\documentclass[openright,10pt]{book}

\usepackage{preamble}
\usepackage{commands}

\begin{document}
\frontmatter
\renewcommand{\next}{\vspace{1em}}

\begin{titlepage}
\begin{center}
\vspace{7cm}
{\bf \LARGE Multiple-Description}\\
\vspace{.7cm}
{\bf \LARGE Lattice Vector Quantization}\\
\vspace{10.7cm}
{\bf \large Proefschrift}\\
\vspace{2cm} ter verkrijging van de graad van doctor\\
aan de Technische Universiteit Delft,\\
op gezag van de Rector Magnificus prof.dr.ir.\ J.\ T.\ Fokkema,\\
voorzitter van het College voor Promoties,\\
in het openbaar te verdedigen op maandag 18 juni 2007 om 12:30 uur\\
door Jan \O STERGAARD\\
Civilingeni\o r van Aalborg Universitet, Denemarken\\
geboren te Frederikshavn.
\end{center}

\newpage\pagestyle{empty}
\noindent Dit proefschrift is goedgekeurd door de promotor:\\
Prof.dr.ir.\ R.\ L.\ Lagendijk \next

\noindent Toegevoegd promotor:
\\ Dr.ir.\ R.\ Heusdens \next

\noindent Samenstelling promotiecommissie:\next

\noindent \hspace{-1em} \begin{tabular}{ll}
Rector Magnificus,              & voorzitter\\
Prof.dr.ir.\ R.\ L.\ Lagendijk,     & Technische Universiteit Delft, promotor\\
Dr.ir.\ R.\ Heusdens,             & Technische Universiteit Delft, toegevoegd\\
                                & promotor\\
Prof.dr.\ J.\ M.\ Aarts,               & Technische Universiteit Delft \\
Prof.dr.ir.\ P.\ Van Mieghem,         & Technische Universiteit Delft \\
Prof.dr.\ V.\ K.\ Goyal,              & Massachusetts Institute of Technology,\\
& Cambridge, United States \\
Prof.dr.\ B.\ Kleijn,                & KTH School of Electrical Engineering,\\
& Stockholm, Sweden  \\
Prof.dr.\ E.\ J.\ Delp,         & Purdue University, Indiana, \\ 
& United States
\end{tabular}
\vspace{1cm}

\includegraphics[width=3cm]{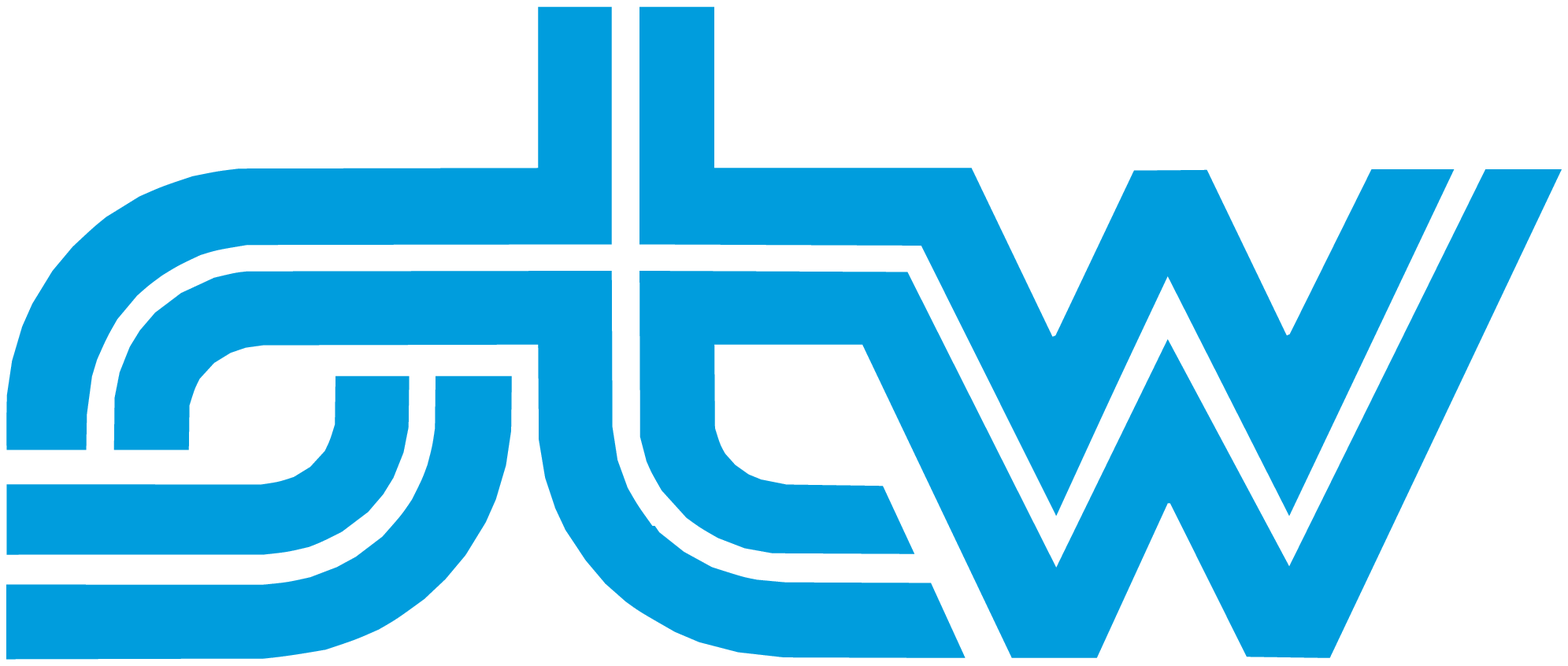}

\noindent The production of this thesis has been financially supported by STW.

\vspace{1cm} \vspace{3em} \noindent ISBN-13: 978-90-9021979-0\next

\noindent Copyright $\copyright$ 2007 by J.\ \O stergaard

\vspace{1em} \noindent All rights reserved. No part of this thesis may be reproduced
or transmitted in any form or by any means, electronic, mechanical, photocopying, any
information storage or retrieval system, or otherwise, without written permission
from the copyright owner.\next

\end{titlepage}


\begin{titlepage}
\begin{center}
\vspace{7cm}
{\bf \LARGE Multiple-Description}\\
\vspace{.7cm}
{\bf \LARGE Lattice Vector Quantization}\\
\vspace{.7cm}
\vspace{12.7cm}

\end{center}
\end{titlepage} \thispagestyle{empty} \ \newpage
\pagenumbering{roman}

\thispagestyle{empty}
%
%
%
%

\thispagestyle{empty}
\chapter*{Preface}
\addcontentsline{toc}{chapter}{Preface}
The research for this thesis was conducted within the STW project DET.5851, \emph{Adapti\-ve Sound Coding (ASC)}. One of the main objectives of the ASC project was to develop a universal audio codec capable of adapting to time-varying characteristics of the input signal, user preferences, and application-imposed constraints or time-varying network-imposed constraints such as bit rate, quality, latency, bandwidth, and bit-error rate (or packet-loss rate).

This research was carried out during the period February 2003 -- February 2007 in the Information and Communication Theory group at Delft University of Technology, Delft, The Netherlands. During the period June 2006 -- September 2006 the research was conducted in the department of Electrical Engineering-Systems at Tel Aviv Univer\-sity, Tel Aviv, Israel.

I would like to take this opportunity to thank my supervisor Richard Heusdens and cosupervisor Jesper Jensen, without whose support, encouragement, and splendid supervision this thesis would not have been possible. 

I owe a lot to my office mates Omar Niamut and Pim Korten.
Omar, your knowled\-ge of audio coding is impressive.
On top of that you are also a gifted musician. A truly marvelous cocktail. 
Pim, you have an extensive mathematical knowledge, which one can only envy. I thank you for the many insightful and delightful discussions we have had over the years. 

I am grateful to Ram Zamir for hosting my stay at Tel Aviv University. I had such a great time and sincerely appreciate the warm welcoming that I received from you, your family and the students at the university. It goes without saying that I will definitely be visiting Tel Aviv again.

I would also like to thank Inald Lagendijk for accepting being my Promotor and
of course cheers go out to the former and current audio cluster fellows; Christoffer R\o dbro, Ivo Shterev, Ivo Batina, Richard Hendriks, and Jan Erkelens for the many insightful discussions we have had.
In addition I would like to acknowledge all members of the ICT group especially Anja van den Berg, Carmen Lai, Bartek Gedrojc and Ana-Ioana Deac.
Finally, the financial support by STW is gratefully acknowledged.

\vspace{1cm}
\noindent J.\ \O stergaard, Delft, January 2007.

\thispagestyle{empty}
\chapter*{Summary}
\addcontentsline{toc}{chapter}{Summary}
Internet services such as voice over Internet protocol (VoIP) and audio/video streaming (e.g.\ video on demand and video conferencing) are becoming more and more popular with the recent spread of broadband networks. These kind of ``real-time'' services often demand low delay, high bandwidth and low packet-loss rates in order to deliver tolerable quality for the end users.
However, the heterogeneous communication infra\-structure of today's packet-switched networks does not provide a guaranteed perfor\-mance in terms of bandwidth or delay and therefore the desired quality of service is generally not achieved. 

To achieve a certain degree of robustness on errorprone channels one can make use of multiple-description (MD) coding, which is a discipline that recently has received a lot of attention. The MD problem is basically a joint source-channel coding problem.
It is about (lossy) encoding of information for transmission over an unreliable $K$-channel com\-munication system. The channels may break down resulting in erasures and a loss of information at the receiving side. 
Which of the $2^{K}-1$ non-trivial subsets of the $K$ channels that are working is assumed known at the receiving side but not at the encoder. The problem is then to design an MD scheme which, for given channel rates (or a given sum rate), minimizes the distortions due to reconstruction of the source using information from any subsets of the channels.

While this thesis focuses mainly on the information theoretic aspects of MD coding, we will also show how the proposed MD coding scheme can be used to construct a perceptually robust audio coder suitable for audio streaming on packet-switched networks. 

We attack the MD problem from a source coding point of view and consider the general case involving $K$ descriptions. 
We make extensive use of lattice vector quantization (LVQ) theory, which turns out to be instrumental in the sense that the proposed MD-LVQ scheme serves as a bridge between theory and practice. In asymptotic cases of high resolution and large lattice vector quantizer dimension, we show that the best known information theoretic rate-distortion MD bounds can be achieved, whereas in non-asymptotic cases of finite-dimensional lattice vector quantizers (but still under high resolution assumptions) we construct practical MD-LVQ schemes, which are comparable and often superior to existing state-of-the-art schemes.

In the two-channel symmetric case it has previously been established that the side descriptions of an MD-LVQ scheme admit side distortions, which (at high resolution conditions) are identical to that of $L$-dimensional quantizers having spherical Voronoi cells. In this case we say that the side quantizers achieve the $L$-sphere bound. Such a result has not been established for the two-channel asymmetric case before. However, the proposed MD-LVQ scheme is able to achieve the $L$-sphere bound for two descriptions, at high resolution conditions, in both the symmetric and asymmetric cases.

The proposed MD-LVQ scheme appears to be among the first schemes in the literature that achieves the largest known high-resolution three-channel MD region in the quadratic Gaussian case.
While optimality is only proven for $K\leq 3$ descriptions we conjecture it to be optimal for any $K$ descriptions.

We present closed-form expressions for the rate and distortion performance for general smooth stationary sources and squared error distortion criterion and at high resolution conditions (also for finite-dimensional lattice vector quantizers). It is shown that the side distortions in the three-channel case is expressed through the dimension\-less normalized second moment of an $L$-sphere independent of the type of lattice used for the side quantizers. This is in line with previous results for the two-description case.

The rate loss when using finite-dimensional lattice vector quantizers is lattice independent and given by the rate loss of an $L$-sphere and an additional term describing the ratio of two dimensionless expansion factors. The overall rate loss is shown to be superior to existing three-channel schemes.

\tableofcontents

\mainmatter
\fancyhead[LE,RO]{\bfseries\thepage}
\fancyhead[LO]{\nouppercase{\bfseries\rightmark}}
\fancyhead[RE]{\nouppercase{\bfseries\leftmark}}
\renewcommand{\headrulewidth}{1pt}

\setcounter{chapter}{0}

\chapter{Introduction}

\section{Motivation}
Internet services such as voice over Internet protocol (VoIP)\index{Voice over IP} and audio/video streaming (e.g.\ video on demand and video conferencing) are becoming more and more popular with the recent spread of broadband networks. These kinds of ``real-time'' services often demand low delay, high bandwidth and low packet-loss rates in order to deliver tolerable quality for the end users.
However, the heterogeneous communication infra\-structure of today's packet-switched networks does not provide a guaranteed perfor\-mance in terms of bandwidth or delay and therefore the desired quality of service is (at least in the authors experience) generally not achieved. 

Clearly, many consumers enjoy the Internet telephony services provided for free through e.g.\ Skype$^{\text{TM}}$. This trend seems to be steadily growing, and more and more people are replacing their traditional landline phones with VoIP compatible systems. On the wireless side it is likely that cell phones soon are to be replaced by VoIP compatible wireless (mobile) phones. A driving impetus is consumer demand for cheaper calls, which sometimes may compromise quality.

The structure of packet-switched networks makes it possible to exploit diversity in order to achieve robustness towards delay and packet losses and thereby improve the quality of existing VoIP services.
For example, at the cost of increased bandwidth (or bit rate), every packet may be duplicated and transmitted over two different paths (or channels) throughout the network. If one of the channels fails, there will be no reduction in quality at the receiving side. Thus, we have a great degree of robustness. 
On the other hand, if none of the channels fail so that both packets are received, there will be no improvement in quality over that of using a single packet. Hence, robustness via diversity comes with a price. 

However, if we can tolerate a small quality degradation on reception of a single packet, we can reduce the bit rates of the individual packets, while still maintaining the good quality on reception of both packets by making sure that the two packets improve upon each other. This idea of trading off bit rate vs.\ quality between a number of packets (or descriptions) is usually referred to as the multiple-description (MD) problem and is the topic of this thesis. 

While this thesis focuses mainly on the information theoretic aspects of MD coding, we will also show how the proposed MD coding scheme can be used to construct a perceptually robust audio coder suitable for audio streaming on packet-switched networks. 
To the best of the author's knowledge the use of MD coding in current state-of-the-art VoIP systems or audio streaming applications is virtually non-existent. We expect, however, that future schemes will employ MD coding to achieve a certain degree of robustness towards packet losses. The research presented in this thesis is a step in that direction.

\section{Introduction to MD Lattice Vector Quantization}
The MD problem is basically a joint source-channel coding\index{joint source-channel coding} problem.
It is about (lossy) encoding of information for transmission over an unreliable $K$-channel com\-munication system. The channels may break down resulting in erasures and a loss of information at the receiving side. 
Which of the $2^{K}-1$ non-trivial subsets of the $K$ channels that are working is assumed known at the receiving side but not at the encoder. The problem is then to design an MD scheme which, for given channel rates (or a given sum rate), minimizes the distortions on the receiver side due to reconstruction of the source using information from any subsets of the channels.

\subsection{Two Descriptions} 
\begin{figure}[ht]
\psfrag{D0}{$\scriptstyle D_0$}
\psfrag{D1}{$\scriptstyle D_1$}
\psfrag{Dc}{$\scriptstyle D_c$}
\psfrag{R0}{$\scriptstyle R_0$}
\psfrag{R1}{$\scriptstyle R_1$}
\psfrag{Description 0}{\scriptsize Description 0}
\psfrag{Description 1}{\scriptsize Description 1}
\begin{center}
\includegraphics[width=7cm]{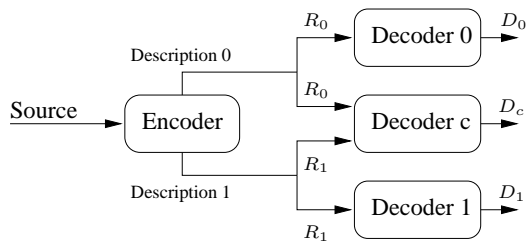}
\caption{The traditional two-channel MD system.}
\label{fig:twochannel_intro}
\end{center}
\end{figure}

The traditional case involves two descriptions as shown in Fig.~\ref{fig:twochannel_intro}. The total bit rate $R_T$, also known as the sum rate, is split between the two descriptions, i.e.\ $R_T=R_0+R_1$, and the distortion observed at the receiver depends on which descriptions arrive. If both descriptions are received, the resulting distortion $(D_c)$ is smaller than if only a single description is received ($D_0$ or $D_1$). It may be noticed from Fig.~\ref{fig:twochannel_intro} that Decoder 0 and Decoder 1 are located on the sides of Decoder $c$ and it is therefore customary to refer to Decoder 0 and Decoder 1 as the side decoders and Decoder $c$ as the central decoder. In a similar manner we often refer to $D_i, i=0,1,$ as the side distortions\index{side distortion} and $D_c$ as the central distortion\index{central distortion}. The situation where $D_0=D_1$ and $R_0=R_1$ is called symmetric MD coding and is a special case of asymmetric MD coding, where we allow unequal side rates\index{description rate} and unequal side distortions.

One of the fundamental problems of MD coding is that if two descriptions both represent the source well, then, intuitively, they must be very similar to the source and therefore also similar to each other. Thus, their joint description is not much better than a single one of them. 
Informally, we may therefore say that the MD problem is about how good one can make the simultaneous representations of the individual descriptions as well as their joint de\-scription.

The two-description MD problem was formalized and presented by Wyner, Witsen\-hausen, Wolf and Ziv at an information theory workshop in September 1979~\cite{goyal:2001a}.\footnote{At that time the problem was already known to several people including Gersho, Ozarow, Jayant, Miller, and Boyle who all made contributions towards its solution, see~\cite{goyal:2001a} for more information.} Formally, the traditional two-description MD problem asks what is the largest possible set of distortions $(D_0,D_1,D_c)$ given the bit rate constraints $(R_0,R_1)$ or alternatively the largest set of bit rates $(R_0,R_1)$ given the distortion constraints $(D_0,D_1,D_c)$? Both these questions were partially answered by El Gamal and Cover who presented an achievable rate-distortion region~\cite{elgamal:1982}, which Ozarow~\cite{ozarow:1980} proved was tight in the case of a memoryless Gaussian source and the squared error distortion measure. Currently, this is the only case where the solution to the MD problem is completely known.

\subsection{Many Descriptions}
Recently, the information theoretic aspects of the general case of $K>2$ descriptions have received a lot of attention~\cite{venkataramani:2001,venkataramani:2003,pradhan:2004,puri:2005,wang:2006}. This case is the natural extension of the two-description case. Given the rate tuple $(R_0,\dotsc,R_{K-1})$, we seek the largest set of simultaneously achievable distortions over all subsets of descriptions. The general $K$-channel MD problem will be treated in greater detail in Chapter~\ref{chap:md_theory}.

With this thesis we show that, at least for the case of audio streaming for lossy packet-switched networks, there seems to be a lot to be gained by using more than two descriptions. 
It is likely that this result carries over to VoIP and video streaming applications. 

\subsection{Scalar vs.\ Vector Quantization}
In the single-description (SD) case it is known that the scalar rate loss (i.e.\ the bit rate increase due to using a scalar quantizer instead of an optimal infinite-dimensional vector quantizer) is approximately $0.2546$ bit/dim.~\cite{gish:1968}. 
For many applications this rate loss is discouraging small and it is tempting to quote Uri Erez:\footnote{Said during a break at the International Symposium on Information Theory in Seattle, July 2006.}
\begin{quote}
\hspace{-5mm}\mbox{\emph{``The problem of vector quantization is that scalar quantization works so well.''}}
\end{quote}

However, in the MD case, the sum (or accumulative) rate loss over many descrip\-tions can be severe. For example, in the two-description case, it is known that the scalar rate loss is about twice that of the SD scalar rate loss\cite{vaishampayan:1998}. 
Therefore, when constructing MD schemes for many descriptions, it is important that the rate loss is kept small. To achieve this, we show in this thesis, that one can, for example, use lattice vector quantizers combined with an index-assignment algorithm.

\section{Contributions}
The MD problem is a joint source-channel coding problem. However, in this work we mainly attack the MD problem from a source coding point of view, where we consider the general case involving $K$ descriptions. 
We make extensive use of lattice vector quantization (LVQ) theory, which turns out to be instrumental in the sense that the proposed MD-LVQ scheme serves as a bridge between theory and practice. In asymptotic cases of high resolution and large lattice vector quantizer dimension, we show that the best known information theoretic rate-distortion MD bounds can be achieved, whereas in non-asymptotic cases of finite-dimensional lattice vector quantizers (but still under high resolution assumptions) we construct practical MD-LVQ schemes, which are comparable and often superior to existing state-of-the-art schemes.

\vspace{2mm}
The main contributions of this thesis are the following:

\begin{enumerate}
\item \textbf{$L$-sphere bound for two descriptions}\par
In the two-channel symmetric case it has previously been established that the side descriptions of an MD-LVQ scheme admit side distortions, which (at high resolution conditions) are identical to that of $L$-dimensional quantizers having spherical Voronoi cells~\cite{servetto:1999,vaishampayan:2001}. In this case we say that the side quantizers\index{side quantizer} achieve the $L$-sphere bound.
Such a result has not been established for the two-channel asymmetric case before. However, the proposed MD-LVQ scheme is able to achieve the $L$-sphere bound for two descriptions, at high resolution conditions, in both the symmetric and asymmetric cases. \\[2mm]
\item \textbf{MD high-resolution region for three descriptions}\par
The proposed MD-LVQ scheme appears to be among the first schemes in the literature that achieves the largest known high-resolution three-channel MD region in the quadratic Gaussian\index{quadratic!Gaussian} case.\footnote{A conference version of the proposed symmetric $K$-channel MD-LVQ scheme appeared in~\cite{ostergaard:2005c} and the full version in~\cite{ostergaard:2004b}. The asymmetric $K$-channel MD-LVQ scheme appeared in~\cite{ostergaard:2005d}. Independently, Chen et al.~\cite{chen:2005a,chen:2005,chen:2006} presented a different design of $K$-channel asymmetric MD coding.}
We prove optimality for $K\leq 3$ de\-scriptions, but conjecture optimality for any $K$ descriptions.
\item \textbf{Exact rate-distortion results for $L$-dimensional LVQ}\par
 We present closed-form expressions for the rate and distortion performance when using $L$-dimensional lattice vector quantizers. These results are valid  for smooth stationary sources and squared-error distortion criterion and at high resolution conditions. 
\item \textbf{Rate loss for finite-dimensional LVQ}\par
The rate loss of the proposed MD-LVQ scheme when using finite-dimensional lattice vector quantizers is lattice independent and given by the rate loss of an $L$-sphere and an additional term describing the ratio of two dimensionless expansion factors. The overall rate loss is shown to be superior to existing three-channel schemes, a result that appears to hold for any number of descriptions.
\item \textbf{$K$-channel asymmetric MD-LVQ}\par
In the asymmetric two-description case it has previously been shown that by introducing weights, the distortion profile of the system can range from successi\-ve refinement\index{successive refinement} to complete symmetric MD coding~\cite{diggavi:2000,diggavi:2002}. We show a similar result for the general case of $K$ descriptions. Furthermore, for any set of weights, we find the optimal number of descriptions and show that the redundancy in the scheme is independent of the target rate\index{target rate}, source distribution and choice of lattices for the side quantizers. Finally, we show how to optimally distribute a given bit budget among the descriptions, which is a topic that has not been addressed in previous asymmetric designs.
\item \textbf{Lattice construction using algebraic $\mathcal{J}$-modules}\par
For the two-description case it has previously been shown that algebraic tools can be exploited to simplify the construction of MD-LVQ schemes~\cite{servetto:1999,vaishampayan:2001,diggavi:2000,diggavi:2002}. We extend these results to $K$-channel MD-LVQ and show that algebraic $\mathcal{J}$-modules provide simple solutions to the problem of constructing the lattices used in MD-LVQ.\\[2mm]
\item \textbf{$K$-channel MD-LVQ based audio coding}\par
We present a perceptually robust audio coder\index{audio coding} based on the modified discrete cosine transform and $K$-channel MD-LVQ. This appears to be the first scheme to consider more than two descriptions for audio coding. Furthermore, we show that using more than two descriptions is advantageous in packet-switched network\index{packet-switched network} environments with excessive packet losses. 
\end{enumerate}

\section{Structure of Thesis}
The main contributions of this thesis are presented in Chapters~\ref{chap:symmetric}--\ref{chap:nac} and the corresponding appendices, i.e.\ Appendices~\ref{app:proofslatticetheory}--\ref{app:nac_results}.

The general structure of the thesis is as follows:
\begin{description}
\item[Chapter 2] The theory of LVQ is a fundamental part of this thesis and in this chapter we describe in detail the construction of lattices and show how they can be used as vector quantizers. A majority of the material in this chapter is known, but the use of $\mathcal{J}$-modules for constructing product lattices based on more than two sublattices is new.
\item[Chapter 3] We consider the MD problem from a source-coding perspective and in this chapter we cover aspects of SD rate-distortion theory, which are also relevant for the MD case.
\item[Chapter 4] In this chapter we present and discuss the existing MD rate-distortion results, which are needed in order to better understand (and to be able to compare to) the new MD results to be presented in the forthcoming chapters.
\item[Chapter 5] Here we present the proposed entropy-constrained $K$-channel symmetric MD-LVQ scheme. We derive closed-form expressions for the rate and distortion performance of MD-LVQ at high resolution and find the optimal lattice parame\-ters, which minimize the expected distortion given the packet-loss probabilities. We further show how to construct practical MD-LVQ schemes and evaluate their numerical performance. This work was presented in part in~\cite{ostergaard:2005c,ostergaard:2004b}.
\item[Chapter 6] We extend the results of the previous chapter to the asymmetric case. In addition we present closed-form expressions for the distortion due to reconstruc\-ting using arbitrary subsets of descriptions. We also describe how to distribute a fixed target bit rate across the descriptions so that the expected distortion is minimized. 
This work was presented in part in~\cite{ostergaard:2005d,ostergaard:2005e}.
\item[Chapter 7] In this chapter we compare the rate-distortion performance of the proposed MD-LVQ scheme to that of existing state-of-the-art MD schemes as well as to known information theoretic high-resolution $K$-channel MD rate-distortion bounds. This work was presented in part in~\cite{ostergaard:2006b,ostergaard:2005e}.
\item[Chapter 8] In this chapter we propose to combine the modified discrete cosine trans\-form with MD-LVQ in order to construct a perceptually robust audio coder. 
Part of the research presented in this chapter represents joint work with O.\ Niamut. This work was presented in part in~\cite{ostergaard:2006a}.
\item[Chapter 9] A summary of results and future research directions are given here.
\item[Appendices] The appendices contain supporting material including proofs of lemmas, propositions, and theorems.
\end{description}

\section{List of Papers}
The following papers have been published by the author of this thesis during his Ph.D.\ studies or are currently under peer review.

\begin{enumerate}
\item J. \O stergaard and R. Zamir, ``Multiple-Description Coding by Dithered Delta Sigma Quantization'', IEEE Proc. Data Compression Conference (DCC), pp. 63 -- 72. March 2007. (Reference~\cite{ostergaard:2007}).

\item J. \O stergaard, R. Heusdens, and J. Jensen, ``Source-Channel Erasure Codes With Lattice Codebooks for Multiple Description Coding'', IEEE Int.\ Sym\-posium on Information Theory (ISIT), pp. 2324 -- 2328, July 2006. (Reference \cite{ostergaard:2006b}).

\item J. \O stergaard, R. Heusdens and J. Jensen ``$n$-Channel Asymmetric Entropy-Constrained Multiple-Description Lattice Vector Quantization'',  Submitted to IEEE Trans.\ Information Theory, June 2006. (Reference~\cite{ostergaard:2005e}).
\item J. \O stergaard, J. Jensen and R. Heusdens,``$n$-Channel Entropy-Constrained Mul\-tiple-Description Lattice Vector Quantization'',  IEEE Trans.\ Information Theory, vol. 52, no. 5, pp. 1956 -- 1973, May 2006. (Reference~\cite{ostergaard:2004b}).

\item J. \O stergaard, O. A. Niamut, J. Jensen and R. Heusdens, ``Perceptual Audio Coding using $n$-Channel Lattice Vector Quantization'', Proc. IEEE Int.\ Conferen\-ce on Audio, Speech and Signal Processing (ICASSP), vol. V, pp. 197 -- 200, May 2006. (Reference~\cite{ostergaard:2006a}).

\item J. \O stergaard, R. Heusdens, and J. Jensen, ``On the Rate Loss in Perceptual Audio Coding'', IEEE Benelux/DSP Valley
Signal Processing Symposium, pp. 27 -- 30, March 2006. (Reference~\cite{ostergaard:2006c}).

\item J. \O stergaard, R. Heusdens, J. Jensen, "$n$-Channel Asymmetric Multiple-Descrip\-tion Lattice Vector Quantization", IEEE Int. Symposium on Informa\-tion Theory (ISIT), pp. 1793 -- 1797. September 2005. (Reference~\cite{ostergaard:2005d}).

\item J. \O stergaard, R. Heusdens, J. Jensen, "On the Bit Distribution in Asymmetric Multiple-Description Coding", 26th Symposium on Information Theory in the Benelux, pp. 81 -- 88, May 2005. (Reference~\cite{ostergaard:2005f}).

\item J. \O stergaard, J. Jensen and R. Heusdens, "$n$-Channel Symmetric Multiple-Description Lattice Vector Quantization", IEEE Proc. Data Compression Con\-fe\-rence (DCC), pp. 378 -- 387, March 2005. (Reference~\cite{ostergaard:2005c}).

\item J. \O stergaard, J. Jensen and R. Heusdens, "Entropy Constrained Multiple De\-scription Lattice Vector Quantization", Proc. IEEE Int.\ Conference on Audio, Speech and Signal Processing (ICASSP), vol. 4, pp. 601 -- 604, May 2004. (Reference~\cite{ostergaard:2004}).
\end{enumerate}

\chapter{Lattice Theory}\label{chap:lattice_theory}
In this chapter we introduce the concept of a lattice and show that it can be used as a vector quantizer. We form subsets (called sublattices) of this lattice, and show that these sublattices\index{lattice!sublattice} can also be used as quantizers. In fact, in later chapters, we will use a lattice as a central quantizer and the sublattices will be used as side quantizers for MD-LVQ. We defer the discussion on rate-distortion properties of the lattice vector quantizer to Chapters~\ref{chap:rd_theory} and~\ref{chap:md_theory}.

We begin by describing a lattice in simple terms and show how it can be used as a vector quantizer. This is done in Section~\ref{sec:latticeintro} and more details can be found in Appendix~\ref{app:lattice_defs}.
Then in Section~\ref{sec:latticealgebra} we show that lattice theory is intimately connected to algebra and it is therefore possible to use existing algebraic tools to solve lattice related problems. For example it is well known that lattices form groups\index{group} under ordinary vector addition and it is therefore possible to link fundamental group theory to lattices. In Section~\ref{sec:latticeconstruction} we then use these algebraic tools to construct lattices and sublattices. It might be a good idea here to consult Appendix~\ref{app:quaternions} for the definition of Quaternions and Appendix~\ref{app:modules} for a brief introduction to module theory. 

We would like to point out that Section~\ref{sec:latticeintro} contains most of the essential lattice theory needed to understand the concept of MD-LVQ. Sections~\ref{sec:latticealgebra} and~\ref{sec:latticeconstruction} are sup\-plementary to Section~\ref{sec:latticeintro}. In these sections we construct lattices and sublattices in an algebraic fashion by using the machinery of module theory\index{module}. This turns out to be a very convenient approach, since it allows simple constructions of lattices. This theory is therefore also very helpful for the practical implementation of MD-LVQ schemes. 
In addition, we would like to emphasize that by use of module theory we are able to prove the existence of lattices which admit the required sublattices and product lattices. In Chapters~\ref{chap:symmetric}--\ref{chap:comparison} we will implicitly assume that all lattices, sublattices, and product lattices are constructed as specified in this chapter.

\section{Lattices}\label{sec:latticeintro}
\begin{figure}[th]
\begin{center}
\subfigure[$\Lambda=\mathbb{Z}^2$]{\includegraphics[width=5.5cm]{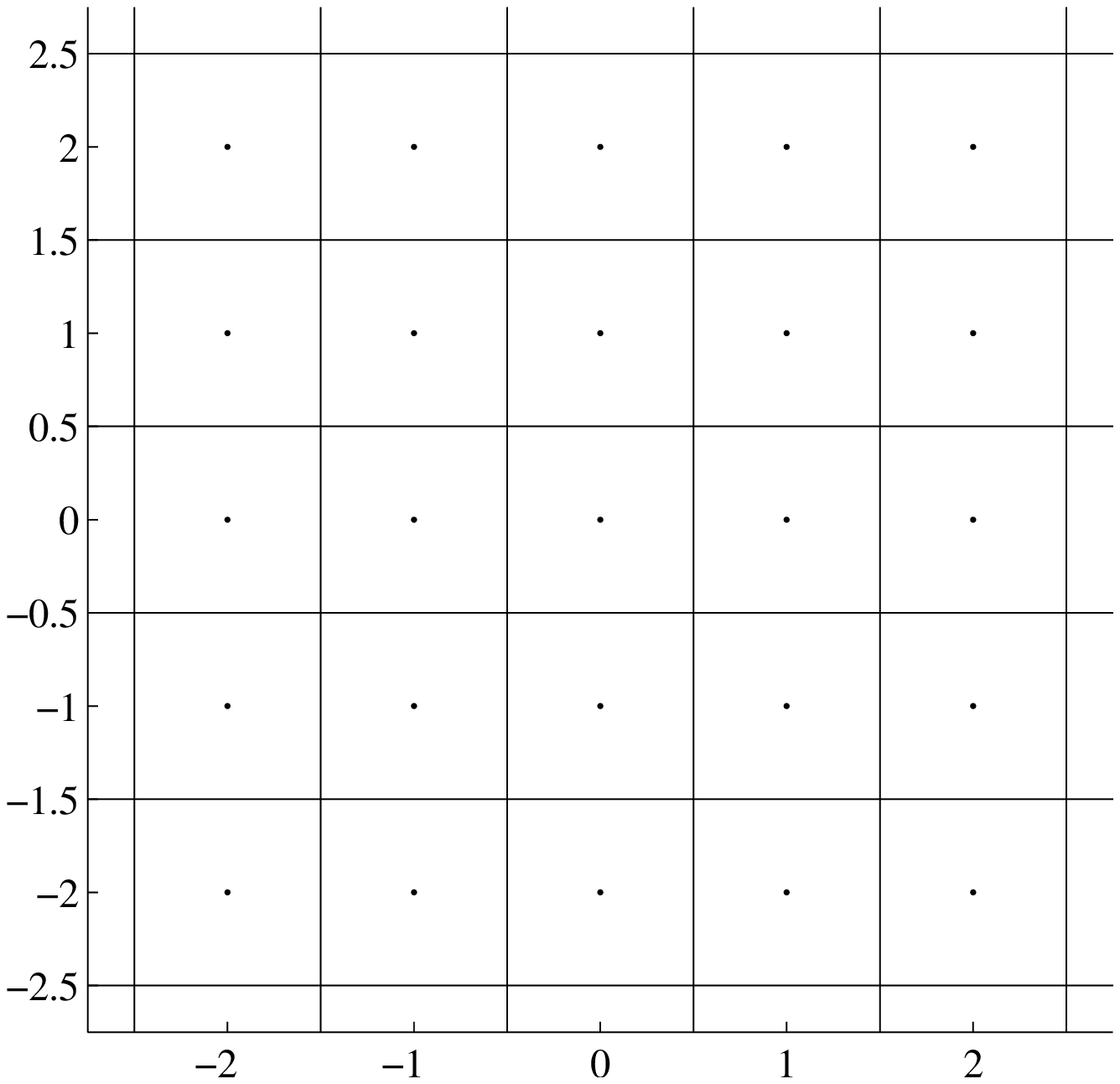}\label{fig:lat_z2}}\quad
\subfigure[Random point set]{\includegraphics[width=5.5cm]{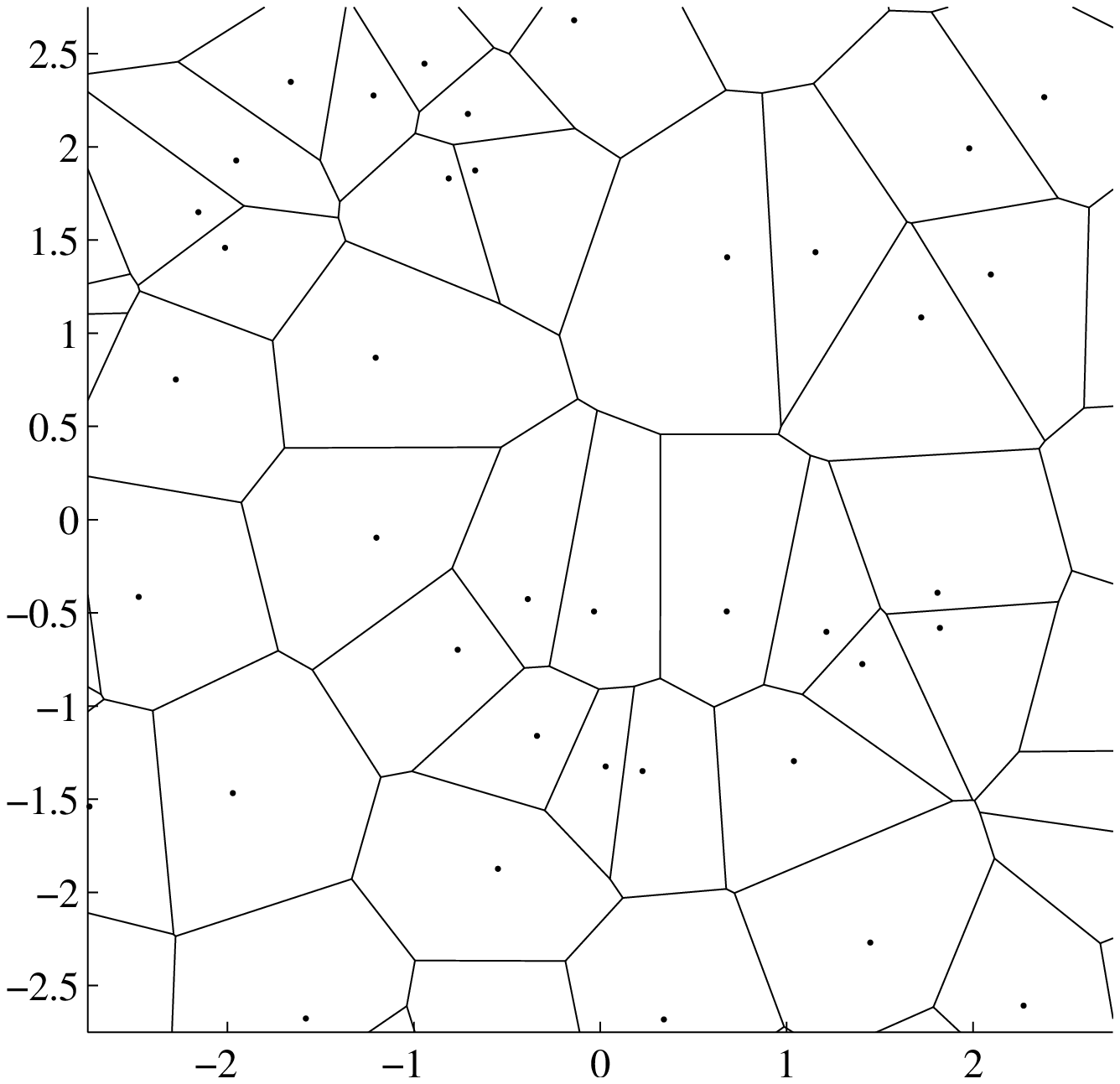}\label{fig:randlat}}
\caption{(a) finite region of the lattice $\Lambda=\mathbb{Z}^2$. (b) randomly selected points of $\mathbb{R}^2$. The solid lines describe the boundaries of the Voronoi cells of the points.}
\label{fig:lat_examples}
\end{center}
\end{figure}

An $L$-dimensional lattice\index{lattice} is a discrete set of equidistantly spaced points in the $L$-dimensional Euclidean vector space $\mathbb{R}^L$. For example, the set of integers $\mathbb{Z}$ forms a lattice in $\mathbb{R}$ and the Cartesian product $\mathbb{Z}\times\mathbb{Z}$ forms a lattice in $\mathbb{R}^2$. More formally, we have the following definition.
\begin{definition}[\!\!\cite{conway:1999}]
A lattice $\Lambda\subset\mathbb{R}^L$ consists of all possible integral linear combinations of a set of basis vectors, that is
\begin{equation}\label{eq:latticedef}
\Lambda = \left\{ \lambda\in \mathbb{R}^L : \lambda = \sum_{i=1}^{L} \xi_i\zeta_i,\ \forall\xi_i \in \mathbb{Z}\right\},
\end{equation}
where $\zeta_i\in \mathbb{R}^L$ are the basis vectors also known as generator vectors\index{generator!vector} of the lattice.
\end{definition}
The generator vectors $\zeta_i, i=1,\dotsc, L,$ (or more correctly their transposes) form the rows of the generator matrix\index{generator!matrix} $M$. Usually there exists several generator matrices which all lead to the same lattice. In Appendix~\ref{app:rootlattices} we present some possible generator matrices for the lattices considered in this thesis. 
\begin{definition}
Given a discrete set of points $S\subset \mathbb{R}^L$, the nearest neighbor region\index{nearest neighbor!region} of $s\in S$ is called a Voronoi cell, Voronoi region or Dirichlet region, and is defined by
\begin{equation}
V(s) \triangleq \{ x\in \mathbb{R}^L : \| x - s\|^2 \leq \|x-s'\|^2,\, \forall\, s' \in S \},
\end{equation}
where $\|x\|$ denotes the usual norm\index{norm!inner product} in $\mathbb{R}^L$, i.e.\ $\|x\|^2=x^Tx$.
\end{definition}
As an example, Fig.~\ref{fig:lat_z2} shows a finite region of the lattice $\Lambda=\mathbb{Z}^2$ consisting of all pairs of integers. For this lattice, the Voronoi cells $V(\lambda), \lambda\in \Lambda$, form squares in the two-dimensional plane.
This lattice is also referred to as the $Z^2$ lattice or the square lattice, cf.\ Appendix~\ref{app:z2}. 
A lattice $\Lambda$ and its Voronoi cells $V(\lambda), \forall \lambda\in\Lambda$, actually form a vector quantizer. When $\Lambda$ is used as a vector quantizer, a point $x$ is mapped (or quantized) to $\lambda\in \Lambda$ if $x\in V(\lambda)$. An example of a non-lattice vector quantizer is shown in Fig.~\ref{fig:randlat}. Here we have randomly picked a set of elements of $\mathbb{R}^2$. Notice that the Voronoi cells are not identical but still their union cover the space.
%
%
On the other hand, in Fig.~\ref{fig:lat_z2}, it may be noticed that the Voronoi cells of $\Lambda$ are all identical, and we say that each one of them describes a fundamental region. A fundamental region of a lattice is a closed region which contains a single lattice point and tessellates\index{tessellating!quantizer} the underlying space. 
\begin{lemma}[\cite{conway:1999}]
All fundamental regions have the same volume $\nu$.
\end{lemma}
\begin{lemma}[\cite{conway:1999}]
The fundamental volume $\nu$ of $\Lambda$ is given by $\nu=\sqrt{\det(A)}$, where $A=MM^T$ is called the Gram matrix\index{Gram matrix}. We sometimes write the volume as $\nu=\det(\Lambda)$\index{determinant of a lattice}.
\end{lemma}
Let us define $V_0 \triangleq V(0)$, i.e.\ the Voronoi cell around the lattice point located at the origin. This region is called a fundamental region of the lattice since it specifies the complete lattice through translations. 
We then have the following definition.
\begin{definition}[\cite{conway:1999}]
The dimensionless normalized second moment of inertia $G(\Lambda)$ of a lattice $\Lambda$ is defined by
\begin{equation}\label{eq:G}
G(\Lambda) \triangleq\frac{1}{L\nu^{1+2/L}}\int_{V_0}\|x\|^2dx,
\end{equation}
where $\nu$ is the volume of $V_0$.
\end{definition}

Applying any scaling or orthogonal transform, e.g.\ rotation or reflection on $\Lambda$ will not change $G(\Lambda)$, which makes it a good figure of merit when comparing different lattices (quantizers). Furthermore, $G(\Lambda)$ depends only upon the shape of $V_0$, and in general, the more sphere-like shape, the smaller normalized second moment~\cite{conway:1999}.

\subsection{Sublattices}
\begin{figure}[th]
\begin{center}
\includegraphics[width=5.5cm]{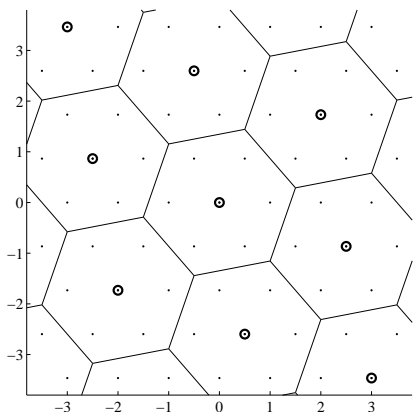}
\caption{The hexagonal lattice $\Lambda$ (small dots) and a sublattice $\Lambda'\subset \Lambda$ (circles) of index $N=7$. The solid lines illustrate the boundaries of the Voronoi cells of $\Lambda'$.}
\label{fig:sublattice_a2}
\end{center}
\end{figure}

If $\Lambda$ is a lattice then a sublattice\index{lattice!sublattice} $\Lambda'\subseteq \Lambda$ is a subset of the elements of $\Lambda$ that is itself a lattice. For example if $\Lambda=\mathbb{Z}$ then the set of all even integers is a sublattice of $\Lambda$. Geometrically speaking, a sublattice $\Lambda'\subset \Lambda$ is obtained by scaling and rotating (and possibly reflecting) the lattice $\Lambda$ so that all points of $\Lambda'$ coincide with points of $\Lambda$. A sublattice $\Lambda'\subset\Lambda$ obtained in this manner is referred to as a geometrically-similar sublattice of $\Lambda$.
Fig.~\ref{fig:sublattice_a2} shows an example of a lattice $\Lambda\subset \mathbb{R}^2$ and a geometrically-similar sublattice $\Lambda'\subset\Lambda$. In this case $\Lambda$ is the hexagonal lattice which is described in Appendix~\ref{app:a2}. It may be noticed from Fig.~\ref{fig:sublattice_a2} that all Voronoi cells of the sublattice $\Lambda'$ contain exactly seven points of $\Lambda$. In general we would like to design a sublattice so that each of its Voronoi cells contains exactly $N$ points of $\Lambda$. We call $N$ the index value of the sublattice and usually write it as $N=|\Lambda/\Lambda'|$. Normalizing $N$ by dimension, i.e.\ $N'=N^{1/L}$, gives what is known as the nesting ratio\index{index!nesting ratio}.
%
%
We call a sublattice $\Lambda'\subset \Lambda$ clean\index{clean sublattice} if no points of $\Lambda$ lies on the boundaries of the Voronoi cells of $\Lambda'$. For example, the sublattice of Fig.~\ref{fig:sublattice_a2} is clean. If $\Lambda'\subset\Lambda$ is a clean sublattice we call the index $N=|\Lambda/\Lambda'|$ an admissible index value. In this work we are mainly interested in clean sublattices and we will further discuss the issue of admissible index values in Section~\ref{sec:admindexvalues}.

\section{$\mathcal{J}$-Lattice}\label{sec:latticealgebra}
We showed in the previous section that, geometrically speaking, an $L$-dimensional lattice $\Lambda\subset \mathbb{R}^L$ is a discrete set of regularly spaced points in $\mathbb{R}^L$. From Appendix~\ref{app:modules} it can be deduced that, algebraically speaking, an $L$-dimensional lattice $\Lambda\subset \mathbb{R}^L$ is a free (torsion-free)\index{torsion!free} discrete $\mathcal{J}$-module of rank $L$ with compact quotient\index{quotient!compact} $\mathbb{R}^L/\Lambda$. In this section we will consider the latter definition of a lattice and construct lattices and sublattices by use of the theory of modules. 

A lattice $\Lambda\subset\mathbb{R}^L$ as defined in~(\ref{eq:latticedef}) forms an additive group\index{group} $(\Lambda,+)$ under ordinary vector addition with the zero-vector being the identity element. If the group further admits left or right multiplication by the ring $\mathcal{J}$ then we call $\Lambda$ a $\mathcal{J}$-module\index{module}. In other words, $\Lambda$ is a $\mathcal{J}$-module if it is closed under addition and subtraction of members of the group and closed under scalar multiplication by members of the ring, see Appendix~\ref{app:modules} for details. Since $\Lambda$ is also a lattice we sometimes prefer the name $\mathcal{J}$-lattice over $\mathcal{J}$-module.

Let $\zeta_i, i=1,\dotsc, L$ be a set of linearly independent vectors in $\mathbb{R}^L$ and let $\mathcal{J}\subset \mathbb{R}$ be a ring. Then a left $\mathcal{J}$-lattice $\Lambda$ generated by $\zeta_i, i=1,\dotsc,L$ consists of all linear combinations
\begin{equation}\label{eq:Jlattice}
\xi_1\zeta_1 + \cdots + \xi_L\zeta_L,
\end{equation}
where $\xi_i\in \mathcal{J}, i=1,\dotsc,L$~\cite{conway:1999}. A right $\mathcal{J}$-lattice is defined similarly with the multiplication of $\zeta_i$ on the right by $\xi_i$ instead.

\begin{figure}[h!]
\begin{center}
\subfigure[$\Lambda=\mathcal{G}\zeta_1$]{\includegraphics[width=5cm]{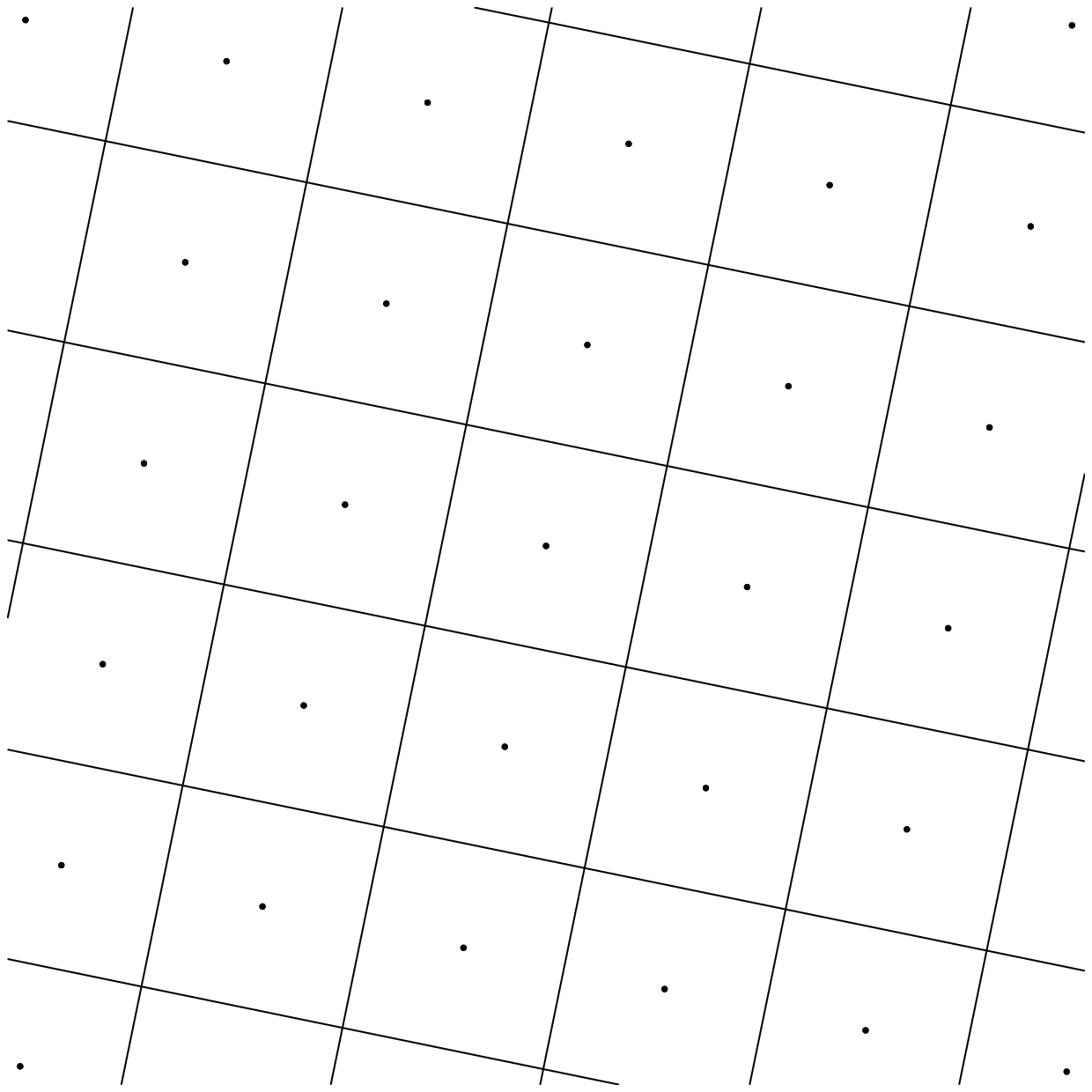}\label{fig:Jlat_G}}\quad
\subfigure[$\Lambda=\mathcal{E}\zeta_1$]{\includegraphics[width=5cm]{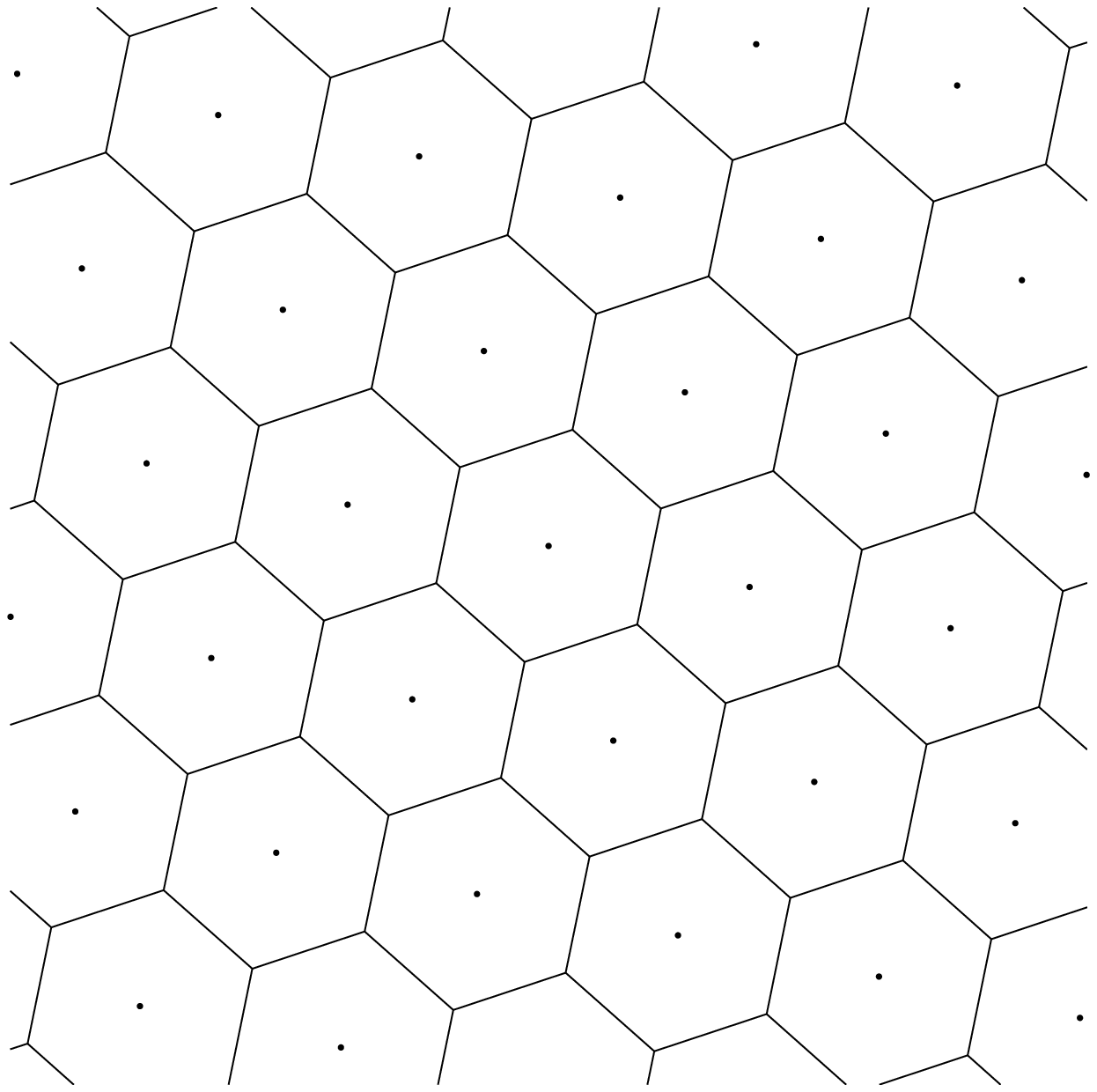}\label{fig:Jlat_E}}\par
\subfigure[$\Lambda=\mathcal{Q}\zeta_1, \omega_1=\sqrt{-5}$]{\includegraphics[width=5cm]{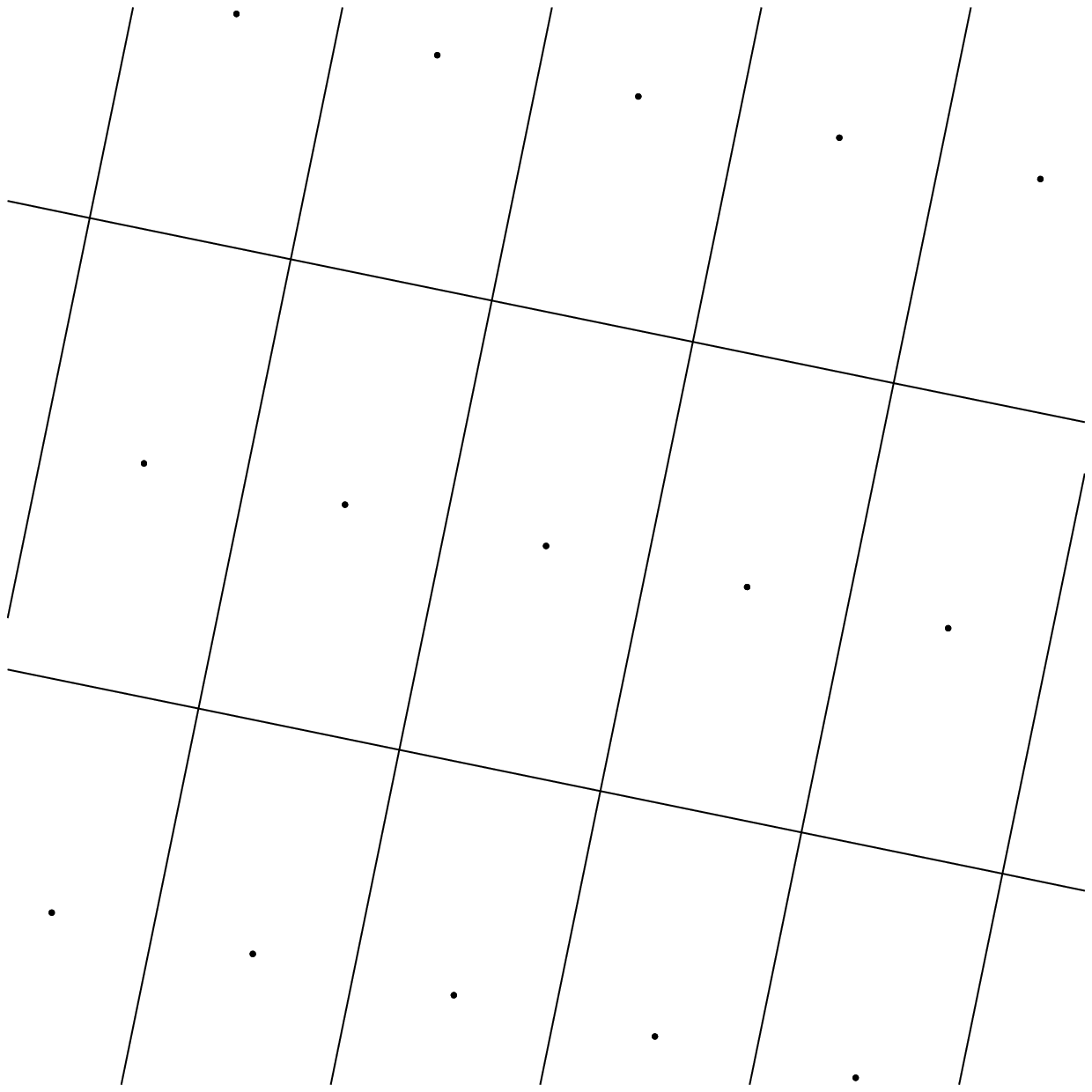}\label{fig:Jlat_Q1}}\quad
\subfigure[$\Lambda=\mathcal{Q}\zeta_1, \omega_1=(-1+\sqrt{-7})/2$]{\includegraphics[width=5cm]{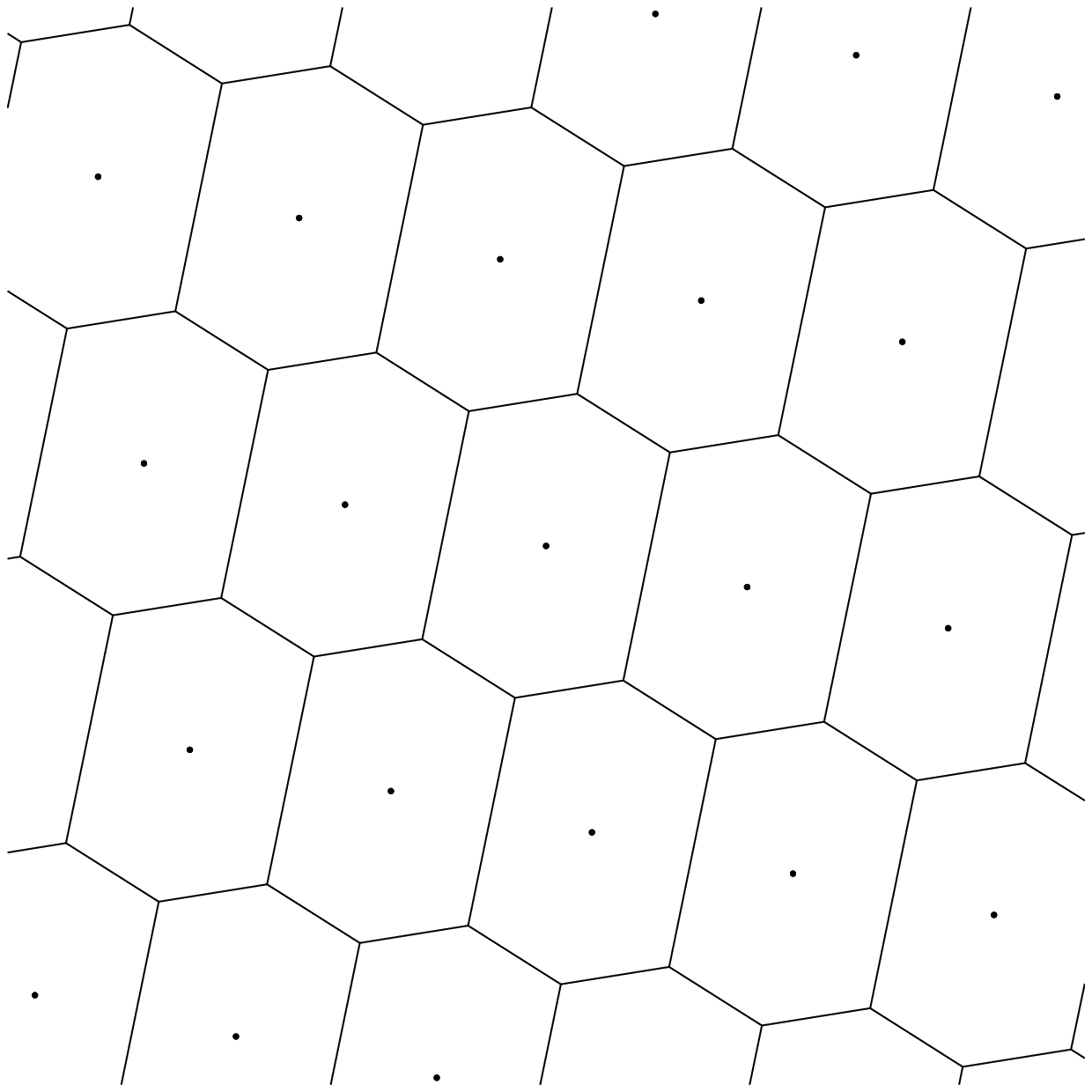}\label{fig:Jlat_Q2}}
\caption{One-dimensional complex $\mathcal{J}$-lattices constructed from different rings of integers by use of the basis $\zeta_1=11.2-2.3i$. The solid lines illustrate the boundaries of the nearest neighbor regions (Voronoi cells) between lattice points.}
\label{fig:J_lattices}
\end{center}
\end{figure}

We have so far assumed that the underlying field is the Cartesian product of the reals, i.e.\ $\mathbb{R}^L$. However, there are other fields which when combined with well defined rings of integers will lead to $\mathcal{J}$-lattices that are good for quantization. Let the field be the complex field $\mathbb{C}$ and let the ring of integers be the Gaussian integers\index{integer!Gaussian} $\mathcal{G}$, where~\cite{conway:1999}
\begin{equation}\label{eq:gaussianint}
\mathcal{G} = \{ \xi_1+i\xi_2 : \xi_1,\xi_2\in \mathbb{Z}\},\quad i=\sqrt{-1}.
\end{equation}
Then we may form a one-dimensional complex lattice (to which there always exists an isomorphism\index{isomorphism} that will bring it to $\mathbb{R}^2$) by choosing any non-zero element (a basis) $\zeta_1\in \mathbb{C}$ and insert in~(\ref{eq:Jlattice}), cf.\ Fig.~\ref{fig:Jlat_G} where we have made the arbitrary choice of $\zeta_1=11.2-2.3i$.
 The lattice described by the set of Gaussian integers is isomorphic to the square lattice $Z^2=\mathbb{Z}^2$. The operation $\mathcal{G}\zeta_1$ then simply rotate and scale the $Z^2$ lattice.
To better illustrate the shape of the $\mathcal{J}$-lattice we have in Fig.~\ref{fig:Jlat_G} also shown the boundaries (solid lines) of the nearest neighbor regions (also called Voronoi cells) between the lattice points. 
Fig.~\ref{fig:Jlat_E} shows an example where the basis $\zeta_1=11.2-2.3i$ has been multiplied by the Eisenstein integers\index{integer!Eisenstein} $\mathcal{E}$, where~\cite{conway:1999}
\begin{equation}\label{eq:eisensteinint}
\mathcal{E}=\{ \xi_1+\omega \xi_2 : \xi_1,\xi_2\in \mathbb{Z}\},\quad \omega = e^{2\pi i/3}.
\end{equation}
The ring of algebraic integers\index{integer!algebraic} $\mathcal{Q}$ is defined by~\cite{conway:1999}
\begin{equation}
\mathcal{Q}=\{\xi_1+\omega_1\xi_2 : \xi_1,\xi_2\in\mathbb{Z}\},
\end{equation}
where $\omega_1$ is, for example, one of 
\begin{equation}
\sqrt{-2}, \sqrt{-5}, \frac{-1+\sqrt{-7}}{2}, \frac{-1+\sqrt{-11}}{2}.
\end{equation}
Figs.~\ref{fig:Jlat_Q1} and~\ref{fig:Jlat_Q2} show examples where $\mathcal{J}$ is the ring of algebraic integers and where $\omega_1=\sqrt{-5}$ and $\omega_1=(-1+\sqrt{-7})/2$, respectively. In both cases we have used the basis $\zeta_1=11.2-2.3i$.
%
%

\subsection{$\mathcal{J}$-Sublattice}
If $\Lambda'$ is a submodule\index{module!sub} of a $\mathcal{J}$-module $\Lambda$ then $\Lambda'$ is simply a sublattice\index{lattice!sublattice} of the lattice $\Lambda$. More formally we have the following lemma.
\begin{lemma}[\cite{adkins:1992}]
Let $\mathcal{J}$ be a ring. If $\Lambda$ is $\mathcal{J}$-module and $\Lambda'\subseteq \Lambda, \Lambda'\neq \emptyset$, then $\Lambda'$ is a $\mathcal{J}$-submodule of $\Lambda$ if and only if $\xi_1\lambda'_1 + \xi_2\lambda'_2 \in \Lambda'$ for all $\lambda'_1,\lambda'_2\in \Lambda'$ and $\xi_1,\xi_2\in \mathcal{J}$.
\end{lemma}

\begin{figure}[ht]
\begin{center}
\includegraphics[width=8cm]{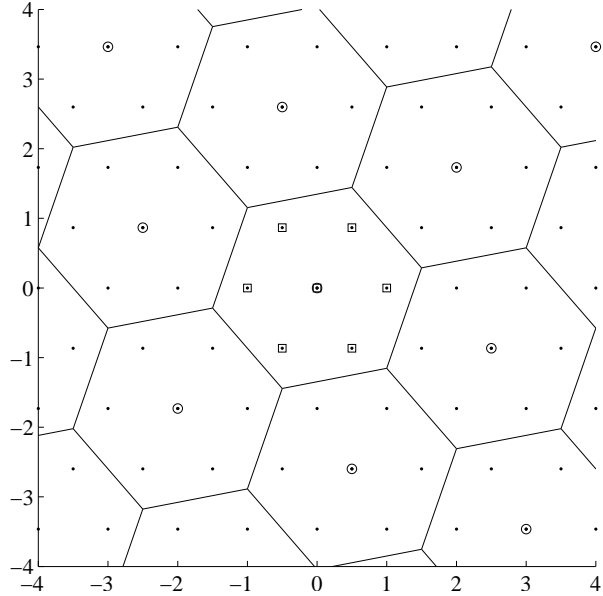}
\caption{The Eisenstein lattice $\Lambda$ is here shown with dots and the circles illustrate points of the sublattice $\Lambda'=\xi\Lambda, \xi=-3-2\omega, \omega=e^{2\pi i/3}$. The solid lines describe the boundaries between neighboring Voronoi cells of the sublattice points. Notice that there are 7 dots in each Voronoi cell. The points marked with squares are the seven coset representatives of the quotient $\Lambda/\Lambda'$.}
\label{fig:submodule_A2}
\end{center}
\end{figure}

Let $\Lambda$ be a $\mathcal{J}$-module. Then we may form the left submodule $\Lambda'=\xi\Lambda$ and the right submodule $\Lambda''=\Lambda\xi$ by left (or right) multiplication of $\Lambda$ by $\xi\in \mathcal{J}$. 
For example let $\Lambda$ be the $\mathcal{J}$-module given by the Eisenstein integers, i.e.\ $\Lambda=\mathcal{E}$. This lattice can be regarded as a two-dimensional real lattice in $\mathbb{R}^2$ in which case it is usually referred to as $A_2$.
Then let us form the submodule $\Lambda'=\xi\Lambda$ where $\xi=-3-2\omega$ and $\omega=e^{2\pi i/3}$, see Fig.~\ref{fig:submodule_A2}. When the modules in question are lattices we will usually call $\Lambda$ a $\mathcal{J}$-lattice and $\Lambda'$ a $\mathcal{J}$-sublattice. Sometimes when the ring $\mathcal{J}$ is clear from the context or irrelevant we will use the simpler terms lattice and sublattice for $\Lambda$ and $\Lambda'$, respectively.
%
%

\subsection{Quotient Modules}\index{module!quotient}
In this section we consider quotient modules and the next section is concerned with group\index{group!action} actions on these quotient modules. 
Although perhaps unclear at this point, we show later that these concepts are important in order to identify or associate a set of sublattice points with a given lattice point. This identification process, which we call either the labeling problem or the problem of constructing an index assignment map, focuses on labeling the coset\index{coset!representative} representatives of $\Lambda/\Lambda'$, i.e.\ the quotient module. It then turns out, as first observed in~\cite{servetto:1999,vaishampayan:2001}, that we actually only need to label the representatives of the orbits\index{orbits} of $\Lambda/\Lambda'/\Gamma_m$ instead of all coset representatives of $\Lambda/\Lambda'$. Further details about quotient modules and group actions are given in Appendix~\ref{app:modules}.

\begin{definition}
Let $\Lambda$ be a $\mathcal{J}$-module and $\Lambda'$ a $\mathcal{J}$-submodule of $\Lambda$.  Then $\Lambda'$ induces a partition $\Lambda/\Lambda'$ of $\Lambda$ into equivalence classes (or cosets) modulo\index{modulo!lattice} $\Lambda'$. We call such a partition the quotient module.
\end{definition}

The order or index $|\Lambda/\Lambda'|$ of the quotient module $\Lambda/\Lambda'$ is finite and each element of $\Lambda/\Lambda'$ is a representative for an infinite set called a coset\index{coset}. For any $\lambda\in\Lambda$ the coset of $\Lambda'$ in $\Lambda$ determined by $\lambda$ is the set $\lambda+\Lambda' = \{ \lambda+\lambda' : \lambda' \in \Lambda' \}$.
In this work we always let the group operation be ordinary vector addition which is a commutative operation so that the left and right cosets coincide. 
As such there is no ambiguity with respect to left and right cosets when referring to the coset $\lambda+\Lambda'$.
We will use the notation $[\lambda]$ when referring to the coset $\lambda + \Lambda'$ and we call $\lambda$ the coset representative. It should be clear that any member of the coset $[\lambda]$ can be the coset representative. 
To be consistent we will always let the coset representative be the unique\footnote{We will later require that $\Lambda'$ is a clean sublattice of $\Lambda$ from which the uniqueness property is evident. If $\Lambda'$ is not clean then we make an arbitrary choice amongst the candidate representatives.} vector of $[\lambda]$ which is in the Voronoi cell of the zero-vector of $\Lambda'$.
For example if $\Lambda$ and $\Lambda'$ are defined as in Fig.~\ref{fig:submodule_A2} then the index $|\Lambda/\Lambda'|=7$ and there is therefore seven distinct cosets in the quotient module $\Lambda/\Lambda'$. The seven cosets representatives are indicated with squares in Fig.~\ref{fig:submodule_A2}.

\subsection{Group Actions on Quotient Modules}\label{sec:actionsonmodules}
\begin{figure}[ht]
\begin{center}
\mbox{%
\subfigure[$\Lambda/\Lambda'/\Gamma_2$]{\includegraphics[width=5cm]{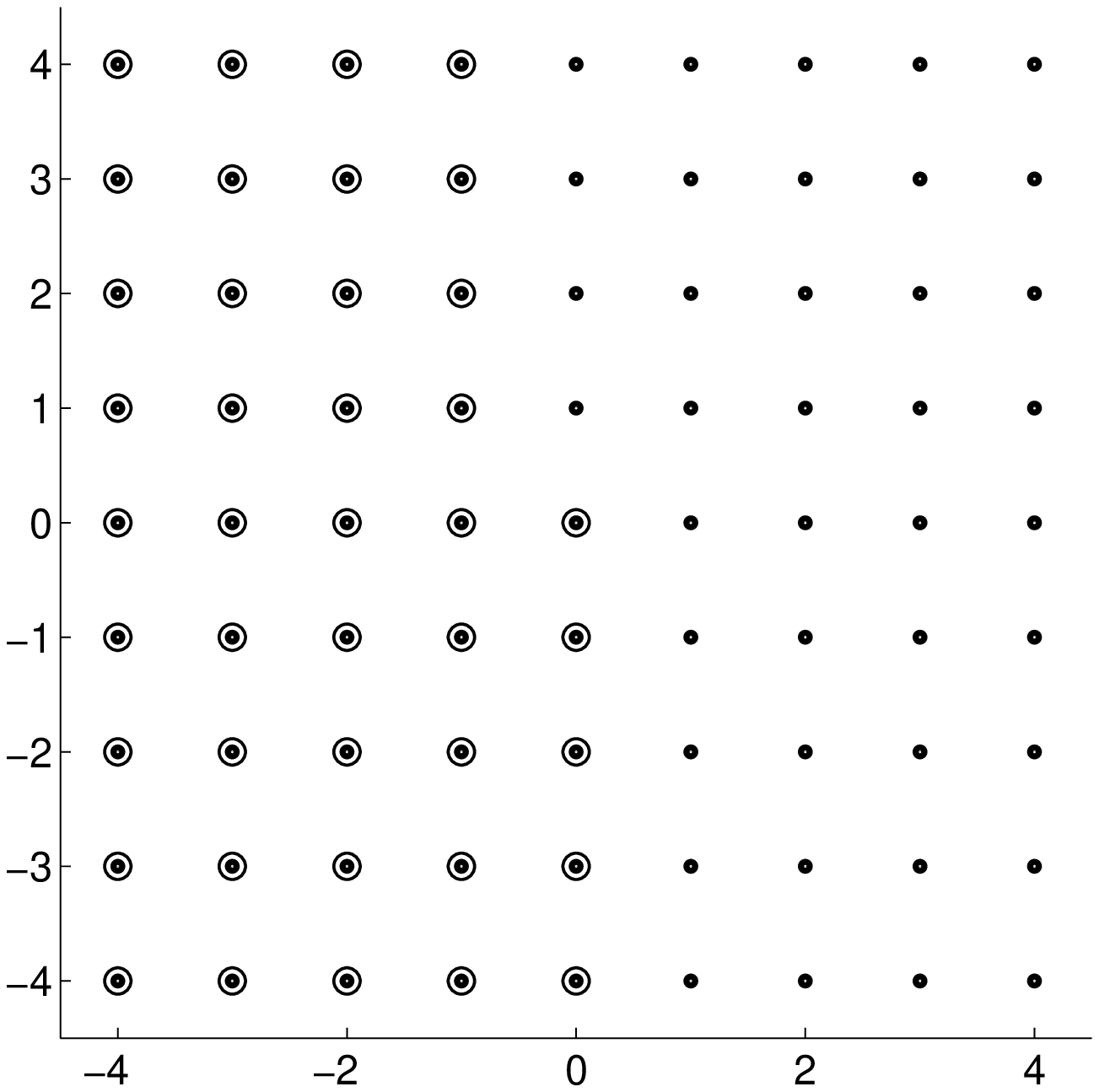}\label{fig:auto1}}\quad
\subfigure[$\Lambda/\Lambda'/\Gamma_4$]{\includegraphics[width=5cm]{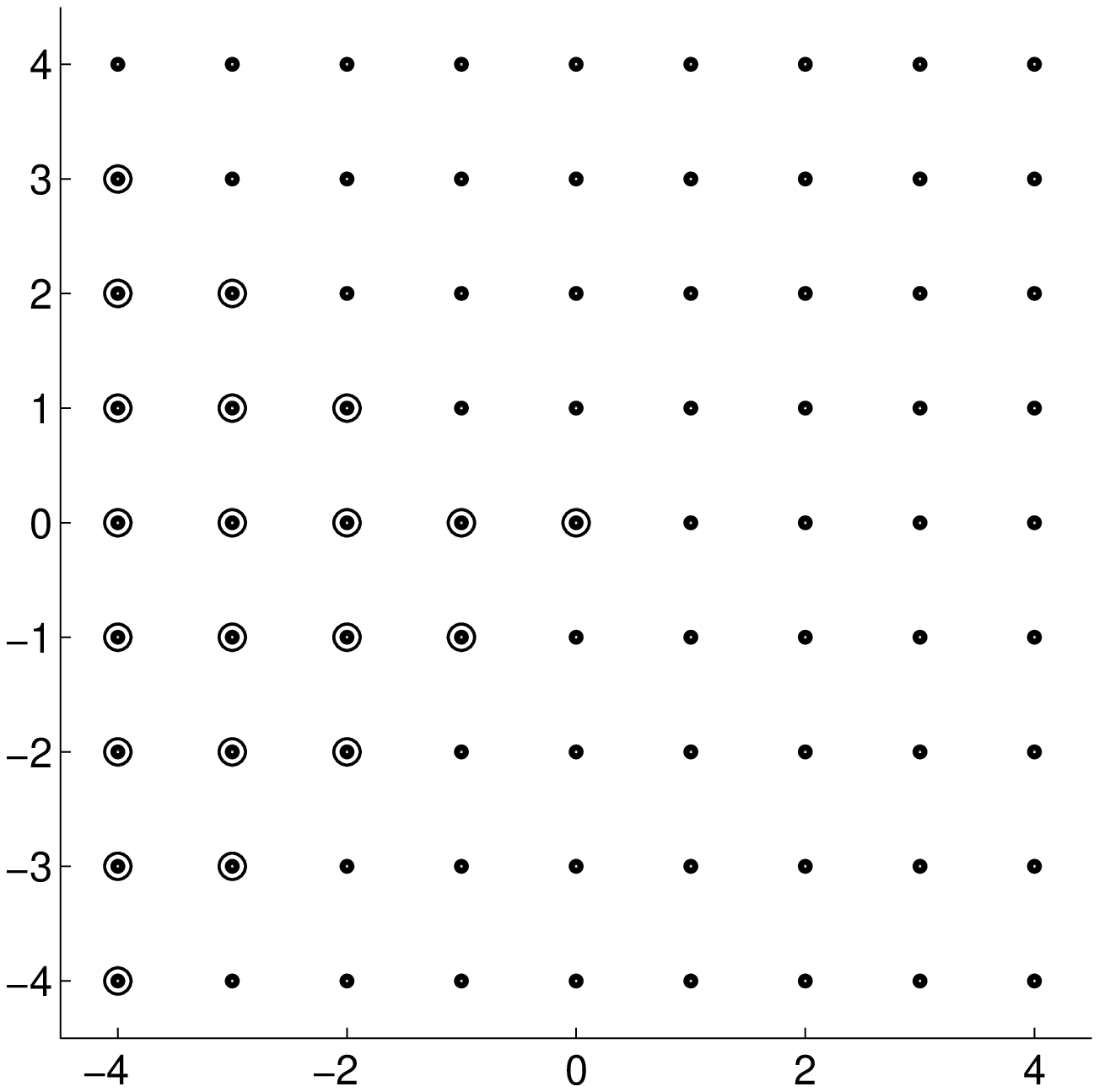}\label{fig:auto2}}}
\caption{The 81 coset representatives for $\Lambda/\Lambda'$ are here shown as dots and representatives for the orbits of (a) $\Lambda/\Lambda'/\Gamma_2$ and (b) $\Lambda/\Lambda'/\Gamma_4$ are shown as circles.}
\label{fig:automorphisms}
\end{center}
\end{figure}

Let $\Gamma_m\subseteq \text{Aut}(\Lambda)$ be a group of order $m$ of automorphisms\index{automorphism!group} of $\Lambda$. 
We then denote the set of orbits\index{orbits} under the action of $\Gamma_m$ on the quotient module\index{module!quotient} $\Lambda/\Lambda'$ by $\Lambda/\Lambda'/\Gamma_m$. 
For example let $\Gamma_2=\{I_2,-I_2\}$ be a group (closed under matrix multiplication) of order 2, where $I_2$ is the two-dimensional identity matrix. Let the $\mathcal{J}$-module $\Lambda$ be identical to $\mathbb{Z}^2$ and let $\Lambda'$ be a submodule of $\Lambda$ of index $N=81$. In other words, there are $N$ coset representatives in the quotient module $\Lambda/\Lambda'$ whereas the set of orbits $\Lambda/\Lambda'/\Gamma_2$ has cardinality $|\Lambda/\Lambda'/\Gamma_2|=41$. 
This is illustrated in Fig.~\ref{fig:auto1} where the coset representatives of $\Lambda/\Lambda'$ are illustrated with dots and representatives of the orbits of $\Lambda/\Lambda'/\Gamma_2$ are marked with circles. 

Next consider the group given by
\begin{equation}\label{eq:gamma4}
\Gamma_4 =\left\{ \pm I_2,
\pm\begin{pmatrix} 0 & -1 \\ 1 & \phantom{-}0 \end{pmatrix}
\right\},
\end{equation}
which has order 4 and includes $\Gamma_2$ as a subgroup\index{group!sub}. Fig.~\ref{fig:auto2} shows coset representatives for $\Lambda/\Lambda'$ and representatives for the set of orbits\index{orbits} $\Lambda/\Lambda'/\Gamma_4$. Notice that $|\Lambda/\Lambda'/\Gamma_4|=21$.
%
%

\section{Construction of Lattices}\label{sec:latticeconstruction}
We now show how to construct the lattices and sublattices that later will be used as quantizers in MD-LVQ.

\subsection{Admissible Index Values}\label{sec:admindexvalues}
For any geometrically-similar sublattice $\Lambda'$ of $\Lambda$, a number of lattice points of $\Lambda$ will be located within each Voronoi cell of $\Lambda'$ and perhaps on the boundaries between neighboring Voronoi cells. 
In the latter case ties must be broken in order to have well defined Voronoi cells. 
To avoid tie breaking it is required that $\Lambda'$ has no lattice points on the boundary of its Voronoi cells. In this case, $\Lambda'$ is said to be clean.
As previously mentioned, we call an index value of a clean sublattice an admissible index value\index{index!admissible}.
In~\cite{conway:1999b} partial answers are given to when $\Lambda$ contains a sublattice $\Lambda'$ of index $N$ that is geometrically-similar to $\Lambda$, and necessary and sufficient conditions are given for any lattice in two dimensions to contain a geometrically-similar and clean sublattice of index $N$. 
These results are extended in~\cite{diggavi:2002} to geometrically-similar and clean sublattices in four dimensions for the $Z^4$ and $D_4$ lattice. In addition, results are given for any $Z^L$ lattice where $L=4k, k\geq1$. Table~\ref{tab:N} briefly summarizes admissible index values for the known cases. 
\begin{table}[ht]\begin{center}
\begin{tabular}{c|c|l}\hline
Lattice & Dim. & Admissible index values \\ \hline\hline
$Z$ & 1  & 1,3,5,7,9,\dots \\
$Z^2$ & 2  & 1,5,9,13,17,25,29,37,41,45,49,\dots \\
$A_2$ & 2  & 1,7,13,19,31,37,43,49,\dots \\
$D_4$ & 4  & 1,25,49,169,289,625,\dots \\
$Z^4$ & 4  & 1,25,49,81,121,169,225,289,361,\dots \\ \hline
\end{tabular}\end{center}
\caption{Admissible index values for geometrically-similar and clean sublattices in one, two and four dimensions. See Appendix~\ref{app:rootlattices} for more information about these sets of index values.}\label{tab:N}
\end{table}
In general $Z^L$ has a geometrically-similar and clean sublattice if and only if $N$ is odd  and
\begin{itemize}
\item[a)] $L$ odd and $N$ an $L^{th}$ power, or
\item[b)] $L=2$ and $N$ of the form $a^2+b^2$, or
\item[c)] $L=4k,\, k\geq 1$ and $N$ of the form $m^{L/2}$ for some integer $m$,
\end{itemize}
see~\cite{diggavi:2002} for details.

It can be shown that squaring an admissible index value yields another admissible index value for all lattices considered in this work. 
We can generalize this even further and show that the product of any number of admissible index values leads to an admissible index value.
\begin{lemma}\label{lem:productindex}
For the lattices $A_2, D_4$ and $Z^L$ where $L=1,2$ or $L=4k$, where $k\geq 1$, the product of two or more admissible index values yields another admissible index value.
\end{lemma} 
\begin{proof}
See Appendix~\ref{app:proofslatticetheory}.
\end{proof}
As noted in~\cite{conway:1999b} it is always possible by e.g.\ exhaustive search to see if a lattice $\Lambda$ contains a sublattice $\Lambda'$ with an index value of $N=c^{L/2}, c\in \mathbb{R}^+$. Let the Gram matrix of $\Lambda$ be $A$. Then search through $\Lambda$ to see if it contains a set of generator vectors\index{generator!vector} with Gram matrix\index{Gram matrix} $cA$. 
In large lattice dimensions this approach easily becomes infeasible. However, for known lattices the admissible index values can be found off-line and then tabulated for later use. 

If two lattices $\Lambda\subset \mathbb{R}^L$ and $\Lambda'\subset \mathbb{R}^{L
'}$ are concatenated (i.e.\ their Cartesian product is formed) then the resulting lattice $\Lambda''$ is of dimension $L''=L+L'$, cf.\ Definition~\ref{def:cartproductlattice}. The set of admissible index values of $\Lambda''$ (when normalized by dimension) might be different than that of $\Lambda$ or $\Lambda'$. For example let $\Lambda=Z^1$ where the admissible index values are the odd integers. Then notice that the four-dimensional $Z^4$ lattice is simply a cascade of four $Z^1$ lattices. However, the admissible index values (normalized per dimension) of $Z^4$ are given by (see Appendix~\ref{app:z4})
\begin{equation}
N'=\{\bs{1},2.24,2.65,\bs{3},3.32,3.61,3.87,4.12,4.36,4.58,4.8,\bs{5},\dots\},
\end{equation}
where we have shown the index values of $Z^1$ in boldface. Thus, by forming a higher dimensional lattice by cascading smaller dimensional lattices it is possible to achieve more (or at least different) index values. 

A different strategy is to change the underlying ring $\mathcal{J}$ as shown in Fig.~\ref{fig:J_lattices} which results in a different lattice of the same dimension that might lead to new index values. In this thesis, however, we will be using the known admissible index values of Table~\ref{tab:N}.

\subsection{Sublattices}
In this section we construct sublattices and primarily focus on a special type of sublattices called product lattices\index{lattice!product lattice}. In~\cite{diggavi:2002} the following definition of a product lattice was presented.
\begin{definition}[\cite{diggavi:2002}]\label{def:productlattice}
Let $\mathcal{J}$ be an arbitrary ring, let $\Lambda=\mathcal{J}$ and form the two sublattices $\Lambda_0=\xi_0\Lambda$ and $\Lambda_1=\Lambda\xi_1,\ \xi_i\in \Lambda, i=0,1$.
Then the lattice $\Lambda_\pi=\xi_0\Lambda\xi_1$ is called a product lattice and it satisfies $\Lambda_\pi\subseteq \Lambda_i, i=0,1$.
\end{definition}
\noindent In this work, however, we will make use of a more general notion of a product lattice which includes Definition~\ref{def:productlattice} as a special case.
\begin{definition}\label{def:generalproductlattice}
A product lattice $\Lambda_\pi$ is any sublattice satisfying $\Lambda_\pi\subseteq \Lambda_i$ where $\Lambda_i=\xi_i\Lambda$ or $\Lambda_i=\Lambda\xi_i, i=0,\dotsc,K-1$.
\end{definition}

The construction of product lattices\index{lattice!product lattice} based on two sublattices as described in Defini\-tion~\ref{def:productlattice} was treated in detail in~\cite{diggavi:2002}.
In this section we extend the existing results of~\cite{diggavi:2002} and construct product lattices based on more than two sublattices for $L=1,2$ and $4$ dimensions for the root lattices $Z^1, Z^2, A_2, Z^4$ and $D_4$, which are described in Appendix~\ref{app:rootlattices}.
Along the same lines as in~\cite{diggavi:2002} we construct sublattices and product lattices by use of the ordinary rational integers\index{integer!rational} $\mathbb{Z}$ as well as the Gaussian integers $\mathcal{G}$, Eisenstein integers $\mathcal{E}$, Lipschitz integral Quaternions\index{integer!Lipschitz} $\mathcal{H}_0$ and the Hurwitz integral Quaternions\index{integer!Hurwitzian} $\mathcal{H}_1$, where $\mathcal{G}$ and $\mathcal{E}$ are given by~(\ref{eq:gaussianint}) and~(\ref{eq:eisensteinint}), respectively, and~\cite{conway:1999}
\begin{align}
\mathcal{H}_0 &= \{\xi_1+i\xi_2+j\xi_3+k\xi_4 : \xi_1,\xi_2,\xi_3,\xi_4\in \mathbb{Z} \}, \\
\mathcal{H}_1 &= \{\xi_1+i\xi_2+j\xi_3+k\xi_4 : \xi_1,\xi_2,\xi_3,\xi_4\ \text{all in}\ \mathbb{Z}\ \text{or all in}\ \mathbb{Z}+1/2 \},
\end{align}
where $i,j$ and $k$ are unit Quaternions, see Appendix~\ref{app:quaternions} for more information. 
For example a sublattice $\Lambda_1$ of $\Lambda=\mathbb{Z}$ is easily constructed, simply by multiplying all points $\lambda\in \Lambda$ by $\xi$ where $\xi\in \mathbb{Z}\backslash\{0\}$.\footnote{Since $\Lambda$ is a torsion free\index{torsion!free} $\mathcal{J}$-module the submodule $\Lambda'=\xi\Lambda$ is a non-trivial cyclic submodule\index{cyclic!submodule} whenever $0\neq \xi \in \Lambda$.}
This gives a geometrically-similar sublattice $\Lambda_1 = \xi\mathbb{Z}$ of index $|\xi|$. This way of constructing sublattices may be generalized by considering different rings of integers. For example, for the square lattice $\Lambda=\mathcal{G}$ whose points lie in the complex plane, a geometrically-similar sublattice of index 2 may be obtained by multiplying all elements of $\Lambda$ by the Gaussian integer $\xi=1+i$.

\subsubsection{Sublattices and product lattices of $Z^1, Z^2$ and $A_2$}
The construction of product lattices\index{lattice!product lattice} based on the sublattices $Z^1, Z^2$ and $A_2$ is a straight forward generalization of the approach taken in~\cite{diggavi:2002}. 
Let the lattice $\Lambda$ be any one of $Z^1=\mathbb{Z}, Z^2= \mathcal{G}$ or $A_2= \mathcal{E}$ and let the geometrically-similar sublattices $\Lambda_i$ be given by $\xi_i\Lambda$ where $\xi_i$ is an element of the rational integers $\mathbb{Z}$, the Gaussian integers $\mathcal{G}$ or the Eisenstein integers $\mathcal{E}$, respectively. 
\begin{lemma}\label{lem:productlattice}
$\Lambda_\pi = \xi_0\xi_1\cdots\xi_{K-1} \Lambda$ is a product lattice.
\end{lemma}
\begin{proof}
See Appendix~\ref{app:proofslatticetheory}.
\end{proof}

Also, as remarked in~\cite{diggavi:2002}, since the three rings considered are unique factorization rings\index{unique factorization ring}, the notion of least common multiple\index{least common multiple} (lcm) is well defined. Let us define $\xi_\cap \triangleq \text{lcm}(\xi_0,\dotsc,\xi_{K-1})$ so that $\xi_i|\xi_\cap$, i.e.\ $\xi_i$ divides $\xi_\cap$. This leads to the following lemma.
\begin{lemma}\label{lem:productlattice_int}
$\Lambda_\pi'=\xi_{\cap}\Lambda$ is a product lattice.
\end{lemma}
\begin{proof}
See Appendix~\ref{app:proofslatticetheory}.
\end{proof}

\begin{figure}[ht]
\begin{center}
\psfrag{Lc}{$\Lambda$}
\psfrag{L0}{$\Lambda_0=\xi_0\Lambda$}
\psfrag{L1}{$\Lambda_1=\xi_1\Lambda$}
\psfrag{LK-1}{$\Lambda_{K-1}=\xi_{K-1}\Lambda$}
\psfrag{Lint}{$\Lambda_\pi'=\xi_{\cap}\Lambda$}
\psfrag{Lpi}{$\Lambda_\pi=\xi_0\xi_1\cdots\xi_{K-1}\Lambda$}
\includegraphics[width=6cm]{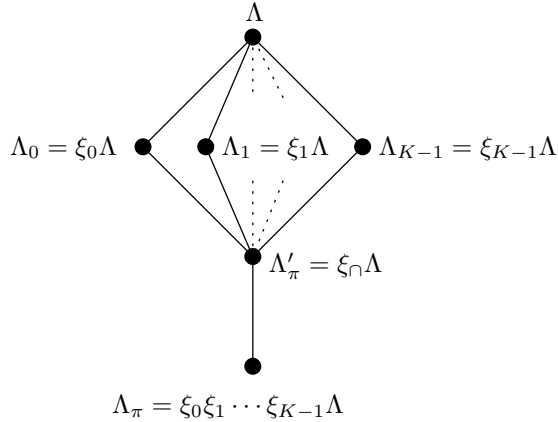}
\caption{The intersection (meet\index{meet}) of $K$ arbitrary sublattices form a product lattice for $Z^1, Z^2$ and $A_2$.}
\label{fig:hasse1}
\end{center}
\end{figure}

\begin{figure}[ht]
\begin{center}
\includegraphics[width=9cm]{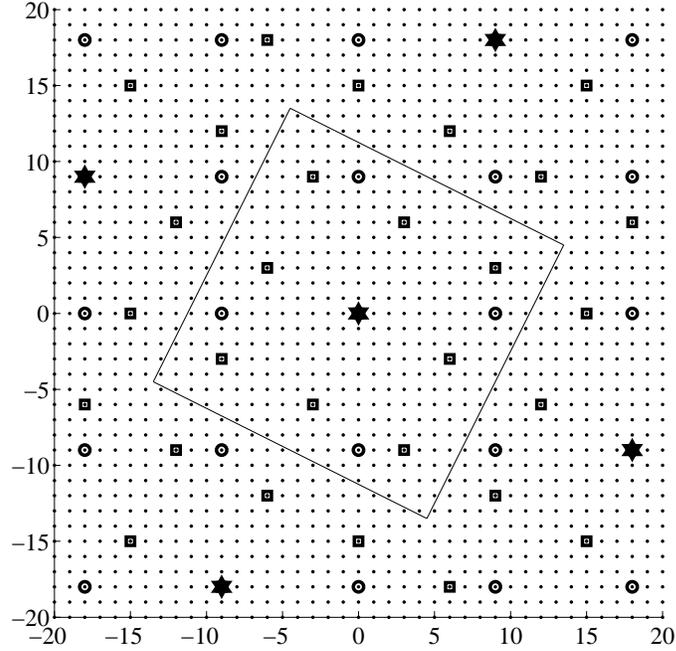}
\caption{The lattice $\Lambda=\mathcal{G}$ is here shown as dots. The two lattices $\Lambda_0=(3+6i)\Lambda$ (squares) and $\Lambda_1=9\Lambda$ (circles) are sublattices of $\Lambda$ and the lattice $\Lambda_\cap=(9+18i)\Lambda$ (stars) is a sublattice of all the lattices $\Lambda_0,\Lambda_1$ and $\Lambda$. The solid lines describe the boundary of the Voronoi cell $V_0$ of the product lattice point located at the origin.}
\label{fig:z2_lcm}
\end{center}
\end{figure}

The relations between $\Lambda,\Lambda_i,\Lambda'_\pi$ and $\Lambda_\pi$ as addressed by Lemmas~\ref{lem:productlattice} and~\ref{lem:productlattice_int} are shown in Fig.~\ref{fig:hasse1}.
For example, let $\Lambda=\mathcal{G} (\equiv\! Z^2)$ and let $N_0=45$ and $N_1=81$. Then we have that $\text{lcm}(45,81)=405$ and $45\cdot 81=3645$. 
We may choose $\xi_0=3+6i, \xi_1=9$ and $\xi_\cap=9+18i$, so that $|\xi_0|^2=45, |\xi_1|^2=81$ and $|\xi_\cap|^2=405$. 
Notice that $\xi_0 | \xi_\cap$ and $\xi_1 | \xi_\cap$, i.e.\ $\frac{\xi_\cap}{\xi_0} = 3\in\mathcal{G}$ and $\frac{\xi_\cap}{\xi_1} = 1+2i\in\mathcal{G}$. 
Since both $\xi_0$ and $\xi_1$ divides $\xi_\cap$, the lattice $\Lambda_\cap=\xi_\cap\Lambda$ will be a sublattice of $\Lambda_0=\xi_0\Lambda$ as well as $\Lambda_1=\xi_1\Lambda$, see Fig.~\ref{fig:z2_lcm}. 

%
%

%
%

\subsubsection{Sublattices and product lattices of $Z^4$}
\begin{figure}[ht]
\begin{center}
\psfrag{Lc}{$\Lambda$}
\psfrag{L0}{$\Lambda_0=\xi_0\Lambda$}
\psfrag{L1}{$\Lambda_1=\Lambda\xi_1$}
\psfrag{Lpi}{$\Lambda_\pi=\xi_0\Lambda\xi_1$}
\includegraphics[width=6cm]{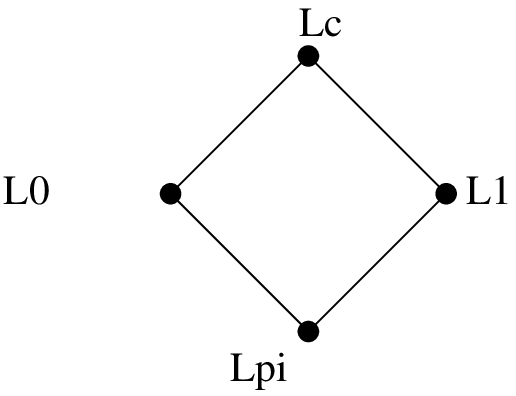}
\caption{Two arbitrary sublattices form a product lattice.}
\label{fig:hasse2}
\end{center}
\end{figure}

As was done in~\cite{diggavi:2002} we will use the Quaternions~\cite{ward:1997,kantor:1989} for the construction of sublattices and product lattices\index{lattice!product lattice} for $Z^4$.
The Quaternions form a non-commutative ring\index{non-commutative ring} and it is therefore necessary to distinguish between left and right multiplication~\cite{ward:1997,kantor:1989}. For the case of two sublattices we adopt the approach of~\cite{diggavi:2002} and construct the sublattice $\Lambda_0$ by multiplying $\Lambda$ on the left, i.e.\ $\Lambda_0=\xi_0\Lambda$ and $\Lambda_1$ is obtained by right multiplication $\Lambda_1=\Lambda\xi_1$, see Fig.~\ref{fig:hasse2}.
More than two descriptions was not considered in~\cite{diggavi:2002}. Let the $K$ sublattices be of index $N_0,\dots, N_{K-1}$ respectively. Then we may form $\Lambda_0=\xi_0\Lambda$ and $\Lambda_1=\Lambda\xi_1$ as above. However, by letting $\Lambda_2=\Lambda\xi_2$ we run into trouble when creating the product lattice. For example, if we define $\Lambda_\pi=\xi_0\Lambda\xi_1\xi_2$ it is clear that $\Lambda_\pi\subseteq \Lambda_0$ and $\Lambda_\pi\subseteq \Lambda_2$. The problem is that in general $\Lambda_\pi\nsubseteq \Lambda_1$ since $\xi_1\xi_2\neq \xi_2\xi_1$ and we therefore have to restrict the set of admissible index values.

\begin{lemma}\label{lem:lipshitz}
Let $N_0$ and $N_1$ be admissible index values for $Z^2$. Then $N_0^2$ and $N_1^2$ (which are admissible index values for $Z^4$), can be associated with a pair of Lipschitz integers\index{integer!Lipschitz} $(\xi_0,\xi_1)$ that commute, i.e.\ $\xi_0\xi_1=\xi_1\xi_0$.
\end{lemma}
\begin{proof}
See Appendix~\ref{app:proofslatticetheory}.
\end{proof}

\begin{figure}[ht]
\begin{center}
\psfrag{Lc}{$\Lambda$}
\psfrag{L0}{$\Lambda_0=\xi_0\Lambda$}
\psfrag{L1}{$\Lambda_1=\Lambda\xi_1$}
\psfrag{LK-1}{$\Lambda_{K-1}=\Lambda\xi_{K-1}$}
\psfrag{Lint}{$\Lambda_\pi'=\xi_0\Lambda\xi'_{\cap}$}
\psfrag{Lpi}{$\Lambda_\pi=\xi_0\Lambda\xi_1\cdots\xi_{K-1}$}
\includegraphics[width=6cm]{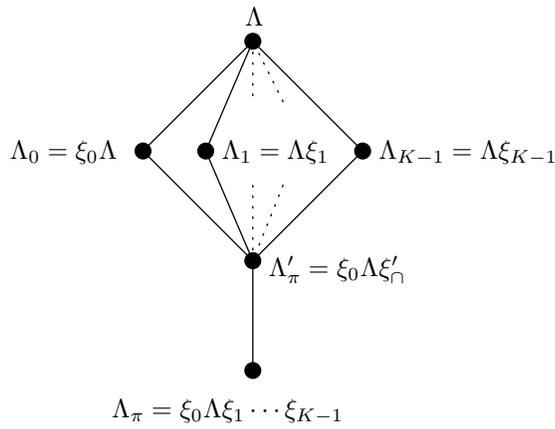}
\caption{The intersection (meet\index{meet}) of $K$ arbitrary sublattices form a product lattice for $Z^4$.}
\label{fig:hasse3}
\end{center}
\end{figure}

From Lemma~\ref{lem:lipshitz} it follows that there exist an infinite number of pairs of admissible index values $(N_0,N_1)$ where $N_0\neq N_1$ such that the Lipschitz integers $\xi_0$ and $\xi_1$ commute. For example, let $N_0=7^2,N_1=13^2,N_2=5^2$ and define $\Lambda_0=\xi_0\Lambda, \Lambda_1=\Lambda\xi_1$ and $\Lambda_2=\Lambda\xi_2$ where $\xi_0=-2 -i -j -k, \xi_1= -2 -i + 0j + 0k$ and $\xi_2=-3 -2i +0j + 0k$. For this example we have $\xi_0\xi_1\neq \xi_1\xi_0$ and $\xi_0\xi_2\neq\xi_2\xi_0$ but $\xi_1\xi_2=\xi_2\xi_1$. 
Letting $\Lambda_\pi=\xi_0\Lambda\xi_1\xi_2$ makes sure that $\Lambda_\pi\subseteq \Lambda_i$ for $i=0,1,2$, since $\Lambda_\pi=(\xi_0\Lambda\xi_1)\xi_2=(\xi_0\Lambda\xi_2)\xi_1$. In general it is possible to construct the product lattice $\Lambda_\pi$ such that $\Lambda_\pi\subseteq \Lambda_i$ for $i=0,\dots, K-1$ as long as any $K-1$ of the $K$ $\xi_i$'s commute, see Fig.~\ref{fig:hasse3}, where $\xi'_\cap=\text{lcm}(\xi_1,\dots,\xi_{K-1})$. If all the pairs $(\xi_i,\xi_j), i,j\in\{0,\dots,K-1\}$ commute the procedure shown in Fig.~\ref{fig:hasse1} is also valid.

\subsubsection{Sublattices and product lattices of $D_4$}
For $D_4$ we use the ring of Hurwitzian integers\index{integer!Hurwitzian}, i.e.\ $\xi_i\in \mathcal{H}_1$. For the case of two sublattices we design the sublattices and product lattices\index{lattice!product lattice} as in~\cite{diggavi:2002} and shown in Fig.~\ref{fig:hasse2}. For more than two sublattices we have to make a restriction on the set of allowable admissible index values. The Quaternions leading to admissible index values for $D_4$ obtained in~\cite{diggavi:2002} are of the form\footnote{With two exceptions being $\xi_i=\frac{1}2+\frac{1}2i+\frac{1}2j+\frac{5}2k$ and $\xi_i=\frac{1}2+\frac{3}2i+\frac{3}2j+\frac{3}2k$ both leading to an index value of $N=49$.} $\xi_i=\frac{a}2(1+i)+\frac{b}2(j+k) \in \mathcal{H}_1$, where $a$ and $b$ are odd positive integers. 
Quaternions of this form do generally not commute since both $a$ and $b$ are nonzero. In fact two Quaternions commute if and only if their vector parts are proportional~\cite{baylis:1989}, i.e.\ linearly-dependent, which rarely happens for the Quaternions of the form $\xi_i=\frac{a}2(1+i)+\frac{b}2(j+k) \in \mathcal{H}_1$. For example we did an exhaustive search based on all admissible index values between 25 and 177241 and found only five pairs (up to permutations) of Quaternions that commute. These are shown in Table~\ref{tab:Qcommute}.

\begin{table}[ht]
\begin{center}
\begin{tabular}{cccc}\hline
$N_0$ & $N_1$ & $\xi_0$ & $\xi_1$ \\ \hline
25 & 15625 & $\frac{1}2+\frac{1}2i+\frac{3}2j+\frac{3}2k$ & $\frac{5}2+\frac{5}2i+\frac{15}2j+\frac{15}2k$ \\
169 & 105625 & $\frac{1}2+\frac{1}2i+\frac{5}2j+\frac{5}2k$ & $\frac{5}2+\frac{5}2i+\frac{25}2j+\frac{25}2k$ \\
625 & 28561 & $\frac{5}2+\frac{5}2i+\frac{5}2j+\frac{5}2k$ & $\frac{13}2+\frac{13}2i+\frac{13}2j+\frac{13}2k$ \\
625 & 83521 & $\frac{5}2+\frac{5}2i+\frac{5}2j+\frac{5}2k$ & $\frac{17}2+\frac{17}2i+\frac{17}2j+\frac{17}2k$ \\
28561 & 83521 & $\frac{13}2+\frac{13}2i+\frac{13}2j+\frac{13}2k$ & $\frac{17}2+\frac{17}2i+\frac{17}2j+\frac{17}2k$ \\ \hline
\end{tabular}
\end{center}
\caption{Each row shows two Quaternions $\xi_0$ and $\xi_1$ which commute, i.e.\ $\xi_0\xi_1=\xi_1\xi_0$.}
\label{tab:Qcommute}
\end{table}

We therefore restrict the set of admissible index values to $N_i\in \{a,b\}$ for $i=0,\dots,K-1$ where $a$ and $b$ are any two admissible index values. 
With this the product lattice, for $K>2$ sublattices, is based on only two integers e.g.\ $\xi_0$ and $\xi_1$ as shown in Fig.~\ref{fig:hasse2} and the index of the product lattice is then $N_\pi=ab$. With this approach it is possible to obtain $\Lambda_\pi\subseteq \Lambda_i$ for $i=0,\dots, K-1$.

\chapter{Single-Description Rate-Distortion Theory}\label{chap:rd_theory}
Source coding\index{source coding} with a fidelity criterion\index{distortion measure!fidelity criterion}, also called rate-distortion theory\index{rate-distortion!theory} (or lossy source coding)\index{source coding!lossy}, was introduced by Shannon in his two landmark papers from 1948~\cite{shannon:1948} and 1959~\cite{shannon:1959} and has ever since received a lot of attention.
For an introduction to rate-distortion theory we refer the reader to the survey papers by Kieffer~\cite{kieffer:1993} and Berger and Gibson~\cite{berger:1998} and the text books by Berger~\cite{berger:1971}, Cisz\'ar and K\"orner~\cite{csiszar:1981} and Cover and Thomas~\cite{cover:1991}.

\section{Rate-Distortion Function}
A fundamental problem of rate-distortion theory is that of describing the rate $R$ required to encode a source $X$ at a prescribed distortion (fidelity) level $D$. Let $X^L=\{X_i\}, i=1,\dotsc,L$ be a sequence of random variables (or letters) of a stationary\footnote{Throughout this work we will assume all stochastic processes to be discrete-time zero-mean weak-sense stationary processes (unless otherwise stated).} random process\index{stationary process} $X$. Let $\hat{X}$ be the reproduction of $X$ and let $x$ and $\hat{x}$ be realizations of $X$ and $\hat{X}$, respectively. 
The alphabets\index{alphabet} $\mathcal{X}$ and $\hat{\mathcal{X}}$ of $X$ and $\hat{X}$, respectively, can be continuous\index{alphabet!continuous} or discrete\index{alphabet!discrete} and in the latter case we distinguish between discrete alphabets of finite or countably infinite cardinality\index{alphabet!countably infinite}\index{alphabet!countably finite}. When it is clear from context we will often ignore the superscript $^L$ which indicates the dimension of the variable or alphabet so that $x\in \mathcal{X}\subset \mathbb{R}^L$ denotes an $L$-dimensional vector or element of the alphabet $\mathcal{X}$ which is a subset of $\mathbb{R}^L$.

\begin{definition}
A fidelity criterion\index{distortion measure!fidelity criterion} for the source $X$ is a family $\rho^{(L)}(X,\hat{X}), L\in\mathbb{N}$ of distortion measures of which $\rho^{(L)}$ computes the distortion when representing $X$ by $\hat{X}$. If $\rho^{(L)}(X,\hat{X})\triangleq\frac{1}{L}\sum_{i=1}^{L}\rho(X_i,\hat{X}_i)$ then $\rho$ is said to be a single-letter fidelity criterion\index{distortion measure!single letter} and we will then use the notation $\rho(X,\hat{X})$. Distortion measures of the form $\rho(X-\hat{X})$ are called difference distortion measures\index{distortion measure!difference}. For example $\rho(X,\hat{X})=\frac{1}{L}\|X-\hat{X}\|^2$ is a difference distortion measure (usually referred to as the squared-error distortion measure)\index{distortion measure!squared error}.
\end{definition}

In this work we will be mainly interested in the squared-error single-letter fidelity criterion which is defined by
\begin{equation}
\rho(X,\hat{X}) \triangleq \frac{1}{L}\sum_{i=1}^{L}(X_i-\hat{X}_i)^2.
\end{equation}

With this, formally stated, Shannon's rate-distortion function $R(D)$ (expressed in bit/dim.) for stationary sources with memory and single-letter fidelity criterion, $\rho$, is defined as~\cite{berger:1971}
\begin{equation}\label{eq:RD}
R(D) \triangleq \lim_{L\rightarrow\infty} R_L(D),
\end{equation}
where the $L^{th}$ order rate-distortion function is given by
\begin{equation}\label{eq:RDLorder}
R_L(D) = \inf\{\frac{1}{L}I(X;\hat{X}) : E\rho(X,\hat{X})\leq D\},
\end{equation}
where $I(X;\hat{X})$ denotes the mutual information\footnote{The mutual information between to continuous-alphabet\index{alphabet!continuous} sources $X$ and $\hat{X}$ with a joint pdf $f_{X,\hat{X}}$ and marginals $\fx$ and $f_{\hat{X}}$, respectively, is defined as~\cite{cover:1991}
\begin{equation*}
I(X;\hat{X}) = \int_{\mathcal{X}}\int_{\hat{\mathcal{X}}}f_{X,\hat{X}}(x,\hat{x})\log_2\left(\frac{f_{X,\hat{X}}(x,\hat{x})}{\fx(x)f_{\hat{X}}(\hat{x})}\right)\,dxd\hat{x}.
\end{equation*}} between $X$ and $\hat{X}$, $E$ denotes the statistical expectation operator and the infimum\index{infimum} is over all\index{mutual information} conditional distributions\index{distribution!conditional} $f_{X|\hat{X}}(\hat{x}|x)$ for which the joint distributions\index{distribution!joint} $f_{X,\hat{X}}(x,\hat{x})=\fx(x)f_{\hat{X}|X}(\hat{x}|x)$ satisfy the expected distortion constraint given by
\begin{equation}
\int_{\mathcal{X}}\int_{\hat{\mathcal{X}}}\fx(x)f_{\hat{X}|X}(\hat{x}|x)\rho(x,\hat{x})d\hat{x}dx\leq D.
\end{equation}
The $L^{th}$ order rate-distortion function $R_L(D)$ can be seen as the rate-distortion function of an $L$-dimensional i.i.d.\ vector source $X$ producing vectors with the distribution\index{distribution!joint} of $X$~\cite{linder:1997}.


Let $h(X)$ denote the differential entropy\index{entropy!differential} (or continuous entropy) of $X$ which is given by~\cite{cover:1991}
\begin{equation*}
h(X) = -\int_{\mathcal{X}}\fx(x)\log_2(\fx(x))\, dx
\end{equation*}
and let the differential entropy rate\index{entropy!rate} $\bar{h}(X)$ be defined by $\bar{h}(X)\triangleq \lim_{L\rightarrow\infty}\frac{1}{L}h(X)$ where for independently and identically distributed (i.i.d.) scalar processes\index{scalar!process} $\bar{h}(X)=\frac{1}{L}h(X)$. With a slight abuse of notation we will also use the notation $\bar{h}(X)$ to indicate the dimension normalized differential entropy of an i.i.d.\ vector source.
If $\rho$ is a difference distortion measure\index{distortion measure!difference}, then~(\ref{eq:RD}) and~(\ref{eq:RDLorder}) can be lower bounded by the Shannon lower bound~\cite{berger:1971}\index{Shannon lower bound}. Specifically, if $E\rho$ is the mean squared error (MSE) fidelity criterion\index{MSE}, then~\cite{berger:1971,linder:1997}
\begin{equation}\label{eq:slb}
R(D) \geq \bar{h}(X) - \frac{1}{2}\log_2(2\pi e D),
\end{equation}
where equality holds at almost all distortion levels $D$ for a (stationary) Gaussian source~\cite{berger:1971}\index{Gaussian process}.\footnote{Eq.~(\ref{eq:slb}) is tight for all $D\leq \text{ess}\inf S_X$, where $S_X$ is the power spectrum of a stationary Gaussian process $X$~\cite{berger:1971}.} In addition it has been shown that~(\ref{eq:slb}) becomes asymptotically tight at high resolution, i.e.\ as $D\rightarrow 0$, for sources with finite differential entropies and finite second moments for general difference distortion measures, cf.~\cite{linder:1997}. 

Recall that the differential entropy\index{entropy!of Gaussian vector} of a jointly Gaussian vector is given by~\cite{cover:1991}
\begin{equation}
h(X) = \frac{1}{2}\log_2((2\pi e)^{L} |\Phi|),
\end{equation}
where $|\Phi|$ is the determinant of $\Phi=EXX^T$, i.e.\ the covariance matrix\index{covariance matrix} of $X$. It follows from~(\ref{eq:slb}) that the rate-distortion function of a memoryless\index{memoryless source} scalar Gaussian process of variance $\sigma_X^2$ is given by
\begin{equation}\label{eq:RDGaussian}
R(D) = \frac{1}{2}\log_2\left(\frac{\sigma_X^2}{D}\right),
\end{equation}
whenever $D\leq \sigma_X^2$ and $R(D)=0$ for $D>\sigma_X^2$ since $R(D)$ is everywhere non-negative.

\begin{figure}[ht]
\psfrag{D}{$\scriptstyle D$ \scriptsize [dB]}
\psfrag{R}{$\scriptstyle R$ \scriptsize [bit/dim.]}
\begin{center}
\includegraphics[width=8cm]{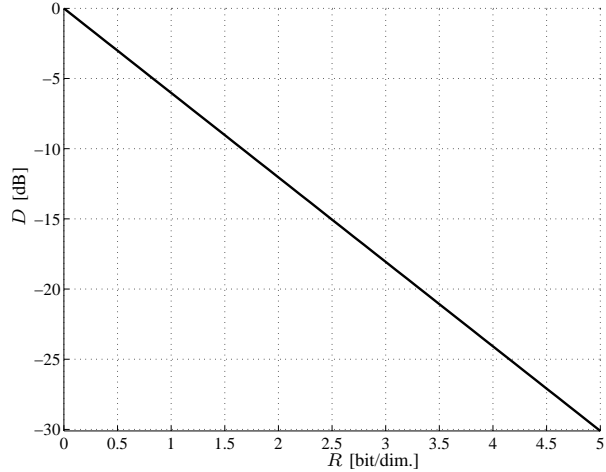}
\caption{$D(R)$ for the unit-variance memoryless Gaussian source.}
\label{fig:DRGaussian}
\end{center}
\end{figure}

The inverse of $R(D)$ is called the distortion-rate function\index{distortion-rate function} $D(R)$ and it basically says that if a source sequence is encoded at a rate $R$ the distortion is at least $D(R)$. From~(\ref{eq:RDGaussian}) we see that the distortion-rate function of the memoryless Gaussian process\index{Gaussian process} is given by
\begin{equation}\label{eq:DRGaussian}
D(R) = \sigma_X^2 2^{-2R},
\end{equation}
which is shown in Fig.~\ref{fig:DRGaussian} for the case of $\sigma_X^2=1$.

\begin{remark}
From~(\ref{eq:DRGaussian}) and also from Fig.~\ref{fig:DRGaussian} it may be seen that each extra bit reduces the distortion by a factor of four --- a phenomena often referred to as the ``6 dB per bit rule''~\cite{gray:1998}\index{6 dB per bit rule}. In fact, the ``6 dB per bit rule'' is approximately true not just for the Gaussian source but for arbitrary sources.
\end{remark}
%
%

The rate-distortion function of a memoryless scalar source\index{memoryless source} and squared-error distortion measure\index{distortion measure!squared error} may be upper and lower bounded by use of the entropy-power inequality\index{inequality!entropy-power}, that is~\cite{berger:1971}
\begin{equation}\label{eq:RDbound_entpower}
\frac{1}{2}\log_2\left(\frac{\sigma_X^2}{D}\right) \geq R(D) \geq \frac{1}{2}\log_2\left(\frac{P_X}{D}\right),
\end{equation}
where $P_X$ is the entropy power.\footnote{The entropy power\index{entropy!power} $P_X\triangleq (2\pi e)^{-1}2^{2h(X)}$ of a source $X$ is defined as the variance of a Gaussian density\index{distribution!Gaussian} that has the same differential entropy as $X$~\cite{berger:1971}.} Similarly, the distortion-rate function is bounded as
\begin{equation}\label{eq:DRbound_entpower}
P_X2^{-2R} \leq D(R) \leq \sigma_X^2 2^{-2R},
\end{equation}
with equalities all the way in both~(\ref{eq:RDbound_entpower}) and~(\ref{eq:DRbound_entpower}) if $X$ is Gaussian. 

\begin{remark}
Inequalities~(\ref{eq:RDbound_entpower}) and~(\ref{eq:DRbound_entpower}) show that, of all sources, the Gaussian source is the hardest to compress.
\end{remark}

\section{Quantization Theory}\index{quantization theory}
A quantizer $Q$ consists of a set of decision cells $\mathcal{S}=\{S_i : i\in \mathcal{I}\}$ where $\mathcal{I}\subseteq \mathbb{N}$ together with a set of reproduction values $\mathcal{C}=\{c_i : i\in \mathcal{I}\}$~\cite{gray:1998}. The operation of quantization is defined as $Q(x)\triangleq c_i$ if $x\in S_i$. We require that $\mathcal{S}$ cover the input space $\mathcal{X}$ which implies that $\bigcup_{i\in\mathcal{I}} S_i \supset \mathcal{X}$ and often we need $\mathcal{S}$ to partition $\mathcal{X}$ so that sets of $\mathcal{S}$ are pairwise disjoint, i.e.\ $S_i\cap S_j=\emptyset, i\neq j$ so that $\bigcup_{i\in\mathcal{I}} S_i = \mathcal{X}$.

\begin{definition}
The decision cells of a nearest neighbor quantizer\index{nearest neighbor!quantizer} are called Voronoi cells, Voronoi regions or Dirichlet regions~\cite{conway:1999}. Given the $i^{th}$ reproduction value $c_i$ the Voronoi cell $V(c_i)$ is defined by
\begin{equation}
V(c_i) \triangleq \{ x\in \mathcal{X} : \rho(x,c_i)\leq \rho(x,c_j),\ \forall j\in \mathcal{I}\},
\end{equation}
where ties (if any) can be arbitrarily broken.\footnote{Two neighboring $L$-dimensional Voronoi cells for continuous-alphabet\index{alphabet!continuous} sources share a common $L'$-dimensional face where $L'\leq L-1$. For discrete-alphabet\index{alphabet!discrete} sources it is also possible that a point is equally spaced between two or more centroids of the codebook, in which case tie breaking is necessary in order to make sure that the point is not assigned to more than one Voronoi cell.}
\end{definition}
It follows that the expected distortion of a quantizer is given by
\begin{equation}
D_Q = \sum_{i\in \mathcal{I}} \int_{x\in S_i}\fx(x)\rho(x,c_i)\, dx.
\end{equation}

Let us for the moment assume that $\mathcal{X}=\mathbb{R}^L$ and $\mathcal{C}=\mathcal{\hat{X}}\subset \mathbb{R}^L$. Then, for the squared error distortion measure\index{distortion measure!squared error}, the Voronoi cells of an $L$-dimensional nearest neighbor quantizer\index{nearest neighbor!quantizer} (vector quantizer) are defined as
\begin{equation}
V(\hat{x}_i) \triangleq \{ x\in \mathbb{R}^L : \|x-\hat{x}_i\|^2 \leq \|x-\hat{x}_j\|^2,\ \forall x_j\in \hat{\mathcal{X}}\},\quad \hat{x}_i\in\hat{\mathcal{X}},
\end{equation}
where $\|\cdot\|$ denotes the $\ell_2$-norm, i.e.\ $\|x\|^2=\sum_{n=1}^{L}x_n^2$.

Vector quantizers are often classified as either entropy-constrained\index{entropy-constrained quantization} quantizers or resolution-constrained quantizers\index{resolution-constrained quantization} or as a mixed class where for example the output of a resolution-constrained quantizer is further entropy coded\index{entropy!coding}.\footnote{Entropy-constrained quantizers (resp.\ resolution-constrained quantizers) are also called variable-rate quantizers (resp.\ fixed-rate quantizers).}
When designing an entropy-constrained quantizer one seeks to form the Voronoi regions $V(\hat{x}_i), \hat{x}_i\in \hat{\mathcal{X}}$, and the reproduction alphabet $\hat{\mathcal{X}}$ such that the distortion $D_Q$ is minimized subject to an entropy constraint $R$ on the discrete entropy $H(\hat{X})$. Recall that the discrete entropy\index{entropy!discrete} of a random variable is given by~\cite{cover:1991}
\begin{equation}
H(\hat{X}) = -\sum_{i\in \mathcal{I}} P(\hat{x}_i)\log_2(P(\hat{x}_i)),
\end{equation}
where $P$ denotes probability and $P(\hat{x}_i) = P(x\in V(\hat{x}_i))$. On the other hand, in resolution-constrained quantization the distortion is minimized subject to a constraint on the cardinality of the reproduction alphabet. 
In this case the elements of $\hat{\mathcal{X}}$ are coded with a fixed rate of $R=\log_2(|\hat{\mathcal{X}}|)/L$. 
For large vector dimensions, i.e.\ when $L\gg 1$, it is very likely that randomly chosen source vectors belong to a typical set\index{typical!set} $\mathcal{A}^{(L)}$ in which the elements are approximately uniformly distributed~\cite{cover:1991}\index{distribution!uniform}. 
As a consequence, in this situation there is not much difference between entropy-constrained and resolution-constrained quantization.

There exists several iterative algorithms for designing vector quantizers. 
One of the earliest such algorithms is the Lloyd algorithm\index{Lloyd algorithm} which is used to construct resolution-constrained scalar quantizers~\cite{lloyd:1957}, see also~\cite{lloyd:1982}. The Lloyd algorithm is basically a cyclic minimizer\index{cyclic!minimizer} that alternates between two phases:
\begin{enumerate}
\item Given a codebook $\mathcal{C}=\hat{\mathcal{X}}$ find the optimal partition of the input space, i.e.\ form the Voronoi cells $V(\hat{x}_i), \forall \hat{x}_i\in \hat{\mathcal{X}}$. 
\item Given the partition, form an optimal codebook, i.e.\ let $\hat{x}_i\in \hat{\mathcal{X}}$ be the centroid\index{centroid} of the set $x\in V(\hat{x}_i)$.
\end{enumerate}
If an analytical description of the pdf\index{probability density function} is unavailable it is possible to estimate the pdf by use of empirical observations~\cite{gersho:1992}. Furthermore, Lloyd's algorithm has been extended to the vector case~\cite{gersho:1992,linde:1980} but has not been explicitly extended to the case of entropy-constrained vector quantization. Towards that end Chou et al.~\cite{chou:1989} presented an iterative algorithm based on a Lagrangian formulation of the optimization problem. In general these empirically designed quantizers are only locally optimal and unless some structure is enforced on the codebooks, the search complexity easily becomes overwhelming (the computational complexity of an unconstrained quantizer increases exponentially with dimension)~\cite{gersho:1992}.
There exists a great deal of different design algorithms and we refer the reader to the text books~\cite{gersho:1992,gray:1990,conway:1999} as well as the in-depth article by Gray and Neuhoff~\cite{gray:1998} for more information about the theory and practice of vector quantization. 

\section{Lattice Vector Quantization}
In this work we will focus on structured vector quantization and more specifically on lattice vector quantization (LVQ)~\cite{conway:1999,gray:1990,gibson:1988}. A family of highly structured quantizers is the tesselating quantizers which includes lattice vector quantizers as a sub family. In a tesselating quantizer all decision cells are translated and possibly rotated and reflected versions of a prototype cell, say $V_0$.
In a lattice vector quantizer all Voronoi cells are translations of $V_0$ which is then taken to be $V_0\triangleq V(0)$, i.e.\ the Voronoi cell of the reproduction point located at the origin (the zero vector) so that $V(\hat{x}_i)=V_0+\hat{x}_i$.\footnote{Notice that not all tesselating quantizers are lattice quantizers. For example, a tesselating quantizer having triangular shaped decision cells is not a lattice vector quantizer.}
In a high-resolution lattice vector quantizer the reproduction alphabet\index{alphabet!reproduction} $\hat{\mathcal{X}}$ is usually given by an $L$-dimensional lattice $\Lambda\subset \mathbb{R}^L$, see Appendices~\ref{app:lattice_defs} and~\ref{app:rootlattices} for more details about lattices.

In order to describe the performance of a lattice vector quantizer it is convenient to make use of high resolution\index{high resolution conditions} (or high rate) assumptions which for a stationary source can be summarized as follows~\cite{gray:1990,gersho:1992}: 
\begin{enumerate}
\item The rate or entropy of the codebook is large, which means that the variance of the quantization error is small compared to the variance of the source. Thus, the pdf of the source can be considered constant within a Voronoi cell, i.e.\ $\fx(x)\approx \fx(\hat{x}_i)$ if $x\in V(\hat{x}_i)$. Hence, the geometric centroids\index{centroid!geometric} of the Voronoi cells are approximately the midpoints of the cells
\item The quantization noise process tends to be uncorrelated even when the source is correlated
\item The quantization error is approximately uncorrelated with the source
\end{enumerate}
Notice that 1) is always true if the source distribution is uniform\index{distribution!uniform}. 
Furthermore, all the above assumptions have been justified rigorously in the limit as the variance of the quantization error tends to zero for the case of smooth sources\index{smooth source} (i.e.\ continuous-alphabet sources having finite differential entropies)~\cite{linder:1994,zamir:1996,viswanathan:2001}. 
If subtractive dither\index{subtractive dither} is used, as is the case of entropy-constrained dithered (lattice) quantization (ECDQ)\index{ECDQ}, the above assumptions are valid at any resolution and furthermore the quantization errors are independent of the source~\cite{ziv:1985,zamir:1992,zamir:1996b}. 
A nice property of ECDQ is that the additive noise model is accurate at any resolution, so that the quantization operation can be modeled as an additive noise process~\cite{zamir:1996b}. 
The dither signal of an ECDQ is an i.i.d.\ random\footnote{The dither signal is assumed known at the decoder so it is in fact a pseudo-random process\index{pseudo-random process}.} process which is uniformly distributed over a Voronoi cell of the quantizer so that for a scalar quantizer the distribution of the quantization errors is uniform. Asymptotically, as the dimension of the ECDQ\index{ECDQ} grows unboundedly, any finite-dimensional marginal\index{marginal!Gaussian distributed} of the noise process becomes jointly Gaussian distributed and the noise process becomes Gaussian distributed in the divergence\footnote{The information divergence (also called Kullback-Leibler distance\index{Kullback-Leibler distance} or relative entropy) between two pdfs $f_X$ and $g_X$ is defined as~\cite{cover:1991}
\begin{equation*} 
\mathbb{D}(f\|g)=\int_{\mathcal{X}} \fx(x)\log_2(\fx(x)/\gx(x))\, dx.
\end{equation*}} 
sense~\cite{zamir:1996}. These properties of the noise process are also valid for entropy-constrained LVQ (without dither) under high resolution assumptions~\cite{zamir:1996}.
It is interesting to see that at very low resolution, i.e.\ as the variance of the quantization error tends to the variance of the source, the performance of an entropy-constrained scalar quantizer is asymptotically as good as any vector quantizer~\cite{marco:2006}.

\subsection{LVQ Rate-Distortion Theory}\label{sec:space_filling}
Let $H$ denote the discrete entropy $H(\hat{\mathcal{X}})$ of the codebook of an entropy-constrained vector quantizer and let the dimension-normalized MSE distortion measure $D_L$ be defined as
\begin{equation}
D_L \triangleq \frac{1}{L}E\|X-\hat{X}\|^2.
\end{equation}
Then by extending previous results of Bennett~\cite{bennett:1948} for high resolution scalar quantization to the vector case it was shown by Zador~\cite{zador:1982} that if $X$ has a probability density\index{probability density function} then\footnote{Later on the precise requirements on the source for~(\ref{eq:zador}) to be valid was formalized by Linder and Zeger~\cite{linder:1994}.}
\begin{equation}\label{eq:zador}
\lim_{H\rightarrow \infty} D_L 2^{2H/L} = a_L 2^{2h(X)/L},
\end{equation}
where $h(X)$ is the differential entropy of $X$ and $a_L$ is a constant that depends only on $L$. In the scalar case where $L=1$ it was shown by Gish and Pierce~\cite{gish:1968} that $a_1=1/12$ and that the quantizer that achieves this value is the unbounded uniform scalar (lattice) quantizer. For the case of $1<L<\infty$ the value of $a_L$ is unknown~\cite{linder:1994}.\footnote{For the case of resolution-constrained quantization both $a_1$ and $a_2$ are known~\cite{gray:2004}.}
It was conjectured by Gersho\index{Gersho's conjecture} in 1979~\cite{gersho:1979} that if the source distribution\index{distribution!uniform} is uniform over a bounded convex set in $\mathbb{R}^L$ then the optimal quantizer will have a partition whose regions are all congruent to some polytope\index{polytope!congruent}.
Today, more than 25 years after, this conjecture remains open. But if indeed it is true then, at high resolution, the optimal entropy-constrained quantizer is a tessellating quantizer independent of the source distribution (as long as it is smooth).

The distortion at high resolution of an entropy-constrained lattice vector quantizer is given by~\cite{gersho:1979,linder:1994}
\begin{equation}\label{eq:DLVQ}
D_L \approx G(\Lambda) \nu^{2/L},
\end{equation}
where $\nu$ (the volume of a fundamental region of the lattice) is given by
\begin{equation}\label{eq:nu2L_rdtheory}
\nu^{2/L} = 2^{2(h(X)-H)/L}.
\end{equation}
Thus, if we assume that Gersho's conjecture is true and furthermore assume that a lattice vector quantizer is optimal then~(\ref{eq:DLVQ}) implies that $a_L = G(\Lambda)$. By inserting (\ref{eq:nu2L_rdtheory}) in~(\ref{eq:DLVQ}) the discrete entropy (again at high resolution) is found to be given by
\begin{equation}\label{eq:discreteentropy}
H(\hat{\mathcal{X}}) \approx h(X) - \frac{L}{2}\log_2\left(\frac{D_L}{G(\Lambda)}\right)\ \text{[bit]}.
\end{equation}

The Shannon lower bound\index{Shannon lower bound} is the most widely used tool to relate the performance of lattice quantizers to the rate-distortion function of a source. For example, at high resolution, the Shannon lower bound is tight for all smooth sources\index{smooth source}, thus 
\begin{equation}
R(D)\approx \bar{h}(X) - \frac{1}{2}\log_2(2\pi e D)\ \text{[bit/dim.]},
\end{equation}
so that the asymptotic rate-redundancy\index{rate redundancy!SD} $R_{\text{red}}$ of a lattice vector quantizer over the rate-distortion function of a smooth source under the MSE distortion measure is\footnote{$R_{\text{red}}$ is in fact the divergence of the quantization noise from Gaussianity in high resolution lattice vector quantization.}
\begin{equation}
R_{\text{red}}=\frac{1}{2}\log_2(2\pi eG(\Lambda))\  \text{[bit/dim.]}
\end{equation}

From~(\ref{eq:DLVQ}) it may be noticed that the distortion of a lattice vector quantizer is source independent and in fact, for fixed $\nu$, the distortion only depends upon $G(\Lambda)$. Furthermore, $G(\Lambda)$ is scale and rotation invariant and depends only upon the shape of the fundamental region $V_0$ of the lattice $\Lambda$~\cite{conway:1999}. In general the more sphere-like shape of $V_0$ the smaller $G(\Lambda)$~\cite{conway:1999}. It follows that $G(\Lambda)$ is lower bounded by $G(S_L)$ the dimensionless normalized second moment\index{second-moment of inertia! of an $L$-sphere} of an $L$-sphere\index{sphere!second-moment} where~\cite{conway:1999}
\begin{equation}
G(S_L) = \frac{1}{(L+2)\pi}\Gamma\left(\frac{L}{2}+1\right)^{2/L},
\end{equation}
and where $\Gamma(\cdot)$ is the Gamma function\index{Gamma function}. For $L\rightarrow \infty$ we have $G(S_\infty)=1/2\pi e$. 
The $L$-fold Cartesian product\index{Cartesian product} of the integers form an $L$-dimensional lattice $Z^L=\mathbb{Z}^L$ which has a hypercubic fundamental region. It can easily be computed that $G(Z^L)=1/12$ which is in fact the largest dimensionless normalized second moment over all admissible fundamental regions~\cite{conway:1999}. Thus,
\begin{equation}
1/12\geq G(\Lambda) \geq G(S_L) \geq \frac{1}{2\pi e},
\end{equation}
 where the first two inequalities become equalities for $L=1$ since in one dimension the only possible lattice is $Z^1$, the scalar uniform lattice, and $G(Z^1)=G(S_1)=1/12$. For $1<L<\infty$ $L$-spheres do not pack the Euclidean space and are therefore not admissible fundamental regions~\cite{conway:1999}. However, for $L\rightarrow \infty$ and with a proper choice of lattice it is known that $G(\Lambda)\rightarrow G(S_\infty)$~\cite{zamir:1996}. Table~\ref{tab:latparm} shows $G(\Lambda)$ for the best known $L$-dimensional lattices with respect to quantization.
\begin{table}[ht]
\begin{center}
\begin{tabular}{lccccc}
Lattice name & Dimension & Notation & $G(\Lambda)$ & $G(S_L)$ \\ \hline
Scalar         &$1$ &$Z^1$ & $0.0833$ & $0.0833$  \\ 
Hexagonal      &$2$ &$A_2$ & $0.0802$ & $0.0796$  \\
BCC            &$3$ &$\tilde{A}_3$& $0.0787$ & $0.0770$ \\
Schl\"{a}fli\index{Schl\"{a}fli lattice}    &$4$ &$D_4$ & $0.0766$ & $0.0750$  \\
---            &$5$ &$\tilde{D}_5$& $0.0756$ & $0.0735$ \\
---            &$6$ &$E_6$ & $0.0743$ & $0.0723$  \\
---            &$7$ &$\tilde{E}_7$ & $0.0731$ & 0.0713 \\
Gosset\index{Gosset lattice}         &$8$ &$E_8$ & $0.0717$ & $0.0705$  \\
Coxeter-Todd\index{Coxeter-Todd lattice}   &$12$ &$K_{12}$      & $0.0701$ & $0.0681$ \\
Barnes-Walls\index{Barnes-Walls lattice}   &$16$ &$BW_{16}$     & $0.0683$ & $0.0666$  \\
Leech\index{Leech lattice}          &$24$ &$\Lambda_{24}$& $0.0658$ & $0.0647$  \\
Poltyrev\footnotemark \index{Poltyrev lattice}&$\infty$ &$\Lambda_{\infty}$ & $0.0585$ & $0.0585$ &  \\ \hline
\end{tabular}
\caption{The dimensionless normalized second moments\index{second-moment of inertia!of a lattice} of the lattice $\Lambda$ and the $L$-sphere are denoted $G(\Lambda)$ and $G(S_L)$, respectively. All figures are obtained from~\cite{conway:1999}.}
\label{tab:latparm}
\end{center}
\end{table}
\footnotetext{The fact that there actually exist lattices in infinite dimensions which are capable of achieving the dimensionless normalized second moment of a sphere was proven in~\cite{zamir:1996}, a proof which was contributed by G.~Poltyrev.}

While all the lattices in Table~\ref{tab:latparm} 
are the best known lattices for quantization in their dimensions it is in fact only $Z^1, A_2,\tilde{A}_3$ and $\Lambda_\infty$ which are known to be optimal among all lattices~\cite{conway:1999} and furthermore, only $Z^1$ and $\Lambda_\infty$ are known to be optimal among all entropy-constrained vector quantizers.


It is interesting to compare the optimal performance of an entropy-constrained scalar quantizer $Z^1$ to that of an optimal entropy-constrained infinite-dimensional lattice vector quantizer $\Lambda_\infty$. From Table~\ref{tab:latparm} it can be seen that the rate loss\index{rate loss!SD} $R_{\text{Loss}}$ (at high resolution or at any resolution for the uniform density) when using $Z^1$ instead of $\Lambda_\infty$ is given by
\begin{equation}\label{eq:Rloss}
R_{\text{Loss}} = \frac{1}{2}\log_2\left(\frac{G(Z^1)}{G(\Lambda_\infty)}\right) = 0.2546\ \text{bit/dim.}
\end{equation}
or equivalently the increase in distortion (also known as the space-filling loss\index{space-filling loss} or space-filling gain when reversed) for using $Z^1$ instead of $\Lambda_\infty$ is given by
\begin{equation}
D_{\text{Loss}}=10\log_{10}\left(\frac{G(Z^1)}{G(\Lambda_\infty)}\right) =1.5329 \ \text{dB}.
\end{equation}

\begin{figure}[ht]
\psfrag{Dloss}{$\scriptstyle D_{\text{Loss}}$ \scriptsize [dB]}
\psfrag{Dimension}{\scriptsize Dimension ($L$)}
\psfrag{Z1}{$\scriptstyle Z^1$}
\psfrag{A2}{$\scriptstyle A_2$}
\psfrag{A3}{$\scriptstyle \tilde{A}_3$}
\psfrag{D4}{$\scriptstyle D_4$}
\psfrag{D5}{$\scriptstyle \tilde{D}_5$}
\psfrag{E6}{$\scriptstyle E_6$}
\psfrag{E7}{$\scriptstyle \tilde{E}_7$}
\psfrag{E8}{$\scriptstyle E_8$}
\psfrag{K12}{$\scriptstyle K_{12}$}
\psfrag{BW16}{$\scriptstyle BW_{16}$}
\psfrag{L24}{$\scriptstyle \Lambda_{24}$}
\begin{center}
\includegraphics[width=10cm]{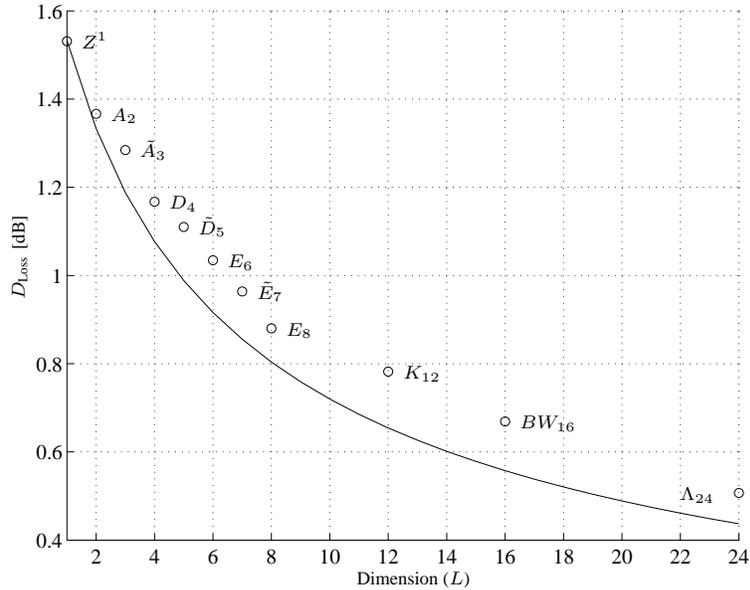}
\caption{Space-filling loss\index{space-filling loss} for the lattices of Table~\ref{tab:latparm}. The solid line describes the space-filling loss of $L$-spheres.}
\label{fig:G}
\end{center}
\end{figure}

Fig.~\ref{fig:G} illustrates the space-filling loss for the lattices of Table~\ref{tab:latparm}. For comparison we also show the space-filling loss of $L$-dimensional ``quantizers'' having spherical Voronoi cells which is given by $D_{\text{Loss}}=10\log_{10}(G(S_L)/G(\Lambda_\infty))$.
%
%

\section{Entropy Coding}
In the previous section we saw that for stationary sources one achieves space-filling gains if vector quantizers are used instead of scalar quantizers. The space-filling gain is independent of the statistical properties of the source. In other words, whether the source is i.i.d.\ or has memory the space-filling gain remains the same.
However, for this to be true, we implicitly assume that any statistical redundancy (correlation) which might be present in the quantized signal is removed by a (lossless) entropy coder\index{entropy!coding}. Recall that the discrete entropy $H(\hat{\mathcal{X}})$ of the quantizer is given by~(\ref{eq:discreteentropy}) and that $P(\hat{x}_i)$ denotes the probability of the symbol $\hat{x}_i$ where $\hat{x}_i\in \hat{\mathcal{X}}$. Assume now that a codeword (of the entropy coder) of length $l_i$ is assigned to the symbol $\hat{x}_i$. Then the average codeword length $\bar{s}$ is given by
\begin{equation}\label{eq:entropycode}
\bar{s}=\sum_{i\in \mathcal{I}}P(\hat{x}_i)l_i.
\end{equation}

The idea of an entropy coder is to assign short codewords to very probable symbols and long codewords to less probable symbols in order to drive $\bar{s}$ towards its minimum. Since we require the (entropy) code to be lossless it means that the code should be a uniquely decodable code. Due to a result of Shannon we can lower bound $\bar{s}$ by the following theorem.
\begin{theorem}~\cite{shannon:1948}
The average codeword length $\bar{s}$ of a uniquely decodable binary code satisfies
\begin{equation}
\bar{s}\geq H(\hat{\mathcal{X}}).
\end{equation}
\end{theorem}
In the same paper Shannon also gave an upper bound on $\bar{s}$, i.e.\ $\bar{s}<H(\hat{\mathcal{X}}) + 1$ and he furthermore showed that if a sequence of, say $s$, symbols is jointly coded then the average number of bits per symbol satisfy
\begin{equation}
H(\hat{\mathcal{X}}) \leq \bar{s} < H(\hat{\mathcal{X}}) + \frac{1}{s},
\end{equation}
which shows that the entropy $H(\hat{\mathcal{X}})$ can be approximated arbitrarily closely by enco\-ding sufficiently long sequences~\cite{shannon:1948}.

In~(\ref{eq:entropycode}) we have $|\mathcal{I}|=|\hat{\mathcal{X}}|$ and we thereby implicitly restrict $\hat{\mathcal{X}}$ to be a discrete alphabet\index{alphabet!discrete} be it finite or countably finite\index{alphabet!countably finite}, but we do in fact not always require that $|\hat{\mathcal{X}}|<\infty$. 
For example it is known that an entropy-constrained vector quantizer (ECVQ) may be recast in a Lagrangian sense~\cite{chou:1989,gyorgy:2002,gray:2002} and that a Lagrangian-optimal\footnote{The operational distortion-rate function\index{operational rate-distortion function} is the infimum\index{infimum} of the set of distortion-rate functions that can be obtained by use of any vector quantizer which satisfies the given entropy constraints. A Lagrangian-optimal ECVQ achieves points on the lower convex hull\index{convex hull} of the operational distortion-rate function and in general any point can be achieved by use of time-sharing~\cite{gyorgy:2002}\index{time sharing}.}
ECVQ\index{Lagrangian optimal quantizer} always exists under general conditions on the source and distortion measure~\cite{gyorgy:2002}.
Furthermore, Gy\"{o}rgy et al.~\cite{gyorgy:2003} showed that, for the squared error distortion measure, a Lagrangian-optimal ECVQ has only a finite number of codewords if the tail of the source distribution\index{distribution!Gaussian} is lighter\index{distribution!light tail} than the tail of the Gaussian distribution (of equal variance), while if the tail is heavier\index{distribution!heavy tail} than that of the Gaussian distribution the Lagran\-gian-optimal ECVQ has an infinite number of codewords~\cite{gyorgy:2003}. If the source distribution is Gaussian then the finiteness of the codebook depends upon the rate of the codebook. 
In addition they also showed that for source distributions with bounded support\index{bounded support} the Lagrangian-optimal ECVQ has a finite number of codewords.\footnote{These quantizers are not unique. For example it was shown by Gray et al.~\cite{gray:2004} that for the uniform density on the unit cube there exists Lagrangian-optimal ECVQs with codebooks of infinite cardinality.}
\footnote{In Chapter~\ref{chap:comparison} we will show that, in certain important cases, the cardinality of lattice codebooks is finite.}

In this work we will not delve into the theory of entropy coding but merely assume that there exist entropy coders which are complex enough so that (at least in theory) the discrete entropies of the quantizers can be reached. 
For more information about entropy coding we refer the reader to Chapter 9 of the text book by Gersho and Gray~\cite{gersho:1992} as well as the references cited in this section.


\chapter{Multiple-Description Rate-Distortion Theory}\label{chap:md_theory}
The MD problem is concerned with lossy encoding of information for transmission over an unreliable $K$-channel communication system. The channels may break down resulting in erasures and a potential loss of information at the receiving side.
Which of the $2^{K}-1$ non-trivial subsets of the $K$ channels that are working is assumed known at the receiving side but not at the encoder.
The problem is then to design an MD system which, for given channel rates or a given sum rate\index{sum rate}, minimizes the distortions due to reconstruction of the source using information from
any subsets of the channels. The compound channel (or composite channel) containing the $K$ subchannels is often described as a packet-switched network where individual packets are either received errorless or not at all. 
In such situations the entire system is identified as a multiple-description system having $K$ descriptions. 

\begin{figure}[ht]
\psfrag{D0}{$\scriptstyle D_0$}
\psfrag{D1}{$\scriptstyle D_1$}
\psfrag{Dc}{$\scriptstyle D_c$}
\psfrag{R0}{$\scriptstyle R_0$}
\psfrag{R1}{$\scriptstyle R_1$}
\psfrag{Description 0}{\scriptsize Description 0}
\psfrag{Description 1}{\scriptsize Description 1}
\begin{center}
\includegraphics[width=7cm]{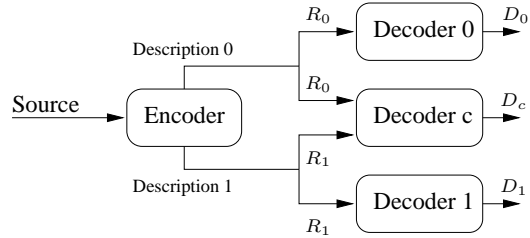}
\caption{The traditional two-channel MD system.}
\label{fig:twochannel}
\end{center}
\end{figure}

\begin{figure}[ht]
\psfrag{X}{$\scriptstyle X$}
\psfrag{Xh}{$\scriptstyle \hat{X}$}
\psfrag{R0}{$\scriptstyle R_0$}
\psfrag{R1}{$\scriptstyle R_1$}
\psfrag{RK}{$\scriptstyle R_{K-1}$}
\psfrag{Description 0}{\hspace{2.5mm}\scriptsize Description 0}
\psfrag{Description 1}{\hspace{2.5mm}\scriptsize Description 1}
\psfrag{Description K-1}{\hspace{2mm}\scriptsize Description $\scriptstyle K-1$}\begin{center}
\includegraphics[width=12cm]{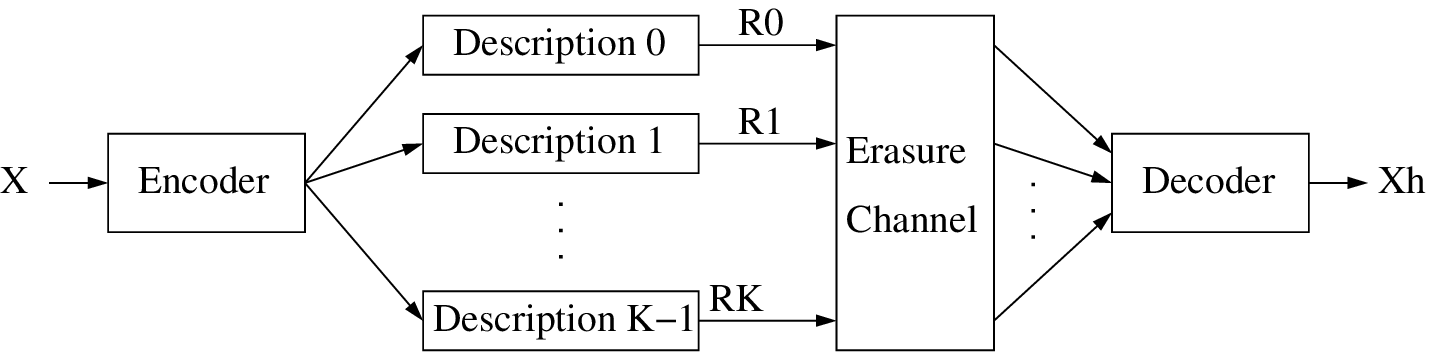}
\caption{General $K$-channel MD system. Descriptions are encoded at an entropy of $R_i$, $i=0,\dots, K-1$. The erasure channel\index{erasure channel} either transmits the $i^{th}$ description errorlessly or not at all.}
\label{fig:Kchannel}
\end{center}
\end{figure}

The classical case involves two descriptions as shown in Fig.~\ref{fig:twochannel}. The total rate $R_T$, also known as the sum rate, is split between the two descriptions, i.e.\ $R_T=R_0+R_1$, and the distortion observed at the receiver depends on which descriptions arrive. If both descriptions are received, the distortion $(D_c)$
is lower than if only a single description is received ($D_0$ or $D_1$). The general $K$-channel MD problem involves $K$ descriptions and is depicted in Fig.~\ref{fig:Kchannel}.

\section{Information Theoretic MD Bounds}
From an information theoretic perspective the MD problem is partly about describing the achievable MD rate-distortion region and partly about designing good practical codes whose performance is (in some sense) near optimum. Before presenting the known information theoretic bounds we need the following definitions.

\begin{figure}[ht]
\psfrag{R0}{\small $R_0$}
\psfrag{R1}{\small $R_1$}
\begin{center}
\includegraphics[width=6cm]{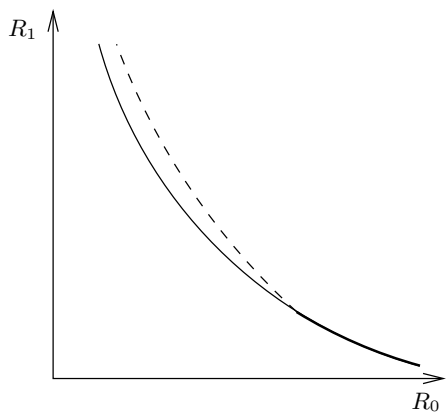}
\caption{The solid thin curve show an example of an outer bound and the dashed curve illustrates an inner bound\index{inner bound}. In the region where the bounds coincide (thick line), the bounds are tight.}
\label{fig:def_bounds}
\end{center}
\end{figure}

\begin{definition}
The \emph{MD rate-distortion region}\index{rate-distortion!MD region} given a source and a fidelity criterion is the closure\index{closure} of the set of simultaneously achievable rates and distortions. 
\end{definition}

\begin{example}
In the two-channel case the MD rate-distortion region is the closure of the set of achievable quintuples\index{quintuple} $(R_0,R_1,D_c,D_0,D_1)$.
\end{example}

\begin{definition}
An \emph{inner bound} to the MD problem is a set of achievable rate-distortion points for a specific source and fidelity criterion.
\end{definition}

\begin{definition}
An \emph{outer bound}\index{outer bound} to the MD problem is a set of rate-distortion points, for a specific source and fidelity criterion, for which it is known that no points outside this bound can be reached.
\end{definition}

\begin{definition}
If the inner and outer bounds coincide they are called \emph{tight}\index{tight bound}.
\end{definition}

\begin{example}
An example of inner and outer bounds for the set of achievable rate pairs $(R_0,R_1)$ given some fixed distortion triple $(D_c,D_0,D_1)$ is shown in Fig.~\ref{fig:def_bounds}. 
In this example there exists a region where the inner and outer bounds meet (coincide) and the bounds are said to be tight within that region.
\end{example}

\begin{remark}
The SD rate-distortion bounds\index{rate-distortion!bounds} form simple outer bounds to the MD problem. For example $R_i \geq R(D_i), i=0,\dotsc,K-1$ and $\sum_{i=0}^{K-1} R_i \geq R(D_c)$ where $R(\cdot)$ describes the SD rate-distortion function.
\end{remark}

\begin{definition}
The term \emph{no excess marginal rates}\index{marginal!no-excess rate} refers to the situation where, for fixed side distortions\index{side distortion} $D_i$,  the side description rates of an MD system meet the SD rate-distortion bounds, i.e.\ $R_i=R(D_i)$. At the other extreme we have the situation of \emph{no excess sum rate}\index{sum rate!no excess} where for a given sum rate the the central distortion $D_c$ achieves its minimum so that $\sum_{i=0}^{K-1} R_i = R(D_c)$.
\end{definition}

An interesting subset of the MD rate-distortion region is the symmetric MD rate-distortion region\index{rate-distortion!function}.\footnote{The lower bound of this symmetric region is the symmetric MD rate-distortion function of the source. With a slight abuse of notation we sometimes call the MD rate-distortion function a region.} The term symmetric relates to the situation where all channel rates (description rates) are equal and the distortion depends only upon the number of working channels (received descriptions) and as such not on which of the channels are working. This is in contrast to the asymmetric case where the description rates as well as side distortions\index{side distortion} are allowed to be unequal. 

Another important subset of the MD rate-distortion region\index{rate-distortion!region} is the high resolution region which refers to an MD rate-distortion region that becomes achievable asymp\-totical\-ly as the description rates of the system become large relative to the variance of the source (or equivalently, asymptotically as the distortions tend to zero).

\subsection{Two-Channel  Rate-Distortion Results}
El Gamal and Cover~\cite{elgamal:1982} obtained inner bounds to the two-channel MD problem (known as the EGC region) and Ozarow~\cite{ozarow:1980} showed that these inner bounds are tight for the memoryless Gaussian source under the squared-error fidelity criterion.\footnote{It is customary in the literature to refer to the case of a memoryless Gaussian source and squared-error fidelity criterion as the quadratic Gaussian case\index{quadratic!Gaussian}.}
Ahlswede~\cite{ahlswede:1985} and Zhang and Berger~\cite{zhang:1987} showed that the EGC region is also tight for general sources and distortion measures in the no excess sum rate case. 
However, in the excess sum rate\index{sum rate!excess} case it was shown by Zhang and Berger~\cite{zhang:1987} that the EGC region is not always tight for the binary memoryless source under the Hamming distortion measure\index{distortion measure!Hamming}. Outer bounds for the binary symmetric source and Hamming distortion have also been obtained by Wolf, Wyner and Ziv~\cite{wolf:1980}, Witsenhausen~\cite{witsenhausen:1980} and Zhang and Berger~\cite{zhang:1995}.
Zamir~\cite{zamir:1999,zamir:2000} obtained inner and outer bounds for smooth stationary sources and the squared-error fidelity criterion and further showed that the bounds become tight at high resolution. High resolution bounds for smooth sources and locally quadratic distortion measures\index{distortion measure!locally quadratic} have been obtained by Linder et al.~\cite{linder:1998}. 
Outer bounds for arbitrary memoryless sources and squared-error distortion measure\index{distortion measure!squared error} were obtained by Feng and Effros~\cite{feng:2005} and Lastras-Monta$\tilde{\text{n}}$o and Castelli~\cite{lastras:2006}. 

To summarize, the achievable MD rate-distortion region is only completely known for the case of two channels, squared-error fidelity criterion and the memoryless Gaussian source~\cite{ozarow:1980,elgamal:1982}. 
This region consists of the convex hull\index{convex hull} of the set of achievable quintuples $(R_0,R_1,D_0,D_1,D_c)$ where the rates satisfy~\cite{ozarow:1980,chen:2006}
\begin{align}
R_0 &\geq R(D_0) = \frac{1}2\log_2\left(\frac{\sigma_X^2}{D_0}\right) \label{eq:2chan_R0}\\
R_1 &\geq R(D_1) = \frac{1}2\log_2\left(\frac{\sigma_X^2}{D_1}\right) \label{eq:2chan_R1}\\ \label{eq:2chan_Rt}
R_0+R_1 &\geq R(D_c) + \frac{1}{2}\log_2\delta(D_0,D_1,D_c) \\
&=
\frac{1}2\log_2\left(\frac{\sigma_X^2}{D_c}\right) + \frac{1}{2}\log_2\delta(D_0,D_1,D_c),
\end{align}
where $\sigma_X^2$ denotes the source variance and $\delta(\cdot)$ is given by~\cite{chen:2006}
\begin{equation}
\delta(D_0,D_1,D_c) = 
\begin{cases}
1, & \hspace{-5cm}D_c< D_0 + D_1 - \sigma_X^2\\
\frac{\sigma_X^2D_c}{D_0D_1}, & \hspace{-5cm} D_c > \left(\frac{1}{D_0} + \frac{1}{D_1} - \frac{1}{\sigma_X^2}\right)^{-1} \\
\frac{(\sigma_X^2-D_c)^2}{(\sigma_X^2-D_c)^2 - \big(\sqrt{(\sigma_X^2-D_0)(\sigma_X^2-D_1)} - \sqrt{(D_0-D_c)(D_1-D_c)}\big)^2}, \text{o.w.},
\end{cases}
\end{equation}
and the distortions satisfy~\cite{ozarow:1980}
\begin{align} 
D_0 &\geq \sigma_X^2 2^{-2R_0} \label{eq:D02chan} \\
D_1 &\geq \sigma_X^2 2^{-2R_1} \label{eq:D12chan} \\ \label{eq:Dc2chan}
D_c &\geq \frac{\sigma_X^2 2^{-2(R_0+R_1)}}{1-\big(\sqrt{\Pi} - \sqrt{\triangle}\big)^2},
\end{align}
where $\Pi = (1-D_0/\sigma_X^2)(1-D_1/\sigma_X^2)$ and $\triangle=(D_0D_1/\sigma_X^4) - 2^{-2(R_0+R_1)}$. In general it is only possible to simultaneously achieve equality in two of the three rate inequalities given by~(\ref{eq:2chan_R0}) -- (\ref{eq:2chan_Rt}). However, in the high side distortion\index{side distortion!high} case, i.e.\ when $\delta(\cdot)=1$, it is in fact possible to have equality in all three~\cite{elgamal:1982}.

\begin{figure}[ht]
\psfrag{R1}{$\scriptstyle R_1$}
\begin{center}
\includegraphics[width=8cm]{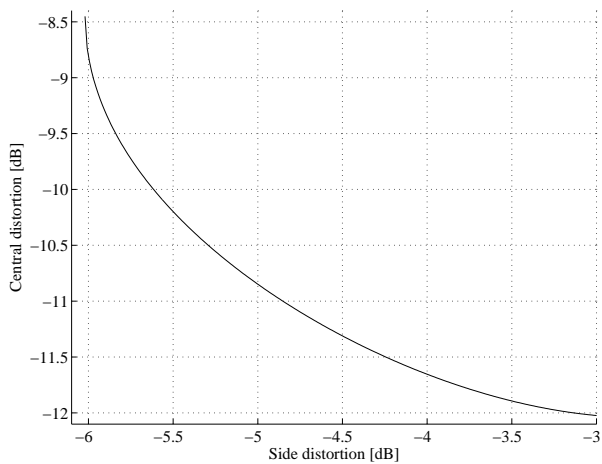}
\caption{Central distortion $(D_c)$ as a function of side distortion $(D_0=D_1)$ in a symmetric setup where $R_0=R_1=1$ bit/dim.\ for the unit-variance Gaussian source and MSE.}
\label{fig:2chan_d0dc}
\end{center}
\end{figure}

Fig.~\ref{fig:2chan_d0dc} shows the central distortion~(\ref{eq:Dc2chan}) as a function of the side distortion~(\ref{eq:D02chan}) in a symmetric setup where $D_0=D_1$ and $R_0=R_1=1$ bit/dim.\ for the unit-variance Gaussian source. 
Notice that at one extreme we have optimal side distortion, i.e.\ $D_0=D(R_0)=-6.02$ dB, which is on the single-channel rate-distortion function of a unit-variance Gaussian source at 1 bit/dim. At the other extreme we have optimal central distortion, i.e.\ $D_c=D(2R_0)=-12.04$ dB, which is on the single-channel rate-distortion function of the Gaussian source at 2 bit/dim. Thus, in this example, the single-channel rate-distortion bounds become effective for the two-channel MD problem only at two extreme points.
%
%

\begin{figure}[ht]
\psfrag{a}{$\scriptstyle \frac{1}{2}$}
\psfrag{b}{$\scriptstyle 1$}
\psfrag{c}{$\scriptstyle 1\frac{1}{2}$}
\psfrag{d}{$\scriptstyle 2$}
\psfrag{e}{$\scriptstyle 2\frac{1}{2}$}
\psfrag{R0}{\small $R_0$}
\psfrag{R1}{\small $R_1$}
\psfrag{R0(d0)}{\small $R(D_0)$}
\psfrag{R1(d1)}{\small $R(D_1)$}
\psfrag{RX}{\small Achievable $(R_0,R_1)$}
\begin{center}
\includegraphics[width=8cm]{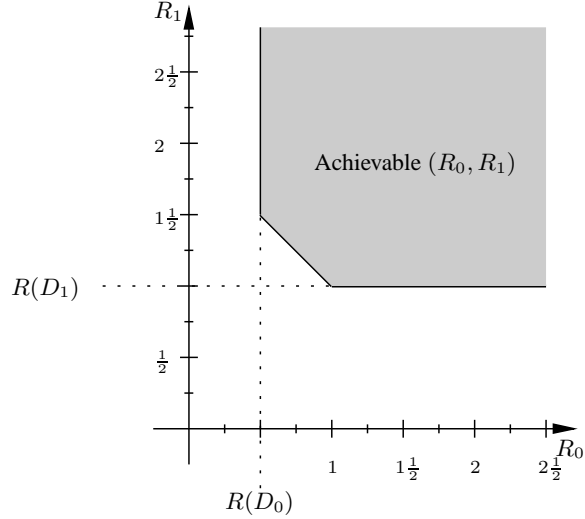}
\caption{Achievable rate pairs $(R_0,R_1)$ for the distortion triplet $(D_c,D_0,D_1)=(\frac{1}{13.9}, \frac{1}{2}, \frac{1}{4})$ in the quadratic Gaussian case.}
\label{fig:2chan_rateregion}
\end{center}\end{figure}

The rate region comprising the set of achievable rate pairs $(R_0,R_1)$ which satisfy (\ref{eq:2chan_R0}), (\ref{eq:2chan_R1}) and~(\ref{eq:2chan_Rt}) is illustrated in Fig.~\ref{fig:2chan_rateregion}. In this example we assume a unit-variance Gaussian source and choose distortions $D_0=\frac{1}{2}, D_1=\frac{1}{4}$ and $D_c=\frac{1}{13.9}$. Notice that $R_0$ and $R_1$ are lower bounded by $0.5$ and $1$ bit/dim., respectively, and the sum rate\index{sum rate} is lower bounded by $R_0+R_1\geq 2$ bit/dim. 

\subsubsection{Ozarow's Double-Branch Test Channel}
\begin{figure}[th]
\psfrag{N1}{$N_1$}
\psfrag{N0}{$N_0$}
\psfrag{X}{$X$}
\psfrag{X1}{$\hat{X}_1$}
\psfrag{X0}{$\hat{X}_0$}
\psfrag{Xc}{$\hat{X}_c$}
\psfrag{a1}{$a_1$}
\psfrag{a0}{$a_0$}
\psfrag{b1}{$b_1$}
\psfrag{b0}{$b_0$}
\begin{center}
\includegraphics[width=5cm]{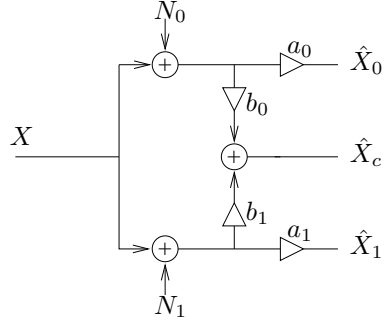}
\caption{The MD optimum test channel of Ozarow~\cite{ozarow:1980}. At high resolution the filters degenerate so in the symmetric case we have $a_i=1$ and $b_i=1/2, i=1,2$ so that $\hat{X}_0=Y_0, \hat{X}_1=Y_1$ and $\hat{X}_c=\frac{1}{2}(\hat{X}_0+\hat{X}_1)$.}
\label{fig:testchannel}
\end{center}
\end{figure}

Ozarow~\cite{ozarow:1980} showed that the double-branch test channel\index{test channel} depicted in Fig.~\ref{fig:testchannel} achieves the complete two-channel MD rate-distortion region in the quadratic Gaussian case\index{quadratic!Gaussian}.
This channel has two additive noise branches $Y_0=X+N_0$ and $Y_1=X+N_1$, where all variables are Gaussian distributed and the noise pair ($N_0,N_1$) is independent of $X$ but jointly Gaussian and \emph{negatively} correlated (except from the case of no-excess marginal rates\index{marginal!no-excess rate}, in which case the noises are independent). In the symmetric case and when the correlation between $N_0$ and $N_1$ is high, i.e.\ near $-1$, the central distortion\index{central distortion} is close to optimum but the side distortions are then generally poor. On the other hand, when the side distortions are optimal, the noise pair becomes independent and the central distortion is not much better than either of the side distortions. The post filters\index{post filter} (Wiener filters) $a_i$ and $b_i, i=0,1$ describe the scalar weights which are required for minimum MSE (MMSE)\index{minimum MSE estimation} estimation of $X$ based on either $Y_0, Y_1$ or both.
At high resolution this test channel\index{test channel} is particularly simple since the filters degenerate. Specifically, in the symmetric case, where the noise variances are equal, the side reconstructions $\hat{X}_0$ and $\hat{X}_1$ become $\hat{X}_0=Y_0$ and
$\hat{X}_1=Y_1$, while the central reconstruction $\hat{X}_c$ becomes a simple average, i.e.\ $\hat{X}_c=(\hat{X}_0+\hat{X}_1)/2$. 

\subsubsection{Rate-Redundancy Region}
\begin{figure}[ht]
\psfrag{r=0.5}{\small $R_c=0.5$}
\psfrag{r=1}{\small $R_c=1$}
\psfrag{r=1.5}{\small $R_c=1.5$}
\psfrag{r=2}{\small $R_c=2$}
\psfrag{r=2.5}{\small $R_c=2.5$}
\psfrag{D1}{\small $D_0$}
\psfrag{rho}{\small $R^*_{\text{red}}$ [bit/dim.]}
\begin{center}
\includegraphics[width=10cm]{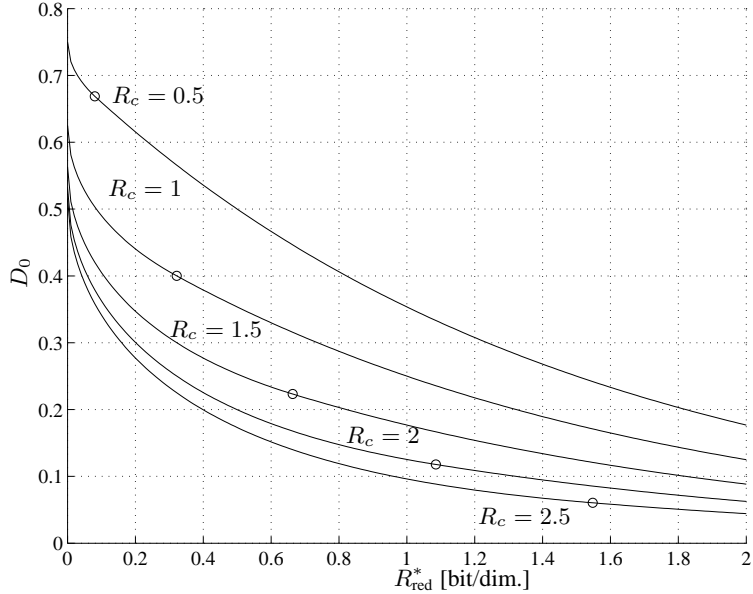}
\caption{Side distortions $D_0=D_1$ as a function of rate redundancy $R^*_{\text{red}}$. For each curve the central distortion is held fixed at $D_c=2^{-2R_c}$. The circles mark the points beyond which the second bound of~(\ref{eq:rateredundancy}) becomes effective. This example is from~\cite{goyal:2001}.}
\label{fig:rate_redundancy}
\end{center}
\end{figure}

The redundancy rate-distortion function (RRD) introduced in~\cite{orchard:1997} for the symmetric case and further developed in~\cite{goyal:2001,wang:2002} describes how fast the side distortion decays with increasing rate redundancy\index{rate redundancy!MD} $R^*_{\text{red}}$ when the central distortion $D_c$ is fixed.\footnote{The rate redundancy $R^*_{\text{Red}}$ is sometimes referred to as the excess sum rate\index{sum rate!no excess}.} Let $R_c = R(D_c)$ be the rate needed for an SD system to achieve the (central) distortion $D_c$. Then consider a symmetric setup where $D_0=D_1$ and $R_0=R_1$ and define $R^*_{\text{red}} \triangleq 2R_0 - R_c$, i.e.\ $R^*_\text{red}$ describes the additional rate needed for an MD system over that of an SD system to achieve the central distortion $D_c$. In order to reduce the side distortion $D_0$ while keeping $D_c$ fixed it is necessary to introduce redundancy such that $2R_0\geq R_c$. 
For a given $R_c$ (or equivalently a given $D_c$) and a given $R^*_{\text{red}}$ the side distortion for the unit-variance Gaussian source is lower bounded by~\cite{goyal:2001}
\begin{equation}\label{eq:rateredundancy}
D_0\geq
\begin{cases}
\frac{1}{2}(1+2^{-2R_c} - (1-2^{-2R_c})\sqrt{1-2^{-2R^*_{\text{red}}}}), & R^*_{\text{red}}\leq \bar{R}^*_{\text{red}} \\
2^{-(R_c+R^*_{\text{red}})}, & R^*_{\text{red}}> \bar{R}^*_{\text{red}}
\end{cases}
\end{equation}
where $\bar{R}^*_{\text{red}}=R_c-1+\log_2(1+2^{-2R_c})$.
If $R^*_{\text{red}}=0$ we have optimum central distortion, i.e.\ no excess sum rate\index{sum rate!no excess}, but the side distortions will then generally be high. As we increase the rate while keeping the central distortion fixed we are able to lower the side distortions. 
Fig.~\ref{fig:rate_redundancy} shows the side distortion $D_0=D_1$ as a function of the rate redundancy $R^*_{\text{red}}$ when the central distortion is fixed at $D_c=2^{-2R_c}, R_c\in\{0.5, 1, 1.5, 2, 2.5\}$. 
%
%
It is interesting to observe that when $D_c$ is optimal, i.e.\ when $R^*_\text{red}=0$, then the gap from $D_1$ to $D(R_0)$ increases with increasing $R_c$. To see this, notice that when $R^*_\text{red}=0$ it follows from the first bound of~(\ref{eq:rateredundancy}) that $D_0\geq (1+2^{-2R_c})/2$. The second bound of~(\ref{eq:rateredundancy}) is actually the SD rate-distortion bound, i.e.\ $2^{-(R_c+R^*_{\text{red}})} = 2^{-2R_0}$, and the gap between these two bounds is shown in Table~\ref{tab:redundancygaps}.
\begin{table}[ht]
\begin{center}
\begin{tabular}{ccccc} 
$R_c$ & $R_0$ & $(1+2^{-2R_c})/2$ & $D(R_0)$ & Gap \\ \hline
0.5   & 0.25  & 0.75             & 0.707    & 0.043 \\
1     & 0.5   & 0.625            & 0.5      & 0.125 \\
1.5   & 0.75  & 0.563            & 0.354    & 0.209 \\
2     & 1     & 0.531            & 0.25     & 0.281 \\
2.5   & 1.25  & 0.516            & 0.177    & 0.339 \\ \hline
\end{tabular}
\caption{The gap between the two bounds of~(\ref{eq:rateredundancy}) when $R^*_\text{red}=0$. In this case $R_0=R_c/2$.}
\label{tab:redundancygaps}
\end{center}
\end{table}

\subsubsection{Two-Channel High-Resolution Results}
Based on the results for the Gaussian source of Ozarow~\cite{ozarow:1980} it was shown by Vaisham\-payan et al.~\cite{vaishampayan:1998c,vaishampayan:1998} that at high resolution and for the symmetric case, if the side distortions satisfy 
\begin{equation}\label{eq:d0approx}
D_0=\sigma_X^2b2^{-2R_0(1-a)},
\end{equation}
for $0< a < 1$ and $b\geq 1$ then the central distortion is lower bounded by
\begin{equation}\label{eq:dcapprox}
D_c \geq \frac{\sigma_X^2}{4b}2^{-2R_0(1+a)},
\end{equation}
which leads to a simple bound on the distortion product\index{distortion product} $D_cD_0$, that is
\begin{equation}\label{eq:distortionproduct}
D_c D_0 \geq \frac{\sigma_X^4}{4}2^{-4R_0}.
\end{equation}
It was further shown that an optimal two-channel scheme achieves equality in~(\ref{eq:dcapprox}) and therefore also in~(\ref{eq:distortionproduct}) at high resolution
and when $D_c\ll D_0$. Since~(\ref{eq:distortionproduct}) is independent of $a$  it serves as a simple means of relating the performance of MD schemes to the information theoretic rate-distortion bounds of~\cite{ozarow:1980}.
It is therefore a standard figure of merit when assessing the performance of two-channel MD schemes at high resolution.
For small ratios of $D_0/D_c$ it is not possible to achieve equality in~(\ref{eq:distortionproduct}). However, at high resolution the more general but less used distortion product\index{distortion product} is also achievable~\cite{vaishampayan:1998c}
\begin{equation}\label{eq:distortionproduct1}
D_c D_0 = \frac{\sigma_X^4}{4}\frac{1}{1-D_c/D_0}2^{-4R_0},
\end{equation}
which meets the lower bound of~(\ref{eq:distortionproduct}) if $D_c/D_0\rightarrow 0$.
If $D_0$ is optimal, i.e.\ if $D_0=D(R_0)$, then it follows from~\cite{ozarow:1980} that $D_c\geq D_0/2$. Using the ratio $D_c/D_0=1/2$ in~(\ref{eq:distortionproduct1}) yields $D_cD_0=\frac{\sigma_X^4}{2} 2^{-4R_0}$ which is twice as large as the lower bound of~(\ref{eq:distortionproduct}).

\begin{figure}[ht]
\psfrag{Dc}{\small $D_c$ [dB]}
\psfrag{D0}{\small $D_0$ [dB]}
\psfrag{R0=1}{\small $R_0=1$}
\psfrag{R0=2}{\small $R_0=2$}
\psfrag{R0=3}{\small $R_0=3$}
\psfrag{R0=4}{\small $R_0=4$}
\psfrag{R0=5}{\small $R_0=5$}
\begin{center}
\includegraphics[width=10cm]{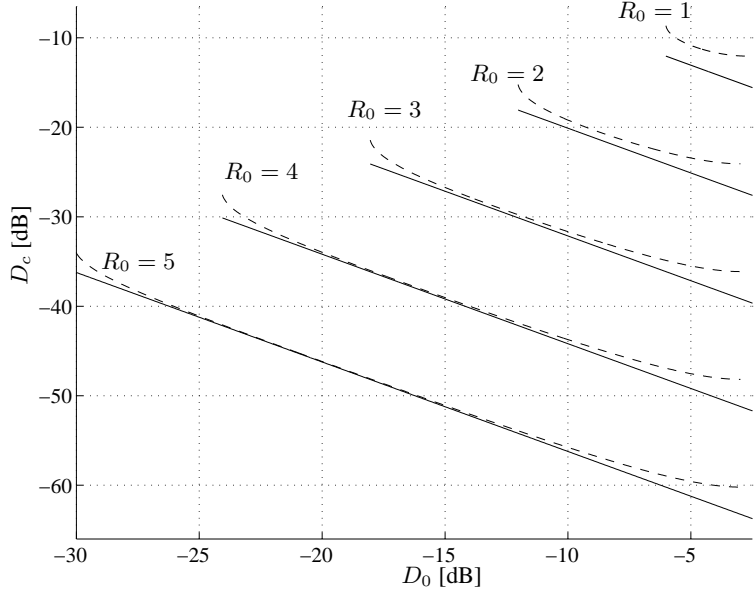}
\caption{The central distortion $D_c$ as a function of side distortions $D_0=D_1$ at different rates $R_0=R_1\in \{1,\dotsc,5\}$. The dashed lines illustrate the true distortion bounds given by~(\ref{eq:D02chan}) and~(\ref{eq:Dc2chan}) and the solid lines represent the high resolution asymptotic bounds given by~(\ref{eq:d0approx}) and~(\ref{eq:dcapprox}).}
\label{fig:d0dc}
\end{center}
\end{figure}

Fig.~\ref{fig:d0dc} compares the high resolution approximations given by~(\ref{eq:d0approx}) and~(\ref{eq:dcapprox}) (solid lines) to the true bounds given by~(\ref{eq:D02chan}) and~(\ref{eq:Dc2chan}) (dashed lines) for the case of a unit-variance memoryless Gaussian source and $b=1$. Notice that the asymptotic expressions meet the true bounds within a growing interval as the rate increases. Since $a$ is positively bounded away from zero and always less than one, the interval where they meet will never include the entire high resolution region. For example only large distortion ratios $D_0/D_c$ are achievable.\footnote{In Section~\ref{sec:comparison_twochannel} Remark~\ref{rem:abounds} we explain in more detail why the asymptotic curves never meet the true curves at the extreme points.} 
%
%

The asymmetric situation is often neglected but it is in fact fairly simple to come up with a distortion product\index{distortion product} in the spirit of~(\ref{eq:distortionproduct}). Let us first rewrite the central and side distortions in Ozarow's solution by use of the entropy power\index{entropy!power} $P_X$ as was done by Zamir~\cite{zamir:1999,zamir:2000}, that is
\begin{equation}\label{eq:2chanD0_entpower}
D_i \geq P_X 2^{-2R_i},\quad i=0,1,
\end{equation}
and 
\begin{equation}\label{eq:2chanDC_entpower}
D_c \geq \frac{P_X2^{-2(R_0+R_1)}}{1- \left(\right|\sqrt{\Pi}-\sqrt{\triangle}\left|^+\right)^{2}},
\end{equation}
where
\begin{equation}
\Pi=(1-D_0/P_X)(1-D_1/P_X)
\end{equation}
and
\begin{equation}
\triangle=D_0D_1/P_X^2 - 2^{-2(R_0+R_1)},
\end{equation}
and where\footnote{\ $|\cdot|^+$ becomes effective only in the high side distortion case, i.e.\ when $D_0+D_1 > \sigma_X^2(1+2^{-2(R_0+R_1)})$~\cite{zamir:2000}.}
\begin{equation}
|x|^{+}\triangleq \begin{cases}
x, & \text{if}\ x>0 \\
0, & \text{otherwise}.
\end{cases}
\end{equation}
An advantage of Zamir's solution is that it acts as an outer bound to the MD problem for general sources under the squared-error distortion measure\index{distortion measure!squared error}. For the memoryless Gaussian source it becomes tight at any resolution, i.e.\ it becomes identical to Ozarow's solution, and for arbitrary smooth stationary sources it becomes asymptotically tight at high resolution.
\begin{lemma}\label{lem:asymproduct}
If $2^{-2(R_0+R_1)}\ll D_0D_1\ll D_i, i=0,1$ then 
\begin{equation}\label{eq:asymlowbound}
D_c(D_0+D_1+2\sqrt{D_0D_1})\geq \left(\frac{2^{2h(X)}}{2\pi e}\right)^{2} 2^{-2(R_0+R_1)}.
\end{equation}
\end{lemma}
\begin{proof}
Let us expand the denominator in~(\ref{eq:2chanDC_entpower}) as\footnote{Here we neglect the high side distortion case.}
{\allowdisplaybreaks
\begin{align}\notag
1-\big(\sqrt{\Pi}-\sqrt{\triangle}\big)^2 &= 1 - \Big(\sqrt{(1-D_0/P_X)(1-D_1/P_X)} \\ \notag
&\quad-\sqrt{D_0D_1/P_X^2 - 2^{-2(R_0+R_1)}}\Big)^2 \\ \notag
&= 1-\Big(
(1-D_0/P_X)(1-D_1/P_X)+D_0D_1/P_X^2-2^{-2(R_0+R_1)} \\ \notag
&\quad-2\sqrt{(1-D_0/P_X)(1-D_1/P_X)(D_0D_1/P_X^2 - 2^{-2(R_0+R_1)})}\Big) \\ \notag
&=D_0/P_X+D_1/P_X-2D_0D_1/P_X^2+2^{-2(R_0+R_1)}\\ \notag
&\phantom{=}+2 \Big(D_0D_1/P_X^2 -D_0^2D_1/P_X^3-D_0D_1^2/P_X^3+(D_0D_1/P_X^2)^2
 \\ \notag
&\phantom{=}-(1-D_0/P_X-D_1/P_X+D_0D_1/P_X^2)2^{-2(R_0+R_1)}\Big)^{\frac{1}2} \\ \label{eq:asymbound}
&\approx \frac{D_0+D_1+2\sqrt{D_0D_1}}{P_X},
\end{align}}
where the approximation follows from the assumption of high resolution, i.e.\ $R_i\rightarrow \infty, i=0,1$, so that we have $2^{-2(R_0+R_1)}\ll D_0D_1\ll D_i$. The inequality $2^{-2(R_0+R_1)}$ $\ll D_0D_1$ is valid when we have excess marginal rates\index{marginal!excess rate},
i.e.\ when at least one of the side decoders is not operating on its lower bound. As such we assume that $D_i$ grows as $\mathcal{O}(2^{-2\tilde{R}_i})$ where $\tilde{R_i}\leq R_i$ and $\tilde{R}_0+\tilde{R}_1< R_0+R_1$ and it follows that the entire expression is dominated by terms that grow as $\mathcal{O}(2^{-2\tilde{R}_i})$ or $\mathcal{O}(2^{-(\tilde{R}_0+\tilde{R}_1)})$. 
Inserting~(\ref{eq:asymbound}) into~(\ref{eq:2chanDC_entpower}) leads to
\begin{equation}
\begin{split}
D_c &\geq \frac{P_X2^{-2(R_0+R_1)}}{1- \left(\sqrt{\Pi}-\sqrt{\triangle}\right)^{2}} \\
 &\approx \frac{P_X^2}{D_0+D_1+2\sqrt{D_0D_1}}2^{-2(R_0+R_1)},
\end{split}
\end{equation}
which completes the proof since $P_X=2^{2h(X)}/(2\pi e)$.
\end{proof}

\begin{remark}
It follows that an optimal asymmetric (or symmetric) MD system achieves equality in~(\ref{eq:asymlowbound}) at high resolution for arbitrary (smooth) sources. Notice that the bound~(\ref{eq:asymlowbound}) holds for arbitrary bit distributions of $R_0$ and $R_1$ as long as their sum remains constant and the inequalities~(\ref{eq:2chanD0_entpower}) and~(\ref{eq:2chanDC_entpower}) are satisfied (or more correctly that the corre\-sponding lower bounds on the individual side rates\index{description rate} and their sum rate are satisfied).
\end{remark}

\subsection{$K$-Channel  Rate-Distortion Results}
Recently, an achievable $K$-channel MD rate-distortion region was obtained by Venka\-taramani, Kramer and Goyal~\cite{venkataramani:2001,venkataramani:2003} for arbitrary memoryless sources and single-letter distortion measures\index{distortion measure!single letter}. 
This region generally takes a complicated form but in the quadratic Gaussian\index{quadratic!Gaussian} case it becomes simpler. 
The region presented in~\cite{venkataramani:2003} describes an asymmetric MD rate-distortion region and includes as a special case the symmetric MD rate-distortion region. The construction of this region relies upon forming layers of conditional random codebooks. 
It was, however, observed by Pradhan, Puri and Ramchandran in a series of papers~\cite{pradhan:2001,pradhan:2001b,puri:2002,puri:2002a,pradhan:2004,puri:2005} that by exploiting recent results on dis\-tributed source coding it is possible to replace the conditional codebooks with universal codebooks whereby the codebook rate can be reduced through random bin\-ning.
While Pradhan et al.\ limited their interests to the symmetric case it can be shown that their results carry over to the asymmetric case as well. This has recently been done by Wang and Viswanath~\cite{wang:2006} who further extended the results to the case of vector Gaussian sources and covariance distortion measure constraints\index{distortion measure!covariance constraints}. 

The largest known achievable rate-distortion region for the $K$-channel MD problem is that of Pradhan et al.~\cite{pradhan:2004,puri:2005}.\footnote{Outer bounds for the $K$-channel quadratic Gaussian problem were presented in~\cite{venkataramani:2003}.}
Common for all the achievable rate-distortion regions is that they represent inner bounds and it is currently not known whether they can be further improved. However, for the quadratic Gaussian case it was con\-jectured in~\cite{puri:2005} that their bound is in fact tight. That conjecture remains open.

The key ideas behind the achievable region obtained by Pradhan et al.\ are well explained in~\cite{pradhan:2004,puri:2005} and we will here repeat some of their insights and results before presenting the largest known $K$-channel achievable rate-distortion region.

Consider a packet-erasure channel\index{erasure channel} with parameters $K$ and $k$, i.e.\ at least $k$ out of $K$ descriptions are received. For the moment being, we assume $k=1$.
Generate $K$ independent random codebooks, say $\mathcal{C}_0,\dots,\mathcal{C}_{K-1}$ each of rate $R$.
The source is now separately and independently quantized using each of the codebooks. The index of the nearest codeword in the $i^{th}$ codebook is transmitted on
the $i^{th}$ channel. A code constructed in this way was dubbed a source-channel erasure code\index{source-channel erasure code} in~\cite{pradhan:2004} which we, for notational convenience, abridge to $(K,k)$ SCEC\@. Notice that since each of the individual codebooks are optimal for the source then if only a single index is received the source is reconstructed with a distortion that is on the distortion-rate function $D(R)$ of the source. However, if more than one index is received, the quality of the reconstructed signal can be strictly improved due to multiple versions of the quantized source.
The above scheme is generalized to $(K,k)$ SCEC for $k> 1$ by making use of random binning.
This is possible since the quantized variables are assumed (symmetrically) correlated\index{distribution!symmetric} so that general results of distributed source coding are applicable. Specifically, due to celebrated results of Slepian and Wolf~\cite{slepian:1973}\index{Slepian-Wolf coding} and Wyner and Ziv~\cite{wyner:1976}\index{Wyner-Ziv coding}, if it is assumed that some $k$ out of the set of $K$ correlated variables are received then (by e.g.\ use of random binning) it is possible to encode at a rate close to the joint entropy of any $k$ variables, in a distributed fashion, so that the encoder does not need to know which $k$ variables that are received. It is usually then not possible to decode on reception of fewer than $k$ variables.

Before presenting the main theorem of~\cite{pradhan:2004} which describes the achievable rate-distortion regions of $(K,k)$ SCECs in the general case of $1\leq k\leq K$, we need some definitions.
\begin{definition}
$D^{(K,k)}$ denotes the distortion when receiving $k$ out of $K$ descriptions. 
\end{definition}

\begin{definition}
A tuple $(R,D^{(K,k)},D^{(K,k+1)},\dots,D^{(K,K)})$ is said to be achievable if for arbitrary $\delta>0$, there exists, for sufficiently large block length $L$, a $(K,k)$ SCEC with parameters $(L,\Theta,\Delta_k,\Delta_{k+1},\dots,\Delta_K)$ with
\begin{equation}
\Theta\leq 2^{L(R+\delta)}\ \text{and}\ \Delta_h \leq D^{(K,h)}+ \delta,\ h=k,k+1,\dots,K.
\end{equation}
\end{definition}
Let $\mathcal{I}_k=\{I:I\subseteq \{0,\dotsc,K-1\}, |I|\geq k\}$. A $(K,k)$ SCEC with parameters $(L,\Theta,\Delta_k,\Delta_{k+1},\dots,\Delta_K)$ is defined by a set of $K$ encoding functions~\cite{pradhan:2004}
\begin{equation}
F_i\colon \mathcal{X}\to\{1,2,\dots,\Theta\},\qquad i=0,\dotsc,K-1,
\end{equation}
and a set of $|\mathcal{I}_k|$ decoding functions
\begin{equation}
G_I\colon \bigotimes_I \{1,2,\dots,\Theta\}\to \hat{\mathcal{X}},\qquad \forall I\in \mathcal{I}_k
\end{equation}
where $\bigotimes$ denotes the Cartesian product\index{Cartesian product} and for all $h \in \{k,k+1,\dots,K\}$ and $X\in \mathcal{X}$ we have
\begin{equation}
\begin{split}
\Delta_h&=E\rho(X,G_I(F_{i_1}(X),\dots,F_{i_h}(X))) \\
I&= \{i_1,\dots,i_h\}, \quad \forall I\in \mathcal{I}_k, |I|=h.
\end{split}
\end{equation}

\begin{theorem}[\cite{pradhan:2004}, Th. 1]\label{theo:scec_partI}
For a probability distribution\footnote{To avoid clutter we omit the subscripts on the probability distributions in this section.}\index{distribution!symmetric}
\begin{equation}
p(x,y_0,\dots,y_{K-1})=q(x)p(y_0,\dots,y_{K-1}|x)
\end{equation}
defined over $\mathcal{X}\bigoplus\mathcal{Y}^K$ where $\mathcal{Y}$ is some finite alphabet\index{alphabet}, $p(y_0,\dots,y_{K-1}|x)$ is symmetric, and a set of decoding functions $\forall I\in \mathcal{I}_k, g_I:\mathcal{Y}^{|I|}\rightarrow \hat{\mathcal{X}}$, if
\begin{equation}
E\rho(X,g_I(Y_I))\leq D^{(K,|I|)}, \qquad \forall I\in \mathcal{I}_k
\end{equation}
and
\begin{equation}
R>\frac{1}{k}H(Y_0,\dots,Y_{k-1})-\frac{1}{K}H(Y_0,\dots,Y_{K-1}|X)
\end{equation}
then $(R,D^{(K,k)},D^{(K,k+1)},\dots,D^{(K,K)})$ is an achievable rate-distortion tuple.
\end{theorem}

\begin{figure}[ht]
\begin{center}
\psfrag{R0}{$\scriptstyle R^{(0)}$}
\psfrag{R1}{$\scriptstyle R^{(1)}$}
\psfrag{R2}{$\scriptstyle R^{(2)}$}
\psfrag{L00}{$\scriptstyle L_{00}$}
\psfrag{L01}{$\scriptstyle L_{01}$}
\psfrag{L02}{$\scriptstyle L_{02}$}
\psfrag{L10}{$\scriptstyle L_{10}$}
\psfrag{L11}{$\scriptstyle L_{11}$}
\psfrag{L12}{$\scriptstyle L_{12}$}
\psfrag{L20}{$\scriptstyle L_{20}$}
\psfrag{L21}{$\scriptstyle L_{21}$}
\psfrag{L22}{$\scriptstyle L_{22}$}
\psfrag{Description 0}{\small Description 0}
\psfrag{Description 1}{\small Description 1}
\psfrag{Description 2}{\small Description 2}
\psfrag{(3,1) SCEC}{\small $(3,1)$ SCEC}
\psfrag{(3,2) SCEC}{\small $(3,2)$ SCEC}
\psfrag{(3,3) SCEC}{\small $(3,3)$ SCEC}
\psfrag{b}{$\underbrace{\hspace{15.5mm}}$}
\includegraphics[width=8cm]{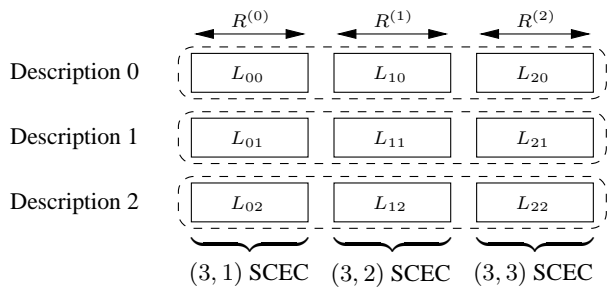}
\caption{Concatenation of (3,1), (3,2) and (3,3) SCECs to achieve the distortion triplet $(D^{(3,1)},D^{(3,2)},D^{(3,3)})$. Each description contains three layers and the rate of each description is $R=R^{(0)}+R^{(1)}+R^{(2)}$.}
\label{fig:scec_partII}
\end{center}
\end{figure}

In~\cite{puri:2005} an achievable rate-distortion region for the $K$-channel MD problem was presented. The region was obtained by constructing a number of layers within each description where the set of all $j^{th}$ layers across the $K$ descriptions corresponds to a $(K,j)$ SCEC.
Assume that it is desired to achieve some distortion triplet $(D^{(3,1)},$ $D^{(3,2)},D^{(3,3)})$ for a three-channel system.
Then first a (3,1) SCEC is constructed using a rate of $R^{(0)}$ bit/dim.\ per description. We use the superscript to distinguish between the rate $R_0$ of encoder 0 in an asymmetric setup and the rate $R^{(0)}$ of layer 0 in a symmetric setup. 
The rate $R^{(0)}$ is chosen such that $D^{(3,1)}$ can be achieved with the reception of any single description. 
If two descriptions are received the distortion is further decreased. However if $D^{(3,2)}$ is not achieved on the reception of any two descriptions, then a (3,2) SCEC is constructed at a rate $R^{(1)}$ bit/dim.\ per description. The rate $R^{(1)}$ is chosen such that $D^{(3,2)}$ can be achieved on the reception of any two descriptions. If $D^{(3,3)}$ is not achieved on the reception of all three descriptions, a refinement layer is constructed at a rate of $R^{(2)}$ bit/dim.\ per description, see Fig.~\ref{fig:scec_partII}. Each description contains three layers, e.g.\ description $i$ consists of a concatenation of $L_{0i}, L_{1i}$ and $L_{2i}$.
It is important to see that the first layer, i.e.\ the (3,1) SCEC is constructed exactly as described by Theorem~\ref{theo:scec_partI}. 
The second layer, i.e.\ the (3,2) SCEC differs from the construction in that of the binning rate. Since, on reception of any two descriptions (say description 0 and 1), we have not only the two second layers ($L_{10}$ and $L_{11}$) but also two base layers ($L_{00}$ and $L_{01}$). This makes it possible to decrease the binning rate $R^{(1)}$ by exploiting correlation across descriptions as well as across layers. The final layer can be a simple refinement layer where bits are evenly split among the three descriptions or for example a (3,3) SCEC\@. The rate of each description is then given by $R=R^{(0)}+R^{(1)}+R^{(2)}$ and the total rate is $3R$.

We are now in a position to introduce the main theorem of~\cite{puri:2005} which describes an achievable rate-distortion region for the concatenations of $(K,k)$ SCECs. 
Let $Y_{ij}$ be a random variable in the $i^{th}$ layer and $j^{th}$ description and let $I_0^{K-1}=\{0,\dotsc,K-1\}$. 
For $i\in I_0^{K-2}$, let $Y_{iI_0^{K-1}} = (Y_{i0},Y_{i1},\dots,Y_{iK-1})$ represent $K$ random variables in the $i^{th}$ layer taking values in alphabet $\mathcal{Y}_i$. Let $Y_{K-1}$ be the last layer refinement variable taking values in the alphabet $\mathcal{Y}_{K-1}$ and
\begin{equation}
Y_{I_0^{K-2}I_0^{K-1}}=(Y_{0I_0^{K-1}},Y_{1I_0^{K-1}},\dots,Y_{(K-2)I_0^{K-1}}).
\end{equation}
A joint distribution $p(y_{I_0^{K-2}I_0^{K-1}},y_{K-1}|x)$ is called symmetric if for all $1\leq r_i\leq K$ where $i\in I_0^{K-2}$, the following is true: the joint distribution of $Y_{K-1}$ and all $(r_0+r_1+\cdots+r_{K-2})$ random variables where any
$r_i$ are chosen from the $i^{th}$ layer, conditioned on $x$, is the same.
\begin{theorem}[\cite{puri:2005}, Th. 2]\label{theo:scec_partII}
For any probability distribution
\begin{equation}
p(x,y_{I_0^{K-2}I_0^{K-1}},y_{K-1})=p(x)p(y_{I_0^{K-2}I_0^{K-1}},y_{K-1}|x)
\end{equation}
where $p(y_{I_0^{K-2}I_0^{K-1}},y_{K-1}|x)$ is symmetric, defined over $\mathcal{X}\times\mathcal{Y}_0^K\times\mathcal{Y}_1^K\times\cdots\times \mathcal{Y}_{K-2}^K\times\mathcal{Y}_{K-1}$, and a set of decoding functions given by\footnote{\ $\mathcal{Y}_i^j$ denotes the $j$ times Cartesian product\index{Cartesian product} of the alphabet $\mathcal{Y}_i$.}
\begin{equation}
\begin{split}
g_{I}&: \mathcal{Y}_0^{|I|}\times\cdots\times\mathcal{Y}_{|I|-1}^{|I|} \rightarrow \hat{\mathcal{X}}\quad \forall I\subset I_0^{K-1} \\
g_{I_0^{K-1}}&: \mathcal{Y}_0^K\times\mathcal{Y}_1^K\times\cdots\times\mathcal{Y}_{K-2}^K\times\mathcal{Y}_{K-1}\rightarrow \hat{\mathcal{X}}
\end{split}
\end{equation}
the convex closure\index{convex hull} of $(R,D^{(K,1)},D^{(K,2)},\dots,D^{(K,K)})$ is achievable where
\begin{equation}
E\rho_I(X,g_I(Y_{I_0^{|I|-1}I}))\leq D^{(K,|I|)}\quad \forall I\subset I_0^{K-1},
\end{equation}
\begin{equation}
E\rho_{I_{0}^{K-1}}(X,g_{I_0^{K-1}}(Y_{I_0^{K-2}I_{0}^{K-1}},Y_{K-1}))\leq D^{(K,K)},
\end{equation}
and
\begin{equation}
\begin{split}
R &\geq H(Y_{00})+\sum_{k=2}^{K-1}\frac{1}{k}H(Y_{{k-1}I_0^{k-1}}|Y_{I_0^{k-2}I_0^{k-1}})\\
&\quad +\frac{1}{K}H(Y_{K-1}|Y_{I_0^{K-2}I_0^{K-1}})-\frac{1}{K}H(Y_{I_0^{K-2}I_0^{K-1}},Y_{K-1}|X).
\end{split}
\end{equation}
\end{theorem}

The main difference between Theorem~\ref{theo:scec_partI} and Theorem~\ref{theo:scec_partII} is that the latter theorem considers the complete $K$-tuple of distortions $(D^{(K,1)},D^{(K,2)},\dots,D^{(K,K)})$ whereas the former theorem considers the $(K-k+1)$-tuple of distortions $(D^{(K,k)},$ $D^{(K,k+1)},\dots,D^{(K,K)})$. Hence, an SCEC based on the construction presented in~\cite{pradhan:2004} is specifically tailored to networks where it is known that at least $k$ channels out $K$ channels are always working.
With the construction presented in~\cite{puri:2005} it is possible to concatenate several SCECs and obtain a code that works for networks where the number of working channels is not known a priori.

\subsection{Quadratic Gaussian $K$-Channel Rate-Distortion Region}
We will now describe the achievable $K$-channel rate-distortion region for the quadratic Gaussian\index{quadratic!Gaussian} case. This appears to be the only case where explicit (and relatively simple) closed-form expressions for rate and distortion have been found.

Consider a unit-variance Gaussian source $X$ and define the random variables $Y_i, i=0,\dotsc,K-1,$ given by
\begin{equation}
Y_i = X + Q_i,
\end{equation}
where the $Q_i$'s are identically distributed jointly Gaussian random variables (independent of $X$) with variance $\sigma_q^2$ and covariance matrix $Q$ given by
\begin{equation}\label{eq:covQ}
Q = \sigma_q^2
\begin{bmatrix}
1 & \rho_q &\rho_q & \cdots & \rho_q \\
\rho_q & 1 & \rho_q & \cdots & \rho_q \\
\rho_q & \rho_q & 1 & \cdots & \rho_q \\
\vdots & \vdots & \vdots & \ddots& \vdots \\
\rho_q & \rho_q & \rho_q & \cdots & 1
\end{bmatrix},
\end{equation}
where, for $K>1$, it is required that the correlation coefficient\index{correlation coefficient} satisfies $-1/(K-1) < \rho_q \leq 1$ to ensure that $Q$ is positive semidefinite~\cite{venkataramani:2003}. In the case of Ozarow's double branch test channel\index{test channel} for $K=2$ descriptions, we only need to consider non positive $\rho_q$'s. This is, in fact, also the case for $K>2$ descriptions~\cite{venkataramani:2003}.

It is easy to show that the MMSE\index{minimum MSE estimation} when estimating $X$ from any set of $m$ $Y_i$'s is given by~\cite{venkataramani:2003,pradhan:2004}
\begin{equation}\label{eq:MMSE}
D^{(K,m)}=\frac{\sigma_q^2( 1+(m-1)\rho_q)}{m+\sigma_q^2(1+(m-1)\rho_q)}.
\end{equation}
We now focus on the $(K,k)$ SCEC as presented in Theorem~\ref{theo:scec_partI}.  The rate of each description is given by~\cite{pradhan:2004}
\begin{equation}\label{eq:Rgaussian}
R = \frac{1}{2}\log_2\left(\frac{k + \sigma_q^2(1+(k-1)\rho_q)}{\sigma_q^2(1-\rho_q)}\right)^{1/k}\left(\frac{1-\rho_q}{1+(K-1)\rho_q}\right)^{1/K}.
\end{equation}
The quantization error variance $\sigma_q^2$ can now be obtained from~(\ref{eq:Rgaussian})
\begin{equation}\label{eq:sigmanorm}
\sigma_q^2 = k\left( (1-\rho_q)2^{2kR}\left(\frac{1+(K-1)\rho_q}{1-\rho_q}\right)^{k/K}-(1+(k-1)\rho_q)\right)^{-1}.
\end{equation}

We will follow~\cite{pradhan:2004} and look at the performance of a $(K,k)$ SCEC in three different situations distinguished by the amount of correlation $\rho_q$ introduced in the quantization noise.
\subsubsection{Independent quantization noise: $\rho_q=0$}
The quantization noise is i.i.d., i.e.\ $\rho_q=0$, hence $Q$ is diagonal. 
Assuming that the quantization noise is normalized such that $\sigma^2_q=k/(2^{2kR}-1)$, we get the following expressions for the distortion
\begin{equation}
D^{(K,k+r)}=\frac{\sigma_q^2}{\sigma_q^2+(k+r)} = \frac{k}{2^{2kR}(k+r)-r}\quad \text{for}\ 0\leq r\leq K-k,
\end{equation}
and
\begin{equation}
D^{(K,m)} = 1\quad \text{for}\ 0\leq m<k.
\end{equation}
The distortion when receiving $k$ descriptions is optimal, i.e.\ $D^{(K,k)}=2^{-2kR}$.

\subsubsection{Correlated quantization noise: $\rho_q=\rho_q^*$}
The amount of correlation $\rho_q^*$ needed in order to be on the distortion-rate
function on the reception of $k=K$ descriptions is given by
\begin{equation}
\rho_q^* = -\frac{2^{2KR}-1}{(K-1)2^{2KR}+1} \approx -\frac{1}{K-1}.
\end{equation}
This leads to the following performance
\begin{equation}
D^{(K,r)} = 1-\frac{r}{K}(1-2^{-2KR})\quad \text{for}\ 0\leq r\leq K.
\end{equation}
Notice that the distortion when receiving $K$ descriptions is optimal, i.e.\ $D^{(K,K)}=2^{-2KR}$.

\subsubsection{Correlated quantization noise: $\rho_q^* < \rho_q < 0$}
Here a varying degree of correlation is introduced and the performance is given by
\begin{equation}\label{eq:scec_kdist}
D^{(K,r)} = \frac{\sigma_q^2(1+(r-1)\rho_q)}{\sigma_q^2(1+(r-1)\rho_q)+r}\quad \text{for}\ k\leq r\leq K,
\end{equation}
and
\begin{equation}
D^{(K,m)} = 1\quad \text{for}\ 0\leq m < k.
\end{equation}

\begin{figure}[ht]
\begin{center}
\includegraphics[width=10cm]{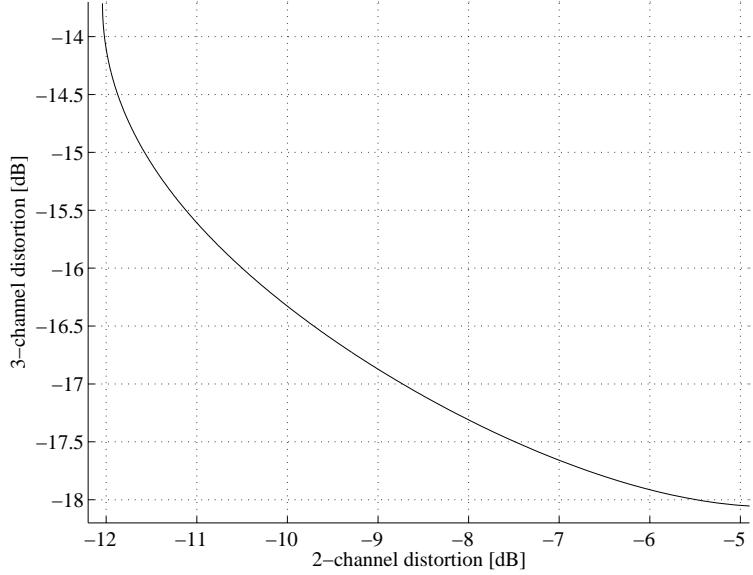}
\caption{Three-channel versus two-channel distortions for a (3,2) SCEC at $R=1$ bit/dim.\ for the unit-variance Gaussian source. This example is from~\cite{pradhan:2004}.}
\label{fig:scec_32}
\end{center}
\end{figure}

Fig.~\ref{fig:scec_32} shows the three-channel distortion $D^{(3,3)}$ as a function of the two-channel distortion $D^{(3,2)}$ when varying $\rho_q$ and keeping the rate constant by use of~(\ref{eq:sigmanorm}). In this example we use a $(3,2)$ SCEC with $R=1$ bit/dim\@. At one end we have $D^{(3,2)}=-12.0412$ dB which is on the distortion-rate function of the source and at the other end we have $D^{(3,3)}=-18.0618$ dB which is also on the distortion-rate function.
%
%

\begin{figure}[ht]
\psfrag{D1}{$\scriptscriptstyle D^{(3,1)}$}
\psfrag{D2}{$\scriptscriptstyle D^{(3,2)}$}
\psfrag{D3-hel}{$\scriptscriptstyle D^{(3,3)}$}
\psfrag{rho}{\small $\rho_q$}
\begin{center}
\includegraphics[width=10cm]{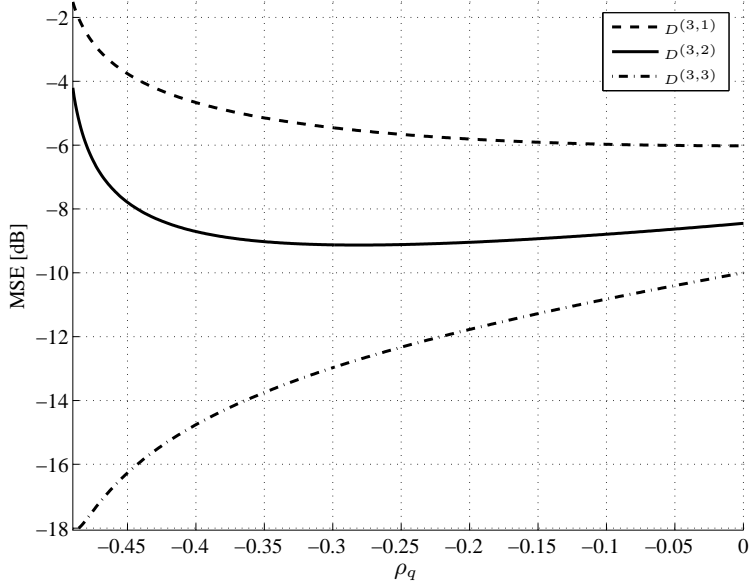}
\caption{The simultaneously achievable one-channel, two-channel and three-channel distortions for the unit-variance Gaussian source at 1 bit/dim.\ for a (3,1) SCEC.}
\label{fig:D1D2D3_R1}
\end{center}
\end{figure}

In Fig.~\ref{fig:D1D2D3_R1} we show the simultaneously achievable one-channel $D^{(3,1)}$, two-channel $D^{(3,2)}$ and three-channel $D^{(3,3)}$ distortions for the unit-variance memoryless Gaussian source at 1 bit/dim.\ when using a $(3,1)$ SCEC. It is interesting to observe that while it is possible to achieve optimum one-channel distortion ($\rho_q=0 \Rightarrow D^{(3,1)}\approx -6$ dB) and optimum three-channel distortion ($\rho_q\rightarrow -1/2 \Rightarrow D^{(3,3)}\approx -18$ dB) it is not possible to drive the two-channel distortion towards its optimum ($\forall\rho_q, D^{(3,2)}< -12$ dB). 
In other words, a $(3,1)$ SCEC can achieve optimal one-channel and three-channel performance and a $(3,2)$ SCEC can achieve optimal two-channel and three-channel performance. 
%
%

\begin{figure}[ht]
\psfrag{D1}{\small $D^{(3,1)}$ [dB]}
\psfrag{D2}{\small $D^{(3,2)}$ [dB]}
\psfrag{D3}{\small $D^{(3,3)}$ [dB]}
\begin{center}
\includegraphics[width=10cm]{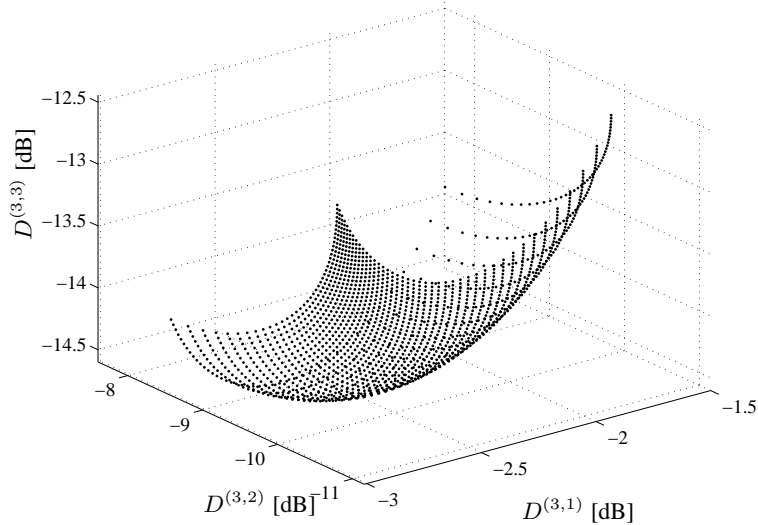}
\caption{Achievable distortion region for $R^{(0)}=R^{(1)}=0.5$ bit/dim.\ per description. The description rate is $R=R^{(0)}+R^{(1)}=1$ bit/dim\@. The dense peak is in the front.}
\label{fig:scec_321region}
\end{center}
\end{figure}

Let us now look at the achievable three-channel region presented in~\cite{puri:2005} for the the memoryless Gaussian source. 
Let $R$ bit/dim.\ per description be the rate of transmission. Let the random variables in the three layers be defined as
\begin{equation}
Y_{0j}=X+Q_{0j}, Y_{1j}=X+Q_{1j},\ \text{and}\ Y_2=X+Q_2\quad \text{for}\ j\in I_0^{2},
\end{equation}
where for $i\in I_0^{1}, Q_{iI_0^{2}}$ are symmetrically distributed\index{distribution!symmetric} Gaussian random variables with variance $\sigma_{q_i}^2$ and correlation coefficient\index{correlation coefficient}
$\rho_{q_i}$ and $Q_2$ is a Gaussian random variable with variance $\sigma^2_{q_2}$. $Q_{0I_0^{2}}, Q_{1I_0^2}$ and $Q_2$ are independent of each other
and $X$. 
By changing the four independent variables $R^{(0)},R^{(1)}, \rho_{q_0}$ and $\rho_{q_1}$ different trade-offs between $D^{(3,1)},D^{(3,2)}$ and $D^{(3,3)}$ can be made. The correlation coefficients are lower bounded by~\cite{puri:2005}
\begin{equation}\label{eq:rhoi}
\rho_{q_i}\geq -\frac{2^{6R^{(i)}}-1}{2\cdot 2^{6R^{(i)}}+1}.
\end{equation}
The variance $\sigma_{q_0}^2$ of the base layer follows from~(\ref{eq:sigmanorm}) by letting $k=1$, that is
\begin{equation}\label{eq:varinvq1}
\sigma_{q_0}^{-2}=2^{2R^{(0)}}(1+2\rho_{q_0})^{\frac{1}3}(1-\rho_{q_0})^{\frac{2}{3}}-1
\end{equation}
and it can be shown that~\cite{puri:2005}
\begin{equation}\label{eq:varinvq2}
\sigma_{q_1}^{-2}=-\frac{1+\rho_{q_1}}{(1+\rho_{q_0})\sigma_{q_0}^2}-\frac{1+\rho_{q_1}}{2}+\frac{2^{4R^{(1)}}(1+\rho_{q_0}+2/\sigma_{q_0}^2)(1+\rho_{q_1}-2\rho_{q_1}^2)^{\frac{2}3}}{2(1+\rho_{q_0})(1-\rho_{q_1})^{\frac{1}3}}.
\end{equation}
Let MSE$_2$ denote the distortion given by any two base-layer random variables,
and MSE$_3$ denote the distortion given by all the random variables in the base
and the second layer. Hence, MSE$_2$ denotes the MMSE obtained in estimating the source $X$ using either ($Y_{00},Y_{01}$), ($Y_{00},Y_{02}$) or ($Y_{01},Y_{02}$). Similarly MSE$_3$ denotes the MMSE in estimating $X$ from ($Y_{00},Y_{01},Y_{02},Y_{10},Y_{11}, Y_{12}$). From~(\ref{eq:MMSE}) it follows that
\begin{equation}
\text{MSE}_2=\frac{\sigma_{q_0}^2(1+\rho_{q_0})}{\sigma_{q_0}^2(1+\rho_{q_0})+2},
\end{equation}
and it can also be shown that~\cite{puri:2005}
\begin{equation}\label{eq:scecmse3}
\begin{split}
\text{MSE}_3&=\frac{\sigma_{q_0}^2\sigma_{q_1}^2(1+2\rho_{q_0}+2\rho_{q_1}+4\rho_{q_0}\rho_{q_1})}{3\sigma_{q_0}^2(1+2\rho_{q_0})+3\sigma_{q_1}^2(1+2\rho_{q_1})
+\sigma_{q_0}^2\sigma_{q_1}^2(1+2\rho_{q_0}+2\rho_{q_1}+4\rho_{q_0}\rho_{q_1})},
\end{split}
\end{equation}
and
\begin{equation}\label{eq:scecsigma3}
\sigma_{q_2}^2=\frac{\text{MSE}_3}{2^{6(R-R^{(0)}-R^{(1)})}-1}.
\end{equation}
Finally, we have~\cite{puri:2005}
\begin{align}
D'^{(3,1)}&=\frac{\sigma_{q_0}^2}{1+\sigma_{q_0}^2}, \label{eq:scecD1}\\
D'^{(3,2)}&=\frac{\sigma_{q_1}^2(1+\rho_{q_1})\text{MSE}_2}{\sigma_{q_1}^2(1+\rho_{q_1})+2\text{MSE}_2}, \label{eq:scecD2} \\ \label{eq:scecD3}
D'^{(3,3)}&=\frac{\sigma_{q_2}^2\text{MSE}_3}{\text{MSE}_3+\sigma_{q_2}^2}.
\end{align}
The lower convex hull\index{convex hull} of $(D'^{(3,1)},D'^{(3,2)},D'^{(3,3)})$ corresponds to an achievable distor\-tion tuple $(D^{(3,1)},D^{(3,2)},D^{(3,3)})$. See Fig.~\ref{fig:scec_321region} for an example of an achievable distortion region for the unit-variance memoryless Gaussian source
for $R^{(0)}=R^{(1)}=0.5$ bit/dim.\ per description and a description rate of $R=R^{(0)}+R^{(1)}=1$ bit/dim\@. In this plot the correlation values are varied throughout the range given by~(\ref{eq:rhoi}).
%
%

\section{Multiple-Description Quantization}
The previous section described known information theoretic bounds. These bounds were shown to be achievable by use of random codebooks. Unfortunately random codebooks are usually not very practical due to e.g.\ high search complexity and large memory requirements. From a practical point of view it is therefore desirable to avoid random codebooks, which is the case for the MD schemes we present in this section.

Existing MD schemes can roughly be divided into three categories: quantizer-based, transform-based and source-channel erasure codes\index{source-channel erasure code} based.
Quantizer-based schemes include \index{scalar!quantization}scalar quantization~\cite{jayant:1981,vaishampayan:1993,vaishampayan:1994,battlo:1997,berger-wolf:2002,tian:2004b,tian:2004,dayan:2002}, trellis coded quantization\index{trellis coded quantization} \cite{vaishampayan:1998c,jafarkhani:1999,wang:2000} 
and vector quantization~\cite{fleming:1999,fleming:2004,gortz:2003,cardinal:2004,servetto:1999,diggavi:2000,vaishampayan:2001,diggavi:2002,tian:2004c,ostergaard:2004,ostergaard:2005c,ostergaard:2005d, ostergaard:2005e,ostergaard:2004b,ostergaard:2007,chen:2005,chen:2006,goyal:2002,kelner:2000,yahampath:2004,koulgi:2003}.
Transform-based approaches include correlating transforms~\cite{orchard:1997,wang:1997,goyal:1998,goyal:2001,goyal:2001b}, overcomplete expansions\index{overcomplete expansion} and filterbanks~\cite{goyal:1998b,goyal:1999,balan:2000,chou:1999,kovacevic:2002,dragotti:2001}. Schemes based on source-channel erasure codes were presented in~\cite{pradhan:2001,pradhan:2001b,puri:2002,puri:2002a,pradhan:2004,puri:2005}.
For further details on many existing MD techniques we refer the reader to the excellent survey article by Goyal~\cite{goyal:2001a}.

The work in this thesis is based on lattice vector quantization and belongs therefore to the first of the catagories mentioned above to which we will also restrict attention.

\subsection{Scalar Two-Channel Quantization with Index Assignments}
\begin{figure}
\begin{center}
\psfrag{Q0:}{\small$Q_0$:}
\psfrag{Q1:}{\small$Q_1$:}
\psfrag{Qc:}{\small$Q_c$:}
\mbox{%
\subfigure[Regular side quantizers]{\includegraphics[width=5.7cm]{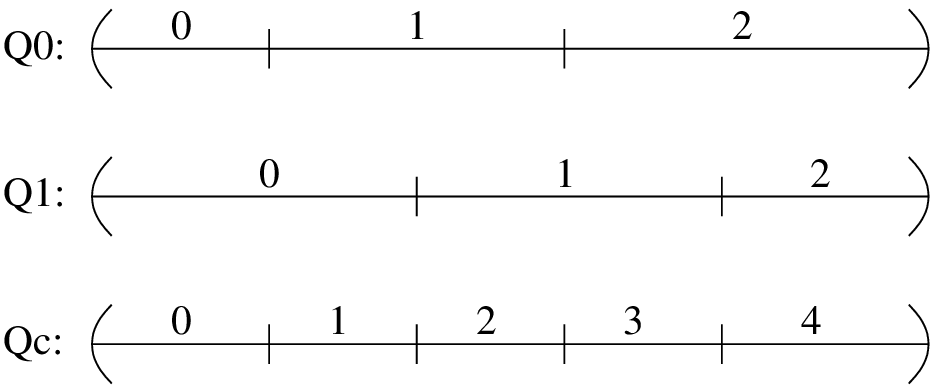}\label{fig:mdsq_a}}\quad
\subfigure[Irregular side quantizers]{\includegraphics[width=5.7cm]{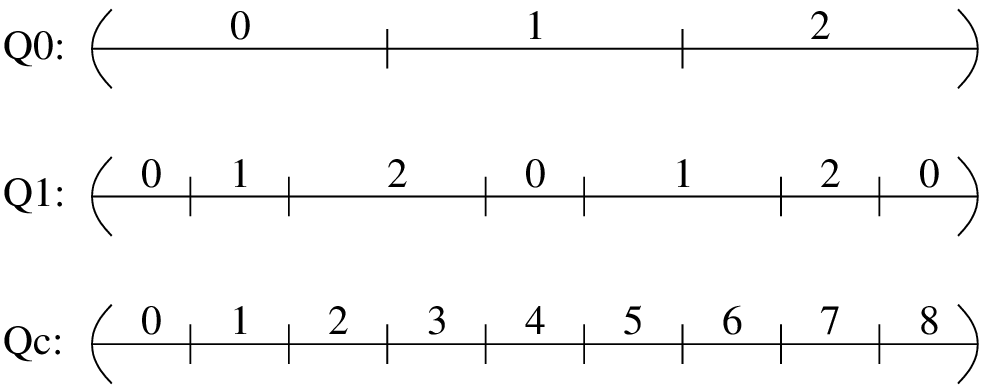}\label{fig:mdsq_b}}}
\caption{Two quantizers are able to refine each other if their cell bounderies do not coincide. In (a) both $Q_0$ and $Q_1$ are good quantizers but $Q_c$ is poor. In (b) quantizer $Q_1$ is poor whereas $Q_0$ and $Q_c$ are both good.}
\label{fig:mdsq}
\end{center}\end{figure}

In some of the earliest MD schemes it was recognized that two separate low-resolution quantizers may be combined to form a high-resolution quantizer. The cells of the high-resolution quantizer are formed as the intersections of the cells of the low-resolution quantizers~\cite{goyal:2001a}. The two low-resolution quantizers are traditionally called the side quantizers\index{side quantizer} and their joint quantizer, the high-resolution quantizer, is called the central quantizer\index{central quantizer}. If the side quantizers are regular quantizers, i.e.\ their cells form connected regions, then the central quantizer is not much better than the best of the two side quantizers. 
However, if disjoint cells are allowed in the side quantizers, then a much better central quantizer can be formed.
According to Goyal's survey article~\cite{goyal:2001a}, the idea of using disjoint cells in the side quantizers\index{side quantizer} seems to originate from some unpublished work of Reudink~\cite{reudink:1980}. Fig~\ref{fig:mdsq_a} shows an example where two regular side quantizers $Q_0$ and $Q_1$ each having 3 cells are combined to form a central quantizer $Q_c$ having 5 cells. Hence, the resolution of the central quantizer is only about twice that of either one of the two side quantizers. 
Fig.~\ref{fig:mdsq_b} shows an example where one of the side quantizers have disjoint cells which makes it possible to achieve a very good joint quantizer. 
In this case both side quantizers have three cells but $Q_1$ has disjoint cells. The central quantizer has 9 cells which is equal to the product of the number of cells of the side quantizers\index{side quantizer}. 
Hence, the resolution of the central quantizer is comparable to an optimal single description scalar quantizer\index{scalar!quantization} operating at the sum rate of the two side quantizers. The price, however, is relatively poor performance of side quantizer $Q_1$.

\begin{figure}
\begin{center}
\psfrag{0}{\small 0}
\psfrag{1}{\small 1}
\psfrag{2}{\small 2}
\psfrag{3}{\small 3}
\psfrag{4}{\small 4}
\psfrag{5}{\small 5}
\psfrag{6}{\small 6}
\psfrag{7}{\small 7}
\psfrag{8}{\small 8}
\psfrag{9}{\small 9}
\psfrag{10}{\small 10}
\psfrag{11}{\small 11}
\psfrag{12}{\small 12}
\psfrag{13}{\small 13}
\psfrag{14}{\small 14}
\psfrag{15}{\small 15}
\mbox{%
\subfigure[Side and central quantizers]{\includegraphics[width=7cm]{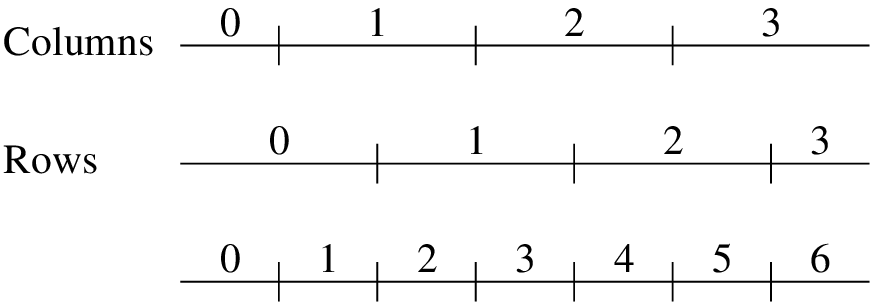}\label{fig:IA_partition_7}}\qquad
\subfigure[Index assignment matrix]{\includegraphics[width=3.5cm]{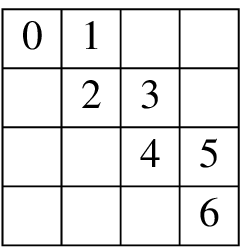}\label{fig:IA_q2}}}
\caption{(a) The two side quantizers each having four cells are offset from each other. Their intersection forms the central quantizer having 7 cells. (b) shows the corresponding index assignment matrix for the quantizers. The columns of the matrix form a side quantizer and the rows also form a side quantizer.} 
\label{fig:IA_7}
\end{center}
\end{figure}

The idea of using two quantizers with disjoint cells as side quantizers and their intersections as a central quantizer was independently discovered by Vaishampayan \cite{vaishampayan:1993} some years after Reudink. Vaishampayan proposed a systematic way to control the redundancy in the two side quantizers by use of an index assignment matrix~\cite{vaishampayan:1993}.
The idea is to first partition the real line into intervals in order to obtain the central quantizer\index{central quantizer} and then assign a set of central cells to each cell in the side quantizers. 
For example let us partition the real line into 7 intervals as shown in Fig.~\ref{fig:IA_partition_7} (the bottom quantizer is the central quantizer). 
We then construct the index assignment matrix as shown in Fig.~\ref{fig:IA_q2}. 
Each column of the matrix represent a cell of the side quantizer shown in the top of Fig.~\ref{fig:IA_partition_7}. Since there are four columns the side quantizer has four cells. Similarly, the four rows of the matrix represent the four cells of the second side quantizer (the middle quantizer of Fig.~\ref{fig:IA_partition_7}). 
The central quantizer in this design, which is based on the two main diagonals and where the side quantizers have connected cells, is known as a staggered quantizer.\footnote{It is often possible to make the second side quantizer a translation of the first side quantizer and use their intersection as the central quantizer. The quality improvement of a central quantizer constructed this way over that of the side quantizers is known as the staggering gain\index{staggering gain}. However, as first observed in~\cite{dayan:2002} and further analyzed in~\cite{tian:2005} the staggering gain dissappears when good high dimensional lattice vector quantizers are used.}
If we only use the main diagonal of the index assignment matrix we get a repetition code\index{repetition code}. In this case the side quantizers are identical and they are therefore not able to refine each other, which means that the central distortion will be equal to the side distortions.

\begin{figure}
\begin{center}
\psfrag{0}{\small 0}
\psfrag{1}{\small 1}
\psfrag{2}{\small 2}
\psfrag{3}{\small 3}
\psfrag{4}{\small 4}
\psfrag{5}{\small 5}
\psfrag{6}{\small 6}
\psfrag{7}{\small 7}
\psfrag{8}{\small 8}
\psfrag{9}{\small 9}
\psfrag{10}{\small 10}
\psfrag{11}{\small 11}
\psfrag{12}{\small 12}
\psfrag{13}{\small 13}
\psfrag{14}{\small 14}
\psfrag{15}{\small 15}
\mbox{%
\subfigure[Side and central quantizers]{\includegraphics[width=8.5cm]{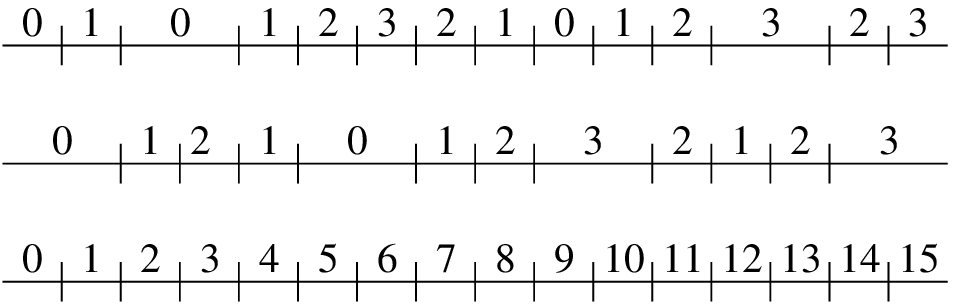}\label{fig:IA_partition_16}}
\subfigure[Index assignment matrix]{\includegraphics[width=3.5cm]{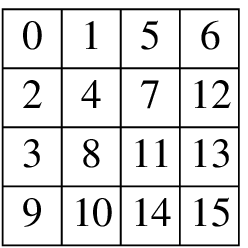}\label{fig:IA_q3}}}
\caption{(a) The two side quantizers each having four cells are not identical. Their intersection forms the central quantizer having 16 cells. (b) shows the corresponding index assignment matrix for the quantizers. The columns of the matrix form a side quantizer and the rows also form a side quantizer.} 
\label{fig:IA_16}
\end{center}
\end{figure}

By placing more elements (numbers) in the index assignment matrix the central quantizer will have more cells and the central distortion can therefore be reduced. There is a trade-off here, since placing more elements in the matrix will usually cause the cells of the side quantizers to be disjoint and the side distortion will then increase. Fig.~\ref{fig:IA_q3} shows an example where the index assignment matrix is full and the central quantizer therefore has 16 cells. Hence, the central distortion is minimized. From Fig.~\ref{fig:IA_partition_16} it is clear that the cells of the side quantizers are disjoint and since each cell is spread over a large region of the central quantizer the side distortion will be large.

The main difficulty of the design proposed by Vaishampayan lies in finding good index assignments, i.e.\ constructing the index assignment matrix. In~\cite{vaishampayan:1993} several heuristic designs were proposed for the case of symmetric resolution-constrained MD scalar quantization. Their performance at high resolution was evaluated in~\cite{vaishampayan:1998} and it was shown that in the quadratic Gaussian case, the distortion product\index{distortion product} $D_cD_0$ was 8.69 dB away from the optimal high resolution distortion product~(\ref{eq:distortionproduct}). Vaishampayan and Do\-maszewicz~\cite{vaishampayan:1994} then proposed an entropy-constrained MD scalar quantizer\index{scalar!quantization} where the index-assignment matrix\index{index-assignment!matrix} was optimized using a generalized Lloyd algorithm. The distortion product of this design was shown to be only 3.06 dB away from the theoretical optimum~\cite{vaishampayan:1998}. Recall that the space-filling loss of an SD scalar quantizer is 1.53 dB so that, quite surprisingly, the gap to the optimal distortion product of a two-description scalar quantizer is twice the scalar space-filling loss\index{space-filling loss}. The design of good index assignments for the scalar case is further considered in~\cite{berger-wolf:2002}.

It is known that the entropy-constrained scalar uniform quantizer is optimal in the SD case, see Chapter~\ref{chap:rd_theory}. This result, however, does not carry over to the MD case. 
Goyal et al.~\cite{kelner:2000,goyal:2002} were the first to recognize that by slightly modifying the central quantizer in a way so that it no longer forms a lattice, it is possible to reduce the distortion product not only in the scalar case but also in the two-dimensional case. This phenomenon was further investigated by Tian et al.~\cite{tian:2004,tian:2004b,tian:2004c} who showed that the scalar distortion product can be further improved by 0.4 dB by modifying the central quantizer.

\subsection{Lattice Vector Quantization for Multiple Descriptions}
Recently, Servetto, Vaishampayan and Sloane~\cite{servetto:1999,vaishampayan:2001} presented a clever construc\-tion based on lattices, which at high resolution and asymptotically in vector dimension is able to achieve the symmetric two-channel MD rate-distortion region. The design of~\cite{servetto:1999,vaishampayan:2001} is again based on index assignments which are non-linear mappings that lead to a curious result. Let $R_s=R_0=R_1$ denote the rate of each of the side quantizers and let $0<a<1$. Then, at high resolution, the central distortion $D_c$ satisfies~\cite{vaishampayan:2001}
\begin{equation}\label{eq:Dc_Vaishampayan}
\lim_{R\rightarrow \infty} D_c 2^{2R_s(1+a)} = \frac{1}{4}G(\Lambda)2^{2h(X)},
\end{equation}
whereas the side distortions $D_0=D_1$ satisfy
\begin{equation}\label{eq:D0_Vaishampayan}
\lim_{R\rightarrow \infty} D_0 2^{2R_s(1-a)} = G(S_L)2^{2h(X)},
\end{equation}
where $G(\Lambda)$ and $G(S_L)$ are the dimensionless normalized second moments of the central lattice $\Lambda$ and an $L$-sphere, respectively. Thus, remarkably, the performance of the side quantizers is identical to that of quantizers having spherical Voronoi cells\index{Voronoi cell!spherical}; note that in the SD case this is not possible for $1<L<\infty$. 

The design presented in~\cite{servetto:1999,vaishampayan:2001} is based on a central lattice $\Lambda$ and a single sublattice $\Lambda_s$ of index $N=|\Lambda/\Lambda_s|$. Each central lattice point $\lambda\in \Lambda$ is mapped to a pair of sublattice points $(\lambda_0,\lambda_1)\in \Lambda_s\times \Lambda_s$ using an index assignment function. 
A pair of sublattice points is called an edge. The edges are constructed by pairing closely spaced (in the Euclidean sense) sublattice points.
By exploiting the direction of an edge (i.e.\ the pair $(\lambda_0,\lambda_1)$ is distinguishable from $(\lambda_1,\lambda_0)$) it is possible to use an edge twice. In order to construct the edges as well as the assignment of edges to central lattice points, geometric properties of the lattices are exploited. Specifically, the edges $(\lambda_0,\lambda_1)$ and $(\lambda_1,\lambda_0)$ are mapped to the central lattice points $\lambda\in\Lambda$ and $\lambda'\in \Lambda$ which satisfy $\lambda+\lambda'= \lambda_0 + \lambda_1$ and furthermore, the distance from the midpoint of an edge and the associated central lattice point should be as small as possible (when averaged over all edges and the corresponding assigned central lattice points). Only a small number of edges and assignments needs to be found, whereafter the symmetry of the lattices can be exploited in order to cover the entire lattice. Examples of edge constructions and assignments are presented in~\cite{servetto:1999,vaishampayan:2001}.

The asymmetric case was considered by Diggavi, Sloane and Vaishampayan~\cite{diggavi:2000,diggavi:2002} who constructed a two-channel scheme also based on index assignments and which, at high resolution and asymptotically in vector dimension, is able to reach the entire two-channel MD rate-distortion region. Specifically, at high resolution, the central distortion satisfies~\cite{diggavi:2002}
\begin{equation}\label{eq:dc_diggavi}
D_c = G(\Lambda)2^{2(h(X)-R_c)},
\end{equation}
where $R_c$ is the rate of the central quantizer and the side distortions satisfy
\begin{equation}\label{eq:d0_diggavi}
D_0 = \frac{\gamma_{1}^2}{(\gamma_0+\gamma_1)^2} G(\Lambda_s)2^{2h(X)}2^{-2(R_0+R_1-R_c)},
\end{equation}
and
\begin{equation}\label{eq:d1_diggavi}
D_1 = \frac{\gamma_{0}^2}{(\gamma_0+\gamma_1)^2} G(\Lambda_s)2^{2h(X)}2^{-2(R_0+R_1-R_c)},
\end{equation}
where $G(\Lambda_s)$ is the dimensionless normalized second moment of a sublattice $\Lambda_s$, which is geometrically-similar to both side lattices $\Lambda_i, i=0,1$, and $\gamma_0,\gamma_1\in \mathbb{R}^+$ are weights which are introduced to control the asymmetry in the side distortions. Notice that in the distortion-balanced case we have $\gamma_0=\gamma_1$ so that $\frac{\gamma_0^2}{(\gamma_0+\gamma_1)^2}=\frac{1}4$ and if $\gamma_0=0$ or $\gamma_1=0$ then the design degenerates to a successive refinement\index{successive refinement} scheme~\cite{diggavi:2002,equitz:1991}. It is worth emphasizing that the side quantizers in the asymmetric design do generally not achieve the sphere bound\index{sphere!bound} in finite dimensions as was the case of the symmetric design. 

\begin{figure}[ht]
\begin{center}
\includegraphics[width=9cm]{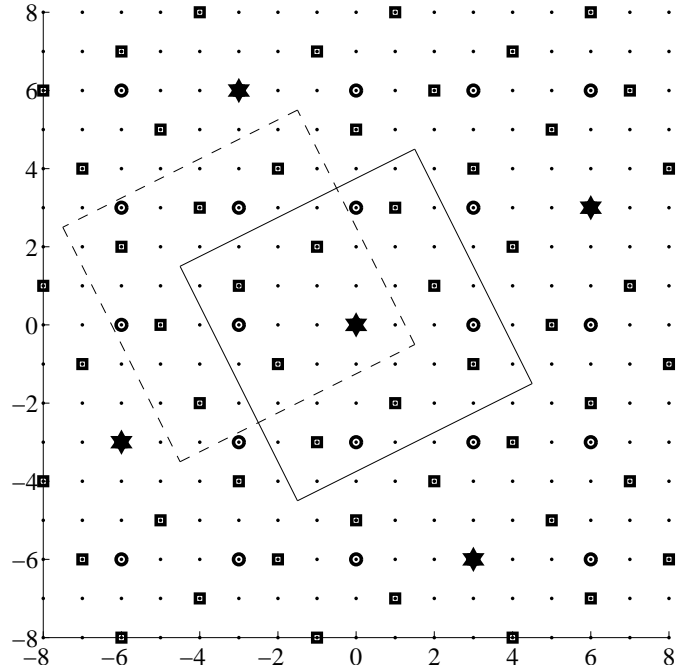}
\caption{A central lattice $\Lambda$ (dots), a sublattice $\Lambda_0$ (squares) of index 5, a sublattice (circles) of index 9, and a product lattice (stars) of index 45. The solid lines denote the Voronoi cell $V_\pi(0)$ of the product lattice point located at the origin. Notice that $V_\pi(0)$ contains 45 central lattice points. The dashed lines denotes $V_\pi(0)$ shifted so it is centered at $(-3,1)$.}
\label{fig:digg_exam}
\end{center}
\end{figure}

The design presented in~\cite{diggavi:2000,diggavi:2002} is based on a central lattice $\Lambda$, two sublattices $\Lambda_0\subset \Lambda$ and $\Lambda_1\subset \Lambda$ of index $N_0=|\Lambda/\Lambda_0|$ and $N_1=|\Lambda/\Lambda_1|$, respectively, and a product lattice $\Lambda_\pi\subset\Lambda_i, i=0,1$, of index $N_\pi = N_0N_1$. The Voronoi cell $V_\pi(\lambda_\pi)$ of the product lattice point $\lambda_\pi\in \Lambda_\pi$ contains $N_1$ sublattice points of $\Lambda_0$ and $N_0$ sublattice points of $\Lambda_1$, see Fig.~\ref{fig:digg_exam} for an example where $N_0=5$ and $N_1=9$. In this example, only the 45 central lattice points located within $V_\pi(0)$ need to have edges assigned. The remaining assignments are done simply by shifting these assignments by $\lambda_\pi\in \Lambda_\pi$. In other words, if the edge $(\lambda_0,\lambda_1)$ is assigned to $\lambda$ then the edge $(\lambda_0+\lambda_\pi,\lambda_1+\lambda_\pi)$ is assigned to $\lambda + \lambda_\pi$. We say that the assignments are shift invariant with respect to the product lattice.

The 45 edges are constructed in the following way. First, create the set $E_{\Lambda_0}$ containing the nine sublattice points of $\Lambda_0$ which are located within $V_\pi(0)$, i.e.\ 
$$
E_{\Lambda_0} = \{(0,0),(-3,1),(-2,-1),(-1,2),(-1,-3),(1,-2),(1,3),(2,1),(3,1)\}.
$$
Let $\lambda_0$ be the first element of $E_{\Lambda_0}$, i.e.\ $\lambda_0=(0,0)$. Pair $\lambda_0$ with the five $\lambda_1$ points located within $V_\pi(0)$. Thus, at this point we have five edges; $\{(0,0),(0,0)\}, \{(0,0),(0,3)\}, \{(0,0),(0,-3)\}, \{(0,0),(3,0)\}$ and $\{(0,0),(-3,0)\}$.
Consider now the second element of $E_{\Lambda_0}$, i.e.\ $\lambda_0=(-3,1)$. Shift $V_\pi(0)$ so that it is centered at $\lambda_0$ (illustrated by the dashed square in Fig.~\ref{fig:digg_exam}). For notational convenience we denote $V_\pi(0)+\lambda_0$ by $V_\pi(\lambda_0)$. We now pair $\lambda_0=(-3,1)$ with the five sublattice points of $\Lambda_1$ which are contained within $V_\pi(\lambda_0)$, i.e.\ $\Lambda_1 \cap V_\pi(\lambda_0) = \{(0,0),(-3,0),(-3,3),(-6,0),(-6,3)\}$. This procedure should be repeated for the remaining points of $E_{\Lambda_0}$ leading to a total of 45 distinct edges. These 45 edges combined with the 45 central lattice points within $V_\pi(0)$ form a bipartite matching problem where the cost of assigning an edge to a central lattice point is given by the Euclidean distance between the mid point (or weighted mid point) of the edge and the central lattice point.

Notice that for large index values, the sublattice point $\lambda_0\in\Lambda_0$ is paired with points of $\Lambda_1$ which are evenly distributed within a region $V_\pi$ centered at $\lambda_0$. If the product lattice $\Lambda_\pi$ is based on the hypercubic lattice $Z^L$ then $V_\pi$ forms a hypercube. In Chapters~\ref{chap:symmetric} and~\ref{chap:asym} we show that it is possible to change the design so that the sublattice points of $\Lambda_1$ which are paired with a given $\lambda_0\in \Lambda_0$ are evenly distributed within an $L$-dimensional hypersphere regardless of the choice of product lattice $\Lambda_\pi$. The purpose of having the points spherically distributed is twofold; first, the side distortion is reduced and second, it allows a simple extension to more than two descriptions.

\subsubsection{Non Index-Assignment Based Designs}\label{sec:nonindexrateloss}
To avoid the difficulty of designing efficient index-assignment maps it was sug\-gested in~\cite{dayan:2002} that the index assignments of a two-description system can be replaced by successive quantization and linear estimation.
More specifically, the two side descrip\-tions can be linearly combined and further enhanced by a refinement layer to yield the central reconstruction.
The design of~\cite{dayan:2002} suffers from a rate loss\index{rate loss!MD} of 0.5 bit/dim.\ at high resolution and is therefore not able to achieve the MD rate-distortion bound.
Recently, however, this gap was closed by Chen et al.~\cite{chen:2005,chen:2006} who showed that by use of successive quantization and source splitting\footnote{Source splitting denotes the process of splitting a source $X$ into two or more source variables, e.g.\ $X\rightarrow (X_1,X_2)$ where $X\rightarrow X_1 \rightarrow X_2$ forms a Markov chain\index{Markov chain} (in that order)~\cite{chen:2006}.}\index{source splitting} it is indeed possible to achieve the two-channel MD rate-distortion bound, at any resolution, without the use of index assignments. Chen et al.\ recognized that the rate region of the MD problem forms a polymatroid\index{polymatroid} and showed that corner points of this rate region can be achieved by successive estimation and quantization.
This design is inherently asymmetric in the description rate since any corner point of a non-trivial rate region will lead to asymmetric rates.
It is therefore necessary to perform source splitting\index{source splitting} in order to
achieve symmetry in the description rate.
When finite-dimensional quantizers are employed there is a space-filling loss due to the fact that the quantizer's Voronoi cells\index{Voronoi cell!spherical} are not completely spherical and each description therefore suffers a rate loss\index{rate loss!MD}. 
The rate loss of the design given in~\cite{chen:2005,chen:2006} is that of $2K-1$ quantizers because source splitting\index{source splitting} is performed by using an additional $K-1$ quantizers besides the conventional $K$ side quantizers. 
In comparison, the designs of the two-channel schemes based on index assignments~\cite{servetto:1999,vaishampayan:2001,diggavi:2000,diggavi:2002} suffer from a rate loss of only that of two quantizers and furthermore, in the symmetric case, they suffer from a rate loss of only that of two spherical quantizers. 
That it indeed is possible to avoid source splitting\index{source splitting} in the symmetric case without the use of index assignments was recently shown by \O stergaard and Zamir~\cite{ostergaard:2007} who constructed a $K$-channel symmetric MD scheme based on dithered Delta-Sigma quan\-tization\index{Delta-Sigma quantization}. The design of~\cite{ostergaard:2007} is able to achieve the entire symmetric two-channel MD rate-distortion region at any resolution and the rate loss when finite-dimensional quantizers are used is that of two lattice quantizers. Hence, in the two-channel case the rate loss when using index assignments is less than or equal to that of the designs which are not using index assignments~\cite{chen:2005,chen:2006,ostergaard:2007}.

\chapter{$K$-Channel Symmetric Lattice Vector Quantization}
\label{chap:symmetric}
In this chapter we consider a special case of the general $K$-channel symmetric MD problem where only a single parameter controls the redundancy tradeoffs between the central and the side distortions. With a single controlling parameter it is possible to describe the entire symmetric rate-distortion region for two descriptions and at high resolution, as shown in~\cite{servetto:1999,vaishampayan:2001}, but it is not enough to describe the symmetric achievable $K$-channel rate-distortion region. 
As such the proposed scheme offers a partial solution to the problem of designing balanced MD-LVQ systems. In Chapter~\ref{chap:comparison} we include more controlling parameters in the design and show that the three-channel MD region given by Theorem~\ref{theo:scec_partI} can be reached at high resolution.

We derive analytical expressions for the central and side quantizers\index{side quantizer} which, under high-resolution assumptions, minimize the expected distortion at the receiving side subject to entropy constraints on the side descriptions for given packet-loss probabilities. 
The central and side quantizers we use are lattice vector quantizers. The central distortion depends upon the lattice type in question whereas the side distortions only depend on the scaling of the lattices but are independent of the specific types of lattices. In the case of three descriptions we show that the side distortions can be expressed through the dimensionless normalized second moment of a sphere as was the case for the two descriptions system presented in~\cite{servetto:1999,vaishampayan:2001}. Furthermore, we conjecture that this is true in the general case of an arbitrary number of descriptions.

In the presented approach the expected distortion observed at the receiving side depends only upon the number of received descriptions, hence the descriptions are mutually refinable and reception of any $\kappa$ out of $K$ descriptions yields equivalent expected distortion. This is different from successive refinement\index{successive refinement} schemes~\cite{equitz:1991} where the individual descriptions often must be received in a prescribed order to be able to refine each other, i.e.\ description number $l$ will not do any good unless descriptions $0,\dots,l-1$ have already been received. We construct a scheme which for given packet-loss probabilities and a maximum bit budget (target entropy) determines the optimal number of descriptions and specifies the corresponding quantizers that minimize the expected distortion.

\section{Preliminaries}\label{sec:prelim}
We consider a central quantizer\index{central quantizer} and $K\geq 2$ side quantizers\index{side quantizer}. The central quantizer is (based on) a lattice $\Lambda_c\subset \mathbb{R}^L$ with a fundamental region of volume $\nu=\det(\Lambda_c)$. The side quantizers are based on a geometrically-similar and clean sublattice\index{sublattice} $\Lambda_s\subseteq \Lambda_c$ of index $N=|\Lambda_c/\Lambda_s|$ and fundamental regions of volume $\nu_s=\nu N$. The trivial case $K=1$ leads to a single-description system, where we would simply use one central quantizer and no side quantizers. 

We will consider the balanced situation, where the entropy $R$ is the same for each description. Furthermore, we consider the case where the contribution $D_i, i=0,\dots,K-1$ of each description to the total distortion is the same. Our design makes sure\footnote{We prove this symmetry property for the asymptotic case of $N\rightarrow\infty$ and $\nu_s\rightarrow 0$. For finite $N$ we cannot guarantee the existence of an exact symmetric solution. However, by use of time-sharing\index{time sharing} arguments, it is always possible to achieve symmetry.}  that the distortion observed at the receiving side depends only on the number of descriptions received; hence reception of any $\kappa$ out of $K$ descriptions yields equivalent expected distortion. 

\subsection{Index Assignments}\label{sec:index}
A source vector $x$ is quantized to a reconstruction point $\lambda_c$ in the central lattice $\Lambda_c$. Hereafter follows index assignments\index{index-assignment!map} (mappings), which uniquely map $\lambda_c$ to one vector (reconstruction point) in each of the side quantizers. This mapping is done through a labeling function $\alpha$, and we denote the individual component functions of $\alpha$ by $\alpha_i$, where $i=0,\dots, K-1$. In other words, the injective map $\alpha$ that maps $\Lambda_c$ into $\Lambda_s\times\dots \times \Lambda_s$, is given by
\begin{align}
\alpha(\lambda_c)&=(\alpha_0(\lambda_c),\alpha_1(\lambda_c),\dots,\alpha_{K-1}(\lambda_c)) \\
&= (\lambda_0,\lambda_1,\dots,\lambda_{K-1}),
\end{align}
where $\alpha_i(\lambda_c)=\lambda_i \in \Lambda_s$ and $i=0,\dots, K-1$. Each $K$-tuple\index{K-tuple@$K$-tuple} $(\lambda_0,\dots,\lambda_{K-1})$ is used only once when labeling points in $\Lambda_c$ in order to make sure that $\lambda_c$ can be recovered unambiguously when all $K$ descriptions are received. At this point we also define the inverse component map, $\alpha_i^{-1}$, which gives the set of distinct central lattice points a specific sublattice point is mapped to. This is given by
\begin{equation}
\alpha_i^{-1}(\lambda_i) = \{\lambda_c \in \Lambda_c : \alpha_i(\lambda_c)=\lambda_i\},\qquad \lambda_i \in \Lambda_s,
\end{equation}
where $|\alpha_i^{-1}(\lambda_i)| \approx N$, since there are $N$ times as many central lattice points as sublattice points within a bounded region of $\mathbb{R}^L$.

Since lattices are infinite arrays of points, we construct a shift invariant\index{shift invariance} labeling function, so we only need to label a finite number of points as is done in~\cite{vaishampayan:2001,diggavi:2002}. Following the approach outlined in Chapter~\ref{chap:lattice_theory} we construct a product lattice $\Lambda_\pi$ which has $N^2$ central lattice points and $N$ sublattice points in each of its Voronoi cells. The Voronoi cells\index{Voronoi cell} $V_\pi$ of the product lattice $\Lambda_\pi$ are all similar so by concentrating on labeling only central lattice points within one Voronoi cell of the product lattice, the rest of the central lattice points may be labeled simply by translating this Voronoi cell throughout $\mathbb{R}^L$. Other choices of product lattices are possible, but this choice has a particular simple construction.  
With this choice of product lattice, we only label central lattice points within $V_\pi(0)$, which is the Voronoi cell of $\Lambda_\pi$ around the origin. With this we get 
\begin{equation}\label{eq:shiftinv}
\alpha(\lambda_c + \lambda_\pi) = \alpha(\lambda_c) + \lambda_\pi,
\end{equation}
for all $\lambda_\pi \in \Lambda_\pi$ and all $\lambda_c \in \Lambda_c$. 

\section{Rate and Distortion Results}
\subsubsection{Central Distortion}\label{sec:centraldist}
Let us consider a scalar process that generates i.i.d.\ random variables with probability density function\index{probability density function} (pdf) $f$. Let $X \in \mathbb{R}^L$ be a random vector made by blocking outputs of the scalar process into vectors of length $L$, and let $x\in \mathbb{R}^L$ denote a realization of $X$. The $L$-fold pdf of $X$ is denoted $f_{X}$ and given by\footnote{It is worth pointing out that we actually only require the individual vectors to be i.i.d.\ and as such correlation within vectors is allowed.}
\begin{equation}
f_X(x) = \prod_{j=0}^{L-1}f(x_j).
\end{equation}
The expected distortion (per dimension) $D_c$ occuring when all packets are received is called the central distortion and is defined as 
\begin{equation}\label{eq:dc}
D_c \triangleq \frac{1}{L}\sum_{\lambda_c \in \Lambda_c}\int_{V_c(\lambda_c)}\|x-\lambda_c\|^2f_X(x)dx,
\end{equation}
where $V_c(\lambda_c)$ is the Voronoi cell of a single reconstruction point $\lambda_c \in \Lambda_c$. 
Using standard high resolution assumptions, cf.\ Chapter~\ref{chap:rd_theory}, we may assume that each Voronoi cell\index{Voronoi cell} is sufficiently small and $f_X(x)$ is smooth and hence approximately constant within each cell. In this case $\lambda_c$ is approximately the centroid\index{centroid} (conditional mean) of the corresponding cell, that is 
\begin{equation}\label{eq:centroid}
\lambda_c \approx \frac{\int_{V_c(\lambda_c)}x f_X(x)dx}{\int_{V_c(\lambda_c)} f_X(x)dx}.
\end{equation}
Since the pdf is approximately constant within a small region we also have that
\begin{equation}
 f_X(x) \approx f_X(\lambda_c), \quad \forall x \in V_c(\lambda_c),
\end{equation}
and we can therefore express the probability, $P$, of a cell as
\begin{equation}
P(V_c(\lambda_c)) = \int_{V_c(\lambda_c)}f_X(x)dx \approx f_X(\lambda_c)\int_{V_c(\lambda_c)}dx = \nu f_X(\lambda_c),
\end{equation}
where $\nu$ is the volume of a Voronoi cell\index{Voronoi cell}. With this, we get
\begin{equation}\label{eq:prob2pdf}
f_X(\lambda_c) \approx \frac{P(V_c(\lambda_c))}{\nu}.
\end{equation}
Inserting~(\ref{eq:prob2pdf}) into~(\ref{eq:dc}) gives
\begin{equation}\label{eq:dcstep1}
D_c \approx \frac{1}{L}\sum_{\lambda_c \in \Lambda_c}P(V_c(\lambda_c))\int_{V_c(\lambda_c)}\frac{\|x-\lambda_c\|^2}{\nu}dx,
\end{equation}
where $\Lambda_c$ is a lattice so all Voronoi cells are congruent and
the integral is similar for all $\lambda_c$'s. Hence, without loss of generality, we let $\lambda_c = 0$ and simplify~(\ref{eq:dcstep1}) as
\begin{equation}\label{eq:dcstep2}
D_c \approx \frac{1}{L}\int_{V_c(0)}\frac{\|x\|^2}{\nu}dx,
\end{equation}
where we used the fact that $\sum_{\lambda_c \in \Lambda_c}P(V_c(\lambda_c)) = 1$.
We can express the average central distortion~(\ref{eq:dcstep2}) in terms of the dimensionless normalized second moment of inertia $G(\Lambda_c)$ by
\begin{align}\label{eq:dcG}
D_c \approx G(\Lambda_c)\nu^{2/L}.
\end{align}

\subsubsection{Side Distortions}
The side distortion for the $i^{th}$ description, i.e.\ the distortion when reconstructing using only the $i^{th}$ description, is given by~\cite{vaishampayan:2001}
{\allowdisplaybreaks
\begin{align} \notag
D_i &= \frac{1}{L}\sum_{\lambda_c \in \Lambda_c}\int_{V_c(\lambda_c)} \|x - \alpha_i(\lambda_c)\|^2f_X(x)dx,\quad  i=0,\dots, K-1,\\ \notag
    &= \frac{1}{L}\sum_{\lambda_c \in \Lambda_c}\int_{V_c(\lambda_c)} \|x - \lambda_c + \lambda_c - \alpha_i(\lambda_c)\|^2f_X(x)dx\\ \notag
    &= \frac{1}{L}\sum_{\lambda_c \in \Lambda_c}\int_{V_c(\lambda_c)}\!\!\! \|x - \lambda_c\|^2f_X(x)dx + 
\frac{1}{L}\sum_{\lambda_c \in \Lambda_c}\int_{V_c(\lambda_c)}\!\!\! \|\lambda_c - \alpha_i(\lambda_c)\|^2f_X(x)dx\\ \notag
&\quad + \frac{2}{L}\sum_{\lambda_c \in \Lambda_c}\int_{V_c(\lambda_c)} 
\langle x-\lambda_c,\lambda_c-\alpha_i(\lambda_c)\rangle f_X(x)dx \\ \notag
    &\approx D_c + 
\frac{1}{L}\sum_{\lambda_c \in \Lambda_c} \|\lambda_c - \alpha_i(\lambda_c)\|^2P(\lambda_c)\\  \notag
&\quad + \frac{2}{L}\sum_{\lambda_c \in \Lambda_c}
\left\langle \int_{V_c(\lambda_c)}x f_X(x)dx-\int_{V_c(\lambda_c)}\lambda_c f_X(x)dx,\lambda_c-\alpha_i(\lambda_c)\right\rangle \\ \label{eq:di}
    &= D_c + \frac{1}{L}\sum_{\lambda_c \in \Lambda_c} \|\lambda_c - \alpha_i(\lambda_c)\|^2P(\lambda_c),
\end{align}}
\!where $P(\lambda_c)$ is the probability that $X$ will be mapped to $\lambda_c$, i.e.\ $Q(X)=\lambda_c$, and the last equality follows since by use of~(\ref{eq:centroid}) we have that 
\begin{equation}
\int_{V_c(\lambda_c)}x f_X(x)dx - \int_{V_c(\lambda_c)}\lambda_c  f_X(x)dx =0.
\end{equation}
We notice from~(\ref{eq:di}) that independent of which labeling function we use, the distortion introduced by the central quantizer\index{central quantizer} is orthogonal (under high-resolution assumptions) to the distor\-tion introduced by the side quantizers.

Exploiting the shift-invariance property of the labeling function~(\ref{eq:shiftinv}) makes it possible to simplify~(\ref{eq:di}) as
\begin{equation}\label{eq:di1}
\begin{split}
D_i&\approx D_c + \frac{1}{L}\sum_{\lambda_\pi \in \Lambda_\pi}\frac{P(\lambda_\pi)}{N^2}
\sum_{\lambda_c \in V_\pi(0)} \|\lambda_c - \alpha_i(\lambda_c)\|^2\\
&= D_c + \frac{1}{N^2}\frac{1}{L}\sum_{\lambda_c \in V_\pi(0)} \|\lambda_c - \alpha_i(\lambda_c)\|^2,\quad i=0,\dots, K-1,
\end{split}
\end{equation}
where we assume the region $V_\pi(0)$ is sufficiently small so $P(\lambda_c) \approx P(\lambda_\pi)/N^2$, for $\lambda_c\in V_\pi(\lambda_\pi)$. Notice that we assume $P(\lambda_\pi)$ to be constant only within each region $V_\pi(\lambda_\pi)$, hence it may take on different values for each $\lambda_\pi \in \Lambda_\pi$.

\subsubsection{Central Rate}\label{sec:rate}
Let $R_c = H(Q(X))/L$ denote the minimum entropy (per dimension) needed for a single-description system to achieve an expected distortion of $D_c$, the central distortion of the multiple-description system as given by~(\ref{eq:dcG}).

The single-description rate $R_c$ is given by
\begin{equation}
R_c = -\frac{1}L\sum_{\lambda_c\in \Lambda_c}\int_{V_c(\lambda_c)}f_X(x)dx\,\log_2\left(\int_{V_c(\lambda_c)}f_X(x)dx\right).
\end{equation}
Using that each quantizer cell has identical volume $\nu$ and assuming that $f_X(x)$ is approximately constant within Voronoi cells of the central lattice $\Lambda_c$, it follows that
\begin{equation}\label{eq:Rc}
\begin{split}
R_c &\approx -\frac{1}L\sum_{\lambda_c\in \Lambda_c}\int_{V_c(\lambda_c)}f_X(x)dx\,\log_2\left( f_X(\lambda_c)\nu\right)\\ 
&=-\frac{1}L\sum_{\lambda_c\in \Lambda_c}\int_{V_c(\lambda_c)}f_X(x)dx\,\log_2\left( f_X(\lambda_c)\right)\\
&\quad -\frac{1}L\sum_{\lambda_c\in \Lambda_c}\int_{V_c(\lambda_c)}f_X(x)dx\,\log_2(\nu)\\ 
&=-\frac{1}L\sum_{\lambda_c\in \Lambda_c}\int_{V_c(\lambda_c)}f_X(x)dx\,\log_2\left( f_X(\lambda_c)\right)-\frac{1}L\log_2(\nu)\\
&= \bar{h}(X) - \frac{1}L\log_2(\nu).
\end{split}
\end{equation}

\subsubsection{Side Rates}
Let $R_i=H(\alpha_i(Q(X)))/L$ denote the entropy (per dimension) of the $i^{th}$ description, where $i=0,\dots,K-1$. Notice that in the symmetric situation we have $R_s=R_i, i\in \{0,\dots,K-1\}$.\index{description rate}

The side descriptions are based on a coarser lattice obtained by scaling (and possibly rotating) the Voronoi cells of the central lattice by a factor of $N$. Assuming the pdf of $X$ is roughly constant within a sublattice cell, the entropy of the $i^{th}$ side description is given by
\begin{equation}\label{eq:Rs}
\begin{split}
R_i=& -\frac{1}L\sum_{\lambda_i\in \Lambda_s}\left(
\sum_{\lambda_c\in \alpha_i^{-1}(\lambda_i)}
\int_{V_c(\lambda_c)}f_X(x)dx\,
\log_2\left(
\sum_{\lambda_c\in \alpha_i^{-1}(\lambda_i)}
\int_{V_c(\lambda_c)}f_X(x)dx \right)\right)\\ 
=&-\frac{1}L\sum_{\lambda_i\in \Lambda_s}\left(
\sum_{\lambda_c\in \alpha_i^{-1}(\lambda_i)}
\int_{V_c(\lambda_c)}f_X(x)dx\,
\log_2\left(
\nu f_X(\lambda_i)N\right)\right)\\ 
=&-\frac{1}L\sum_{\lambda_i\in \Lambda_s}\left(
\sum_{\lambda_c\in \alpha_i^{-1}(\lambda_i)}
\int_{V_c(\lambda_c)}f_X(x)dx\,
\log_2\left(f_X(\lambda_i)\right)\right) \\ 
&-\frac{1}L\sum_{\lambda_i\in \Lambda_s}\left(
\sum_{\lambda_c\in \alpha_i^{-1}(\lambda_i)}
\int_{V_c(\lambda_c)}f_X(x)dx\,
\log_2(\nu N)\right) \\
=& \bar{h}(X) - \frac{1}L\log_2(N\nu).
\end{split}
\end{equation}
The entropy of the side descriptions is related to the entropy of the single-description system by
\begin{equation}
R_i = R_c - \frac{1}L\log_2(N).
\end{equation}

\section{Construction of Labeling Function}
\label{sec:label}
The index assignment\index{index-assignment!labeling function} is done by a labeling function $\alpha$, that maps central lattice points to sublattice points. An optimal index assignment minimizes a cost functional when $0< \kappa < K$ descriptions are received. In addition, the index assignment should be invertible so the central quantizer can be used when all descriptions are received. Before defining the labeling function we have to define the cost functional to be minimized. To do so, we first describe how to approximate the source sequence when receiving only $\kappa$ descriptions and how to determine the expected distortion in that case.
Then we define the cost functional\index{cost functional} to be minimized by the labeling function $\alpha$ and describe how to minimize it.

\subsection{Expected Distortion}
\label{sec:reconstruction}
At the receiving side, $X\in \mathbb{R}^L$ is reconstructed to a quality that is determined only by the number of received descriptions. If no descriptions are received we reconstruct using the expected value, $EX$, and if all $K$ descriptions are received we reconstruct using the inverse map $\alpha^{-1}(\lambda_0,\dotsc,\lambda_{K-1})$, hence obtaining the quality of the central quantizer. 

In this work we use a simple reconstruction rule which applies for arbitrary sources.\footnote{We show in Chapter~\ref{chap:comparison} that this simple reconstruction rule is, at high resolution, optimal in the quadratic Gaussian case, i.e.\ we show that the largest known three-channel MD region can be achieved in that case. This is in line with Ozarow's double-branch test-channel, where the optimum post filters are trivial at high resolution.} When receiving $1\leq \kappa<K$ descriptions we reconstruct using the average of the $\kappa$ descriptions. We show later (Theorem~\ref{theo:sums}) that using the average of received descriptions as reconstruction rule makes it possible to split the distortion due to reception of any number of descriptions into a sum of squared norms between pairs of lattice points. Moreover, this lead to the fact that the side quantizers' performances approach those of quantizers having spherical Voronoi cells\index{Voronoi cell!spherical}. 
There are in general several ways of receiving $\kappa$ out of $K$ descriptions. Let $\mathcal{L}^{(K,\kappa)}$ denote an index set consisting of all possible $\kappa$ combinations out of $\{0,\dots, K-1\}$. Hence $|\mathcal{L}^{(K,\kappa)}| = \binom{K}{\kappa}$. We denote an element of $\mathcal{L}^{(K,\kappa)}$ by $l=\{l_0, \dots , l_{\kappa-1}\}\in \mathcal{L}^{(K,\kappa)}$. Upon reception of any $\kappa$ descriptions we reconstruct $\hat{X}$ using 
\begin{equation}
\hat{X}=\frac{1}{\kappa}\sum_{j=0}^{\kappa-1}\lambda_{l_j}.
\end{equation}

Our objective is to minimize some cost functional subject to entropy constraints on the description rates. We can, for example, choose to minimize the distortion when receiving any two out of three descriptions. Another choice is to minimize the weighted distortion over all possible description losses. 
In the following we will assume that the cost functional to be minimized is the expected weighted distortion over all description losses and we further assume that the weights are given by the packet-loss probabilities\index{packet-loss probability}. We discuss the case where the weights are allowed to be chosen almost arbitrarily in Chapter~\ref{chap:asym}. 
 
Assuming the packet-loss probabilities, say $p$, are independent and are the same for all descrip\-tions, we may use~(\ref{eq:di1}) and write the expected distortion when receiving $\kappa$ out $K$ descriptions as
\begin{equation}\label{eq:expdist}
\begin{split}
D_a^{(K,\kappa)}\approx (1-p)^{\kappa}p^{K-\kappa}\!
\times\!\! \left(\!\!
\binom{K}{\kappa}D_c 
+ \frac{1}{N^2}\frac{1}{L}
\sum_{l\in \mathcal{L}^{(K,\kappa)}}\sum_{\lambda_c\in V_\pi(0)}\!
\left\|\lambda_c - \frac{1}{\kappa}\sum_{j=0}^{\kappa-1} \lambda_{l_j}\right\|^2\right)\!\!,
\end{split}
\end{equation}
where $\lambda_{l_j}=\alpha_{l_j}(\lambda_c)$ and the two cases $\kappa\in \{0,K\}$, which do not involve the index-assignment map, are given by $D_a^{(K,0)}\approx p^K E\|X\|^2/L$ and $D_a^{(K,K)}\approx (1-p)^K D_c$.

\subsection{Cost Functional}
From~(\ref{eq:expdist}) we see that the distortion expression may be split into two terms, one describing the distortion occurring when the central quantizer is used on the source, and one that describes the distortion due to the index assignment. 
An optimal index assignment jointly minimizes the second term in~(\ref{eq:expdist}) over all $1\leq \kappa\leq K-1$ possible descriptions.
The cost functional\index{cost functional} $J^{(K)}$ to be minimized by the index assignment algorithm is then given by
\begin{equation}\label{eq:IAcost}
J^{(K)}=\sum_{\kappa=1}^{K-1}J^{(K,\kappa)},
\end{equation}
where 
\begin{equation}\label{eq:costfunctional}
J^{(K,\kappa)}= \frac{(1-p)^{\kappa}p^{K-\kappa}}{L N^2}
\sum_{l\in \mathcal{L}^{(K,\kappa)}}\sum_{\lambda_c\in V_\pi(0)}
\left\| \lambda_c - \frac{1}{\kappa}\sum_{j=0}^{\kappa-1} \lambda_{l_j}\right\|^2.
\end{equation}
The cost functional should be minimized subject to an entropy constraint on the side descriptions. We remark here that the side entropies depend solely on $\nu$ and $N$ and as such not on the particular choice of $K$-tuples. In other words, for fixed $N$ and $\nu$ the index assignment problem is solved if~(\ref{eq:IAcost}) is minimized. The problem of choosing $\nu$ and $N$ such that the entropy constraint is satisfied is independent of the assignment problem and deferred to Section~\ref{sec:opt}.

The following theorem makes it possible to rewrite the cost functional in a way that brings more insight into which $K$-tuples to use.\footnote{Notice that Theorem~\ref{theo:sums} is very general. We do not even require $\Lambda_c$ or $\Lambda_s$ to be lattices, in fact, they can be arbitrary sets of points.}

\begin{theorem}\label{theo:sums}
For $1\leq \kappa \leq K$ we have
\begin{equation*}
\begin{split}
\sum_{l\in\mathcal{L}^{(K,\kappa)}}
\sum_{\lambda_c}
\left\| \lambda_c - \frac{1}{\kappa}\sum_{j=0}^{\kappa-1} \lambda_{l_j}\right\|^2
&= \sum_{\lambda_c}\binom{K}{\kappa}\Bigg(\left\| \lambda_c - 
\frac{1}{K}\sum_{i=0}^{K-1}\lambda_i\right\|^2 \\
&\quad+ \left(\frac{K-\kappa}{K^2\kappa(K-1)}\right)
\sum_{i=0}^{K-2}\sum_{j=i+1}^{K-1}\| \lambda_i - \lambda_j\|^2\Bigg).
\end{split}
\end{equation*}
\end{theorem}

\begin{proof}
See Appendix~\ref{app:sums}.
\end{proof}
From Theorem~\ref{theo:sums} it is clear that~(\ref{eq:costfunctional}) can be written as
\begin{equation}\label{eq:costfunctional2}
\begin{split}
J^{(K,\kappa)}&=\frac{(1-p)^{\kappa}p^{K-\kappa}}{L N^2}
\binom{K}{\kappa}\left(
\sum_{\lambda_c\in V_\pi(0)}\left\| \lambda_c - 
\frac{1}{K}\sum_{i=0}^{K-1}\lambda_i\right\|^2 \right.\\
&\quad +
\left.\sum_{\lambda_c\in V_\pi(0)}\left(
\frac{K-\kappa}{K^2\kappa(K-1)}\right)
\sum_{i=0}^{K-2}\sum_{j=i+1}^{K-1}\| \lambda_i - \lambda_j\|^2\right).
\end{split}
\end{equation}

The first term in~(\ref{eq:costfunctional2}) describes the distance from a central lattice point to the centroid of its associated $K$-tuple. The second term describes the sum of pairwise squared distances (SPSD)\index{sum of pairwise squared distances} between elements of the $K$-tuples. In Section~\ref{sec:highrate} (by Proposition~\ref{prop:growthriemann2}) we show that, under a high-resolution assumption, the second term in~(\ref{eq:costfunctional2}) is dominant, from which we conclude that 
in order to minimize~(\ref{eq:IAcost}) we have to choose
the $K$-tuples with the lowest SPSD\@. 
These $K$-tuples are then assigned to central lattice points in such a way, that the first term in~(\ref{eq:costfunctional2}) is minimized. 

Independent of the packet-loss probability, we always minimize the second term in~(\ref{eq:costfunctional2}) by using those $K$-tuples that have the smallest SPSD\@. This means that, at high resolution, the optimal $K$-tuples are independent of packet-loss probabilities and, consequently, the optimal assignment is independent\footnote{Given the central lattice and  the sublattice, the optimal assignment is independent of $p$. However, we show later that the optimal sublattice index $N$ depends on $p$.} of the packet-loss probability.

\subsection{Minimizing Cost Functional}
\label{sec:ktuples}
\begin{figure}[ht]
\psfrag{sub}{$\in \Lambda_s$}
\psfrag{prod}{$\in \Lambda_\pi$}
\psfrag{vtilde}{$\tilde{V}$}
\psfrag{V0}{$V_\pi$}
\psfrag{Vn}{$V_\pi(0)$}
\psfrag{K=3}{$K=3$}
\psfrag{N=25}{$N=25$}
\psfrag{Nt=5}{$\tilde{N}=5$}
\begin{center}
\includegraphics{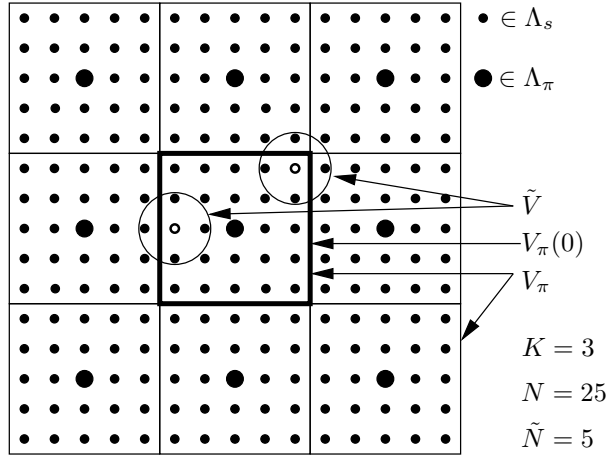}
\end{center}
\caption{The region $\tilde{V}$ (big circles) is here shown centered at two different sublattice points within $V_\pi(0)$. Small dots represents sublattice points of $\Lambda_s$ and large dots represents product lattice points $\lambda_\pi \in \Lambda_\pi$. Central lattice points are not shown here. $V_\pi$ (shown as squares) contains 25 sublattice points centered at product lattice points. In this example $\tilde{V}$ contains 5 sublattice points.}
\label{fig:vtilde}
\end{figure}

In order to make sure that $\alpha$ is shift-invariant, a given $K$-tuple of sublattice reconstruc\-tion points is assigned to only one central lattice point $\lambda_c\in \Lambda_c$. Notice that two $K$-tuples which are translates of each other by some $\lambda_\pi \in \Lambda_\pi$ must not both be assigned to central lattice points located within the same region $V_\pi(\lambda_\pi)$, since this causes assignment of the same $K$-tuples to multiple central lattice points.
The region $V_\pi(0)$ will be translated throughout $\mathbb{R}^L$ and centered at $\lambda_\pi \in \Lambda_\pi$, so there will be no overlap between neighboring regions, i.e.\ $V_\pi(\lambda'_\xi) \cap V_\pi(\lambda''_\xi) = \emptyset$, for $\lambda'_\xi,\lambda''_\xi \in \Lambda_\pi$ and $\lambda'_\xi \neq \lambda''_\xi$. 
One obvious way of avoiding assigning $K$-tuples to multiple central lattice points is then to exclusively use sublattice points located within $V_\pi(0)$. 
However, sublattice points located close to but outside $V_\pi(0)$, might be better  candidates than sublattice points within $V_\pi(0)$ when labeling central lattice points close to the boundary. A consistent way of constructing $K$-tuples, is to center a region $\tilde{V}$ at all sublattice points $\lambda_0 \in \Lambda_s \cap V_\pi(0)$, and construct $K$-tuples by combining sublattice points $\lambda_i\in \Lambda_s, i=1,\dots,K-1$ within $\tilde{V}(\lambda_0)$ in all possible ways and select the ones that minimize~(\ref{eq:costfunctional2}). This is illustrated in Fig.~\ref{fig:vtilde}.
For a fixed $\lambda_i\in \Lambda_s$, the expression $\sum_{\lambda_j\in \Lambda_s \cap \tilde{V}(\lambda_i)}\|\lambda_i - \lambda_j\|^2$ is minimized when $\tilde{V}$ forms a sphere centered at $\lambda_i$. Our construction allows for $\tilde{V}$ to have an arbitrary shape, e.g.\ the shape of $V_\pi$ which is the shape used for the two-description system presented in~\cite{diggavi:2002}. However, if $\tilde{V}$ is not chosen to be a sphere, the SPSD\index{sum of pairwise squared distances} is in general not minimized.

For each $\lambda_0\in \Lambda_s\cap V_\pi(0)$ it is possible to construct $\tilde{N}^{K-1}$ $K$-tuples, where $\tilde{N}$ is the number of sublattice points within the region $\tilde{V}$. 
This gives a total of $N\tilde{N}^{K-1}$ $K$-tuples when all $\lambda_0\in \Lambda_s \cap V_\pi(0)$ are used.
However, only $N^2$ central lattice points need to be labeled ($V_\pi(0)$ only contains $N^2$ central lattice points). When $K=2$, we let $\tilde{N}=N$, so the number of possible $K$-tuples is equal to $N^2$, which is exactly the number of central lattice points in $V_\pi(0)$. 
In general, for $K>2$, the volume $\tilde{\nu}$ of $\tilde{V}$ is smaller than the volume of $V_\pi(0)$ and as such $\tilde{N}<N$. We can approximate $\tilde{N}$ through the volumes $\nu_s$ and $\tilde{\nu}$, i.e.\ $\tilde{N}\approx \tilde{\nu}/\nu_s$. To justify this approximation let $\Lambda \subset \mathbb{R}^L$ be a real lattice and let $\nu=\det(\Lambda)$ be the volume of a fundamental region. Let $S(c,r)$ be a sphere in $\mathbb{R}^L$ of radius $r$ and center $c\in \mathbb{R}^L$. 
According to Gauss' counting principle, the number $A_{\mathbb{Z}}$ of integer lattice points in a convex body $\mathcal{C}$ in $\mathbb{R}^L$ equals the volume Vol$(\mathcal{C})$ of $\mathcal{C}$ with a small error term~\cite{mazo:1990}. In fact if $\mathcal{C}=S(c,r)$ then by use of a theorem due to Minkowski it can be shown that, for any $c\in\mathbb{R}^L$ and asymptotically as $r\rightarrow \infty$, $A_{\mathbb{Z}}(r)=\text{Vol}(S(c,r))=\omega_L r^L$, where $\omega_L$ is the volume of the $L$-dimensional unit sphere~\cite{fricker:1982}, see also~\cite{erdos:1989,bokowski:1973,vinograd:1963,gruber:1987,kratzel:1988}. 
It is also known that the number of lattice points $A_\Lambda(n)$ in the first $n$ shells\index{lattice!shell} of the lattice $\Lambda$ satisfies, asymptotically as $n\rightarrow \infty$, $A_\Lambda(n) = \omega_L n^{L/2}/\nu$~\cite{vaishampayan:2001}. 
Hence, based on the above we approximate the number of lattice points in $\tilde{V}$ by $\tilde{\nu}/\nu_s$, which is an approximation that becomes exact as the number of shells $n$ within $\tilde{V}$ goes to infinity\footnote{For the high-resolution analysis given in Section~\ref{sec:highrate} it is important that $\tilde{\nu}$ is kept small as the number of lattice points within $\tilde{V}$ goes to infinity. This is easily done by proper scaling of the lattices, i.e.\ making sure that $\nu_s\rightarrow 0$ as $N\rightarrow \infty$.} (which corresponds to $N\rightarrow\infty$). Our analysis is therefore only exact in the limiting case of $N\rightarrow \infty$. With this we can lower bound $\tilde{\nu}$ by 
\begin{equation}\label{eq:vtilde}
\lim_{N\rightarrow\infty}
\tilde{\nu}\geq \nu_s\, N^{1/(K-1)}.
\end{equation}
Hence, $\tilde{V}$ contains $\tilde{N}\geq N^{1/(K-1)}$ sublattice points so that the total number of possible $K$-tuples is $N\tilde{N}^{K-1}\geq N^2$.

In Fig.~\ref{fig:vtilde} is shown an example of $\tilde{V}$ and $V_\pi$ regions for the two-dimensional $Z^2$ lattice. In the example we used $K=3$ and $N=25$, hence there are 25 sublattice\index{sublattice} points within $V_\pi$. There are $\tilde{N}=N^{1/(K-1)}=5$ sublattice points in $\tilde{V}$ which is exactly the minimum number of points required, according to~(\ref{eq:vtilde}). 

With equality in~(\ref{eq:vtilde}) we obtain a region that contains the exact number of sublattice points required to construct $N$ tuples for each of the $N$ $\lambda_0$ points in $V_\pi(0)$. 
According to~(\ref{eq:costfunctional2}), a central lattice point should be assigned that $K$-tuple where a weighted average of any subset of the elements of the $K$-tuple is as close as possible to the central lattice point. 
The optimal assignment of $K$-tuples to central lattice points can be formulated and solved as a linear assignment problem~\cite{west:2001}.  

\subsubsection{Shift-Invariance by use of Cosets}
By centering $\tilde{V}$ around each $\lambda_0\in\Lambda_s\cap V_\pi(0)$, we make sure that the map $\alpha$ is shift-invariant. However, this also means that all $K$-tuples have their first coordinate (i.e.\ $\lambda_0$) inside $V_\pi(0)$. To be optimal this restriction must be removed which is easily done by considering all cosets of each $K$-tuple. The coset\index{coset} of a fixed $K$-tuple, say $t=(\lambda_0,\lambda_1,\dots,\lambda_{K-1})$ where $\lambda_0\in \Lambda_s\cap V_\pi(0)$ and $(\lambda_1,\dots,\lambda_{K-1}) \in \Lambda_s^{K-1}$, is given by $\mathrm{Coset}(t)=\{t+\lambda_\pi : \forall \lambda_\pi\in\Lambda_\pi \}$ . $K$-tuples in a coset are distinct modulo $\Lambda_\pi$ and by making sure that only one member from each coset is used, the shift-invariance property is preserved. In general it is sufficient to consider only those $\lambda_\pi$ product lattice points that are close to $V_\pi(0)$, e.g.\ those points whose Voronoi cell touches $V_\pi(0)$. The number of such points is given by the kissing-number\index{kissing-number} $\mathfrak{K}(\Lambda_\pi)$ of the particular lattice~\cite{conway:1999}.

\subsubsection{Dimensionless Expansion Factor $\psi_L$}\label{sec:psiL}
\begin{figure}[ht]
\psfrag{l0}{$\lambda_0$}
\psfrag{l1}{$\lambda_1$}
\psfrag{l2}{$\lambda_2$}
\psfrag{l1p}{$\lambda_1'$}
\psfrag{l2p}{\raisebox{-1mm}{$\lambda_2'$}}
\psfrag{vtilde}{$\tilde{V}$}
\psfrag{K=3}{$K=3$}
\psfrag{N=81}{$N=81$}
\psfrag{Nt=9}{$\tilde{N}=9$}
\begin{center}
\includegraphics{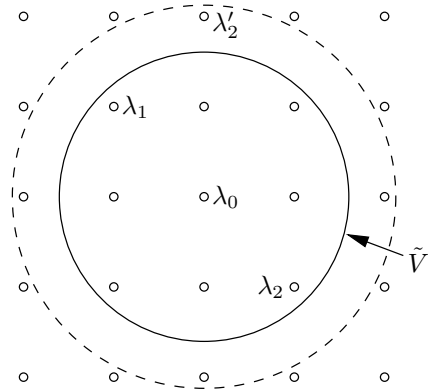}
\end{center}
\caption{The region $\tilde{V}$ is here centered at the point $\lambda_0$. Notice that the distance between $\lambda_1$ and $\lambda_2$ is about twice the maximum distance from $\lambda_0$ to any point in $\Lambda_s\cap \tilde{V}$. The dashed circle illustrates an enlargement of $\tilde{V}$.}
\label{fig:tuples}
\end{figure}

Centering $\tilde{V}$ around $\lambda_0$ points causes a certain asymmetry in the pairwise distances of the elements within a $K$-tuple. Since the region is centered around $\lambda_0$ the maximum pairwise distances between $\lambda_0$ and any other sublattice point will always be smaller than the maximum pairwise distance between any two sublattice points not including $\lambda_0$. 
This can be seen more clearly in Fig.~\ref{fig:tuples}. Notice that the distance between the pair of points labeled $(\lambda_1,\lambda_2)$ is twice the distance of that of the pair $(\lambda_0,\lambda_1)$ or $(\lambda_0,\lambda_2)$. However, by slightly increasing the region $\tilde{V}$ to also include $\lambda'_2$ other tuples may be made, which actually have a lower pairwise distance than the pair $(\lambda_1,\lambda_2)$. For this particular example, it is easy to see that the $3$-tuple $t=(\lambda_0,\lambda_1,\lambda_2)$ has a greater SPSD\index{sum of pairwise squared distances} than the $3$-tuple $t'=(\lambda_0,\lambda_1,\lambda'_2)$.

For each $\lambda_0\in V_\pi(0)$ we center a region $\tilde{V}$ around the point, and choose those $N$ $K$-tuples, that give the smallest SPSD\@. By expanding $\tilde{V}$ new $K$-tuples can be constructed that might have a lower SPSD than the SPSD of the original $N$ $K$-tuples. However, the distance from $\lambda_0$ to the points farthest away increases as $\tilde{V}$ increases. Since we only need $N$ $K$-tuples, it can be seen that $\tilde{V}$ should never be larger than twice the lower bound in~(\ref{eq:vtilde}) because then the distance from the center to the boundary of the enlarged $\tilde{V}$ region is greater than the maximum distance between any two points in the $\tilde{V}$ region that reaches the lower bound. 
In order to theoretically describe the performance of the quantizers, we introduce a dimensionless expansion factor\index{$\psi_L$} $1\leq \psi_L <2$
which describes how much $\tilde{V}$ must be expanded from the theoretical lower bound~(\ref{eq:vtilde}), to make sure that $N$ optimal $K$-tuples can be constructed by combining sublattice points within a region $\tilde{V}$. 

For the case of $K=2$ we always have $\psi_L=1$ independent of the dimension $L$ so it is only in the case $K\geq 3$ that we need to find expressions for $\psi_L$.
\begin{theorem}\label{theo:psiLK3}
For the case of $K=3$ and any odd $L$, the dimensionless expansion factor is given by 
\begin{equation}
\psi_L=\left(\frac{\omega_L}{\omega_{L-1}}\right)^{1/2L}\left(\frac{L+1}{2L}\right)^{1/2L}\beta_L^{-1/2L},
\end{equation}
where $\omega_L$ is the volume of an $L$-dimensional unit sphere and $\beta_L$ is given by
\begin{equation}\label{eq:betaL}
\begin{split}
\beta_L&=
\sum_{n=0}^{\frac{L+1}2}\binom{\frac{L+1}2}{n}2^{\frac{L+1}2-n}(-1)^n 
\sum_{k=0}^{\frac{L-1}2} \frac{\left(\frac{L+1}2\right)_k \left(\frac{1-L}2\right)_k}{\left(\frac{L+3}2\right)_k\, k!}\\
&\quad \times\sum_{j=0}^k\binom{k}{j}\left(\frac{1}2\right)^{k-j}(-1)^j\left(\frac{1}{4}\right)^j \frac{1}{L+n+j}.
\end{split}
\end{equation}
\end{theorem}
\begin{proof}
See Appendix~\ref{app:theo:psiLK3}.
\end{proof}

For the interesting case of $L\rightarrow \infty$ we have the following theorem.
\begin{theorem}\label{theo:psiLinf}
For $K=3$ and $L\rightarrow \infty$ the dimensionless expansion factor\index{$\psi_L$} $\psi_L$ is given by
\begin{equation}\label{eq:psiinfty}
\psi_{\infty} = \left(\frac{4}{3}\right)^{1/4}.
\end{equation}
\end{theorem}
\begin{proof}
See Appendix~\ref{app:theo:psiLinf}.
\end{proof}

Table~\ref{tab:psiL} lists\footnote{Theorem~\ref{theo:psiLK3} is only valid for $L$ odd. However, in the proof of Theorem~\ref{theo:psiLK3} it is straightforward to replace the volume of spherical caps by standard expressions for circle cuts in order to obtain $\psi_2$.} $\psi_L$\index{$\psi_L$} for $K=3$ and different values of $L$ and it may be noticed that $\psi_\infty = \sqrt{\psi_1}$. 
\begin{table}[ht]
\begin{center}
\begin{tabular}{c|l}\hline
$L$   & $\psi_L$ \\ \hline
1     & $1.1547005\cdots$ \\
2     & $1.1480804\cdots$ \\
3     & $1.1346009\cdots$ \\
5     & $1.1240543\cdots$ \\
7     & $1.1172933\cdots$ \\
9     & $1.1124896\cdots$ \\
11    & $1.1088540\cdots$ \\
13    & $1.1059819\cdots$ \\ \hline
\end{tabular}\quad
\begin{tabular}{c|l}\hline
$L$   & $\psi_L$ \\ \hline
15    & $1.1036412\cdots$ \\
17    & $1.1016878\cdots$ \\
19    & $1.1000271\cdots$ \\
21    & $1.0985938\cdots$ \\
51    & $1.0883640\cdots$ \\
71    & $1.0855988\cdots$ \\
101   & $1.0831849\cdots$ \\
$\infty$ & $1.0745699\dots$ \\ \hline
\end{tabular}
\end{center}
\caption{$\psi_L$ values obtained by use of Theorems~\ref{theo:psiLK3} and~\ref{theo:psiLinf} for $K=3$.}
\label{tab:psiL}
\end{table}

\begin{remark}
In order to extend these results to $K>3$ it follows from the proof of Theorem~\ref{theo:psiLK3} that we need closed-form expressions for the volumes of all the different convex regions that can be obtained by $K-1$ overlapping spheres. With such expressions it should be straightforward to find $\psi_L$ for any $K$. However, the analysis of $\psi_L$ for the case of $K=3$ (as given in the proof of Theorem~\ref{theo:psiLK3}) is constructive in the sense that it reveals how $\psi_L$ can be numerically estimated for any $K$ and $L$, see Appendix~\ref{app:estimating_psi}.
\end{remark}

\begin{remark}
In order to achieve the shift-invariance property of the index-assignment algorithm, we impose a restriction upon $\lambda_0$ points. Specifically, we require that $\lambda_0\in V_\pi(0)$ so that the first coordinate of any $K$-tuple is within the region $V_\pi(0)$. To avoid excluding $K$-tuples that have their first coordinate outside $V_\pi(0)$ we form cosets of each $K$-tuple and allow only one member from each coset\index{coset} to be assigned to a central lattice point within $V_\pi(0)$. This restriction, which is only put on $\lambda_0\in \Lambda_s$, might cause a bias towards $\lambda_0$ points. 
However, it is easy to show that, asymptotically as $N\rightarrow \infty$, any such bias can be removed. For the case of $K=2$ we can use similar arguments as used in~\cite{diggavi:2002}, and for $K>2$, as shown in Chapter~\ref{chap:asym}, the number of $K$-tuples affected by this restriction is small compared to the number of $K$-tuples not affected. 
So for example this means that we can enforce similar restrictions on all sublattice points, which, asymptotically as $N\rightarrow\infty$, will only reduce the number of $K$-tuples by a neglectable amount. And as such, any possible bias towards the set of points $\lambda_0\in\Lambda_s$ is removed. 

As mentioned above, the $K$-tuples need to be assigned to central lattice points within $V_\pi(0)$. This is a standard linear assignment problem where a cost measure is minimized. However, solutions to linear assignment problems are generally not unique. Therefore, there might exist several labelings, which all yield the same cost, but exhibit a different amount of asymmetry. Theoretically, exact symmetry may then be obtained by e.g.\ time sharing\index{time sharing} through a suitable mixing of labelings. 
In practice, however, any scheme would use a finite $N$ (and finite rates). In addition, for many applications, time sharing\index{time sharing} is inconvenient. In these non-asymptotic cases we cannot guarantee exact symmetry. To this end, we have provided a few examples that assess the distortions obtained from practical experiments, see Section~\ref{sec:num} (Tables~\ref{tab:kappa1} and~\ref{tab:kappa2}).
\end{remark}

\section{High-Resolution Analysis}\label{sec:highrate}
In this section we derive high-resolution approximations for the expected distortion. For this high-resolution analysis we let $N\rightarrow \infty$ and $\nu_s\rightarrow 0$. Thus, the index $N$ of the sublattices increases, but the actual volumes of the Voronoi cells shrink.

\subsection{Total Expected Distortion}
We wish to obtain an analytical expression for the expected distortion given by~(\ref{eq:expdist}). In order to achieve this we first relate the sum of distances between pairs of sublattice points to $G(S_L)$, the dimensionless normalized second moment of an $L$-sphere. This is done by Proposition~\ref{prop:riemann2}.

\begin{proposition}\label{prop:riemann2}
For $K=2$ and asymptotically as $N\rightarrow \infty$ and $\nu_s\rightarrow 0$, as well as for $K=3$ and asymptotically as $N,L\rightarrow \infty$ and $\nu_s\rightarrow 0$, we have for any pair $(\lambda_i,\lambda_j),\ i,j=0,\dots,K-1,\ i\neq j$,
\begin{equation*}
\frac{1}{L}\sum_{\lambda_c\in V_\pi(0)}\!\!\!\| \alpha_i(\lambda_c)-\alpha_j(\lambda_c)\|^2
= G(S_L)\psi_L^{2}N^2N^{2K/L(K-1)}\nu^{2/L}.
\end{equation*}
\end{proposition}
\begin{proof}
See Appendix~\ref{app:riemann2}.
\end{proof}

\begin{conjecture}\label{con:riemann2}
Proposition~\ref{prop:riemann2} is true also for $K>3$ asymptotically as $N,L\rightarrow\infty$ and $\nu_s\rightarrow 0$. 
\end{conjecture}

\begin{remark}
Arguments supporting conjecture~\ref{con:riemann2} are given in Appendix \ref{app:riemann2}. 
\end{remark}
\begin{remark}
In Appendix~\ref{app:riemann2} we also present an exact expression for Proposition \ref{prop:riemann2} for $K=3$ and finite $L$.
\end{remark}

Recall that we previously showed that by use of Theorem~\ref{theo:sums} it is possible to split~(\ref{eq:expdist}) into two terms; one that describes the distance from a central lattice point to the centroid of its associated $K$-tuple and another which describes the sum of pairwise squared distances (SPSD)\index{sum of pairwise squared distances} between elements of the $K$-tuples. To determine which of the two terms that are dominating we present the following proposition:

\begin{proposition}\label{prop:growthriemann2}
For $N\rightarrow \infty$ and $2\leq K<\infty$ we have
\begin{equation}
\frac{\sum_{\lambda_c\in V_\pi(0)}\left\| \lambda_c - \frac{1}{K}\sum_{i=0}^{K-1}\lambda_i\right\|^2}
{\sum_{\lambda_c\in V_\pi(0)}\sum_{i=0}^{K-2}\sum_{j=i+1}^{K-1}\| \lambda_i - \lambda_j\|^2} \rightarrow 0.
\end{equation}
\end{proposition}

\begin{proof}
See Appendix~\ref{app:growthriemann2}.
\end{proof}

The expected distortion~(\ref{eq:expdist}) can by use of Theorem~\ref{theo:sums} be written as
\begin{equation}\label{eq:da}
\begin{split}
D_a^{(K,\kappa)}&\approx (1-p)^{\kappa}p^{K-\kappa}\!
\times\!\!\left(\!\!\!
\binom{K}{\kappa}D_c 
+ \frac{1}{L}\frac{1}{N^2}
\sum_{l\in \mathcal{L}^{(K,\kappa)}}\sum_{\lambda_c\in V_\pi(0)}
\left\|\lambda_c - \frac{1}{\kappa}\sum_{j=0}^{\kappa-1} \lambda_{l_j}\right\|^2\right) \\ 
&=
(1-p)^{\kappa}p^{K-\kappa}\binom{K}{\kappa}\times\Bigg(D_c +
\frac{1}{L}\frac{1}{N^2}\sum_{\lambda_c\in V_\pi(0)}
\left(
\left\| \lambda_c - \frac{1}{K}\sum_{i=0}^{K-1}\lambda_i\right\|^2
\right.
\\
&\quad+\left.
\left(\frac{K-\kappa}{K^2\kappa(K-1)}\right)
\sum_{i=0}^{K-2}\sum_{j=i+1}^{K-1}\| \lambda_i - \lambda_j\|^2 
\right)\Bigg).
\end{split}
\end{equation}

By use of Proposition~\ref{prop:riemann2} (as an approximation that becomes exact for $L\rightarrow \infty$), Proposition~\ref{prop:growthriemann2} and Eq.~(\ref{eq:dcG}) it follows that~(\ref{eq:da}) can be written as
\begin{align}\notag
D_a^{(K,\kappa)}&\approx (1-p)^{\kappa}p^{K-\kappa}\binom{K}{\kappa} \\
&\quad \times\!\!\left(\!D_c +
\frac{1}{L}\frac{1}{N^2}\!\!\sum_{\lambda_c\in V_\pi(0)}\!\!
\left(\frac{K-\kappa}{K^2\kappa(K-1)}\right)
\sum_{i=0}^{K-2}\sum_{j=i+1}^{K-1}\| \lambda_i - \lambda_j\|^2 
\right)\\ \notag
&\approx (1-p)^{\kappa}p^{K-\kappa}\binom{K}{\kappa} \\ \label{eq:distkleqK}
&\quad\times\left(G(\Lambda_c)\nu^{2/L} +
\left(\frac{K-\kappa}{2K\kappa}\right)
G(S_L)\psi_L^{2}N^{2K/L(K-1)}\nu^{2/L}
\right).
\end{align}
The second term in~(\ref{eq:distkleqK}), that is
\begin{equation}\label{eq:sidedistkappa}
\left(\frac{K-\kappa}{2K\kappa}\right)
G(S_L)\psi_L^{2}N^{2K/L(K-1)}\nu^{2/L}
\end{equation}
is the dominating term for $\kappa<K$ and $N\rightarrow \infty$ and describes the side distortion\index{side distortion} due to reception of any $\kappa<K$ descriptions. Observe that this term is only dependent upon $\kappa$ through the coefficient $\frac{K-\kappa}{2K\kappa}$.

The total expected distortion $D_a^{(K)}$ is obtained from~(\ref{eq:distkleqK}) by summing over $\kappa$ including the cases where $\kappa=0$ and $\kappa=K$, which leads to
\begin{equation}\label{eq:adistopt}
\begin{split}
D_a^{(K)} &\approx \hat{K}_1G(\Lambda_c)\nu^{2/L} +\hat{K}_2 
G(S_L)\psi_L^{2}N^{2K/L(K-1)}\nu^{2/L} + p^K E\|X\|^2/L,
\end{split}
\end{equation}
where $\hat{K}_1$ is given by
\begin{equation}
\begin{split}
\hat{K}_1&=\sum_{\kappa=1}^K\binom{K}{\kappa}p^{K-\kappa}(1-p)^\kappa\\
&= 1-p^K,
\end{split}
\end{equation}
and $\hat{K}_2$ is given by
\begin{equation}\label{eq:K2}
\hat{K}_2=\sum_{\kappa=1}^K\binom{K}{\kappa}p^{K-\kappa}(1-p)^\kappa\frac{K-\kappa}{2\kappa K}.
\end{equation}

Using~(\ref{eq:Rc}) and~(\ref{eq:Rs}) we can write $\nu$ and $N$ as a function of differential entropy and side entropies, that is
\begin{equation}\label{eq:nu2L}
\nu^{2/L}=2^{2(\bar{h}(X)-R_c)},
\end{equation}
and
\begin{equation}\label{eq:N2KL}
N^{2K/L(K-1)}=2^{\frac{2K}{K-1}(R_c-R_s)},
\end{equation}
where $R_s=R_i,i=0,\dotsc,K-1$ denotes the side description rate.
Inserting~(\ref{eq:nu2L}) and~(\ref{eq:N2KL}) in~(\ref{eq:adistopt}) makes it possible to write the expected distortion as a function of entropies
\begin{equation}\label{eq:daopt}
\begin{split}
D_a^{(K)} &\approx \hat{K}_1G(\Lambda_c)2^{2(\bar{h}(X)-R_c)}\\
&\quad+ \hat{K}_2\psi_L^{2}G(S_L) 2^{2(\bar{h}(X)-R_c)}2^{\frac{2K}{K-1}(R_c-R_s)} 
+ p^K E\|X\|^2/L,
\end{split}
\end{equation}
where we see that the distortion due to the side quantizers depends only upon the scaling (and dimension) of the sublattice but not upon which sublattice is used. Thus, the side distortions\index{side distortion} can be expressed through the dimensionless normalized second moment of a sphere.

\subsection{Optimal $\nu$, $N$ and $K$}\label{sec:opt}
We now derive expressions for the optimal $\nu$, $N$ and $K$. Using these values we are able to construct the lattices $\Lambda_c$ and $\Lambda_s$. The optimal index assignment is hereafter found by using the approach outlined in Section~\ref{sec:label}. These lattices combined with their index assignment completely specify an optimal entropy-constrained MD-LVQ system.

In order for the entropies of the side descriptions to be equal to the target entropy\index{target rate} $R_T/K$, we rewrite~(\ref{eq:Rs}) and get
\begin{equation}\label{eq:tau}
N\nu = 2^{L(\bar{h}(X)-R_T/K)} \triangleq \tau,
\end{equation}
where $\tau$ is constant. 
The expected distortion $D_a^{(K)}$~(\ref{eq:daopt}) may now be expressed as a function of $\nu$,
\begin{equation}
\begin{split}
D_a^{(K)}&=\hat{K}_1G(\Lambda_c)\nu^{2/L}\\  
&+\hat{K}_2\psi_L^{2}G(S_L) \nu^{2/L}\nu^{-\frac{2K}{L(K-1)}}\tau^{\frac{2K}{L(K-1)}}+ p^K E\|X\|^2/L.
\end{split}
\end{equation}
Differentiating w.r.t.\ $\nu$ and equating to zero gives,
\begin{equation}\label{eq:diffda}
\begin{split}
0&=\frac{\partial D_a^{(K)}}{\partial \nu} \\
&=
\frac{2}{L}\hat{K}_1G(\Lambda_c)\frac{\nu^{2/L}}{\nu}
+ \left(\frac{2}{L} - \frac{2}{L}\frac{K}{K-1}\right)\hat{K}_2\psi_L^{2}G(S_L) \frac{\nu^{2/L}}{\nu}\nu^{-\frac{2K}{L(K-1)}}\tau^{\frac{2K}{L(K-1)}},
\end{split}
\end{equation}
from which we obtain the optimal value of $\nu$
\begin{equation}\label{eq:optnu}
\nu = \tau\left(\frac{1}{K-1} \frac{\hat{K}_2}{\hat{K}_1} \frac{G(S_L)}{G(\Lambda_c)}\psi_L^{2}\right)^{\frac{L(K-1)}{2K}}.
\end{equation}
The optimal $N$ follows easily by use of~(\ref{eq:tau})
\begin{equation}\label{eq:optN}
N = \left((K-1)\frac{\hat{K}_1}{\hat{K}_2} \frac{G(\Lambda_c)}{G(S_L)}\frac{1}{\psi_L^{2}}\right)^{\frac{L(K-1)}{2K}}.
\end{equation}
Eq.~(\ref{eq:optN}) shows that the optimal redundancy $N$ is, for fixed $K$, independent of the sublattice as well as the target entropy\index{target rate}.

For a fixed $K$ the optimal $\nu$ and $N$ are given by~(\ref{eq:optnu}) and~(\ref{eq:optN}), respectively, and the optimal $K$ can then easily be found by evaluating~(\ref{eq:adistopt}) for various values of $K$, and choosing the one that yields the lowest expected distortion. The optimal $K$ is then given by
\begin{equation}\label{eq:optK}
K_\text{opt}=\arg\,\min_{K} D_a^{(K)},\quad K=1,\dots, K_\text{max},
\end{equation}
where $K_\text{max}$ is a suitable chosen positive integer. In practice $K$ will always be finite and furthermore limited to a narrow range of integers, which makes the complexity of the minimization approach, given by~(\ref{eq:optK}), negligible.

\section{Construction of Practical Quantizers}
\label{sec:construction}

\subsection{Index Values}
Eqs.~(\ref{eq:optnu}) and~(\ref{eq:optN}) suggest that we are able to continuously trade off central versus side-distortions by adjusting $N$ and $\nu$ according to the packet-loss probability. 
This is, however,  not the case, since certain constraints must be imposed on $N$. 
First of all, since $N$ denotes the number of central lattice points within each Voronoi cell of the sublattice, it must be integer and positive. 
Second, we require the sublattice to be geometrically similar to the central lattice. Finally, we require the sublattice to be a clean sublattice, so that no central lattice points are located on boundaries of Voronoi cells of the sublattice. This restricts the amount of admissible index values for a particular lattice to a discrete set, cf.\ Section~\ref{sec:admindexvalues}.

\begin{figure}[ht]
\begin{center}
\psfrag{Index values [N]}{\small Index values $[N]$}
\psfrag{Packet loss probability [\%]}{\small Packet-loss probability $[\%]$}
\psfrag{K=2: Theoretical N}{\scriptsize $K\!\!=\!2$: Theoretical $N$}
\psfrag{K=3: Theoretical N}{\scriptsize $K\!\!=\!3$: Theoretical $N$}
\psfrag{K=4: Theoretical N}{\scriptsize $K\!\!=\!4$: Theoretical $N$}
\psfrag{Admissible N}{\scriptsize Admissible $N$}
\includegraphics[width=10cm]{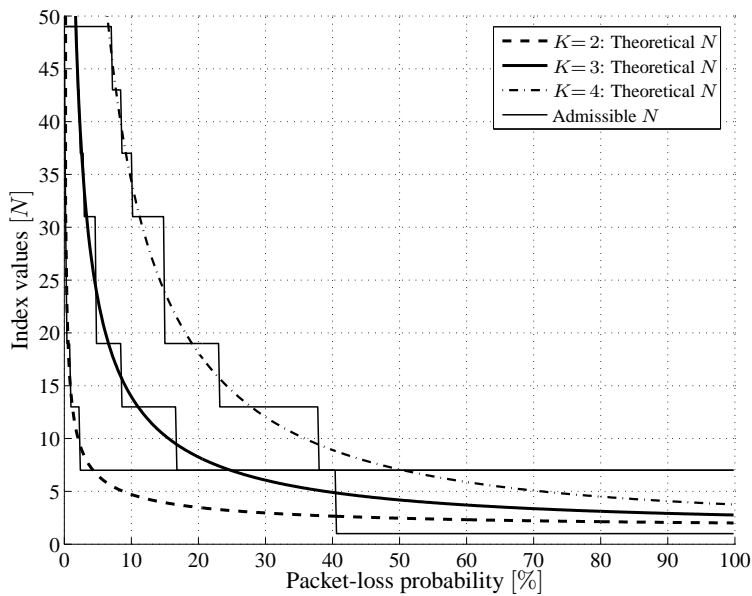}
\caption{Theoretical optimal index values for the $A_2$ quantizer as a function of the packet-loss probability. The thin solid lines are obtained by restricting the theoretical optimal index values given by~(\ref{eq:optN}) to be (optimal) admissible values given by the set $N=\{1,7,13,19,31,37,43,49,\dotsc\}$. The optimal admissible index values are those that minimize~(\ref{eq:adistopt}) for a given $p$.}
\label{fig:index_values}
\end{center}\end{figure}

Fig.~\ref{fig:index_values} shows the theoretically optimal index values (i.e.\ ignoring the fact that $N$ belongs to a discrete set) for the $A_2$ quantizer, given by~(\ref{eq:optN}) for $\psi_L=1,1.1481$ and $1.1762$ corresponding to $K=2,3$ and $4$, respectively.\footnote{The value $\psi_L=1.1762$ for $K=4$ is estimated numerically by using the method outlined in Appendix~\ref{app:estimating_psi}.} Also shown are the theoretical optimal index values when restricted to admissible index values. 
Notice that the optimal index value $N$ increases for increasing number of descriptions. 
This is to be expected since a higher index value leads to less redundancy;
this redundancy reduction, however, is balanced out by the redundancy increase resulting from the added number of descriptions.
%
%
In \cite{ostergaard:2004} we observed that for a two-description system, usually only very few index values would be used (assuming a certain minimum packet-loss probability). Specifically, for the two-dimensional $A_2$ quantizer, only $N\in\{1,7,13\}$ was used, while for higher dimensional quantizers greater index values would be used. However, here we see that by increasing the number of descriptions beyond $K=2$, it is optimal to use greater index values which adds more flexibility to the scheme.

From Fig.~\ref{fig:index_values} it can be seen that when the continuous optimal index value is rounded to the optimal admissible index value it is always the closest one from either below or above. This means that, at least for the $A_2$ lattice, the optimal admissible index value is found by considering only the two values closest to the continuous index value, and using the one that minimizes~(\ref{eq:adistopt}).

\subsection{Constructing $K$-tuples}\label{sec:summarize_ktuples}
The design procedure for constructing $K$-tuples as described in Section~\ref{sec:ktuples} can be summarized as follows:
\begin{enumerate}
\item Center a sphere $\tilde{V}$ at each $\lambda_0\in \Lambda_s\cap V_\pi(0)$ and construct all possible $K$-tuples $(\lambda_0,\dotsc,\lambda_{K-1})$ where $\lambda_i\in\Lambda_s, i=1,\dotsc,K-1$. This makes sure that all $K$-tuples have their first coordinate $(\lambda_0)$ inside $V_\pi(0)$ and they are therefore shift invariant\index{shift invariance}. We will only use $K$-tuples whose elements satisfy $\|\lambda_i-\lambda_j\|\leq r, \forall i,j\in \{0,\dotsc,K-1\}$, where $r$ is the radius of $\tilde{V}$. Make $\tilde{V}$ large enough so that at least $N$ distinct $K$-tuples are found for each $\lambda_0$.
\item Construct cosets\index{coset} of each $K$-tuple.
\item The $N^2$ central lattice points in $\Lambda_c\cap V_\pi(0)$ must now be matched to distinct $K$-tuples. This is a standard linear assignment problem where only one member from each coset is (allowed to be) matched to a central lattice point in $V_\pi(0)$.
\end{enumerate}
The restriction $\|\lambda_i-\lambda_j\|\leq r$ from step 1) which is used to avoid bias towards any of the sublattices, reduces the number of $K$-tuples that can be constructed within the sphere $\tilde{V}$. To be able to form $N$ $K$-tuples it is therefore necessary to use a sphere $\tilde{V}$ with a volume larger than the lower bound~(\ref{eq:vtilde}). 
This enlargement is exactly given by $\psi_L$. As such, for each $\lambda_0$, we form (at least) $N$ $K$-tuples and these $K$-tuples are the ones having minimum norm. We show later (see Lemma~\ref{lem:equivktuples} and its proof) that we actually form all such $K$-tuples of minimal norm which implies that no other $K$-tuples can improve the SPSD\index{sum of pairwise squared distances}.

\subsection{Assigning $K$-Tuples to Central Lattice Points}
In order to assign the set of $K$-tuples to the $N^2$ central lattice points we solve a linear assignment problem\index{linear assignment}. 
However, for large $N$, the problem becomes difficult to solve in practice.
To solve a linear assignment problem or more specifically a bipartite matching problem\index{bipartite matching problem}, one can make use of the Hungarian\index{Hungarian method} method~\cite{kuhn:1955}, which has complexity of cubic order. Hence, if the Hungarian method is used to solve the assignment problem the complexity is of order $O(N^6)$. 
We would like point out that letting $N_\pi=N^2$ is a convenient choice, which is valid for any lattice. However, it is possible to let $N_\pi=N\xi$, where both $N$ and $\xi$ are admissible index values. In this case $N_\pi$ is also guaranteed to be an admissible index value by Lemma~\ref{lem:productindex}. If $\xi=1$ then $N_\pi=N$, which is a special case where $V_\pi(0)$ contains a single sublattice point $\lambda_0$ of $\Lambda_s$.\footnote{Practical experiments have shown that having too few sublattice points in $V_\pi(0)$ leads to a poor index assignment. Theoretically, we do not exclude the possibility that $N_\pi=N$, since we only require that $\tilde{V}$ contains a large number of sublattice points but such a contraint is not imposed on $V_\pi(0)$. However, in the following chapter, where we consider the asymmetric case (so there are several index values), the special case is not allowed.} 
With $N_\pi=N\xi$, the complexity is reduced to $\mathcal{O}(N^3)$. 

Vaishampayan et al.\ observed in~\cite{servetto:1999,vaishampayan:2001} that the number of central lattice points to be labeled can be reduced by exploiting symmetries in the lattices. For example, one can form the quotient $\mathcal{J}$-module\index{quotient!module} $\Lambda/\Lambda_\pi$ and only label representatives of the orbits of $\Lambda/\Lambda_\pi/\Gamma$, where $\Gamma$ is a group of automorphisms, cf.\ Chapter~\ref{chap:lattice_theory}.
While only two descriptions were considered in~\cite{vaishampayan:2001} it is straight-forward to show that their idea also works in our design for an arbitrary number of descriptions. This is because we design the sublattices and product lattices as described in Chapter~\ref{chap:lattice_theory}, hence the notion of quotient modules and group action are well defined.
Since the order of the group $\Gamma$ depends on the lattices but is independent of $N$, the complexity reduction by exploiting the symmetry of the quotient module is a constant multiplicative factor which disappears in the order notation. 

Alternatively, Huang and Wu~\cite{huang:2006} recently showed that for certain low dimensional lattices it is possible to avoid the linear assignment problem by applying a greedy algorithm\index{index-assignment!greedy}, without sacrificing optimality. 
The complexity of the greedy approach is on the order of $\mathcal{O}(N)$, which is a substantial improvement for large $N$.

In the present work we show that the assignment problem can always be posed and solved as a bipartite matching problem. This holds for any lattice in any dimension and it also holds in the asymmetric case to be discussed in Chapter~\ref{chap:asym}. 
In a practical situation it might, however, be convenient to compute the assignments offline and tabulate for further use.

\subsection{Example of an Assignment}
\begin{figure}[ht]
\begin{center}
\includegraphics[width=7cm]{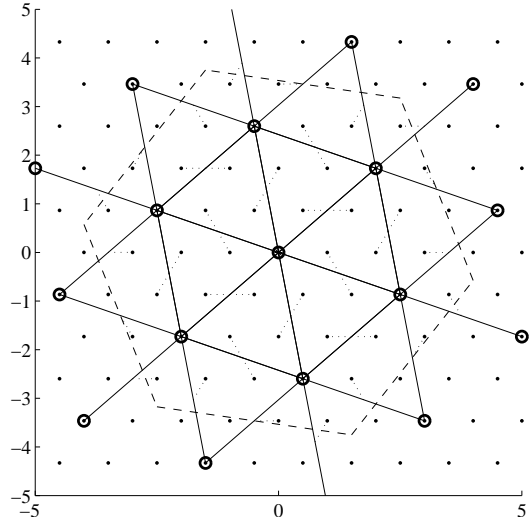}
\end{center}
\caption{A central lattice $\Lambda_c$ (dots) based on the $A_2$ lattice and a sublattice $\Lambda_s$ (circles) of index 7. The hexagonal region (dashed lines) describes $V_\pi(0)$. The solid lines connect pairs of sublattice points (also called 2-tuples or edges) and the dotted lines connect each edge to a central lattice point. A total of 49 edges (some overlaping each other) are shown and these edges are associated with the 49 central lattice points contained within $V_\pi(0)$.}
\label{fig:assign_example}
\end{figure}

In the following we show a simple assignment for the case of $K=2,N=7$ and the $A_2$ lattice. Since $N=7$ we have $N_\pi=49$ and as such there is 49 central lattice points within $V_\pi(0)$, see Fig.~\ref{fig:assign_example}. The individual assignments are also shown in Table~\ref{tab:assign_example}.

\begin{center}
\begin{table}[ht]
\mbox{%
\small
\begin{tabular}{ccc} \hline
$\lambda_c$ & $\lambda_0$ & $\lambda_1$ \\ \hline
(-1, 3.46) &  (-1, 5.2) &  (-0.5, 2.6) \\
(0, 3.46)  &  (1.5, 4.33)&  (-0.5, 2.6)\\
(-1.5, 2.6) & (-3, 3.46)&  (-0.5, 2.6)\\
(-0.5, 2.6) & (-0.5, 2.6)&  (-0.5, 2.6) \\
(0.5, 2.6) & (2, 1.73)&  (-0.5, 2.6) \\
(1.5, 2.6) & (1.5, 4.33)&   (2, 1.73) \\
(2.5, 2.6)  & (4, 3.46)&   (2, 1.73) \\
(-3, 1.73) & (-3, 3.46)&  (-2.5, 0.87) \\
(-2, 1.73) & (-2.5, 0.87)&  (-0.5, 2.6)\\
(-1, 1.73) & (-0.5, 2.6)&  (-2.5, 0.87)\\
(0, 1.73)  & (0, 0)          &  (-0.5, 2.6)\\
(1, 1.73)  & (-0.5, 2.6) &   (2, 1.73) \\
(2, 1.73)  & (2, 1.73) &   (2, 1.73) \\
(3, 1.73)  &  (4.5, 0.87) &   (2, 1.73) \\
(-3.5, 0.87) &(-5, 1.73) &  (-2.5, 0.87) \\
(-2.5, 0.87) &(-2.5, 0.87) &  (-2.5, 0.87) \\
(-1.5, 0.87)& (0, 0)      &  (-2.5, 0.87) \\
(-0.5, 0.87)& (-0.5, 2.6) &       (0, 0)          \\
(0.5, 0.87)&  (0, 0)       &   (2, 1.73) \\
(1.5, 0.87)&   (2, 1.73) &       (0, 0)          \\
(2.5, 0.87)&   (2.5, -0.87) &   (2, 1.73) \\
(-3, 0 )&  (-4.5, -0.87) &  (-2.5, 0.87) \\
(-2, 0)&  (-2, -1.73)  &  (-2.5, 0.87) \\
(-1, 0)&  (-2.5, 0.87) &       (0, 0)          \\ \hline
\end{tabular}
\hspace{2mm}
\raisebox{-2.1mm}{%
\begin{tabular}{ccc} \hline
$\lambda_c$ & $\lambda_0$ & $\lambda_1$ \\ \hline
  (0, 0) &   (0, 0)     &    (0, 0)              \\
  (1, 0) &  (2.5, -0.87) &       (0, 0)           \\
  (2, 0) &  (2, 1.73) &  (2.5, -0.87)\\
  (3, 0) &  (4.5, 0.87) &  (2.5, -0.87)\\
  (-2.5, -0.87) & (-2.5, 0.87) & (-2, -1.73)\\
  (-1.5, -0.87) & (-2, -1.73)  &  (0, 0)          \\
  (-0.5, -0.87) & (0,0)        & (-2, -1.73)\\
  (0.5, -0.87) &  (0.5, -2.6) &       (0, 0)          \\
  (1.5, -0.87) &  (0,0)       &  (2.5, -0.87)\\
  (2.5, -0.87) &  (2.5, -0.87) &  (2.5, -0.87)\\
  (3.5, -0.87) &  (5, -1.73) &  (2.5, -0.87)\\
  (-3, -1.73) &  (-4.5, -0.87) & (-2, -1.73)\\
  (-2, -1.73) &  (-2, -1.73) & (-2, -1.73)\\
  (-1, -1.73) &  (0.5, -2.6) & (-2, -1.73)\\
  (0, -1.73) &       (0, 0)           &  (0.5, -2.6)\\
  (1, -1.73) &  (0.5,  -2.6) &  (2.5, -0.87)\\
  (2, -1.73) &  (2.5, -0.87)  &  (0.5, -2.6)\\
  (3, -1.73) &  (3, -3.47)  &  (2.5, -0.87)\\
  (-2.5, -2.6) & (-4, -3.47) & (-2, -1.73)\\
  (-1.5, -2.6) & (-1.5, -4.33) & (-2, -1.73)\\
  (-0.5, -2.6) & (-2, -1.73) &  (0.5, -2.6)\\
  (0.5, -2.6) &  (0.5 - 2.6) &  (0.5, -2.6)\\
  (1.5, -2.6) &  (3, -3.46) &  (0.5, -2.6)\\
  (0, -3.46) &  (-1.5, -4.33) &  (0.5, -2.6)\\
  (1, -3.46) &  (1, -5.2) &  (0.5, -2.6)\\ \hline
\end{tabular}}}
\caption{A complete assignment for the 49 central lattice points contained within $V_\pi(0)$ for the case of $K=2$ and $N=7$.}
\label{tab:assign_example}
\end{table}
\end{center}

The assignments shown in Table~\ref{tab:assign_example} are obtained by using the procedure outlined in Section~\ref{sec:summarize_ktuples}. Since we have $N_\pi=49$ and $K=2$ it follows that we have 7 sublattice points of $\Lambda_s$ within $V_\pi(0)$ (one of them is the origin). Let us denote this set of sublattice points by $E_{\Lambda_s}$.

\begin{enumerate}
\item Center a sphere $\tilde{V}$ at the first element of $E_{\lambda_s}$, i.e.\ the origin. Pick the candidate sublattice points of $\Lambda_s$, i.e.\ those which are contained within $\tilde{V}\cap\Lambda_s$. 
We make sure that the radius of the sphere is so large that it contains $V_\pi(0)$. Thus, we have at least as many sublattice points in $\tilde{V}$ as in $V_\pi(0)$. Then form all possible distinct edges (2-tuples) having the origin ($\lambda_0$) as first coordinate and $\lambda_1\in V_\pi(\lambda_0)\cap \Lambda_s$ as second coordinate. Notice that we have at least $N$ edges. Repeat this for the remaining elements of $E_{\Lambda_s}$ so that we end up having at least $N_\pi$ edges in total. 
\item Form the coset of each edge. Specifically, construct the following set of edges:
\begin{equation}
\text{Coset}(\lambda_0,\lambda_1)=\{ (\lambda_0 + \lambda_\pi,\lambda_1 + \lambda_\pi) : \lambda_\pi \in \Lambda_\pi\}.
\end{equation}
In practice we restrict each coset to contain a finite number of elements. In fact, we usually only require that the cardinality of the cosets is greater than $\mathfrak{K}(\Lambda_s)$, the kissing number of the lattice. The product lattice points we use when constructing the cosets are then the $\mathfrak{K}(\Lambda_s)+1$ points of smallest norm.
\item If we have more edges than central lattice points we introduce ``dummy'' central lattice points. In this way we have an equal amount of edges and central lattice points. 
The assignment of edges to central lattice points is now a straight forward bipartite problem, where the costs of the dummy nodes are set to zero, so that the optimal solutions are not affected. We note that only one element of each coset is used. In this way we preserve the shift invariance property of the assignments. We only keep the $N_\pi$ assignments belonging to \emph{true} central lattice points and as such we discard the assignments (if any) that belong to ``dummy'' nodes.
\end{enumerate}

In Appendix~\ref{app:assignment_example} we show part of a complete assignment of a more complicated example.

\section{Numerical Results}\label{sec:num}
In this section we compare the numerical performances of two-dimensional entropy-constrained MD-LVQ (based on the $A_2$ lattice) to their theoretical prescribed perfor\-man\-ces. 

\subsection{Performance of Individual Descriptions}
In the first experiment we design three-channel MD-LVQ based on the $A_2$ quantizer. We quantize an i.i.d.\ unit-variance Gaussian source which has been blocked into two-dimensional vectors. The number of vectors used in the experiment is $2\cdot10^6$. The entropy of each side description is 5 bit/dim.\ and we vary the index value in the range $31$ -- $67$. The dimensionless expansion factor $\psi_L$ is set to $1.14808$, see Table~\ref{tab:psiL}. The numerical and theoretical distortions when receiving only a single description out of three is shown in Table~\ref{tab:kappa1}. Similarly, Table~\ref{tab:kappa2} shows the distortions of the same system due to reception of two out of three descriptions and Table~\ref{tab:kappa3} shows the performance of the central quantizer when all three descriptions are received. The column labeled ``Avg.'' illustrates the average distortion of the three numerically measured distortions and the column labeled ``Theo.'' describes the theoretical distor\-tions given by~(\ref{eq:sidedistkappa}). It is clear from the tables that the system is symmetric; the achieved distortion depends on the number of received descriptions but is essentially independent of \emph{which} descriptions are used for reconstruction. The numerically measu\-red discrete entropies of the side descriptions are shown in Table~\ref{tab:rates5bit}.
%
%
%
\begin{table}[ht]
\begin{center}
\begin{tabular}{cccccc}
{\small $N$} & {\small $\lambda_0$} &  {\small $\lambda_1$} &  {\small $\lambda_2$} & {\small Avg.} & {\small Theo.} \\ \hline
31 & $-25.6918$ & $-25.6875$ & $-25.6395$ & $-25.6729$ & $-24.8853$ \\
37 & $-24.5835$ & $-24.5324$ & $-24.5404$ & $-24.5521$ & $-24.5011$ \\
43 & $-24.5772$ & $-24.5972$ & $-24.5196$ & $-24.5647$ & $-24.1748$ \\
49 & $-24.2007$ & $-24.2837$ & $-24.2713$ & $-24.2519$ & $-23.8911$ \\
61 & $-23.8616$ & $-23.9011$ & $-23.8643$ & $-23.8757$ & $-23.4155$ \\
67 & $-23.7368$ & $-23.7362$ & $-23.7655$ & $-23.7462$ & $-23.2118$ \\ \hline
\end{tabular}
\caption{Distortion (in dB) due to reception of a single description out of three.}
\label{tab:kappa1}
\end{center}
\end{table}
\begin{table}[ht]
\begin{center}
\begin{tabular}{p{1mm}p{12mm}p{12mm}p{12mm}cc}
{\small $N$} & \mbox{{\small $\frac{1}2(\lambda_0+\lambda_1)$}} & \mbox{{\small $\frac{1}2(\lambda_0+\lambda_2)$}} & \mbox{{\small $\frac{1}2(\lambda_1+\lambda_2)$}} & {\small Avg.} & {\small Theo.} \\ \hline
31 & $-30.7792$ & $-30.7090$ & $-30.7123$ & $-30.7335$ & $-30.9059$ \\
37 & $-29.8648$ & $-29.8430$ & $-29.9472$ & $-29.8850$ & $-30.5217$ \\
43 & $-29.9087$ & $-29.8749$ & $-29.9641$ & $-29.9159$ & $-30.1954$ \\
49 & $-29.6290$ & $-29.5577$ & $-29.6662$ & $-29.6176$ & $-29.9117$ \\
61 & $-29.3076$ & $-29.2185$ & $-29.3715$ & $-29.2992$ & $-29.4361$ \\
67 & $-29.1752$ & $-29.2128$ & $-29.2151$ & $-29.2010$ & $-29.2324$ \\ \hline
\end{tabular}
\caption{Distortion (in dB) due to reception of two descriptions out of three.}
\label{tab:kappa2}
\end{center}
\end{table}

\begin{table}[ht]
\begin{center}
\begin{tabular}{ccc}
{\small $N$} & {\small $\lambda_c$} & {\small Theo.} \\ \hline
31  & $-43.6509$ & $-43.6508$ \\ 
37  & $-44.4199$ & $-44.4192$ \\
43  & $-45.0705$ & $-45.0719$ \\
49  & $-45.6401$ & $-45.6391$ \\
61  & $-46.5879$ & $-46.5905$ \\
67  & $-46.9992$ & $-46.9979$ \\ \hline
\end{tabular}
\caption{Distortion (in dB) due to reception of all three descriptions.}
\label{tab:kappa3}
\end{center}
\end{table}

\begin{table}[ht]
\begin{center}
\begin{tabular}{cccccc}
{\small $N$} & {\small $\lambda_0$} &  {\small $\lambda_1$} &  {\small $\lambda_2$} \\ \hline
31 & $5.0011$ & $5.0012$ & $5.0012$ \\
37 & $4.9925$ & $4.9982$ & $4.9988$ \\
43 & $4.9967$ & $5.0006$ & $5.0006$ \\
49 & $4.9993$ & $5.0004$ & $5.0004$ \\
61 & $5.0018$ & $5.0017$ & $5.0017$ \\
67 & $5.0023$ & $5.0022$ & $5.0022$ \\ \hline
\end{tabular}
\caption{Numerically measured discrete entropies [bit/dim.] for the individual descriptions. Here the target description rate is set to 5 bit/dim.}
\label{tab:rates5bit}
\end{center}
\end{table}

\begin{figure}[ht]
\psfrag{D1}{$\scriptscriptstyle D^{(3,1)}$}
\psfrag{D2}{$\scriptscriptstyle D^{(3,2)}$}
\psfrag{D3-hel}{$\scriptscriptstyle D^{(3,3)}$}
\psfrag{rho}{\small $\rho_q$}
\begin{center}
\includegraphics[width=10cm]{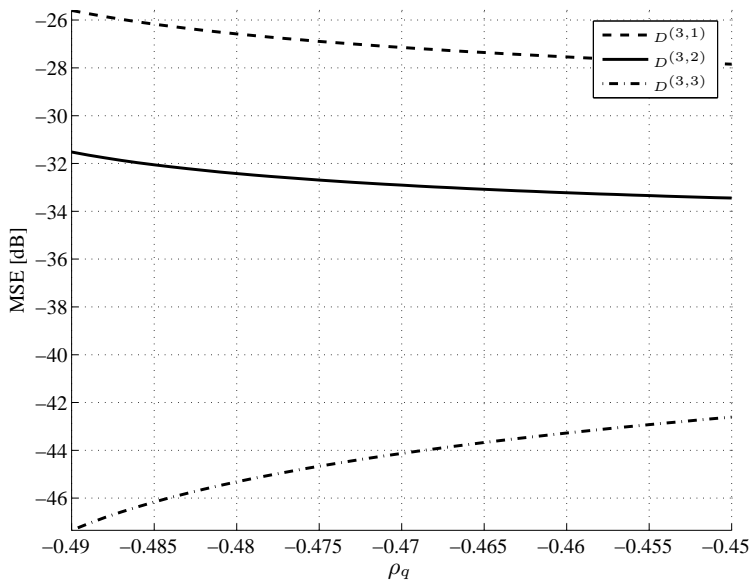}
\caption{The simultaneously achievable one-channel, two-channel and three-channel distortions for the unit-variance Gaussian source at 5 bit/dim.\ for $-0.49<\rho_q<-0.45$ for a $(3,1)$ SCEC.}
\label{fig:D1D2D3_R5}
\end{center}
\end{figure}

The distortions shown in Tables~\ref{tab:kappa1} to~\ref{tab:kappa3} correspond to the case where we vary the index value $N$ throughout the range $67\geq N \geq 31$ for three-channel MD-LVQ operating at $R_s= 5$ bit/dim.\ per description. To achieve similar performance with a (3,1) SCEC we need to vary the correlation $\rho_q$ within the interval $-0.49\leq\rho_q\leq-0.45$, as shown in Fig.~\ref{fig:D1D2D3_R5}.

%
%

\subsection{Distortion as a Function of Packet-Loss Probability}
\begin{figure}[ht]
\begin{center}
\psfrag{K=1}{$\scriptscriptstyle K=1$}
\psfrag{K=2}{$\scriptscriptstyle K=2$}
\psfrag{K=3}{$\scriptscriptstyle K=3$}
\includegraphics[width=9cm]{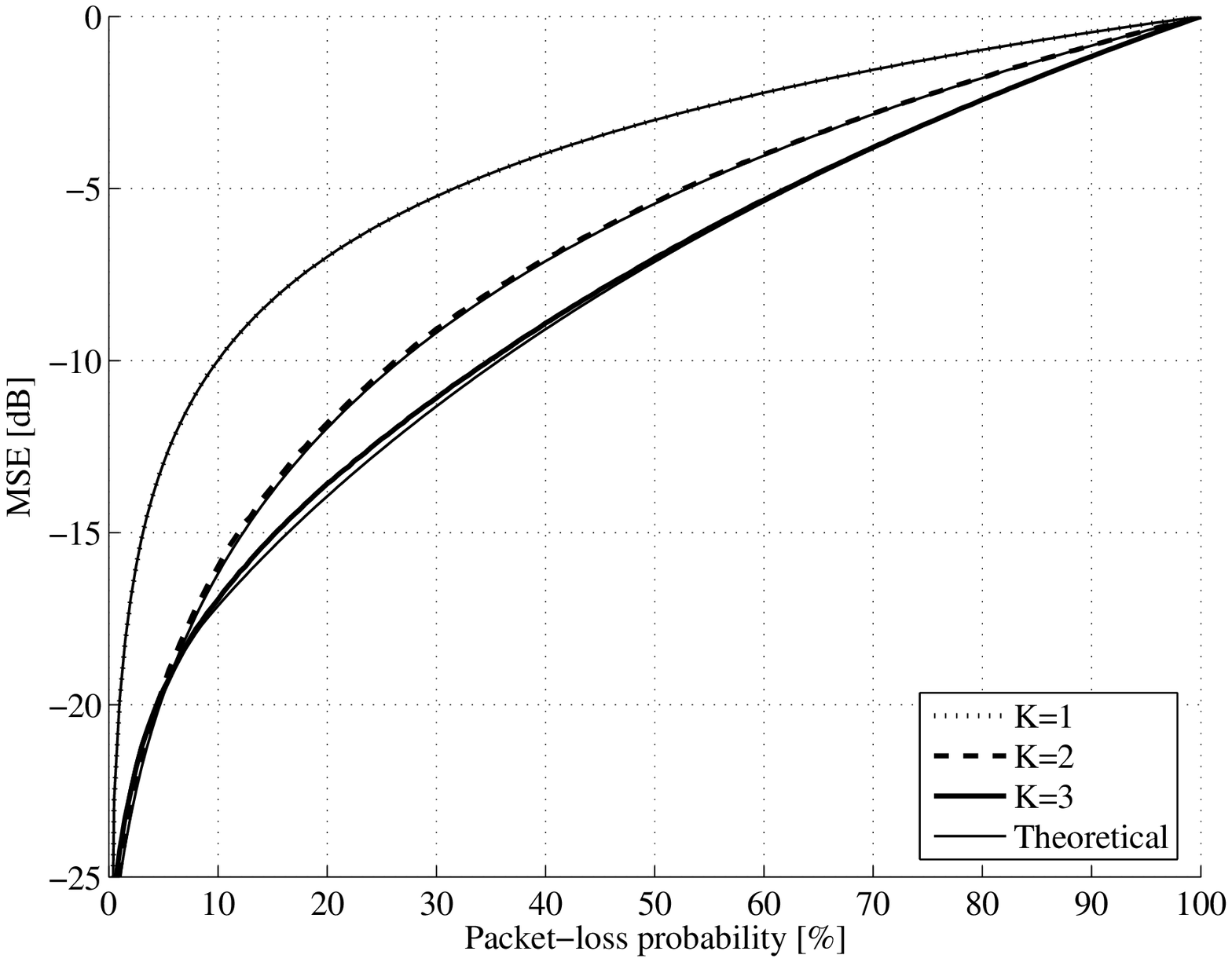}
\caption{Distortion as a function of the packet-loss probability for the $A_2$ quantizer. The target entropy is 6 bit/dim., so each description gets 6/K bit/dim. Thick lines show numerical performance and thin solid lines show theoretical performance. The two curves at the top (coinciding) illustrate the case of $K=1$, the two curves in the middle illustrate the case of $K=2$, and the bottom two curves illustrate the case of $K=3$ descriptions.}
\label{fig:numdistpp}
\end{center}\end{figure}

We now show the expected distortion as a function of the packet-loss probability for $K$-channel MD-LVQ where $K=1,2,3$. We block the i.i.d.\ unit-variance Gaussian source into $2\cdot 10^6$ two-dimensional vectors and let the total target entropy be 6 bit/dim. The expansion factor is set to $\psi_2=1$ for $K=1,2$ and $\psi_2=1.14808$ for $K=3$. We sweep the packet-loss probability $p$ in the range $p\in [0;1]$ in steps of 1/200 and for each $p$ we measure the distortion for all admissible index values and use that index value which gives the lowest distortion. This gives rise to an operational lower hull (OLH)\index{operational lower hull} for each quantizer. This is done for the theoretical curves as well by inserting admissible index values in~(\ref{eq:adistopt}) and use that index value that gives the lowest distortion. 
In other words we compare the numerical OLH with the theoretical OLH and not the ``true''\footnote{A lattice is restricted to a set of admissible index values. This set is generally expanded when the lattice is used as a product quantizer, hence admissible index values closer to the optimal values given by~(\ref{eq:optN}) can in theory be obtained, cf.\ Section~\ref{sec:admindexvalues}.} lower hull that would be obtained by using the unrestricted index values given by~(\ref{eq:optN}).
The target entropy\index{target rate} is evenly distributed over $K$ descriptions. For example, for $K=2$ each description uses 3 bit/dim., whereas for $K=3$ each description uses only 2 bit/dim. The performance is shown in Fig.~\ref{fig:numdistpp}. The practical performance of the scheme is described by the lower hull of the $K$-curves. Notice that at higher packet-loss probabilities ($p>5\%$) it becomes advantageous to use three descriptions instead two. 
%
%

\begin{figure}[ht]
\begin{center}
\includegraphics[width=9cm]{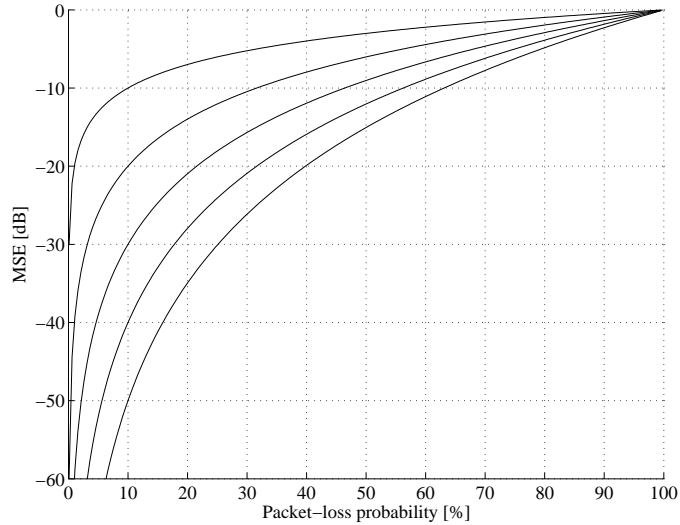}
\caption{Estimation error $\frac{1}{L}E\|X\|^2 p^K$ as a function of the packet-loss probability for different number of descriptions. The top curve is for $K=1$, the second from the top is for $K=2$, and so on. The bottom curve is for $K=5$.}
\label{fig:estimation_pp}
\end{center}\end{figure}

It is important to see that when the distortion measure is the expected distortion based on the packet-loss probability, then the notion of high resolution is slightly misleading. For example, if we let the rate go to infinity, then for a given fixed packet-loss probability $p$ the only contributing factor to the expected distortion is the distortion due to the estimation of the source when all packets are lost. This term is given by $\frac{1}{L}E\|X\|^2p^{K}$ so that for a unit-variance source, in the asymptotic case of $R\rightarrow \infty$, the expected distortion is simply given by $K10\log_{10}(p)$ dB. In other words, with a packet-loss probability of 10\%, if the number of packets is increased by one, then the corresponding decrease in distortion is exactly 10 dB. We have illustrated this in Fig.~\ref{fig:estimation_pp} for $K=1,\dotsc, 5$.
%
%

\section{Conclusion}
We derived closed-form expressions for the central and side quantizers which, at high-resolution conditions, minimize the expected distortion of a symmetric $K$-channel MD-LVQ scheme subject to entropy constraints on the side descriptions for given packet-loss probabilities.
The expected distortion observed at the receiving side depends only upon the number of received descriptions but is independent of which descriptions are received.
We focused on a special case of the symmetric MD problem where only a single parameter (i.e.\ $N$) controls the redundancy tradeoffs between the central and the side distortions. 
We showed that the optimal amount of redundancy is in independent of the source distribution, the target rate and the type of lattices used for the side quantizers. 

The practical design allows an arbitrary number of descriptions and the optimal number of descriptions depends (among other factors) upon the packet-loss probability. The theoretical rate-distortion results were proven for the case of $K\leq 3$ descriptions and conjectured to be true in the general case of arbitrary $K$ descriptions.

\chapter{$K$-Channel Asymmetric Lattice Vector Quantization}
\label{chap:asym}
\begin{figure}[ht]
\psfrag{X}{$\scriptstyle X$}
\psfrag{Xh}{$\scriptstyle \hat{X}$}
\psfrag{R0}{$\scriptstyle R_0$}
\psfrag{R1}{$\scriptstyle R_1$}
\psfrag{RK}{$\scriptstyle R_{K-1}$}
\psfrag{Description 0}{\scriptsize Description 0}
\psfrag{Description 1}{\scriptsize Description 1}
\psfrag{Description K-1}{\scriptsize Description $\scriptstyle K-1$}
\begin{center}
\includegraphics[width=10cm]{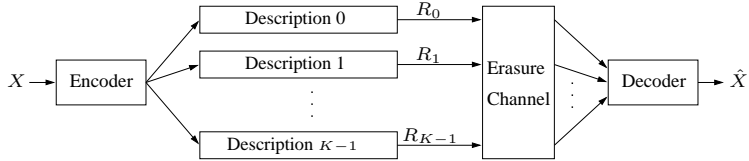}
\caption{General $K$-channel system. Descriptions are encoded at an entropy of $R_i$, $i=0,\dots, K-1$. The erasure channel either transmits the $i^{th}$ description errorless or not at all.}
\label{fig:nchannel}
\end{center}
\end{figure}

In this chapter we will focus on asymmetric MD-LVQ for $K\geq 2$ descriptions, see Fig.~\ref{fig:nchannel}. 
Asymmetric schemes offer additional flexibility over the symmetric schemes, since the bit distribution is also a design parameter and different weights are introduced in order to control the distortions. In fact, symmetric MD-LVQ is a special case of asymmetric MD-LVQ.

In~\cite{diggavi:2000,diggavi:2002} asymmetric two-channel MD-LVQ systems are derived subject to entropy constraints on the individual side entropies. 
However, since these schemes are subject to individual side entropy constraints and not subject to a single constraint on the sum of the side entropies, the problem of how to distribute a total bit budget among the two descriptions is not addressed. In this chapter we derive MD quantizer parameters subject to individual side entropy constraints and/or subject to a total entropy constraint on the sum of the side entropies. We then show that the optimal bit distribution among the descriptions is not unique but is in fact characterized by a set of solutions, which all lead to minimal expected distortion.

For the case of $K=2$ our design admits side distortions which are superior to the side distortions of~\cite{diggavi:2000,diggavi:2002} while achieving identical central distortion\index{central distortion}. 
Specifically, we show that the side distortions\index{side distortion} of our design can be expressed through the dimensionless normalized second moment $G(S_L)$ of an $L$-sphere whereas the side distortions of previous asymmetric designs~\cite{diggavi:2000,diggavi:2002} depend on the dimensionless normalized second moment $G(\Lambda)$ of the $L$-dimensional lattices. More accurately, the difference in side distortions between the two schemes is given by the difference between $G(S_L)$ and $G(\Lambda)$. 
Notice that $G(S_L)\leq G(\Lambda)$ with equality for $L=1$ and for $L\rightarrow\infty$ by a proper choice of lattice~\cite{zamir:1996}, cf.\ Section~\ref{sec:space_filling}.
We also show that, for the case of $K=3$ and asymptotically in lattice vector dimension, the side distortions can again be expressed through $G(S_L)$ and we further conjecture this to be true for $K>3$ descriptions.

\section{Preliminaries}\label{sec:prelim_asym}
To be consistent with the previous chapter we will here introduce the set of lattices required for the asymmetric design and emphasize how they differ from the symmetric design.

Just as in the symmetric case we use a single lattice $\Lambda_c$ as the central quantizer. However, we will make use of several sublattices $\Lambda_i, i=0,\dots, K-1$ for the side quantizers. In fact, we use one side quantizer (sublattice) for each description.
We assume that all sublattices are geometrically-similar to $\Lambda_c$ and clean.
The sublattice index of the $i^{th}$ sublattice $\Lambda_i$ is given by $N_i=|\Lambda_c/\Lambda_i|, N_i\in \mathbb{Z}^+$.
The volume $\nu_i$ of a sublattice Voronoi cell in the $i^{th}$ sublattice is given by $\nu_i=N_i\nu$, where $\nu$ is the volume of a Voronoi cell of $\Lambda_c$. As in the symmetric case we will also here make use of a product lattice $\Lambda_\pi \subseteq \Lambda_i \subseteq \Lambda_c$ of index $N_\pi = |\Lambda_c / \Lambda_\pi|$ in the design of the index-assignment map.

The general framework of asymmetric MD-LVQ is similar to the symmetric case. We use a single index-assignment map $\alpha$, which maps central lattice points to $K$-tuples of sublattice points. The main difference is that in the asymmetric case the sublattice index values, $N_i, i=0,\dotsc,K-1,$ are not necessarily equal, which means that the side descriptions rates $R_i$ are not necessarily equal either. Furthermore, the weights for the case of receiving $\kappa$ out of $K$ descriptions depend upon which $\kappa$ descriptions are considered. This was not so in the symmetric case.

\subsection{Index Assignments}\label{sec:index_asym}
The index assignment map (or labeling function) differs from the symmetric case in that it maps from a single lattice to several distinct sublattices. Specifically, let $\alpha$ denote the labeling function\index{index-assignment!labeling function} and let the individual component functions of $\alpha$ be denoted by $\alpha_i$. The injective map $\alpha$ that maps $\Lambda_c$ into $\Lambda_0 \times \dots \times \Lambda_{K-1}$, is then given by
\begin{align}
\alpha(\lambda_c)&=(\alpha_0(\lambda_c),\alpha_1(\lambda_c),\dots,\alpha_{K-1}(\lambda_c)) \\
&= (\lambda_0,\lambda_1,\dots,\lambda_{K-1}),
\end{align}
where $\alpha_i(\lambda_c)=\lambda_i \in \Lambda_i$ and $i=0,\dots, K-1$. 

We generalize the approach of the previous chapter and construct a product lattice $\Lambda_\pi$ which has $N_\pi$ central lattice points and $N_\pi/N_i$ sublattice points from the $i^{th}$ sublattice in each of its Voronoi cells. The Voronoi cells $V_\pi$ of the product lattice $\Lambda_\pi$ are all similar so by concentrating on labeling only central lattice points within one cell, the rest of the central lattice points may be labeled simply by translating this cell throughout $\mathbb{R}^L$. 
Without loss of generality we let $N_\pi=\prod_{i=0}^{K-1}N_i$, i.e.\ by construction we let $\Lambda_\pi$ be a geometrically-similar and clean sublattice of $\Lambda_i$ for all $i$.\footnote{From Lemma~\ref{lem:productindex} it follows that the product of admissible index values leads to an admissible index value.}
With this choice of $\Lambda_\pi$, we only label central lattice points within $V_\pi(0)$, which is the Voronoi cell of $\Lambda_\pi$ around the origin. 

\subsection{Rate and Distortion Results}
The central distortion $D_c$ is identical to that of a symmetric system, which is given by~(\ref{eq:dcG}). It also follows from the symmetric case see~(\ref{eq:di}) that the side distortion for the $i^{th}$ description is given by
\begin{equation}\label{eq:di1_asym}
D_i = D_c + \frac{1}{L}\frac{1}{N_\pi}\sum_{\lambda_c \in V_\pi(0)} \|\lambda_c - \alpha_i(\lambda_c)\|^2,\quad i=0,\dots, K-1.
\end{equation}
\begin{definition}
$R_i$ denotes the entropy of the individual descriptions. The entropy of the $i^{th}$ description is defined as $R_i\triangleq H(\alpha_i(Q(X)))/L$.
\end{definition}
The side descriptions are based on a coarser lattice obtained by scaling the Voronoi cells of the central lattice by a factor of $N_i$. Assuming the pdf of $X$ is roughly constant within a sublattice cell, the entropies of the side descriptions are given by
\begin{equation}\label{eq:Ri}
R_i\approx \bar{h}(X) - \frac{1}L\log_2(N_i\nu).
\end{equation}
The entropies of the side descriptions are related to the entropy $R_c$ of the central quantizer, given by~(\ref{eq:Rc}), by
\begin{equation*}
R_i \approx R_c - \frac{1}L\log_2(N_i).
\end{equation*}

\section{Construction of Labeling Function}
\label{sec:label_asym}
In this section we construct the index-assignment map $\alpha$, which takes a single vector $\lambda_c$ and maps it to a set of $K$ vectors $\{\lambda_i\}, i=0,\dots, K-1$, where $\lambda_i\in \Lambda_i$. The mapping is invertible so that we have $\lambda_c=\alpha^{-1}(\lambda_0,\dots,\lambda_{K-1})$. 

In asymmetric MD-LVQ weights are introduced in order to control the amount of asymmetry between the side distortions. We will in the following assume that these weights are based on the packet-loss probabilities of the individual descriptions. However, it should be clear that the weights are not limited to represent packet-loss probabilities but can in fact be almost arbitrarily chosen. We will consider the case where the index-assignment map is constructed such that the expected distortion, given by the sum of the distortions due to all possible description losses weighted by their corresponding loss probabilities, is minimized. 

In addition to knowing the weighted distortion over all description losses it is also interesting to know the distortion of any subset of the $K$ descriptions. This issue is considered in Section~\ref{sec:networkcoding}.

\subsection{Expected Distortion}
\label{sec:reconstruction_asym}
At the receiving side, $X\in \mathbb{R}^L$ is reconstructed to a quality that is determined by the received descriptions. If no descriptions are received we reconstruct using the expected value, $EX$, and if all $K$ descriptions are received we reconstruct using the inverse map $\alpha^{-1}(\lambda_0,\dotsc,\lambda_{K-1})$, hence obtaining the quality of the central quantizer. 
In all other cases, we reconstruct to the average of the received descriptions as was done in the symmetric case.

There are in general several ways of receiving $\kappa$ out of $K$ descriptions. Let $\mathcal{L}^{(K,\kappa)}$ denote an index set consisting of all possible $\kappa$ combinations out of $\{0,\dots, K-1\}$ so that $|\mathcal{L}^{(K,\kappa)}| = \binom{K}{\kappa}$. We denote an element of $\mathcal{L}^{(K,\kappa)}$ by $l=\{l_0, \dots , l_{\kappa-1}\}$. The complement $l^c$ of $l$ denotes the $K-\kappa$ indices not in $l$, i.e.\ $l^c = \{0,\dots, K-1\}\backslash l$.
We will use the notation $\mathcal{L}_{i}^{(K,\kappa)}$ to indicate the set of all $l\in \mathcal{L}^{(K,\kappa)}$ that contains the index $i$, i.e., $\mathcal{L}_i^{(K,\kappa)}=\{ l : l \in \mathcal{L}^{(K,\kappa)}\ \text{and}\ i\in l\}$ and similarly $\mathcal{L}_{i,j}^{(K,\kappa)} = \{ l : l \in \mathcal{L}^{(K,\kappa)}\ \text{and}\ i,j\in l\}$.
Furthermore, let $p_i$ be the packet-loss probability for the $i^{th}$ description and let $\mu_i=1-p_i$ be the probability that the $i^{th}$ description is received. Finally, let $p(l)=\prod_{i\in l}\mu_i \prod_{j\in l^c}p_j$, $p(\mathcal{L}^{(K,\kappa)}) = \sum_{l\in\mathcal{L}^{(K,\kappa)}} p(l)$, $p(\mathcal{L}_i^{(K,\kappa)}) = \sum_{l\in\mathcal{L}_i^{(K,\kappa)}} p(l)$ and $p(\mathcal{L}_{i,j}^{(K,\kappa)}) = \sum_{l\in\mathcal{L}_{i,j}^{(K,\kappa)}} p(l)$. For example, for $K=3$ and $\kappa=2$ we have $\mathcal{L}^{(3,2)}=\{ \{0,1\}, \{0,2\}, \{1,2\}\}$ and hence $p(\mathcal{L}^{(3,2)})=\mu_0\mu_1 p_2 + \mu_0\mu_2 p_1 + \mu_1\mu_2 p_0$. In a similar manner for $K=6$ and $\kappa=3$ we have 
\begin{equation*}
\mathcal{L}_{1,2}^{(6,3)}=\{ \{0,1,2\}, \{1,2,3\}, \{1,2,4\}, \{1,2,5\} \},
\end{equation*}
and
\begin{equation*}
p(\mathcal{L}_{1,2}^{(6,2)})=\mu_0\mu_1\mu_2 p_3 p_4 p_5 + \mu_1\mu_2\mu_3 p_0 p_4 p_5 + \mu_1\mu_2\mu_4 p_0 p_3 p_5 + \mu_1\mu_2\mu_5 p_0 p_3 p_4.
\end{equation*}

As in the symmetric case, upon reception of any $\kappa$ out of $K$ descriptions we reconstruct to $\hat{X}$ using
\begin{equation*}
\hat{X}=\frac{1}{\kappa}\sum_{j\in l}\lambda_j.
\end{equation*}
The distortion when receiving a set of descriptions can be derived in a similar way as was done in the symmetric case. Thus, by use of~(\ref{eq:di}) and~(\ref{eq:di1_asym}) it can be shown that the norm of~(\ref{eq:di1_asym}), when receiving descriptions $i$ and $j$, should read $\|\lambda_c-0.5(\alpha_i(\lambda_c)+\alpha_j(\lambda_c))\|^2$. It follows that the expected distortion when receiving $\kappa$ out of $K$ descriptions is given by
\begin{equation}\label{eq:expdist_asym}
\begin{split}
D_a^{(K,\kappa)} &\approx \sum_{l\in \mathcal{L}^{(K,\kappa)}} p(l)\left(D_c + \frac{1}{L}\frac{1}{N_\pi}\sum_{\lambda_c\in V_{\pi}(0)}\left\| \lambda_c - \frac{1}{\kappa}\sum_{j=0}^{\kappa-1}\lambda_{l_j} \right\|^2\right) \\
&= p(\mathcal{L}^{(K,\kappa)})D_c + \frac{1}{L}\frac{1}{N_\pi}\sum_{\lambda_c\in V_{\pi}(0)}\sum_{l\in \mathcal{L}^{(K,\kappa)}}p(l)\left\| \lambda_c - \frac{1}{\kappa}\sum_{j=0}^{\kappa-1}\lambda_{l_j} \right\|^2,
\end{split}
\end{equation}
where $\lambda_{l_j}=\alpha_{l_j}(\lambda_c)$ and the two special cases $\kappa\in \{0,K\}$ are given by $D_a^{(K,0)}\approx \frac{1}{L}E\|X\|^2\prod_{i=0}^{K-1}p_i$ and $D_a^{(K,K)}\approx D_c\prod_{i=0}^{K-1}\mu_i$.

\subsection{Cost Functional}
From~(\ref{eq:expdist_asym}) we see that the distortion $D_a^{(K,\kappa)}$ may be split into two terms, one describing the distortion occurring when the central quantizer is used on the source, and one that describes the distortion due to the index assignment. An optimal index assignment minimizes the second term in~(\ref{eq:expdist_asym}) for all possible combinations of descriptions. The cost functional\index{cost functional} $J^{(K)}$ to be minimized by the index-assignment algorithm can then be written as
\begin{equation}\label{eq:costfunctional_asym}
J^{(K)} = \sum_{\kappa=1}^{K-1} J^{(K,\kappa)},
\end{equation}
where 
\begin{equation}
J^{(K,\kappa)}= 
\frac{1}{L}\frac{1}{N_\pi}\sum_{\lambda_c\in V_{\pi}(0)}\sum_{l\in \mathcal{L}^{(K,\kappa)}}p(l)\left\| \lambda_c - \frac{1}{\kappa}\sum_{j=0}^{\kappa-1}\lambda_{l_j} \right\|^2.
\end{equation}
The cost functional should be minimized subject to some entropy-constraints on the side descriptions or on e.g.\ the sum of the side entropies. We remark here that the side entropies depend solely on $\nu$ and $N_i$ (see~(\ref{eq:Ri})) but not on the particular choice of $K$-tuples. In other words, for fixed $N_i$'s and a fixed $\nu$ the index assignment problem is solved if~(\ref{eq:costfunctional_asym}) is minimized. The problem of choosing $\nu$ and $N_i$ such that certain entropy constraints are not violated is independent of the assignment problem and deferred to Section~\ref{sec:optquantRi}.

\begin{theorem}\label{theo:sums_asym}
For any $1\leq \kappa \leq K$ we have
\begin{equation*}
\begin{split}
\sum_{\lambda_c}&\sum_{l\in \mathcal{L}^{(K,\kappa)}}p(l)\left\| \lambda_c - \frac{1}{\kappa}\sum_{j=0}^{\kappa-1}\lambda_{l_j} \right\|^2 \\
&=
\sum_{\lambda_c}\bigg(p(\mathcal{L}^{(K,\kappa)})\left\| \lambda_c - \frac{1}{\kappa p(\mathcal{L}^{(K,\kappa)})}\sum_{i=0}^{K-1}p(\mathcal{L}_i^{(K,\kappa)})\lambda_i \right\|^2 \\
&+\frac{1}{\kappa^2}\sum_{i=0}^{K-2}\sum_{j=i+1}^{K-1}
\left(\frac{p(\mathcal{L}_i^{(K,\kappa)})p(\mathcal{L}_j^{(K,\kappa)})}{p(\mathcal{L}^{(K,\kappa)})}-p(\mathcal{L}_{i,j}^{(K,\kappa)})\right)\|\lambda_i - \lambda_j\|^2\bigg).
\end{split}
\end{equation*}
\end{theorem}
\begin{proof}
See Appendix~\ref{app:proofsums_asym}.
\end{proof}

The cost functional~(\ref{eq:costfunctional_asym}) can by use of Theorem~\ref{theo:sums_asym} be written as
\begin{equation}\label{eq:costfunctional2_asym}
\begin{split}
J^{(K,\kappa)}&= \frac{1}{L}\frac{1}{N_\pi}
\sum_{\lambda_c\in V_\pi(0)}\bigg(p(\mathcal{L}^{(K,\kappa)})\left\| \lambda_c - \frac{1}{\kappa p(\mathcal{L}^{(K,\kappa)})}\sum_{i=0}^{K-1}\lambda_i p(\mathcal{L}_i^{(K,\kappa)})\right\|^2\\
&\quad+
\frac{1}{\kappa^2}\sum_{i=0}^{K-2}\sum_{j=i+1}^{K-1}
\|\lambda_i - \lambda_j\|^2
\left(\frac{p(\mathcal{L}_i^{(K,\kappa)})p(\mathcal{L}_j^{(K,\kappa)})}{p(\mathcal{L}^{(K,\kappa)})}-p(\mathcal{L}_{i,j}^{(K,\kappa)})\right)\bigg).
\end{split}
\end{equation}
The first term in~(\ref{eq:costfunctional2_asym}) describes the distance from a central lattice point to the weighted centroid of its associated $K$-tuple. The second term describes the weighted sum of pairwise squared distances (WSPSD)\index{weighted sum of pairwise squared distances} between elements of the $K$-tuple. In Section~\ref{sec:highrate_asym} (Proposition~\ref{prop:growthriemann2_asym}) we show that, under a high-resolution assumption, the second term in~(\ref{eq:costfunctional2_asym}) is dominant, from which we conclude that 
in order to minimize~(\ref{eq:costfunctional_asym}) we must use $K$-tuples with the smallest WSPSD\@. 
These $K$-tuples are then assigned to central lattice points in such a way, that the first term in~(\ref{eq:costfunctional2_asym}) is minimized. 

\subsection{Minimizing Cost Functional}
\label{sec:ktuples_asym}
We follow the approach of the symmetric case and center a region $\tilde{V}$ at all sublattice points $\lambda_0 \in \Lambda_0 \cap V_\pi(0)$, and construct $K$-tuples by combining sublattice points from the other sublattices (i.e.\ $\Lambda_i, i=1,\dots,K-1$) within $\tilde{V}(\lambda_0)$ in all possible ways and select the ones that minimize~(\ref{eq:costfunctional_asym}). For each $\lambda_0\in \Lambda_0\cap V_\pi(0)$ it is possible to construct $\prod_{i=1}^{K-1}\tilde{N}_i$ $K$-tuples, where $\tilde{N}_i$ is the number of sublattice points from the $i^{th}$ sublattice within the region $\tilde{V}$. 
This gives a total of $(N_\pi/N_0)\prod_{i=1}^{K-1}\tilde{N}_i$ $K$-tuples when all $\lambda_0\in \Lambda_0 \cap V_\pi(0)$ are used. 
The number $\tilde{N}_i$ of lattice points within a connected region $\tilde{V}$ of $\mathbb{R}^L$ may be approximated by $\tilde{N}_i\approx \tilde{\nu}/\nu_i$ where $\tilde{\nu}$ is the volume of $\tilde{V}$, which is an approximation that becomes exact as the number of shells\index{lattice!shell} of the lattice within $\tilde{V}$ goes to infinity, cf.\ Section~\ref{sec:ktuples}. Therefore, our analysis is only exact in the asymptotic case of $N_i\rightarrow \infty$ and $\nu_i\rightarrow 0$. 
Since $\tilde{N}_i\approx\tilde{\nu}/\nu N_i$ and we need $N_0$ $K$-tuples for each $\lambda_0\in V_\pi(0)$ we see that
\begin{equation*}
N_0 = \prod_{i=1}^{K-1}\tilde{N}_i\approx \frac{\tilde{\nu}^{K-1}}{\nu^{K-1}}\prod_{i=1}^{K-1}N_i^{-1},
\end{equation*}
so in order to obtain at least $N_0$ $K$-tuples, the volume of $\tilde{V}$ must satisfy 
\begin{equation}\label{eq:vtilde_asym}
\lim_{N_i\rightarrow\infty, \forall i}\tilde{\nu}\geq \nu\prod_{i=0}^{K-1}N_i^{1/(K-1)}.
\end{equation}
For the symmetric case, i.e.\ $N=N_i$, $i=0,\dots,K-1$, we have $\tilde{\nu}\geq \nu N^{K/(K-1)}$, which is in agreement with the results obtained in Chapter~\ref{chap:symmetric}.

Before we outline the design procedure for constructing an optimal index assignment we remark that in order to minimize the WSPSD\index{weighted sum of pairwise squared distances} between a fixed $\lambda_i$ and the set of points $\{\lambda_j \in \Lambda_j\cap \tilde{V}\}$ it is required that $\tilde{V}$ forms a sphere centered at $\lambda_i$. The design procedure can be outlined as follows:
\begin{enumerate}
\item Center a sphere $\tilde{V}$ at each $\lambda_0\in \Lambda_0\cap V_\pi(0)$ and construct all possible $K$-tuples $(\lambda_0,\lambda_1,\dots,\lambda_{K-1})$ where $\lambda_i\in \Lambda_i\cap \tilde{V}(\lambda_0)$ and $i=1,\dots, K-1$. This ensures that all $K$-tuples have their first coordinate ($\lambda_0$) inside $V_\pi(0)$ and they are therefore shift-invariant. We will only use $K$-tuples whose elements satisfy $\|\lambda_i-\lambda_j\|\leq r, \forall i,j\in 0,\dots K-1$, where $r$ is the radius of $\tilde{V}$. Make $\tilde{V}$ large enough so at least $N_0$ distinct $K$-tuples are found for each $\lambda_0$.
\item Construct cosets\index{coset} of each $K$-tuple.
\item The $N_\pi$ central lattice points in $\Lambda_c\cap V_\pi(0)$ must now be matched to distinct $K$-tuples. As in the symmetric case, this is a standard linear assignment problem \cite{west:2001} where only one member from each coset is (allowed to be) matched to a central lattice point in $V_\pi(0)$.
\end{enumerate}

The restriction $\|\lambda_i-\lambda_j\|\leq r$ from step 1), which is used to avoid bias towards any of the sublattices, reduces the number of distinct $K$-tuples that can be constructed within the region $\tilde{V}$. To be able to form $N_0$ $K$-tuples it is therefore necessary to use a region $\tilde{V}$ with a volume larger than the lower bound in~(\ref{eq:vtilde_asym}). In order to theoretically describe the performance of the quantizers we need to know the optimal $\tilde{\nu}$, i.e.\ the smallest volume which (asymptotically for large $N_i$) leads to exactly $N_0$ $K$-tuples. 
In Section~\ref{sec:psiL} a dimensionless expansion factor $\psi_L$\index{$\psi_L$} which only depends on $K$ and $L$ was introduced. $\psi_L$ was used to describe how much $\tilde{V}$ had to be expanded from the theoretical lower bound~(\ref{eq:vtilde_asym}), to make sure that $N_0$ optimal $K$-tuples could be constructed by combining sublattice points within a region $\tilde{V}$. 
\begin{lemma}
The dimensionless expansion factor $\psi_L$ for the asymmetric case is identical to the one for the symmetric case.
\end{lemma}
\begin{proof}
Follows by replacing the constant $\nu_s$ by $\nu_i$ in the proof of Theorem~\ref{theo:psiLK3}.
\end{proof}
Adopting this approach leads to
\begin{equation*}
\tilde{\nu}=\psi_L^L\nu\prod_{i=0}^{K-1}N_i^{1/(K-1)}.
\end{equation*}

\begin{remark}
It might appear that the shift invariance\index{shift invariance} restriction enforced by using only one member from each coset will unfairly penalize $\Lambda_0$. However, the next two lemmas prove that, asymptotically as $N_i\rightarrow \infty$, there is no bias towards any of the sublattices. 
We will consider here the case of $K>2$ (for $K=2$ we can use similar arguments as given in~\cite{diggavi:2002}).
\end{remark}

\begin{lemma}\label{lem:nobias}
For $K>2$ the number of $K$-tuples that is affected by the coset restriction is (asymptotically as $N_i\rightarrow \infty, \forall i$) neglectable compared to the number of $K$-tuples which are not affected. 
\end{lemma}
\begin{proof}
See Appendix~\ref{app:lemmas}.
\end{proof}

\begin{lemma}\label{lem:equivktuples}
The set of $N_\pi$ $K$-tuples that is constructed by centering $\tilde{V}$ at each $\lambda_0\in V_\pi(0)\cap \Lambda_0$ is asymptotically identical to the set constructed by centering $\tilde{V}$ at each $\lambda_i\in V_\pi(0)\cap \Lambda_i$, for any $i\in \{1,\dots,K-1\}$.
\end{lemma}
\begin{proof}
See Appendix~\ref{app:lemmas}.
\end{proof}

\begin{remark}
The $K$-tuples need to be assigned to central lattice points within $V_\pi(0)$. This is a standard linear assignment problem where a cost measure is minimized. However, solutions to linear assignment problems are generally not unique. Therefore, there might exist several labelings, which all yield the same cost, but exhibit a different amount of asymmetry. To achieve the specified distortions it may then be necessary to e.g.\ use time sharing\index{time sharing} through a suitable mixing of labelings.
\end{remark}

\subsection{Comparison to Existing Asymmetric Index Assignments}
In this section we have presented a new design for asymmetric MD-LVQ based on the asymmetric design of Diggavi et al.~\cite{diggavi:2002}. The main difference between the existing design of Diggavi et al.\ and the proposed design is that of the shape of the region within which sublattice points are distributed. More specifically, in the design of Diggavi et al., a given sublattice point $\lambda_0\in \Lambda_0$ is paired with a set of sublattice points of $\Lambda_1$ which are all evenly distributed within a Voronoi cell of $\Lambda_\pi$, the product lattice. However, in the proposed design, a sublattice point $\lambda_0\in \Lambda_0$ is paired with a set of sublattice points of $\Lambda_1$ which are all evenly distributed within an $L$-dimensional hypersphere.

Let us emphasize some of the advantages as well as weaknesses of the proposed design.

\begin{itemize}
\item Advantages
  \begin{enumerate}
  \item The side distortion is reduced (compared to the previous design) when finite dimensional lattice vector quantizers are used (when the dimension is strictly greater than one). To see this, notice that the side distortion is a function of the dimensionless normalized second moment of the region over which the sublattice points are distributed. For $L=1$ as well as $L\rightarrow \infty$ spheres pack space and it is possible to have spherical Voronoi cells of $\Lambda_\pi$ by a proper choice of product lattice. 
  \item To simplify the design it is often convenient to base the product lattice upon the simple hypercubic $Z^L$ lattice. 
In this case, the side distortion of the design of Diggavi et al.\ is independent of the vector dimension of the lattices, whereas with the proposed design the distortion steadily decreases as the dimension increases. The reduction in side distortion is upper bounded by approximately 1.53 dB per description.
  \item The proposed design scales easily to more than two descriptions. It is not clear how to obtain more than two descriptions with the previous designs.
\end{enumerate}
\item Weaknesses
\begin{enumerate}
\item The design of Diggavi et al.\ exploits several geometric properties of the underlying lattices to ensure that any single sublattice point of $\Lambda_0$ is paired with exactly $N_0$ sublattice points of $\Lambda_1$. On the other hand, the proposed design guarantees such a symmetry property only in asymptotic cases. Thus, in practice, if such a symmetry property is desired, one might need to search within a set of candidate solutions. 
\end{enumerate}
\end{itemize}

\section{High-Resolution Analysis}\label{sec:highrate_asym}
In this section we derive high-resolution approximations for the expected distortion. In line with the high-resolution analysis presented in Chapter~\ref{chap:symmetric} we let $N_i\rightarrow \infty$ and $\nu_i\rightarrow 0$, i.e.\ for each sublattice the index increase, while the volume of their Voronoi cell shrink.

\subsection{Total Expected Distortion}
Using Theorem~\ref{theo:sums_asym}, the expected distortion~(\ref{eq:expdist_asym}) when $\kappa$ out of $K$ descriptions are received can be written as
\begin{equation}\label{eq:da_asym}
\begin{split}
D_a^{(K,\kappa)}&\approx p(\mathcal{L}^{(K,\kappa)})\,D_c + \frac{1}{L}\frac{1}{N_\pi}\sum_{\lambda_c\in V_{\pi}(0)}\sum_{l\in \mathcal{L}^{(K,\kappa)}}p(l)\left\| \lambda_c - \frac{1}{\kappa}\sum_{j=0}^{\kappa-1}\lambda_{l_j} \right\|^2 \\
&=
p(\mathcal{L}^{(K,\kappa)})\,D_c \\
&\quad+ 
\frac{1}{L}\frac{1}{N_\pi}\sum_{\lambda_c\in V_\pi(0)}\bigg(p(\mathcal{L}^{(K,\kappa)})\left\| \lambda_c - \frac{1}{\kappa p(\mathcal{L}^{(K,\kappa)})}\sum_{i=0}^{K-1}p(\mathcal{L}_i^{(K,\kappa)})\lambda_i \right\|^2 \\
&\quad +
\frac{1}{\kappa^2}\sum_{i=0}^{K-2}\sum_{j=i+1}^{K-1}
\left(\frac{p(\mathcal{L}_i^{(K,\kappa)})p(\mathcal{L}_j^{(K,\kappa)})}{p(\mathcal{L}^{(K,\kappa)})}-p(\mathcal{L}_{i,j}^{(K,\kappa)})\right)\|\lambda_i - \lambda_j\|^2\bigg).
\end{split}
\end{equation}

\begin{proposition}\label{prop:riemann2_asym}
For $K=2$ and asymptotically as $N_i\rightarrow \infty, \nu_i\rightarrow 0$ as well as for $K=3$ and asymptotically as $N_i, L\rightarrow \infty$ and $\nu_i\rightarrow 0$, we have
for any pair of sublattices, $(\Lambda_i,\Lambda_j),\ i,j=0,\dots,K-1,\ i\neq j$,
\begin{equation*}
\frac{1}{L}\sum_{\lambda_c\in V_\pi(0)}
\| \alpha_i(\lambda_c)-\alpha_j(\lambda_c)\|^2 = \psi_L^{2}\nu^{2/L} G(S_L) N_\pi\prod_{m=0}^{K-1}N_m^{2/L(K-1)}.
\end{equation*}
\end{proposition}

\begin{proof}
See Appendix~\ref{app:riemann2_asym}.
\end{proof}

\begin{conjecture}
Proposition~\ref{prop:riemann2_asym} is true for any $K$ asymptotically as $L,N_i\rightarrow \infty$ and $\nu_i\rightarrow 0, \forall i$. 
\end{conjecture}

\begin{proposition}\label{prop:growthriemann2_asym}
For $N_i\rightarrow \infty$ we have
\begin{equation*}
\frac{\displaystyle\sum_{\lambda_c\in V_\pi(0)}\left\| \lambda_c - \frac{1}{\kappa p(\mathcal{L})}\sum_{i=0}^{K-1}p(\mathcal{L}_i^{(K,\kappa)})\lambda_i \right\|^2}
{\displaystyle\sum_{\lambda_c\in V_\pi(0)}\sum_{i=0}^{K-2}\sum_{j=i+1}^{K-1}
\left(\frac{p(\mathcal{L}_i^{(K,\kappa)})p(\mathcal{L}_j^{(K,\kappa)})}{p(\mathcal{L}^{(K,\kappa)})}-p(\mathcal{L}_{i,j}^{(K,\kappa)})\right)\| \lambda_i - \lambda_j\|^2 } \rightarrow 0.
\end{equation*}
\end{proposition}

\begin{proof}
See Appendix~\ref{app:growthriemann2_asym}.
\end{proof}

By use of Propositions~\ref{prop:riemann2_asym} and~\ref{prop:growthriemann2_asym} and~(\ref{eq:dcG}) it follows that~(\ref{eq:da_asym}) can be written as
\begin{equation*}
\begin{split}
&D_a^{(K,\kappa)}\approx p(\mathcal{L}^{(K,\kappa)})\,D_c\\
& + 
\frac{1}{L}\frac{1}{N_\pi}\!\sum_{\lambda_c\in V_\pi(0)}\!\!\left(
\frac{1}{\kappa^2}\sum_{i=0}^{K-2}\sum_{j=i+1}^{K-1}\!
\left(\!\frac{p(\mathcal{L}_i^{(K,\kappa)})p(\mathcal{L}_j^{(K,\kappa)})}{p(\mathcal{L}^{(K,\kappa)})}-p(\mathcal{L}_{i,j}^{(K,\kappa)})\!\right)\|\lambda_i - \lambda_j\|^2\!\!\right) \\
&\approx G(\Lambda_c)\nu^{2/L}p(\mathcal{L}^{(K,\kappa)}) + 
\psi_L^{2}\nu^{2/L} G(S_L)\beta^{(K,\kappa)}\prod_{m=0}^{K-1}N_m^{2/L(K-1)},
\end{split}
\end{equation*}
where $\beta^{(K,\kappa)}$ depends on the packet-loss probabilities and is given by
\begin{equation*}
\beta^{(K,\kappa)}=\frac{1}{\kappa^2}
\sum_{i=0}^{K-2}\sum_{j=i+1}^{K-1}
\left(\frac{p(\mathcal{L}_i^{(K,\kappa)})p(\mathcal{L}_j^{(K,\kappa)})}{p(\mathcal{L}^{(K,\kappa)})}-p(\mathcal{L}_{i,j}^{(K,\kappa)})\right).
\end{equation*}
The total expected distortion $D_a^{(K)}$ is obtained by summing over $\kappa$ including the cases where $\kappa=0$ and $\kappa=K$,
\begin{equation}\label{eq:adistopt_asym}
\begin{split}
D_a^{(K)} &\approx G(\Lambda_c)\nu^{2/L}\hat{p}(\mathcal{L}^{(K)}) +
\psi_L^{2}\nu^{2/L} G(S_L)\prod_{m=0}^{K-1}N_m^{2/L(K-1)}\hat{\beta}^{(K)} \\
&\quad+ \frac{1}{L}E\|X\|^2\prod_{i=0}^{K-1}p_i,
\end{split}
\end{equation}
where
\begin{equation*}
\hat{p}(\mathcal{L}^{(K)})=\sum_{\kappa=1}^{K}p(\mathcal{L}^{(K,\kappa)})
\end{equation*}
and
\begin{equation*}
\hat{\beta}^{(K)} = \sum_{\kappa=1}^{K}\beta^{(K,\kappa)}.
\end{equation*}

Using~(\ref{eq:Rc}) and~(\ref{eq:Ri}) we can write $\nu$ and $N_i$ as a function of differential entropy and side entropies, that is
\begin{equation*}
\nu^{2/L}=2^{2(\bar{h}(X)-R_c)},
\end{equation*}
and
\begin{equation*}
\prod_{i=0}^{K-1}N_i^{2/L(K-1)}=2^{\frac{2K}{K-1}\left(R_c-\frac{1}{K}\sum_{i=0}^{K-1}R_i\right)}.
\end{equation*}
Inserting these results in~(\ref{eq:adistopt_asym}) leads to 
\begin{equation}\label{eq:daopt_asym}
\begin{split}
D_a^{(K)} &\approx G(\Lambda_c)2^{2(\bar{h}(X)-R_c)}\hat{p}(\mathcal{L}^{(K)}) \\
&\quad +\psi_L^{2} G(S_L)2^{2(\bar{h}(X)-R_c)}2^{\frac{2K}{K-1}\left(R_c - \frac{1}{K}\sum_{i=0}^{K-1}R_i\right)}\hat{\beta}^{(K)}
+ \frac{1}{L}E\|X\|^2\prod_{i=0}^{K-1}p_i,
\end{split}
\end{equation}
where we see that the distortion due to the side quantizers is independent of the type of sublattices. 

\section{Optimal Entropy-Constrained Quantizers }\label{sec:optquantRi}
In this section we first derive closed-form expressions for the optimal scaling factors $\nu$ and $N_i$ subject to entropy constraints on the $K$ side descriptions. With these scaling factors we are able to construct a central lattice and $K$ sublattices. The index assign\-ments are then found using the approach outlined in Section~\ref{sec:label_asym}. The central lattice and the $K$ side lattices combined with their index assignment map completely specify an optimal scheme for asymmetric entropy-constrained MD-LVQ. We then consider the situation where the total bit budget is constrained, i.e.\ we find the optimal scaling factors subject to entropy constraints on the sum of the side entropies $\sum_i R_i \leq R^*$, where $R^*$ is the target entropy\index{target rate}. We also find the optimal bit distribution among the $K$ descriptions. 

\subsection{Entropy Constraints Per Description}
We assume $K$ descriptions are to be used. Packet-loss probabilities $p_i, i=0,\dots, K-1,$ are given as well as entropy-constraints on the side descriptions, i.e.\ $R_i \leq R_i^*$, where $R_i^*$ are known target entropies\index{target rate}. To be optimal, the entropies of the side descrip\-tions must be equal to the target entropies, hence by use of~(\ref{eq:Ri}) we must have that
\begin{equation*}
R_i = \bar{h}(X) - \frac{1}{L}\log_2(N_i\nu) = R_i^*,
\end{equation*}
from which we get
\begin{equation}\label{eq:Ninu}
N_i\nu = 2^{L(\bar{h}(X)-R_i^*)} = \tau_i,
\end{equation}
where $\tau_i$ are constants. It follows that $N_i=\tau_i/\nu$ and since $\prod_{i=0}^{K-1}N_i^{2/L(K-1)} = \nu^{-2K/L(K-1)}\prod_{i=0}^{K-1}\tau_i^{2/L(K-1)}$ we can express~(\ref{eq:adistopt_asym}) as a function of $\nu$, i.e.\
\begin{equation*}
\begin{split}
D_a^{(K)} &\approx G(\Lambda_c)\nu^{2/L}\hat{p}(\mathcal{L}^{(K)})\\
&\quad +
\psi_L^{2}\nu^{2/L} G(S_L)\nu^{-2K/L(K-1)}\tau^{2/L(K-1)}
\hat{\beta}^{(K)}
+ \frac{1}{L}E\|X\|^2\prod_{i=0}^{K-1}p_i \\
&=G(\Lambda_c)\nu^{2/L}\hat{p}(\mathcal{L}^{(K)}) +
\psi_L^{2} G(S_L)\nu^{-2/L(K-1)}\tau^{2/L(K-1)}
\hat{\beta}^{(K)} \\
&\quad+ \frac{1}{L}E\|X\|^2\prod_{i=0}^{K-1}p_i,
\end{split}
\end{equation*}
where $\tau=\prod_{i=0}^{K-1}\tau_i$.

Differentiating w.r.t.\ $\nu$ and equating to zero gives,
\begin{equation*}
\begin{split}
\frac{\partial D_a^{(K)}}{\partial \nu}&=
\frac{2}{L}G(\Lambda_c)\nu^{2/L-1}\hat{p}(\mathcal{L}^{(K)}) \\
&\quad -
\frac{2}{L(K-1)}\psi_L^{2} G(S_L)\nu^{-2/L(K-1)-1}\tau^{2/L(K-1)}\hat{\beta}^{(K)}
= 0,
\end{split}
\end{equation*}
from which we obtain the optimal value of $\nu$
\begin{equation}\label{eq:optnu_asym}
\begin{split}
\nu &= \tau^{1/K}\left(\psi_L^{2}\frac{1}{K-1}\frac{G(S_L)}{G(\Lambda_c)}
\frac{\hat{\beta}^{(K)}} 
{\hat{p}(\mathcal{L}^{(K)})}\right)^{\frac{L(K-1)}{2K}}  \\
& = 2^{L(\bar{h}(X) - \frac{1}{K}\sum_i R_i^*)}\left(\psi_L^{2}\frac{1}{K-1}\frac{G(S_L)}{G(\Lambda_c)}
\frac{\hat{\beta}^{(K)}} 
{\hat{p}(\mathcal{L}^{(K)})}\right)^{\frac{L(K-1)}{2K}}.
\end{split}
\end{equation}
The optimal $N_i$'s follow easily by use of (\ref{eq:Ninu}):
\begin{equation}\label{eq:optN_asym}
N_i = \frac{\tau_i}{\nu}=
\tau_i\tau^{-1/K}\left(\frac{1}{\psi_L^{2}}(K-1)\frac{G(\Lambda_c)}{G(S_L)}
\frac{\hat{p}(\mathcal{L}^{(K)})}{\hat{\beta}^{(K)}}\right)^{\frac{L(K-1)}{2K}}.
\end{equation}
Eq.~(\ref{eq:optN_asym}) shows that the optimal redundancies $N_i$'s are, for fixed $K$, independent of the sublattices. Moreover, since $\tau_i\tau^{-1/K}=2^{-L(R_i^*-\frac{1}{K}\sum_jR_j^*)}$ the source-dependent term $\bar{h}(X)$ is eliminated and it follows that the redundancies $N_i$ are independent of the source but also of actual values of target entropies ($N_i$ depends only upon the difference between the average target entropy and $R_i^*$).

\subsection{Total Entropy Constraint}\label{sec:optquantRt}
First we observe from~(\ref{eq:daopt_asym}) that the expected distortion depends upon the \emph{sum} of the side entropies and not the individual side entropies. In order to be optimal it is necessary to achieve equality in the entropy constraint, i.e.\ $R^*=\sum_i R_i$. From~(\ref{eq:Ri}) we have
\begin{equation*}
R^*=\sum_{i=0}^{K-1}R_i = K\bar{h}(X)-\frac{1}{L}\sum_{i=0}^{K-1}\log_2(N_i\nu).
\end{equation*}
This equation can be rewritten as
\begin{equation}\label{eq:tmptau}
\prod_{i=0}^{K-1}(N_i\nu) = 2^{L(K\bar{h}(X)-R^*)} = \tau_*,
\end{equation}
where $\tau_*$ is constant for fixed target entropy\index{target rate} and differential entropies. Writing~(\ref{eq:tmptau}) as
\begin{equation*}
\prod_{i=0}^{K-1}N_i^{2/L(K-1)}=\nu^{-2K/L(K-1)}\tau_*^{2/L(K-1)},
\end{equation*}
and inserting in~(\ref{eq:adistopt_asym}) leads to
\begin{equation*}
\begin{split}
D_a^{(K)} &\approx G(\Lambda_c)\nu^{2/L}\hat{p}(\mathcal{L}^{(K)}) +
\psi_L^{2}\nu^{-2/L(K-1)}\tau_*^{2/L(K-1)} G(S_L)\hat{\beta}^{(K)} \\
&\quad+ \frac{1}{L}E\|X\|^2 \prod_{i=0}^{K-1} p_i.
\end{split}
\end{equation*}
Differentiating w.r.t.\ $\nu$ and equating to zero gives
\begin{equation*}
\begin{split}
\frac{\partial D_a^{(K)}}{\partial \nu}&=
\frac{2}{L}G(\Lambda_c)\nu^{2/L-1}\hat{p}(\mathcal{L}^{(K)}) \\
&\quad -
\frac{2}{L(K-1)}\psi_L^{2} G(S_L)\nu^{-2/L(K-1)-1}\tau_*^{2/L(K-1)}\hat{\beta}^{(K)} =0,
\end{split}
\end{equation*}
from which we obtain the optimal value of $\nu$, that is
\begin{equation}\label{eq:optnuRt}
\nu = 2^{L(\bar{h}(X)-\frac{1}{K}R^*)}\left(\psi_L^{2}\frac{1}{K-1}\frac{G(S_L)}{G(\Lambda_c)}\frac{\hat{\beta}^{(K)}}
{\hat{p}(\mathcal{L}^{(K)})}\right)^{\frac{L(K-1)}{2K}}.
\end{equation}
We note that this expression is identical to~(\ref{eq:optnu_asym}). The results of this section show that the optimal $\nu$ is the same whether we optimize subject to entropy constraints on the individual side entropies or on the sum of the side entropies as long as the total bit budget is the same. 

At this point we still need to find expressions for the optimal $R_i$ (or equivalently optimal $N_i$ given $\nu$). Let $R_i = a_iR^*$, where $\sum_i a_i=1, a_i\geq 0$, hence $R^*=\sum_i R_i$. From~(\ref{eq:Ri}) we have
\begin{equation*}
R_i=\bar{h}(X)-\frac{1}L\log_2(N_i\nu)=a_iR^*,
\end{equation*}
which can be rewritten as
\begin{equation*}
N_i=\nu^{-1}2^{L(\bar{h}(X)-a_iR^*)}.
\end{equation*}
Inserting~(\ref{eq:optnuRt}) leads to an expression for the optimal index value $N_i$, that is
\begin{equation}\label{eq:optNiRt}
N_i = 2^{\frac{L}{K}(1-a_i)R^*}
\left(\psi_L^{-2}(K-1)\frac{G(\Lambda_c)}{G(S_L)}\frac{\hat{p}(\mathcal{L}^{(K)})}{\hat{\beta}^{(K)}}\right)^{\frac{L(K-1)}{2K}}.
\end{equation}

It follows from~(\ref{eq:Rc}) and~(\ref{eq:Ri}) that $R_c\geq a_iR^*$ so that $a_i\leq R_c/R^*$. In addition, since the rates must be positive, we obtain the following inequalities:
\begin{equation}\label{eq:ai}
0< a_iR^*\leq R_c, \quad i=0,\dots, K-1.
\end{equation}

Thus, when we only have a constraint $R^*$ on the sum of the side entropies, the individual side entropies $R_i=a_iR^*$ can be arbitrarily chosen (without loss of per\-for\-mance) as long as they satisfy~(\ref{eq:ai}) and $\sum_i a_i=1$. We remark that $R_i$ is bounded away from zero by a positive constant, cf.~(\ref{eq:2chan_R0}) and~(\ref{eq:2chan_R1}).
For example, for the two-channel case we have $R_0=a_0R^*$ and $R_1=a_1R^*=(1-a_0)R^*$, so that $R_c\geq (1-a_0)R^*$ which implies that $R^*-R_c\leq R_0 \leq R_c$.\footnote{Recall that $R_c$ is fixed, since it depends on $\nu$ which is given by~(\ref{eq:optnuRt}).}

This result leads to an interesting observation. 
Given a single entropy constraint on the sum of the side entropies, the optimal bit distribution among the two descriptions is not unique but contains in fact a set of solutions (i.e.\ a set of quantizers) which all lead to minimal expected distortion.\footnote{In retrospect, this is not a surprising result since, for the two-description case, we already saw that for a fixed distortion tuple $(D_c,D_0,D_1)$ the lower bound of the rate region is piece-wise linear, cf.\ Fig.~\ref{fig:2chan_rateregion}. Furthermore, when the sum rate is minimum, this line segment has a 45 degree (negative) slope. Hence, any choice of rate pairs on this line segment satisfies the sum rate\index{sum rate}. The new observation here, however, is that now we have a practical scheme, which for any number of descriptions, also satisfies this property.}
This allows for additional constraints to be imposed on the quantizers without sacrificing optimality with respect to minimal expected distortion. 
For example, in some mobile wireless environments, it might be beneficial to use those quantizers from the set of optimal quantizers that require the least amount of power. 

\subsection{Example With Total Entropy Constraint}
Let us show by an example some interesting aspects resulting from the fact that we obtain a set of candidate solutions, which all minimize the expected distortion. For example, consider IP-telephony applications, which with the recent spread of broadband networks are being used extensively throughout the world today. 
More specifically, let us consider a packet-switched network where a user has access to two different channels both based on the unreliable user datagram protocol~\cite{udp}. 
Channel 0 is a non priority-based channel whereas channel 1 is a priority-based channel or they are both priority-based channels but of different priorities. 
Equivalently this network can be thought of as a packet-switched network where the individual packets are given priorities; low or high priority. In any case, we assume that only a single packet is transmitted on each channel for each time instance (this can be justified with e.g.\ tight delay constraints).
The priority-based channel favor packets with higher priority and the packet-loss probability $p_1$ on channel 1 is therefore lower than that of channel 0, i.e.\ $p_1<p_0$. Assume the Internet telephony service provider (ITSP) in question charges a fixed amount of say \$1 (\$2) per bit transmitted via channel 0 (channel 1). 
If we then use say 6 bits on channel 1 the quality is better than if we use the 6 bits on channel 0. 
It is therefore tempting to transmit all the bits through channel 1 (or equivalently send both packets with high priority) since it offers better quality than channel 0. 
However, our results reveal that it is often beneficial to make use of both channels (or equivalently send two packets simultaneously of low and high priority). The importance of exploiting two channels is illustrated in Table~\ref{tab:bitalloc1} for the examples given above for a total bit budget of 6 bits and packet-loss probabilities $p_0=5\%$ and $p_1=2\%$. Notice the peculiarity that since the total bit budget is limited to 6 bits then even if the user is willing to pay more than \$8 the performance would be no better than what can be achieved when paying exactly \$8.
%
\begin{table}[t]
\begin{center}
\begin{tabular}{cccccc}
Network & $R_0$ & $R_1$ & Price & Quality & Expected distortion \\ \hline
Single-channel & 6 & 0 & \$6 & Poor & -12.98 dB\\
Single-channel &  0 & 6& \$12 & Good & -16.91 dB\\
Two-channel   &  2 & 4 & \$10 & Optimal & -22.20 dB\\
\textbf{Two-channel}   &  \textbf{4} & \textbf{2} & \textbf{\$8} & \textbf{Optimal} & \textbf{-22.20 dB}\\ \hline
\end{tabular}
\caption{A total bit budget of 6 bits is spent in four different ways. The bottom row shows the most economical way of spending the bits and still achieve optimal performance. The packet-loss probabilities are $p_0=5\%$ and $p_1=2\%$.}
\label{tab:bitalloc1}
\end{center}
\end{table}
The last column of Table~\ref{tab:bitalloc1} describes the expected distortion occurring when quantizing a unit-variance Gaussian source which has been scalar quantized at a total entropy of 6 bit/dim.
The packet-loss probabilities are $p_0=0.05$ and $p_1=0.02$. The quantization error (hence not taking packet losses into account) for an optimal entropy-constrained SD system is $-34.59$ dB but the expected distortion is dominated by the estimation error due to description losses, i.e.\ $10\log_{10}(p_0)=-13.01$ dB and $10\log_{10}(p_1)=-16.99$. It follows that the expected distortion for channel 0 and channel 1 is given by $-12.98$ dB and $-16.91$ dB, respectively. 
For the two-description system the expected distortion is found by use of~(\ref{eq:daopt_asym}) to be $-22.20$ dB, hence a gain of more than $5$ dB is possible when using both channels.

\section{Distortion of Subsets of Descriptions}\label{sec:networkcoding}
We have so far considered the expected distortion occurring when all possible combi\-nations of $K$ descriptions are taken into account. In a sense this corresponds to having only a single receiver. In this section we consider a generalization to multiple receivers that have access to non-identical subsets of the $K$ descriptions and where no packet losses occur. For example one receiver has access to descriptions $\{0,3\}$ whereas another has access to descriptions $\{0,1,2\}$. A total of $2^K-1$ non-trivial subsets are possible. 
We note that the design of the index-assignment map is assumed unchanged. We are still minimizing the cost functional given by~(\ref{eq:costfunctional_asym}). The only difference is that the weights do not necessarily reflect packet-loss probabilities but can be (almost) arbitrarily chosen to trade off distortion among different subsets of descriptions. For example, in a two-description system it is possible to decrease the distortion of description 0 by increasing the distortion of description 1 without affecting the rates.

The main result of this section is given by Theorem~\ref{theo:sidedist}.

\begin{theorem}\label{theo:sidedist}
The side distortion\index{side distortion} $D^{(K,l)}$ due to reception of descriptions $\{l\}$, where $l\in \mathcal{L}^{(K,\kappa)}$ for any $1\leq \kappa\leq K\leq 3$ is, asymptotically as $L,N_i\rightarrow \infty$ and $\nu_i \rightarrow 0$, given by
\begin{equation*}
D^{(K,l)} = \omega^{(K,l)}
\psi_L^{2}\nu^{2/L} G(S_L)\prod_{i=0}^{K-1}N_i^{2/L(K-1)},
\end{equation*}
where 
\begin{equation*}
\begin{split}
\omega^{(K,l)}&=\frac{1}{p(\mathcal{L}^{(K,\kappa)})^2\kappa^2}
\times\bigg(p(\mathcal{L}^{(K,\kappa)})^2\kappa^2-p(\mathcal{L}^{(K,\kappa)})^2\binom{\kappa}{2}\\ 
&\quad - p(\mathcal{L}^{(K,\kappa)})\sum_{j\in l} p(\mathcal{L}_j^{(K,\kappa)})  -
\sum_{i=0}^{K-2}\sum_{j=i+1}^{K-1}p(\mathcal{L}_i^{(K,\kappa)})p(\mathcal{L}_j^{(K,\kappa)})\bigg)
\end{split}
\end{equation*}
and $\binom{\kappa}{2}=0$ for $\kappa=1$.
\end{theorem}

\begin{proof}
See Appendix~\ref{app:sidedist}.
\end{proof}

\begin{conjecture}
Theorem~\ref{theo:sidedist} is true for $K>3$ as $L,N_i\rightarrow\infty$ and 
$\nu_i\rightarrow 0$. 
\end{conjecture}

\begin{remark}
For $K=2$ Theorem~\ref{theo:sidedist} is true also for finite $L$.\footnote{This follows since Proposition~\ref{prop:riemann2_asym} is true for any $L$ for $K=2$.} For $K=3$ it should be seen as an approximation for finite $L$.\footnote{It is in fact possible to find an exact expression for finite $L$. See Remark~\ref{rem:optfiniteL}.} 
\end{remark}

In Theorem~\ref{theo:sidedist} the term $\omega^{(K,l)}$ is a weight factor that depends on the particular subset of received descriptions. For example for $K=2$ we let $\gamma_0=\mu_0 p_1$ and $\gamma_1=\mu_1 p_0$ then for $\kappa=1$ the weights for description 0 and 1 are given by
\begin{equation}\label{eq:weightsK2}
\omega^{(2,0)} = \frac{\gamma_1^2}{(\gamma_0 + \gamma_1)^2}\quad
\text{and}\quad
\omega^{(2,1)} = \frac{\gamma_0^2}{(\gamma_0 + \gamma_1)^2},
\end{equation}
which are in agreement with the results obtained for the two-channel system in~\cite{diggavi:2002}. 

For $K=3$ and $\kappa=1$ we let $\gamma_0 = \mu_0 p_1 p_2$, $\gamma_1=\mu_1 p_0 p_2$ and $\gamma_2 = \mu_2 p_0 p_1$ and the weight for description 0 is then given by
\begin{equation*}
\omega^{(3,0)} = \frac{\gamma_1^2 + \gamma_2^2 + \gamma_1\gamma_2}{(\gamma_0 + \gamma_1 + \gamma_2)^2},
\end{equation*}
whereas for $\kappa=2$ we use the notation $\gamma_{01}=\mu_0\mu_1 p_2$, $\gamma_{02}=\mu_0\mu_2 p_1$ and $\gamma_{12}=\mu_1\mu_2 p_0$ from which we find the weight when receiving description 0 and 1 to be
\begin{equation*}
\omega^{(3,\{0,1\})} = \frac{\gamma_{02}^2 + \gamma_{12}^2 + \gamma_{02}\gamma_{12}}{4(\gamma_{01} + \gamma_{02} + \gamma_{12})^2}.
\end{equation*}

\subsection{Asymmetric Assignment Example}
\begin{figure}[ht]
\begin{center}
\includegraphics[width=7cm]{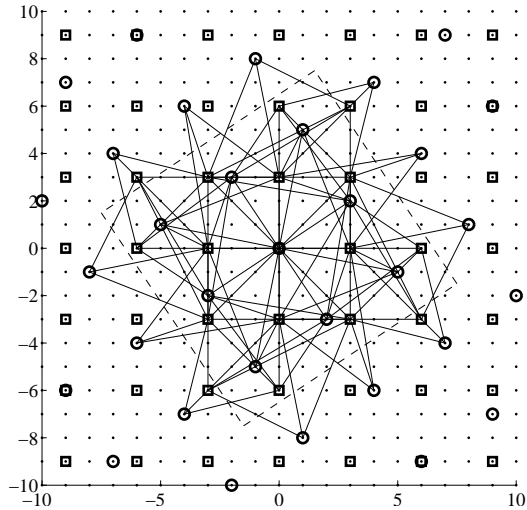}
\caption{A central lattice based on $Z^2$ (dots) and two geometrically-similar sublattices of index 13 (circles) and 9 (squares), respectively. The dashed square illustrates the boundary of $V_\pi(0)$. The solid lines illustrate the 117 edges (where some are overlapping). The weight ratio is here set to $\gamma_0/\gamma_1=1$.}
\label{fig:assign_example_sym_weight}
\end{center}
\end{figure}

In this section we illustrate by an example how one can achieve asymmetric distortions for the case of $K=2$ and the $Z^2$ lattices. Let $N_0=13$ and $N_1=9$ so that $N_\pi=117$. Thus, within $V_\pi(0)$ we have 117 central lattice points, 9 sublattice points of $\Lambda_0$, and 13 sublattice points of $\Lambda_1$. This is illustrated in Fig.~\ref{fig:assign_example_sym_weight}. We first let the weight ratio\footnote{The term weight ratio can be related to the ratio of the side distortions by use of~(\ref{eq:ourd0}) and~(\ref{eq:ourd1}). Specifically, it can be shown that $D_1/D_0 \approx \gamma_0^2/\gamma_1^2$.} \index{weight ratio} be $\gamma_0/\gamma_1=1$ so that the two side distortions are identical. 
In this case several sublattice points of $\Lambda_0$ located outside $V_\pi(0)$ will be used when labeling central lattice points inside $V_\pi(0)$. The solid lines in Fig.~\ref{fig:assign_example_sym_weight} illustrate the 117 edges that are assigned to the 117 central lattice points. 

\begin{figure}[ht]
\begin{center}
\includegraphics[width=7cm]{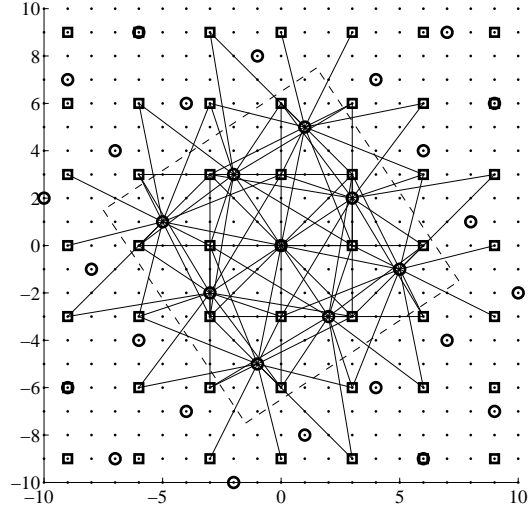}
\caption{A central lattice based on $Z^2$ (dots) and two geometrically-similar sublattices of index 13 (circles) and 9 (squares), respectively. The dashed square illustrates the boundary of $V_\pi(0)$. The solid lines illustrate the 117 edges (where some are overlapping). The weight ratio is here set to $\gamma_0/\gamma_1=4$.}
\label{fig:assign_example_asym_weight}
\end{center}
\end{figure}

If we let the weight ratio\index{weight ratio} be $\gamma_0/\gamma_1=4$ we favor $\Lambda_0$ over $\Lambda_1$. In this case the edge assignments are chosen such that for a given edge, the sublattice point belonging to $\Lambda_0$ is closer to the central lattice point than the sublattice point belonging to $\Lambda_1$. This is illustrated in Fig.~\ref{fig:assign_example_asym_weight}. Notice that in this case the sublattice points of $\Lambda_0$ used for the edges that labels central lattice points within $V_\pi(0)$ are all located within $V_\pi(0)$. Furthermore, in order to construct the required 117 edges, sublattice points of $\Lambda_1$ at greater distance from $V_\pi(0)$ need to be used.

In practice, large index values are required in order to achieve large weight ratios $\gamma_0/\gamma_1$ or $\gamma_1/\gamma_0$. Notice that we can achieve asymmetric side distortions even in the case where the sublattices are identical (so that $N_0=N_1$ and the rates are therefore identical) simply by letting $\gamma_0\neq \gamma_1$. Moreover, we can achieve symmetric side distortions by letting $\gamma_0=\gamma_1$ even when $N_0\neq N_1$ (i.e.\ $R_0\neq R_1$). In the case where either $\gamma_0=0$ and $\gamma_1\neq 0$ or $\gamma_1=0$ and $\gamma_0\neq 0$ the scheme degenerates to a successive refinement scheme, where the side distortion corresponding to the zero weight cannot be controlled. In practice this happens if either $\gamma_0 \gg \gamma_1$ or $\gamma_1\gg \gamma_0$.

\section{Numerical Results}\label{sec:results}
To verify theoretical results we present in this section experimental results obtained by computer simulations. 
In all simulations we have used $2\cdot 10^6$ unit-variance independent Gaussian vectors constructed by blocking an i.i.d.\ scalar Gaussian process into two-dimensional vectors. We first assess the two-channel performance of our scheme. This is interesting partly because it is the only case where the complete achievable MD rate-distortion region is known and partly because it makes it possible to compare to existing schemes. 
We end this section by showing the expected distortion~(\ref{eq:daopt_asym}) in an asymmetric setup using three descriptions.

\subsection{Assessing Two-Channel Performance}
\begin{figure}[ht]
\psfrag{Theo: d0}{\scriptsize Theo: $D_0$}
\psfrag{Theo: d1}{\scriptsize Theo: $D_1$}
\psfrag{Num: d0}{\scriptsize Num: $D_0$}
\psfrag{Num: d1}{\scriptsize Num: $D_1$}
\psfrag{Theo: dddd0}{\scriptsize Theo: $\bar{D}_0$}
\psfrag{Theo: dd1}{\scriptsize Theo: $\bar{D}_1$}
\psfrag{R1[bit/dim.]}{\scriptsize $R_1$ [bit/dim.]}
\begin{center}
\includegraphics[width=10cm]{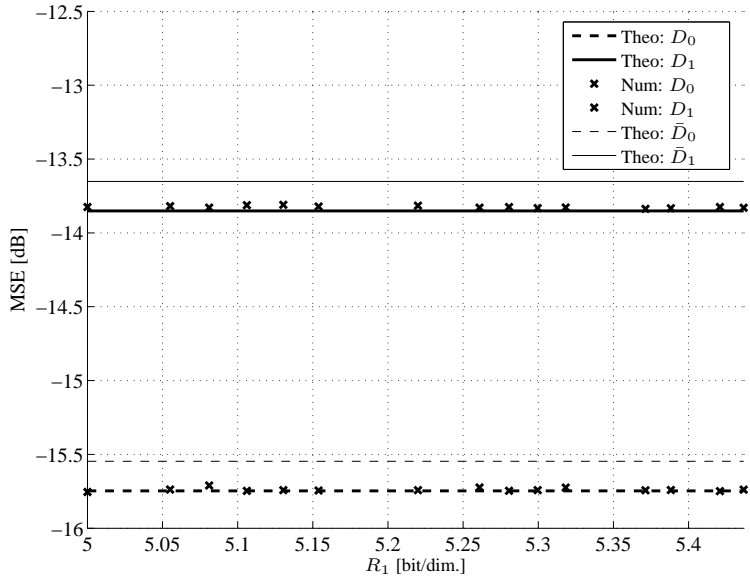}
\caption{The side distortions are here kept fixed as the rate is increased. Notice that the numerically obtained side distortions $D_0$ and $D_1$ (crosses) are strictly smaller than the theoretical $\bar{D}_0$ and $\bar{D}_1$ (thin lines).}
\label{fig:2chan_fix_z2}
\end{center}
\end{figure}

The side distortions $\bar{D}_0$ and $\bar{D}_1$ of the two-channel asymmetric MD-LVQ system presented in~\cite{diggavi:2000,diggavi:2002} are given by~(\ref{eq:d0_diggavi}) and~(\ref{eq:d1_diggavi}) and the central distortion is given by 
\begin{equation}\label{eq:centraldist}
\bar{D}_c\approx G(\Lambda_c)2^{2(\bar{h}(X)-R_c)}. 
\end{equation}
The asymmetric scheme presented in this paper satisfies
\begin{equation}\label{eq:ourd0}
D_0 \approx \frac{\gamma_1^2}{(\gamma_0+\gamma_1)^2}G(S_L)2^{2\bar{h}(X)}2^{-2(R_0+R_1-R_c)},
\end{equation}
and
\begin{equation}\label{eq:ourd1}
D_1 \approx \frac{\gamma_0^2}{(\gamma_0+\gamma_1)^2}G(S_L)2^{2\bar{h}(X)}2^{-2(R_0+R_1-R_c)},
\end{equation}
and the central distortion is identical to~(\ref{eq:centraldist}). It follows that the only difference between the pair of side distortions $(\bar{D}_0, \bar{D}_1)$ and $(D_0,D_1)$ is that the former depends upon $G(\Lambda_\pi)$ and the latter upon $G(S_L)$. In other words, the only difference in distortion between the schemes is the difference between $G(S_L)$ and $G(\Lambda_\pi)$.
For the two dimensional case it is known that $G(S_2)=1/4\pi$ whereas if $\Lambda_\pi$ is similar to $Z^2$ we have $G(\Lambda_\pi)=1/12$ which is approximately $0.2$ dB worse than $G(S_2)$. Fig.~\ref{fig:2chan_fix_z2} shows the performance when quantizing a unit-variance Gaussian source using the $Z^2$ quantizer for the design of~\cite{diggavi:2000,diggavi:2002} as well as for the proposed system. 
In this setup we have fixed $R_0=5$ bit/dim.\ but $R_1$ is varied in the range $5$ -- $5.45$ bit/dim. To do so we fix $N_1=101$ and let $N_0$ step through the following sequence of admissible index values:
\begin{equation*}
\{101,109,113,117,121,125,137,145,149,153,157,169,173,181,185\},
\end{equation*}
and for each $N_0$ we scale $\nu$ such that $R_0$ remains constant. When $N_0=101$ then $R_0=R_1=5$ bit/dim.\ whereas when $N_0>N_1$ then $R_1>R_0$.
We have fixed the ratio $\gamma_0/\gamma_1=1.55$ and we keep the side distortions fixed and change the central distortion. Since the central distortion is the same for the two schemes we have not shown it. Notice that $D_0$ (resp.\ $D_1$) is strictly smaller (about $0.2$ dB) than $\bar{D}_0$ (resp.\ $\bar{D}_1$). This is to be expected since $G(S_2)$ is approximately $0.2$ dB smaller than $G(\Lambda_\pi)$.
%
%

\subsection{Three Channel Performance}
\begin{figure}[ht]
\begin{center}
\psfrag{p3}{$p_3$ \small $[\%]$}
\includegraphics[width=10cm]{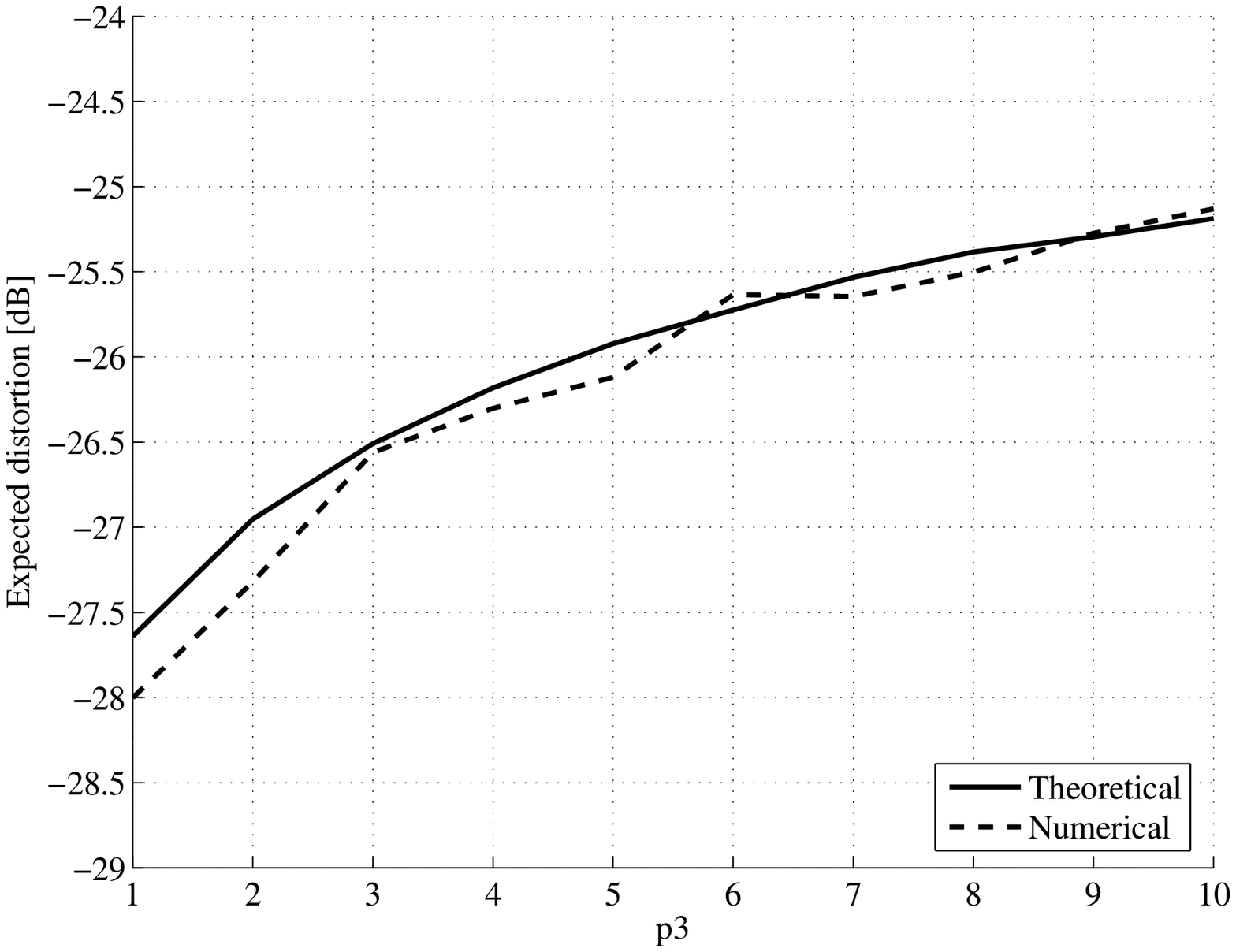}
\label{fig:a2K4perf}
\caption{Expected distortion as a function of packet-loss probabilities for $K=3$ packets and an entropy of 3 bit/dim.\ per description. The packet-loss probabilities are $p_0=2.5\%, p_1=7.5\%, 1\%\leq p_2\leq 10\%$ and $\psi_L=1.14808$.}
\label{fig:a2K4_perf}
\end{center}
\end{figure}

In this setup we let $\psi_L=1.4808$ and the packet-loss probabilities are fixed at $p_0=2.5\%, p_1=7.5\%$ except for $p_2$ which is varied in the range $[1,10]\%$. 
As $p_2$ is varied we update $\nu$ according to~(\ref{eq:optnuRt}) and pick the index values $N_i$ such that $\sum_i R_i\leq R^*$. Since index values are restricted to a certain set of integers, cf.\ Section~\ref{sec:admindexvalues}, the side entropies might not sum exactly to $R^*$. To make sure the target entropy is met with equality we then re-scale $\nu$ as $\nu=2^{L(\bar{h}(X)-\frac{1}K R^*)}\prod_{i=0}^{K-1}N_i^{-1/K}$. We see from Fig.~\ref{fig:a2K4_perf} a good correspondence between the theoretically and numerically obtained results.
%

\section{Conclusion}
We presented a design for high-resolution $K$-channel asymmetric MD-LVQ. 
Along the lines of the previous chapter, closed-form expressions for the optimal central and side quantizers based on packet-loss probabilities and subject to target entropy constraints were derived and practical quantizers were constructed to verify theoretical results. 
For the two-channel case we compared the proposed MD-LVQ scheme to a state-of-the-art two-channel asymmetric scheme and showed that the performance of the central quantizer was equivalent to that of the state-of-the-art scheme whereas the side quantizers were strictly superior in finite dimensions greater than one. 
The problem of distributing bits among the $K$ descriptions was analyzed and it was shown that the optimal solution was not unique. 
In fact, it turned out that bits could be almost arbitrarily distributed among the $K$ descriptions without loss of performance. 
As was the case for the symmetric design, the practical design of asymmetric MD-LVQ allows an arbitrary number of descriptions but the theoretical rate-distortion results were only proven for the case of $K\leq 3$ descriptions and conjectured to be true in the general case of arbitrary $K$ descriptions.

\chapter{Comparison to Existing High-Resolution MD Results}\label{chap:comparison}
In this chapter we compare the rate-distortion performance of the proposed MD-LVQ scheme to that of existing state-of-the-art schemes as well as to known information theoretic high-resolution $K$-channel MD rate-distortion bounds.

\section{Two-Channel Performance}\label{sec:comparison_twochannel}
We will first consider the symmetric case and show that, while the proposed design is different than the design of Vaishampayan et al.~\cite{vaishampayan:2001}, the two-channel performance is, in fact, identical to the results of~\cite{vaishampayan:2001}. Then we consider the asymmetric case and show that the asymmetric distortion product given by Lemma~\ref{lem:asymproduct} is achievable.

\subsection{Symmetric Case}
Let $K=2$ so that $\psi_L^2=1$. From Theorem~\ref{theo:sidedist} (see also~(\ref{eq:sidedistkappa})) we see that the side distortion (i.e.\ for $k=1$) for the symmetric case, i.e.\ $D_0=D_1$ and $R_s=R_i, i=0,1$, is given by (asymptotically as $N\rightarrow\infty$ and $\nu_s\rightarrow 0$)
\begin{equation}\label{eq:sidedista}
D_0 = \frac{1}{4}G(S_L)N^{4/L}\nu^{2/L}.
\end{equation}
In order to trade off the side rate for the central rate we use an idea of~\cite{vaishampayan:2001} and let $2^{-2aR_s} = 4N^{-2/L}$ where $0<a<1$, which implies that
\begin{equation}\label{eq:Na}
N=2^{L(aR_s + 1)}.
\end{equation}
Let us insert~(\ref{eq:Na}) into~(\ref{eq:Rs}) in order to express $\nu$ as a function of $R_s$ and $a$,
\begin{equation}\label{eq:nua}
\nu = 2^{L(h(X) - aR_s - R_s -1)}.
\end{equation}
From~(\ref{eq:centraldist}) we know that the two-channel central distortion $D_c$ is given by $D_c = G(\Lambda_c)\nu^{2/L}$ which by use of~(\ref{eq:nua}) can be rewritten as
\begin{equation}
D_c = G(\Lambda_c) 2^{2(h(X) - aR_s - R_s -1)},
\end{equation}
which leads to
\begin{equation}\label{eq:DcOpt2chan}
\lim_{R_s\rightarrow\infty} D_c 2^{2R_s(1+a)} = \frac{1}{4}G(\Lambda_c)2^{2h(X)}.
\end{equation}
By inserting~(\ref{eq:Na}) and~(\ref{eq:nua}) in~(\ref{eq:sidedista}) we find
\begin{equation}
D_0 = \frac{1}{4}G(S_L)2^{4(aR_s +1) + 2(h(X) - aR_s - R_s -1)},
\end{equation}
which leads to
\begin{equation}\label{eq:D0Opt2chan}
\lim_{R_s\rightarrow \infty} D_0 2^{2R_s(1-a)} = G(S_L) 2^{2h(X)}.
\end{equation}
Comparing~(\ref{eq:DcOpt2chan}) and~(\ref{eq:D0Opt2chan}) with those of Vaishampayan~(\ref{eq:Dc_Vaishampayan}) and~(\ref{eq:D0_Vaishampayan}) reveals that the performance of the proposed two-channel design achieve the same performance as the two-channel design of Vaishampayan et al.~\cite{vaishampayan:2001}. 
Furthermore, let $b=1$ and $L\rightarrow\infty$ and notice that in the memoryless Gaussian case $G(S_\infty)2^{2h(X)}=\sigma_X^2$ so that, by comparing~(\ref{eq:DcOpt2chan}) and~(\ref{eq:D0Opt2chan}) with~(\ref{eq:dcapprox}) and~(\ref{eq:d0approx}), we see that the high-resolution two-channel symmetric rate-distortion function of Ozarow can be achieved.

\begin{remark}\label{rem:abounds}
It is important to see that $a$ in~(\ref{eq:DcOpt2chan}) and~(\ref{eq:D0Opt2chan}) is bounded away from zero and one. 
In the extreme case where $a=0$ the ratio of side distortion over central distortion is small and $N$ cannot be made arbitrarily large as is required for the asymptotic expressions to be valid. On the other hand, when $a=1$ we can no longer force the cells of the side quantizers to be small compared to the variance of the source and the high resolution assumptions are therefore not satisfied. 
This is also true for the general case of $K>2$ descriptions. 
\end{remark}
\begin{remark}
We would like to point out an error in~\cite{ostergaard:2006b} where we overlooked the requirement that $a<1$. In~\cite{ostergaard:2006b} we showed that the high resolution performance of $(3,2)$ SCECs\index{source-channel erasure code} can be achieved by use of lattice codebooks and index assigments (which is true) but we also wrongly claimed that in the extreme case where $a=1$, lattice codebooks achieve rate-distortion points that cannot be achieved by random codebooks, obviously, this cannot be true since, for $a=1$, the high resolution assump\-tions are not satisfied (Remark~\ref{rem:abounds}).
\end{remark}

\subsection{Asymmetric Case}
We already showed in Section~\ref{sec:results} that the performance of the asymmetric two-channel scheme by Diggavi et al.~\cite{diggavi:2000,diggavi:2002} can be achieved. In fact, in finite dimensions greater than one, the performance of the proposed scheme was strictly superior to that of Diggavi et al.
Furthermore, it is easy to show that the high resolution asymmetric distortion product presented by Lemma~\ref{lem:asymproduct} can be achieved. To see this note that by use of~(\ref{eq:ourd0}),~(\ref{eq:ourd1}) and~(\ref{eq:centraldist}) we get
\begin{align}
D_c(D_0+D_1+2\sqrt{D_0D_1}) &= G(\Lambda_c) 2^{2(h(X) - R_c)} \\ \notag
&\quad\times\bigg(
\frac{\gamma_0^2+\gamma_1^2}{(\gamma_0+\gamma_1)^2}G(S_L)2^{2h(X)}2^{-2(R_0+R_1-R_c)} \\ \notag
&\quad +2\sqrt{\frac{\gamma_0^2\gamma_1^2}{(\gamma_0+\gamma_1)^4}G(S_L)^22^{4h(X)}2^{-4(R_0+R_1-R_c)}} \bigg) \\ \notag
&= G(\Lambda_c)G(S_L)2^{4h(X)}2^{-2(R_0+R_1)},
\end{align}
which, asymptotically as $L\rightarrow \infty$, leads to Lemma~\ref{lem:asymproduct}.

\section{Achieving Rate-Distortion Region of $(3,1)$ SCECs}
We will now consider the symmetric three-channel case and show that the rate-distortion performance of $(3,1)$ SCECs can be achieved at high resolution. 

We are interested in the three-channel case, i.e.\ $K=3$, and in the limit of $L\rightarrow\infty$ so that 
\begin{equation}
G(S_L) \rightarrow \frac{1}{2\pi e}
\end{equation}
and
\begin{equation}
\psi_\infty^2 = \sqrt{\frac{4}{3}}.
\end{equation}
Furthermore, without any loss of generality, we assume that the source has unit variance. Thus, from~(\ref{eq:sidedistkappa}) we see that
\begin{equation}\label{eq:sidedist1}
D^{(3,1)} = \frac{1}3\psi_\infty^{2}N'2^{-2R_s},
\end{equation}
since $R_c = R_s + \log_2(N')$,
\begin{equation}\label{eq:sidedist2}
D^{(3,2)} = \frac{1}{12}\psi_\infty^{2}N'2^{-2R_s},
\end{equation}
and the central distortion $D_c=D^{(3,3)}$ given by~(\ref{eq:centraldist}) can be written as
\begin{equation}\label{eq:centraldist1}
D^{(3,3)} = \left(\frac{1}{N'}\right)^2 2^{-2R_s}.
\end{equation}

The following lemma shows that symmetric three-channel MD-LVQ can achieve the rate-distortion region of $(3,1)$ SCECs at high resolution.
\begin{lemma}\label{lem:31scecdist}
At high resolution, the one, two and three-channel distortions of $(3,1)$ SCECs are identical to~(\ref{eq:sidedist1}) -- (\ref{eq:centraldist1}) in the quadratic Gaussian case\index{quadratic!Gaussian}.
\end{lemma}
\begin{proof}
See Appendix~\ref{app:comparison_proofs_lemmas}.
\end{proof}

\begin{remark}
The notion of a large sublattice index $N$ in $K$-channel MD-LVQ corresponds to a large (negative) codebook correlation $\rho_q$ for $(K,1)$ SCECs and in the limit of $N\rightarrow\infty$ we actually have $\rho_q\rightarrow -1/(K-1)$. Thus, for $K=3$ we have $\rho_q\rightarrow -1/2$ as $N\rightarrow\infty$. 
\end{remark}

\section{Achieving Rate-Distortion Region of $(3,2)$ SCECs}
We will now show that the rate-distortion performance of $(3,2)$ SCECs can be achieved by extending the proposed design of three-channel MD-LVQ to include random binning\index{random!binning} on the side codebooks. Specifically, we show that the achievable two-channel versus three-channel distortion region of $(3,2)$ SCECs for the memoryless Gaussian source and MSE can be achieved under high-resolution assumptions. Since the performance of a $(3,3)$ SCEC is identical to that of a single description scheme, it is clear that we can also achieve such perfor\-man\-ce simply by letting $K=1$ and only use the central quantizer.
Explicit bounds for $K>3$ descriptions were not derived in~\cite{pradhan:2004,puri:2005} but we expect that these (non-derived) bounds are also achievable with the proposed $K$-channel MD-LVQ scheme.

We will begin by considering the general situation where we allow finite dimensio\-nal lattice vector quantizers and asymmetric rates and distortions. 
Then, at the end of the section, we focus on the symmetric case and infinite-dimensional lattice vector quantizers in order to compare the performance to the existing bounds. 


Recall that the proposed design of $K$-channel MD-LVQ is able to vary the redun\-dancy by changing the number of descriptions $K$ as well as the index values $N_i$.
In addition, it is possible to trade off distortion among subsets of descriptions, without affecting the rates, simply by varying the weights. 
Increasing $N_i$ and at the same time decreasing $\nu$ so that $\nu_i=N_i\nu$ remains constant does not affect the rate $R_i$. However, the distortion due to the $i^{th}$ description is affected (unless counteracted by the weights). For example in the symmetric setup where $N=N_i$ for all $i$ and the weights are also balanced, the side distortion due to reception of only a subset of descriptions is increased as $N$ is increased and $\nu N$ is kept constant. 
However, in this case, the central distortion due to reception of all descriptions is decreased. In other words, in the symmetric case, for a given $K$, the degree of redundancy is controled by the single parameter $N$.

\subsection{Random Binning on Side Codebooks of MD-LVQ Schemes}\label{sec:scec}
In order to achieve the performance of general $(K,k)$ SCECs we need to introduce more controlling parameters into the design of $K$-channel MD-LVQ. 
To do so we follow an idea of Pradhan et al.~\cite{pradhan:2004} and exploit recent results on distributed source coding. More specifically, we apply random binning\index{random!binning} on the side codebooks of the $K$-channel MD-LVQ scheme. This corresponds in some sense to replacing the random codebooks\index{random!codebook} of $(K,k)$ SCECs with structured lattice codebooks except that we also have a central quantizer and an index assignment map to consider.

Random binning\index{random!binning} is usually applied on (in principle infinite-dimensional) random codebooks\index{random!codebook}. The idea is to exploit the fact that for a given codevector, say $\lambda_0$, of codebook $\mathcal{C}_0$, only a small set of the codevectors in codebook $\mathcal{C}_1$ is jointly typical\index{typical!jointly} with $\lambda_0$.
Then by randomly distributing the codevectors of $\mathcal{C}_1$ over $M_1$ bins, it is unlikely that two or more codevectors, which are all jointly typical with $\lambda_0$, end up in the same bin (at least this is true if $M_1$ is large enough). Thus, if the binning rate $R_{b,1}=\log_2(M_1)$ is less than the codebook rate $R_1$ then it is possible to reduce the description rate by sending the bin indices instead of the codevector indices. 

The rate and distortion performance of lattice vector quantizers are often described using high-resolution assumptions, i.e.\ the rate of the quantizer is sufficiently high and the source pdf sufficiently smooth, so that the pdf can be considered constant within Voronoi regions of the code vectors. Under these assumptions the theoretical performance of lattice vector quantizers can be derived for arbitrary vector dimension. This is in contrast to the asymptotics used when deriving theoretical expressions for the performance of random codebooks\index{random!codebook}. 
For random codebooks the theoretical performance is usually derived based on asymptotically high vector dimension but arbitrary rates. The theory behind random binning relies upon asymptotically high vector dimension and as such when using random binning in $K$-channel MD-LVQ we make use of both asymptotics, i.e.\ high vector dimension and high rates. It is also worth mentioning that we consider memoryless sources with infinite alphabets\index{alphabet!continuous} such as e.g.\ the Gaussian source, whereas the analysis of SCECs relies upon strong typicality\index{typical!strongly} and as such only discrete alphabet memoryless sources are valid.\footnote{However, in~\cite{pradhan:2004} the authors remark that the analysis of SCECs can be generalized to continuous-alphabet memoryless sources by using the techniques of~\cite[Ch.7]{gallager:1968}.}

In lattice codebooks, the code vectors are generally not jointly typical and the concept of random binning is therefore not directly applicable.
It is, however, possible to simulate joint typicality by for example some distance metric, so that code vectors close together (in e.g.\ Euclidean sense) are said to be ``jointly typical''. The index assignments of MD-LVQ is another example of how to simulate joint typicality. We use the term admissible $K$-tuple for any set of $K$ code vectors $(\lambda_0,\dots,\lambda_{K-1})$ which is obtained by applying the index-assignment map on a code vector $\lambda_c$, i.e.\ $\alpha(\lambda_c)=(\lambda_0,\dots,\lambda_{K-1})$ for all $\lambda_c\in \Lambda_c$. 
So for lattice code vectors we exploit that only a subset of all $K$-tuples are admissible $K$-tuples which, in some sense, corresponds to the fact that only a subset of all $K$-tuples of random code vectors are jointly typical.

Let $J\subseteq \{0,\dots,K-1\}$ denote an index set, where $|J|=k$. 
A $k$-tuple is a set of $k$ elements $\{\lambda_j\}, j\in J$ where $\lambda_{j} \in \Lambda_{j}$. The $k$-tuple given by $\{\lambda_j\}=\{\alpha_j(\lambda_c)\}, j\in J$ for any $\lambda_c\in \Lambda_c$ is said to be an admissible $k$-tuple.
Each lattice $\Lambda_i$ contains an infinite number of lattice points (or reproduction vectors) but we show by Lemma~\ref{lem:boundCi} that only finite sets of these points are needed for the codebooks of the side quantizers and we denote these sets by $\mathcal{C}_i \subset \Lambda_i$, where $|\mathcal{C}_i|< \infty$.
The set 
\begin{equation}
\begin{split}
\{\lambda_2|&\{\lambda_1,\lambda_0\}\}= \\ 
&\{\lambda_2\in \Lambda_2 : \lambda_2=\alpha_2(\lambda_c)\ \text{and}\ (\alpha_1(\lambda_c),\alpha_0(\lambda_c))=(\lambda_1,\lambda_0),\ \forall \lambda_c\in \Lambda_c\},
\end{split}
\end{equation}
denotes the set of $\lambda_2$'s which are in admissible $k$-tuples that also contain the specific elements $\lambda_0$ and $\lambda_1$. 

Since we consider the asymmetric case some ambiguity is present in the term $D^{(K,k)}$, because it is not specified which $k$ out of the $K$ descriptions that are to be considered. To overcome this technicality we introduce the notation $D^{(K,J)}, J\in \mathcal{K}$, where $\mathcal{K}$ denotes the set of combinations of descriptions for which the distortion is specified. For example, if we are only interested in the distortion when receiving descriptions $\{0,1\},\{0,2\}$ or $\{0,1,2\}$ out of all subset of $\{0,1,2\}$, then $\mathcal{K}=\{\{0,1\},$ $\{0,2\},\{0,1,2\}\}$ and nothing is guaranteed upon reception of either a single description or the pair of descriptions $\{1,2\}$.

We will now outline the construction of $(K,\mathcal{K})$ MD-LVQ. It can be seen that the construction of $(K,\mathcal{K})$ MD-LVQ resembles the construction of $(K,k)$ SCECs given in~\cite{pradhan:2004}.

\paragraph{Construction of lattice codebooks}
Construct a $K$-channel MD-LVQ system with one central quantizer and $K$ side quantizers of rate $R_i$.
Let $\mathcal{C}_c$ be the codebook of the central quantizer and let $\lambda_c(j_c)\in \mathcal{C}_c$ denote the $j_c^{th}$ element of $\mathcal{C}_c$. Similarly, let $\mathcal{C}_i$ where $i=0,\dots K-1$ denote the codebook of the $i^{th}$ side quantizer and let $\lambda_i(j_i)\in \mathcal{C}_i$ denote the $j_i^{th}$ codeword of $\mathcal{C}_i$. Finally, let $\alpha$ be the index-assignment function that maps central lattice points to sublattice points.

\paragraph{Random binning}
Perform random binning\index{random!binning} on each of the side codebooks $\mathcal{C}_i$ to reduce the side description rate from $R_i$ to $R_{b,i}$ bit/dim., where we assume $R_i>R_{b,i}$.
Let $\xi_i=2^{L(R_i-R_{b,i}+\gamma_i)}$ where $\gamma_i>0$. Assign $\xi_i$ codewords to each of the $2^{LR_{b,i}}$ bins of each codebook. The codewords for a given bin of codebook $\mathcal{C}_0$ is found by randomly extracting $\xi_0$ codewords from $\mathcal{C}_0$ uniformly, independently and with replacement. This procedure is then repeated for all the remaining codebooks $\mathcal{C}_i, i=1,\dots,K-1$.

\paragraph{Encoding}
Given a source word $X\in \mathbb{R}^L$, find the closest element $\lambda_c\in \mathcal{C}_c$ and use $\alpha$ to obtain the corresponding $K$-tuple, i.e.\ $\alpha(\lambda_c)=(\lambda_0(j_0),\dots,\lambda_{K-1}(j_{K-1}))$. If the codeword $\lambda_i\notin \mathcal{C}_i$ then set $j_i$ equal to a fixed special symbol\footnote{The rate increase caused by the introduction of the additional symbol $\vartheta$ is vanishing small for large $L$.}, say $j_i=\vartheta$. For $i=0,\dots, K-1$ define the function $f_i(\lambda_i(j_i))$ which indicates the index of a bin containing the codeword $\lambda_i(j_i)$. If $\lambda_i(j_i)$ is found in more than one bin, set $f_i(\lambda_i(j_i))$ equal to the least index of these bins. If $\lambda_i(j_i)$ is not in any bin, set $f_i(\lambda_i(j_i))=\vartheta$. The bin index $f_i(\lambda_i(j_i))$ is sent over channel $i$.

\paragraph{Decoding}
The decoder receives some $m$ bin indices and searches through the corre\-sponding bins to identify a unique admissible $m$-tuple. 

\begin{remark}
For $(K,\mathcal{K})$ MD-LVQ the notion of large block length, i.e.\ $L\rightarrow\infty$, is introduced in order to make sure that standard binning arguments can be applied. However, it should be clear that the quantizer dimension is allowed to be finite.
If finite quantizer dimension is used it must be understood that (finite length) codewords from consecutive blocks are concatenated to form an $L$-sequence of codewords. The dimension of the $L$-sequence becomes arbitrarily large as $L\rightarrow \infty$, but the quantizer dimension remains fixed. As such this will not affect the binning rate but the distortion tuple $\{D^{(K,J)}\}_{J\in\mathcal{K}}$ is affected in an obvious way. 
\end{remark}

\begin{theorem}\label{theo:scec-lc}
Let $X\in \mathbb{R}^L$ be a source vector constructed by blocking an arbitrary i.i.d.\ source with finite differential entropy into sequences of length $L$. Let $J\subseteq\{0,\dots,K-1\}$ and let $\lambda_{J}$ denote the set of codewords indexed by $J$. The set of decoding functions is denoted $g_J\colon \bigotimes_{j\in J}\Lambda_j \to \mathbb{R}^L$. Then, under high-resolution assumptions, if 
\begin{equation*}
E[\rho(X,g_{J}(\lambda_J))] \leq D^{(K,J)},\quad \forall J\in \mathcal{K},
\end{equation*}
where $\rho(\cdot,\cdot)$ is the squared-error distortion measure and for all $S\subseteq J$
\vspace{-2mm}
\begin{equation}\label{eq:sumbinrateR}
\sum_{i\in S} R_{b,i} > \sum_{i\in S}\gamma_i + \frac{1}{L}\log_2(|\{\lambda_S|\lambda_{J-S}\}|),
\end{equation}
the rate-distortion tuple $(R_{b,0},\dots,R_{b,(K-1)},\{D^{(K,J)}\}_{J\in\mathcal{K}})$ is achievable.
\end{theorem}
\begin{proof}
See Appendix~\ref{app:theo:scec-lc}.
\end{proof}

We have the following corollary for the symmetric case:
\begin{corollary}[Symmetric case]\label{cor:scec-lc}
Let $X\in \mathbb{R}^L$ be a source vector constructed by blocking an arbitrary i.i.d.\ source with finite differential entropy into sequences of length $L$. For any $J\subseteq \{0,\dots,K-1\}$ let $\lambda_{J}$ denote the set of received codewords and let $g_J\colon \bigotimes_{j\in J}\Lambda_j \to \mathbb{R}^L$ be the set of decoding functions. Then, under high-resolution assumptions, if 
\begin{equation*}
E[\rho(X,g_{J}(\lambda_J))] \leq D^{(K,|J|)},\quad \forall J \subseteq \{0,\dots,K-1\}, |J|\geq k,
\end{equation*}
where $\rho(\cdot,\cdot)$ is the squared-error distortion measure and for all $S\subseteq J$
\vspace{-2mm}
\begin{equation}\label{eq:binrateR}
R_b > \gamma + \frac{1}{|S|L}\log_2(|\{\lambda_S|\lambda_{J-S}\}|),
\end{equation}
the tuple $(R_b,D^{(K,k)},D^{(K,k+1)},\dots,D^{(K,K)})$ is achievable.
\end{corollary}
\begin{proof}
Follows immediately from Theorem~\ref{theo:scec-lc}.
\end{proof}

\subsection{Symmetric Case}
To actually apply Theorem~\ref{theo:scec-lc} we need to find a set of binning rates $\{R_{b,i}\},i=0,\dots, K-1$, such that~(\ref{eq:sumbinrateR}) is satisfied for all subsets $S$ of $J$ and for all elements $J$ of $\mathcal{K}$. Let us consider the symmetric case where $K=3$ and design a $(3,2)$ MD-LVQ system. We then have $J=\{i_0,i_1\}$ and it suffices to check the two cases where $S=i_0$ and $S=\{i_0,i_1\}$. Without loss of generality we assume that $i_0=0$ and $i_1=1$. 
The number of distinct $\lambda_1$'s that is paired with a given $\lambda_0$ can be approximated\footnote{Recall that this approximation becomes exact as $N\rightarrow\infty$.} by $(\psi_L \sqrt{N'})^L$, where $N'=N^{1/L}$ is the dimension normalized index value describing the index (redundancy) per dimension. 
Let $S=\{0,1\}$ and notice that $|\{\lambda_S\}|\leq |\{\lambda_1|\lambda_0\}|\cdot |\mathcal{C}_0|$. Then, asymptotically, as $N\rightarrow\infty$, it follows that $|\{\lambda_1|\lambda_0\}|=(\psi_L\sqrt{N'})^L$. Let us now bound the codebook cardinality. 

\begin{lemma}\label{lem:boundCi}
$|\mathcal{C}_i| = 2^{LR_i}$. Furthermore, the entropy of the quantizer indices is upper bounded by $R_i$.\footnote{For large $L$ there is really no loss by assuming that $2^{LR_i}$ is an integer.} 
\end{lemma}
\begin{proof}
See Appendix~\ref{app:comparison_proofs_lemmas}.
\end{proof}

We are now able to find $R_b$ by considering the two cases $|S|=1,2$. For $|S|=1$ we have from~(\ref{eq:binrateR}) that
\vspace{-1mm}
\begin{equation}\label{eq:RI}
R_{b,I} > \gamma+\log_2(\psi_L\sqrt{N'}),
\end{equation}
whereas for $|S|=2$ 
\vspace{-1mm}
\begin{equation}\label{eq:RII}
R_{b,II} > \frac{1}{2}R_s+\gamma+\frac{1}{2}\log_2(\psi_L\sqrt{N'}).
\end{equation}
The dominant $R_b$ is then given by $R_b=\max(R_{b,I},R_{b,II})$. 
Since~(\ref{eq:RI}) and~(\ref{eq:RII}) depends upon $N'$ the dominating binning rate depends upon $N'$. To resolve this problem, we form the inequality $R_{b,II}\geq R_{b,I}$ and find that $N'\leq 2^{2R_s}/\psi_L^2$.
So for $N'\leq 2^{2R_s}/\psi_L^2$ we have $R_b=R_{b,II}$. It is interesting to see that when inserting $N'=2^{2R_s}/\psi_L^2$ in (\ref{eq:RI}) we get $R_{b,I}=\gamma+R_s$. Coincidently, $R_{b,I}$ becomes effective when the binning rate $R_b$ is equal to the codebook rate $R_s$, which violates the assumption that $R_b>R_s$.

It is clear that if we set $R_b$ equal to the lower bound in~(\ref{eq:RII}) we get
\begin{equation*}
R_b = \frac{1}2 R_s + \frac{1}{4}\log_2(N') + \frac{1}{2}\log_2(\psi_L),
\end{equation*}
from which we can express $N'$ and $R_s$ as functions of each other and $R_b$, that is
\begin{equation}
N' = 2^{4R_b-2R_s}\psi_L^{-2},
\end{equation}
and
\begin{equation}\label{eq:RsNR}
R_s = 2R_b - \log_2(\psi_L)-\frac{1}{2}\log_2(N').
\end{equation}
It follows that when varying the redundancy per dimension $N'$, the binning rate $R_b$ can be kept constant by adjusting $R_s$ according to~(\ref{eq:RsNR}). 

In order to compare these results to the existing bounds we let $L\rightarrow\infty$ so that by inserting~(\ref{eq:RsNR}) in~(\ref{eq:sidedistkappa}) we get
\begin{equation}\label{eq:D2barbin}
\begin{split}
D^{(3,2)} &= \frac{1}{12}\psi_\infty^2 N' 2^{-2R_s} \\
 &= \frac{1}{12}\psi_\infty^4 (N')^2 2^{-4R_b}.
\end{split}
\end{equation}
The central distortion ($D_c = D^{(3,3)}$) in MD-LVQ is given by
\begin{equation}\label{eq:Dcbin2}
D^{(3,3)} = 2^{-2R_c},
\end{equation}
where $R_c = R_s + \log_2(N')$ which leads to
\begin{equation}\label{eq:Rcbin2}
R_c = 2R_b - \log_2(\psi_\infty) + \frac{1}{2}\log_2(N').
\end{equation}
Inserting~(\ref{eq:Rcbin2}) into~(\ref{eq:Dcbin2}) leads to
\begin{equation}\label{eq:Dcbin2a}
D^{(3,3)} = \frac{\psi_\infty^2}{N'}2^{-4R_b}.
\end{equation}

\begin{lemma}\label{lem:32scecdist}
At high resolution, the two and three-channel distortions of $(3,2)$ SCECs are identical to~(\ref{eq:D2barbin}) and~(\ref{eq:Dcbin2a}) in the quadratic Gaussian case\index{quadratic!Gaussian}.
\end{lemma}
\begin{proof}
See Appendix~\ref{app:comparison_proofs_lemmas}.
\end{proof}

\subsection{Asymmetric Case}
For the asymmetric case, $K=3$ and where $\mathcal{K}=\{\{0,1\},\{0,2\},\{1,2\},\{0,1,2\}\}$, i.e.\ reconstruction is possible when any two or more descriptions are received, it can be shown (similar to the symmetric case) that the binning rate $R_{b,i}$ is lower bounded by $R_{b,i}=\text{max}(R_{b,i_I},R_{b,i_{II}})$ where
\begin{equation}\label{eq:RiI}
R_{b,i_I} = \log_2(\psi_L) + \frac{1}{2}\log_2(N_\pi') - \log_2(N_i')
\end{equation}
and
\begin{equation}\label{eq:RiII}
R_{b,i_{II}} = \frac{1}{2}\log_2(\psi_L) + \frac{1}{4}\log_2(N_\pi') - \frac{1}{2}\log_2(N_i')+\frac{1}{2}R_i,
\end{equation}
where $N_\pi'=N'_0N'_1N'_2$.

To see this note that if $\lambda_i$ and $\lambda_j$ both are in the same admissible $K$-tuple, then $\lambda_i$ must be within a sphere $\tilde{V}$ centered at $\lambda_j$. The volume of $\tilde{V}$ is $\tilde{\nu}$, which implies that the maximum number of distinct $\lambda_i$ points within $\tilde{V}$ is approximately $\tilde{\nu}/\nu_i$.  In other words,
\begin{equation}
\begin{split}
|\{\lambda_i|\lambda_j\}| &\approx \tilde{\nu}/\nu_i \\
& = (\psi_L\sqrt{N_0'N_1'N_2'}/N_i)^L,
\end{split}
\end{equation}
where the approximation becomes exact for large index values. With this it is easy to see that 
\begin{equation}
\begin{split}
R_{b,i} &> \frac{1}{L}\log_2(|\{\lambda_i|\lambda_j\}|) \\
&\approx \log_2(\psi_L) + \frac{1}{2}\log_2(N_0'N_1'N_2') -\log_2(N_i),
\end{split}
\end{equation}
which is identical to~(\ref{eq:RiI}). From Theorem~\ref{theo:scec-lc} we can also see that the pair-wise sum rate\index{sum rate} must satisfy
\begin{equation}
\begin{split}
R_{b,i} + R_{b,j} &> \frac{1}{L}\log_2(|C_i||\{\lambda_j|\lambda_i\}|) \\
&\approx R_i +  \log_2(\psi_L) + \frac{1}{2}\log_2(N_0'N_1'N_2') -\log_2(N_j),
\end{split}
\end{equation}
which can equivalently be expressed as $R_{b,i} + R_{b,j} > \frac{1}{L}\log_2(|C_j||\{\lambda_i|\lambda_j\}|)$ from which the individual rate requirements can be found to be given by~(\ref{eq:RiII}).

\section{Comparison to $K$-Channel Schemes}
In the asymptotic case of large lattice vector quantizer dimension and under high resolution conditions, we showed in the previous sections that existing MD bounds can be achieved. However, it is also of interest to consider the rate-distortion perfor\-mance that can be expected in practical situations. Towards that end we presented some numerical results obtained through computer simulations in Sections~\ref{sec:num} and~\ref{sec:results}. 

In this section we will compare the theoretical performance of the proposed MD-LVQ scheme to existing state-of-the-art $K$-channel MD schemes~\cite{chen:2006,ostergaard:2007}.
While the schemes~\cite{chen:2006,ostergaard:2007} (as well as the proposed scheme) can be shown to be optimal, under certain asymptotic conditions, 
they are not without their disadvantages when used in practical situations. 
We will, however, refrain from comparing implementation specific factors such as computational complexity as well as scalability in dimension, description rate and number of descriptions. Such comparisons, although relevant, are often highly application specific. 

The above mentioned schemes are all based on LVQ and it is therefore possible to compare their theoretical rate-distortion performance when finite-dimensional lattice vector quantizers are used. Recall from Section~\ref{sec:nonindexrateloss} that the scheme of Chen et al.~\cite{chen:2006} has a rate loss of $(2K-1)$ $L$-dimensional lattice vector quantizers.\footnote{In the asymmetric case where corner points of the rate region are desired, the rate loss of~\cite{chen:2006} is only that of $K$ lattice vector quantizers. However, in the symmetric case, source splitting\index{source splitting} is necessary and there is an additional rate loss.}
The scheme of \O stergaard and Zamir~\cite{ostergaard:2007} was, for the case of $K=2$, shown to have a rate loss\index{rate loss!MD} of only two $L$-dimensional lattice vector quantizers. While this design was shown to permit an arbitrary number of descriptions, the rate loss for $K>2$ descriptions was not assessed.
The rate loss of the proposed scheme, on the other hand, has a somewhat peculiar form. In the case of two descriptions, the rate loss is given by that of two $L$-dimensional quantizers having spherical Voronoi cells.\footnote{This is true in the symmetric case as well as in the asymmetric case.}
However, in the case of $K>2$ descriptions, there is an additional term which influences the rate loss. 

\subsection{Rate Loss of MD-LVQ}
\begin{figure}[ht]
\psfrag{L}{\small $L$}
\begin{center}
\includegraphics[width=10cm]{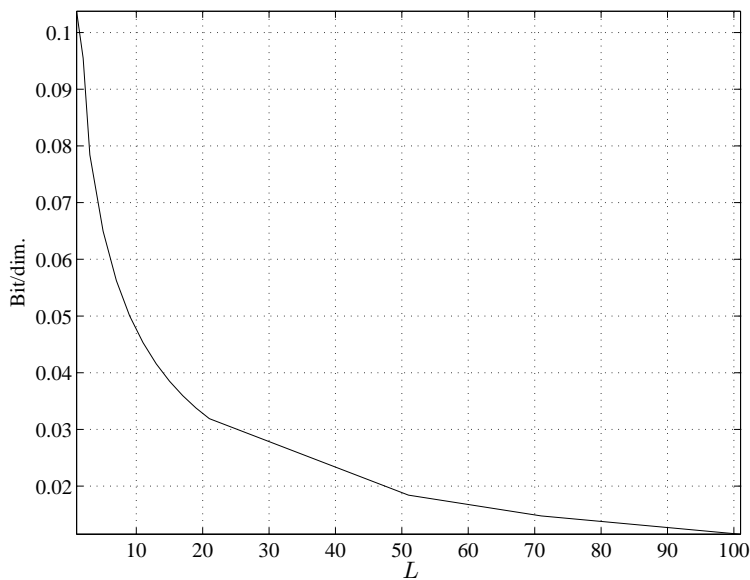}
\end{center}
\caption{The rate loss due to the term $\log_2(\psi_L/\psi_\infty)$ is here expressed in bit/dim.\ as a function of the dimension $L$.}
\label{fig:psi_loss}
\end{figure}

To be able to assess the rate loss\index{rate loss!MD-LVQ} of MD-LVQ when using finite-dimensional quantizers and more than two descriptions, we let $R_f$ denote the description rate (where the subscript ${}_f$ indicates that finite-dimensional quantizers are used). Then the distortion when receiving a single description out of $K=3$ can be found by use of~(\ref{eq:sidedistkappa}) to be given by
\begin{equation}\label{eq:finitequant31}
D^{(3,1)} = \frac{1}{3}G(S_L)(2\pi e) \psi_L^2 N' 2^{-2R_f}.
\end{equation}
Equalizing~(\ref{eq:sidedist1}) and~(\ref{eq:finitequant31}) reveals that the rate loss ($R_f-R_s$), for $K=3$, is given by (at high resolution)
\begin{equation}\label{eq:rateloss31}
R_f-R_s = \frac{1}{2}\log_2(G(S_L)(2\pi e)) + \log_2(\psi_L/\psi_\infty).
\end{equation}
Since $\psi_L\leq \psi_1 = \psi_\infty^2$ (at least for $K=3$) we can upper bound the second term of~(\ref{eq:rateloss31}) by
\begin{equation}\label{eq:volumeloss}
\log_2(\psi_L/\psi_\infty) \leq \log_2(\psi_\infty) = 0.1038\ \text{bit/dim.}
\end{equation}
Fig.~\ref{fig:psi_loss} shows $\log_2(\psi_L/\psi_\infty)$ for $1 \leq L \leq 101$ for $K=3$ using the values of $\psi_L$ from Table~\ref{tab:psiL}.

\begin{remark}
For $K=2$ we have $\psi_L=1, \forall L$, and~(\ref{eq:rateloss31}) is true. Furthermore, if the $K$-channel MD-LVQ scheme is optimal also for $K>3$, as we previously conjectured, then~(\ref{eq:rateloss31}) is true for any $K\geq 2$ (at high resolution).
\end{remark}

\begin{figure}[ht]
\psfrag{[1]}{\small Chen et al.\cite{chen:2006}}
\psfrag{[2]}{\small $(3/2)\log_2(G(\Lambda)2\pi e)$}
\psfrag{3*1/2*log2(2pie/12)}{\small $(3/2)\log_2(2\pi e/12)$}
\psfrag{3*1/2*log2(G(SL)***2pie)}{\small $(3/2)\log_2(G(S_L)2\pi e)$}
\psfrag{L}{\small $L$}
\begin{center}
\includegraphics[width=10cm]{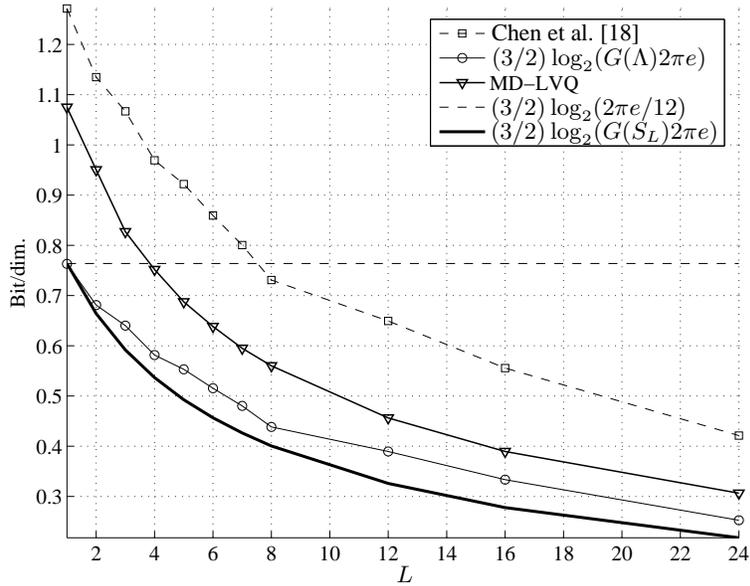}
\end{center}
\caption{Rate loss of the different three-channel MD schemes. For comparison we have included a lower bound (thick solid curve), which corresponds to the sum rate\index{sum rate} loss of three $L$-dimensional quantizers having spherical Voronoi cells. The horizontal dashed line corresponds to the sum rate loss of three lattice vector quantizers with hypercubic cells and the circles illustrate the performance of a scheme having a sum rate loss of three optimal lattice vector quantizers.}
\label{fig:rateloss31}
\end{figure}

It is interesting to observe that both terms in~(\ref{eq:rateloss31}) are independent of the particular type of lattice being used. For example, if the product lattice $\Lambda=Z^{\infty}$ is used for the side quantizers, then the rate loss vanishes (it becomes identical to zero) even though $G(\Lambda)=1/12$.\footnote{Recall that the central distortion depends upon the type of lattice being used. However, at high resolution conditions, the index value is large (i.e.\ $N\rightarrow\infty$) and as such the central distortion is very small compared to the side distortion and can therefore be neglected.} This is not the case with the other two schemes mentioned above, i.e.\ \cite{chen:2006,ostergaard:2007}.
Fig.~\ref{fig:rateloss31} shows the rate loss of the different schemes for $K=3$ descriptions. The lattices used are those of Table~\ref{tab:latparm}. Since we do not have $\psi_L$ values for all even $L$ we have in Fig.~\ref{fig:rateloss31} simply used the average of the two neighboring values, that is
\begin{equation}
\psi_L = 
\begin{cases}
\displaystyle\frac{\psi_{L-1}+\psi_{L+1}}{2}, & L\ \text{even} \\
\psi_L, & L\ \text{odd}.
\end{cases}
\end{equation}

%
%

\section{Conclusion}
In the previous two chapters we initially used a single index-assignment map to control the redundancy between descriptions. 
In this chapter we then showed that by use of random binning\index{random!binning} on the side codebooks in addition to the index-assignment map it was possible to introduce more rate-distortion controlling parameters into the design. While the use of random binning is standard procedure in distributed source coding and has previously been applied to MD schemes based on random codebooks as well, it appears to be the first time it is used in connection with index-assignment based MD schemes. 

We showed that the proposed design of MD-LVQ, asymptotically in rate and lattice vector quantizer dimension, achieves existing rate-distortion MD bounds in the quadratic Gaussian case for two and three descriptions. 

In finite lattice vector quantizer dimensions, we showed that the rate-loss of the proposed design is superior to existing state-of-the-art schemes.

\chapter{Network Audio Coding}\label{chap:nac}
In this chapter we apply the developed MD coding scheme to the practical problem of network audio coding. Specifically, 
we consider the problem of reliable distribution of audio over packet-switched networks\index{packet-switched network} such as the Internet or general ad hoc networks.\footnote{Part of the research presented in this chapter represents joint work with O.\ Niamut.}
Thus, in order to combat (excessive) audio packet losses we choose to transmit multiple audio packets. 

Many state-of-the-art audio coding\index{audio coding} schemes perform time-frequency analysis\index{time-frequency analysis} of the source signal, which makes it possible to exploit perceptual models in both the time and the frequency domain in order to discard perceptually-irrelevant information.
This is done in e.g.\ MPEG-1 (MP3)~\cite{MPEG1MP3}\index{MP3}, MPEG-2\index{MPEG-2} advanced audio coding (AAC)~\cite{MPEG2AAC}, Lucent PAC~\cite{sinha:1998}\index{PAC} and Ogg Vorbis~\cite{vorbis:2006}\index{Ogg Vorbis}. 
The time-frequency analysis is often done by a transform coder which is applied to blocks of the input signal. A common approach is to use the modified discrete cosine transform\index{modified discrete cosine transform} (MDCT)~\cite{malvar:1992} as was done in e.g.\ MPEG-2 AAC~\cite{MPEG2AAC}, Lucent PAC~\cite{sinha:1998} and Ogg Vorbis~\cite{vorbis:2006}. In this chapter we combine the MDCT with $K$-channel MD-LVQ in order to obtain a perceptual transform coder, which is robust to packet losses.

MD coding of audio has to the best of the author's knowledge so far only been considered for two descriptions~\cite{arean:2000,schuller:2005}. However, here we propose a scheme that is able to use an arbitrary number of descriptions without violating the target entropy. We show how to distribute the bit budget among the MDCT coefficients and present closed-form expressions for the rate and distortion performance of the $K$-channel MD-LVQ system which minimize the expected distortion based on the packet-loss probabilities. Theoretical results are verified with numerical computer simulations and it is shown that in environments with excessive packet losses it is advantageous to use more than two descriptions. We verify the findings that more than two descriptions are needed by subjective listening tests, which further proves that acceptable audio quality can be obtained even when the packet-loss rate is as high as 30\%.

\section{Transform Coding}
In this section we describe the MDCT and we define a perceptual distortion measure\index{distortion measure!perceptual} in the MDCT domain.

\subsection{Modified Discrete Cosine Transform}
The MDCT\index{modified discrete cosine transform} is a modulated lapped transform~\cite{malvar:1992} which is applied on overlapping blocks of the input signal. A window of $2M$ time-domain samples is transformed into $M$ MDCT coefficients, whereafter the window is shifted $M$ samples for the next $M$ MDCT coefficients to be calculated. 

Given a block $s\in \mathbb{R}^{2M}$, the set of $M$ MDCT coefficients is given by~\cite{malvar:1992}
\begin{equation}
x_k=\frac{1}{\sqrt{2M}}\sum_{n=0}^{2M-1} h_ns_n \cos\left( \frac{(2n+M+1)(2k+1)\pi}{4M}\right),\quad k=0,\dotsc,M-1,
\end{equation}
where $x_k,h_n\in \mathbb{R}$ and $h$ is an appropriate analysis window, e.g.\ the symmetric sine window~\cite{malvar:1992}
\begin{equation}
h_n=\sin\left(\bigg(n+\frac{1}2\bigg)\bigg(\frac{\pi}{2M}\bigg)\right), \quad n=0,\dotsc, 2M-1.
\end{equation}

The inverse MDCT is given by~\cite{malvar:1992}\footnote{Notice that the MDCT is not an invertible transform on a block-by-block basis since $2M$ samples are transformed into only $M$ samples. We therefore use the tilde notation to indicate that, at this point, the reconstructed samples $\tilde{s}_n$ are not identical to the original samples $s_n$. In order to achieve perfect reconstruction we need to perform overlap-add of consecutive reconstructed blocks~\cite{malvar:1992}.}
\begin{equation}
\tilde{s}_n = h_n\frac{1}{\sqrt{2M}}\sum_{k=0}^{M-1}x_k\cos\left( \frac{(2n+M+1)(2k+1)\pi}{4M}\right),\quad n=0,\dots,2M-1.
\end{equation}

\subsection{Perceptual Weighting Function}
On each block a psycho-acoustic analysis\index{psycho-acoustic analysis} is performed which leads to a masking curve that describes thresholds in the frequency domain below which distortions are inaudible. In our work the masking curve\index{masking curve} is based on a $2nM$-point DFT where $n\in \mathbb{N}$ and the computation of the masking curve is described in detail in~\cite{par:2002}. Let us denote the masking curve by $\Sigma$. We then define a perceptual weight $\mu$ as the inverse of the masking threshold $\Sigma$ evaluated at the center frequencies of the MDCT basis functions, that is
\begin{equation}\label{eq:perceptweight}
\mu_k=\Sigma^{-1}_{2nk+1},\quad k=0,\dots,M-1.
\end{equation}

We require $\mu$ to be a multiplicative weight but otherwise arbitrary. We will not go into more details about $\mu$ except mentioning that we assume it can be efficiently encoded at e.g.\ $4$ kpbs as was done in~\cite{niemeyer:2005}.

\subsection{Distortion Measure}
Let $X\in \mathbb{R}^M$ denote a random vector process\footnote{In fact it is the output of the MDCT of a random vector process $S\in \mathbb{R}^{2M}$.} and let $x\in \mathbb{R}^M$ be a realization of $X$. By $X_k$ and $x_k$ we denote the $k^{th}$ components of $X$ and $x$, respectively, and we will use $\mathcal{X}_k$ to denote the alphabet of $X_k$. The pdf of $X$ is denoted $f_{X}$ with marginals $f_{X_k}$.

We define a perceptual distortion measure\index{distortion measure!perceptual} in the MDCT domain between $x$ and a quantized version $\hat{x}$ of $x$ to be the single-letter distortion measure given by\footnote{Strictly speaking this is not a single-letter distortion measure since the perceptual weight depends upon the entire vector.}
\begin{equation}\label{eq:perceptdist}
\rho(x,\hat{x})\triangleq \frac{1}{M}\sum_{k=0}^{M-1}\mu_k|x_k-\hat{x}_k|^2,
\end{equation}
where $\mu_k$ is given by~(\ref{eq:perceptweight}). The expected perceptual distortion follows from~(\ref{eq:perceptdist}) simply by taking the expectation over $x$, that is
\begin{equation}\label{eq:expperceptdist}
D(x,\hat{x}) = \frac{1}{M}\sum_{k=0}^{M-1}\int_{\mathcal{X}_k}\mu_k|x_k-\hat{x}_k|^2f_{X_k}(x_k)dx_k,
\end{equation}
where we remark that $\mu$ depends on $s$ through $x$.

\subsection{Transforming Perceptual Distortion Measure to $\ell_2$}
For the traditional MSE distortion measure which is also known as the $\ell_2$ distortion measure,\index{distortion measure!squared error} it is known that, under high-resolution assumptions, a lattice vector quantizer is good (even optimal as $L\rightarrow\infty$) for smooth sources, see Chapter~\ref{chap:rd_theory}. 
The MSE distortion measure is used mainly due its mathematical tractability. However, in applications involving a human observer it has been noted that distortion measures which include some aspects of human auditory perception generally perform better than the MSE\@.
A great number of perceptual distortion measures are non-difference distortion measures\index{distortion measure!non difference} and unfortunately even for simple sources their corresponding rate-distortion functions are not known.
For example, the perceptual distortion measure given by~(\ref{eq:expperceptdist}) is an input-weighted MSE (because $\mu$ is a function of $s$), hence it is a non-difference distortion measure.

In certain cases it is possible to derive the rate-distortion functions for general sources under non-difference distortion measures. For example, for the Gaussian pro\-cess with a weighted squared error criterion, where the weights are restricted to be linear time-invariant operators, the complete rate-distortion function was found in~\cite{sakrison:1968}. Other examples include the special case of locally quadratic distortion measures\index{distortion measure!locally quadratic} for fixed rate vector quantizers and under high-resolution assumptions~\cite{gardner:1995}, results which are extended to variable-rate vector quantizers in~\cite{li:1999,linder:1999a}. With regards to the MD problem,~\cite{linder:1998} presents a high-resolution rate-distortion region for smooth sources and locally quadratic distortion measures for the case of two descriptions. The case of vector sources and more than two descriptions remains unsolved.

\begin{remark}
In the SD case it has been shown that it is sometimes possible to apply a function (called a multidimensional compressor\index{compressor}) on the source signal in order to transform it into a domain where a lattice vector quantizer is good. This approach was first considered by Bennett in~\cite{bennett:1948} for the case of a scalar compressor followed by uniform scalar quantization. The general case of a multidimensional compressor followed by lattice vector quantization was considered in~\cite{linder:1999b}.
In general an $L$-dimensional source vector $X$ is ``compressed'' by an invertible mapping $F$.\footnote{The invertible mapping $F$ is for historically reasons called a compressor and said to compress the signal. However, $F$ is allowed to be any invertible mapping (also an expander) but we will use the term compressor to be consistent with earlier literature.} Hereafter $F(X)$ is quantized by a lattice vector quantizer. To obtain the reconstructed signal $\hat{X}$, the inverse mapping $F^{-1}$ (the expander) is applied, that is
\begin{equation}
X\rightarrow F(\cdot)\rightarrow Q(\cdot) \rightarrow F^{-1}(\cdot)\rightarrow \hat{X},
\end{equation}
where $Q$ denotes a lattice vector quantizer.
It was shown in~\cite{linder:1999b} that an optimal compressor $F$ is independent of the source distribution and only depends upon the distortion measure. However, it was also shown that an optimal compressor does not always exists.\footnote{In the scalar case an optimal compressor always exists for a wide range of distortion measures.} In the MD case, results on optimal compressors are very limited. However, it was suggested in~\cite{linder:1998}, that a compressor obtained in a similar way as for the SD case, might perform well also in the two-description case for smooth scalar processes. 
Unfortunately, we have been unsuccessful in finding an analytical expression for such a vector compressor for our distortion measure~(\ref{eq:perceptdist}). 
\end{remark}

In this chapter we will assume that the decoder has access to the perceptual weight $\mu$, which makes it possible to exploit $\mu$ also at the encoder when quantizing the MDCT coefficients. This has been done before by e.g.\ Edler et al.~\cite{edler:2000}. In addition, in the perceptual MD low delay audio coder presented in~\cite{schuller:2005} a post filter\index{post filter}, which resembles the auditory masking curve, was transmitted as side information. The input signal was first pre filtered by a perceptual filter which transforms the input signal into a perceptual domain that approximates an $\ell_2$ domain. A lattice vector quantizer is used in this domain and at the decoder the signal is reconstructed by applying the post filter. 

We adopt the approach of normalizing the input signal by the perceptual weight. First we show that, under a mild assumption on the masking curve, this procedure transforms the perceptual distortion measure\index{distortion measure!perceptual} into an $\ell_2$ distortion measure\index{distortion measure!squared error}. From~(\ref{eq:expperceptdist}) we have that
\begin{align}\label{eq:perceptnorm}
D(x,\hat{x}) &= \frac{1}{M}\sum_{k=0}^{M-1}\int_{\mathcal{X}_k}\mu_k|x_k-\hat{x}_k|^2f_{X_k}(x_k)dx_k \\ \notag
&\overset{(a)}{=} \frac{1}{M}\sum_{k=0}^{M-1}\sum_{j}\int_{\mathcal{X}_{k}\cap V_j'}\mu_k|x_k-\hat{x}_k|^2f_{X_k}(x_k)dx_k \\ \label{eq:ysubx}
&= \frac{1}{M}\sum_{k=0}^{M-1}\sum_j\int_{\mathcal{X}_{k}\cap V_j'}|y_k-\hat{y}_k|^2f_{X_k}(x_k)dx_k,
\end{align}
where $y_k=x_k\sqrt{\mu_k}, \hat{y}_k=\hat{x}_k\sqrt{\mu_k}$ and $(a)$ follows by breaking up the integral into disjoint partial integrals over each Voronoi cell $V_j'$ of the quantizer. In order to perform the necessary variable substitution in the integral given by~(\ref{eq:ysubx}) we write
\begin{equation}\label{eq:dydx}
\frac{dy_k}{dx_k}=x_k\frac{d}{dx_k}(\sqrt{\mu_k})+\sqrt{\mu_k}.
\end{equation}
At this point we enforce the following condition on the masking curve.
Within each quantization cell, the first derivative of the masking curve with respect to the source signal is assumed approximately zero so that from~(\ref{eq:dydx}) $dx_k\approx \sqrt{1/\mu_k}dy_k$.\footnote{To justify this assumption notice that we can approximate the masking curve by piece-wise flat regions (since the masking curve also needs to be coded), which means that small deviations of the source will not affect the masking curve.}
Inser\-ting this in~(\ref{eq:ysubx}) leads to
\begin{equation}
\begin{split}
D(x,\hat{x}) &\approx \frac{1}{M}\sum_{k=0}^{M-1}\sum_j\int_{\mathcal{Y}_{k}\cap V_j}|y_{k}-\hat{y}_k|^2f_{X_k}(x_k)\sqrt{1/\mu_k}dy_k \\ 
&= \frac{1}{M}\sum_{k=0}^{M-1}\sum_j\int_{\mathcal{Y}_{k}\cap V_j}|y_k-\hat{y}_k|^2f_{Y_k}(y_k)dy_k, \\
&= \frac{1}{M}E \sum_{k=0}^{M-1}|y_k-\hat{y}_k|^2,
\end{split}
\end{equation}
since it can be shown that $f_{Y_k}(y_k)=f_{X_k}(x_k)\sqrt{1/\mu_k}$ cf.~\cite[p.100]{stark:1986}. 
In other words, simply by normalizing the input signal $x$ by the root of the input-dependent weight $\mu$, the perceptual distortion measure for $x$ is transformed into an $\ell_2$ distortion measure for $y$. Therefore, when quantizing $y$, the distortion is approximately the same when measuring the $\ell_2$-distortion i.e.\ $E\|y-\hat{y}\|^2/M$ or transforming $y$ and $\hat{y}$ back into $x$ and $\hat{x}$, respectively, and measuring the perceptual distortion given by~(\ref{eq:expperceptdist}).

\subsection{Optimal Bit Distribution}\label{sec:bitalloc}
Each block $s$ leads to $M$ MDCT coefficients, which we first normalize by $\sqrt{\mu}$ and then vector quantize using $K$-channel MD-LVQ. Since, the number of coefficients in the MDCT is quite large, e.g.\ $M=1024$ in our case, it is necessary to split the sequence of $M$ coefficients into smaller vectors to make the quantization problem practically feasible. Any small number of coefficients can be combined and jointly quantized. For example, if the set of $M$ coefficients is split into $M'$ bands (vectors) of length $L_k$ where $k=0,\dots, M'-1$ it can be deduced from~(\ref{eq:daopt}) that the total distortion is given by\footnote{The distortion over individual normalized MDCT coefficients is additive in the MDCT domain (recall that we are using a single-letter distortion measure). However, adding the entropies of a set of MDCT coefficients is suboptimal unless the coefficients are independent. Futhermore, the individual MDCT coefficients will generally be correlated over consecutive blocks. For example, overlapping blocks of an i.i.d.\ process yields a Markov process\index{Markov process}. For simplicity, we do not exploit any correlation across blocks nor between the vectors of MDCT coeffficients (but only within the vectors).}
\begin{equation}\label{eq:daopttot}
\begin{split}
D_a&=\frac{1}{M'}\sum_{k=0}^{M'-1}\hat{K}_{1,k}G(\Lambda_k)2^{2(\bar{h}(Y_k)-R_{c_k})} \\
&\quad + \hat{K}_{2,k}\psi_{L_k}^{2}G(S_k) 2^{2(\bar{h}(Y_k)-R_{c_k})}2^{\frac{2K_k}{K_k-1}(R_{c_k}-R_k)} + \frac{p^{K_k}}{L_k} E\|Y_k\|^2,
\end{split}
\end{equation}
where we allow the quantizers $\Lambda_k$ and the number of packets $K_k$ to vary among the $M'$ bands as well as from block to block.
For a given target entropy $R^*$ we need to find the individual entropies $R_k$ for the $M'$ bands, such that $\sum R_k=R^*/K$ and in addition we need to find the entropies $R_{c_k}$ of the central quantizers. For simplicity we assume in the following that the $M'$ bands are of equal dimension $L'$, that similar central lattices $\Lambda_{c}$ are used, and that the number of packets $K$ is fixed for all $k$. 

We now use the fact that~(\ref{eq:optnu}) and~(\ref{eq:optN}) hold for any bit distribution, hence we may insert~(\ref{eq:optnu}) and~(\ref{eq:optN}) into~(\ref{eq:daopttot}) which leads to individual distortions given by
\begin{equation}
\begin{split}
D_k&=\hat{K}_1G(\Lambda_c)2^{2(\bar{h}(Y_k)-R_k)}
\left(\frac{1}{K-1}\frac{\hat{K}_2}{\hat{K}_1}\frac{G(S_{L'})}{G(\Lambda_c)}\psi_{L'}^2\right)^{\frac{K-1}{K}} \\
 &+ \hat{K}_2G(S_{L'}) 2^{2(\bar{h}(Y_k)-R_k)}\left((K-1)\frac{\hat{K}_1}{\hat{K}_2}\frac{G(\Lambda_c)}{G(S_{L'})}\right)\left(\frac{1}{K-1}\frac{\hat{K}_2}{\hat{K}_1}\frac{G(S_{L'})}{G(\Lambda_c)}\psi_{L'}^2\right)^{\frac{K-1}{K}} \\
&\quad + \frac{p^K}{L'} E\|Y_k\|^2 \\
&= a_02^{2(\bar{h}(Y_k)-R_k)}+\frac{p^K}{L'} E\|Y_k\|^2,
\end{split}
\end{equation}
where $a_0$ is independent of $k$ and given by
\begin{equation}
a_0 = \hat{K}_1G(\Lambda_c)\left(\frac{1}{K-1}\frac{\hat{K}_2}{\hat{K}_1}\frac{G(S_{L'})}{G(\Lambda_c)}\psi_{L'}^2\right)^{\frac{K-1}{K}}.
\end{equation}
In order to find the optimal bit distribution among the $M'$ bands subject to the entropy constraint $\sum_{k=0}^{M'-1} R_k=R^*/K$ we take the common approach of turning the con\-strained optimization problem into an unconstrained problem by introducing a Lagran\-gian cost functional\index{cost functional} of the form
\begin{equation}\label{eq:costfunc}
J = \sum_{k=0}^{M'-1}D_k + \lambda\sum_{k=0}^{M'-1}R_k.
\end{equation}
Differentiating~(\ref{eq:costfunc}) w.r.t.\ $R_k$ leads to 
\begin{equation}\label{eq:costfuncpartial}
\frac{\partial J}{\partial R_k} = -2\ln(2)a_02^{2(\bar{h}(Y_k)-R_k)}+\lambda.
\end{equation}
After equating~(\ref{eq:costfuncpartial}) to zero and solving for $R_k$ we get
\begin{equation}\label{eq:Rk}
R_k = -\frac{1}{2}\log_2\left(\frac{\lambda}{2\ln(2)a_0}\right)+\bar{h}(Y_k).
\end{equation}
In order to eliminate $\lambda$ we invoke the sum-rate constraint $\sum_{k=0}^{M'-1} R_k=R^*/K$ and get
\begin{equation}
\sum_{k=0}^{M-1}R_k = -\frac{M'}{2}\log_2\left(\frac{\lambda}{2\ln(2)a_0}\right)+\sum_{k=0}^{M'-1}\bar{h}(Y_k) = R^*/K,
\end{equation}
from which we obtain
\begin{equation}\label{eq:lambda}
\lambda = 2\ln(2)a_02^{-\frac{2}{M'}(R^*/K-\sum_{k=0}^{M'-1} \bar{h}(Y_k))}.
\end{equation}
We can now eliminate $\lambda$ by inserting~(\ref{eq:lambda}) into~(\ref{eq:Rk}), that is
\begin{equation}\label{eq:optRk}
R_k = \frac{R^*/K-\sum_{k=0}^{M'-1} \bar{h}(Y_k)}{M'}+\bar{h}(Y_k).
\end{equation}

With the simple Lagrangian approach taken here there is no guarantee that the entropies $R_k$ given by~(\ref{eq:optRk}) are all non-negative. It might be possible to extend the Lagrangian cost functional~(\ref{eq:costfunc}) by $M'$ additional Lagrangian weights\index{Lagrangian weights} (also called ``complementary slackness''\index{complementary slackness variables} variables~\cite{sundaram:1999}) in order to obtain $M'$ inequality constraints making sure that $R_k\geq 0$ in addition to the single equality constraint $\sum R_k=R^*/K$. 
While the resulting problem can be solved using numerical techni\-ques, it does not appear to lead to a closed-form expression for the individual entropies $R_k$. 
It is not possible either to simply set negative entropies equal to zero since this will most likely violate the constraint $\sum R_k=R^*/K$. 
Instead we propose a sequential procedure where we begin by considering all $M'$ bands and then one-by-one eliminate bands having negative entropies. 
We assign entropies to each band using~(\ref{eq:optRk})
and then find the one having the largest negative entropy and exclude that one from the optimization process. This procedure continues until all entropies are positive or zero as shown in Table~\ref{tab:bitallocalg}.
\begin{table}[ht]
\begin{center}
\begin{boxedminipage}{11cm}
\begin{enumerate}
\item $\mathcal{I}=\{0,\dots,M'-1\}$
\item $h = \sum_{k\in\mathcal{I}}\bar{h}(Y_k)$
\item $c = \frac{R^*/K - h}{|\mathcal{I}|}$
\item $\mathcal{R}=\{R_k : R_k=c+\bar{h}(Y_k)\ \text{and}\ R_k<0, k\in \mathcal{I} \}$
\item If $|\mathcal{R}|>0$ then goto 2 and set $\mathcal{I}:= \mathcal{I}\backslash j$, where $R_j\leq R_k, \forall k\in \mathcal{I}$
\item $R_k = 
\begin{cases} c+\bar{h}(Y_k), & k\in \mathcal{I} \\
0, & \text{otherwise}
\end{cases}$
\end{enumerate}
\end{boxedminipage}
\caption{Bit-allocation algorithm.\index{bit-allocation algorithm}}
\label{tab:bitallocalg}
\end{center}
\end{table}

The motivation for this approach is that ultimately we would like the contribution of each band to the total distortion to be equal, since they are all approximately equally sensitive to distortion after being flattened by the masking curve\index{masking curve}. However, the normalized MDCT coefficients in some bands have variances which are smaller than the average distortion, hence assigning zero bits to these bands leads to distortions which are lower than the average distortion over all bands. Therefore, the bit budget should only be distributed among the higher variance components.

\section{Robust Transform Coding}
\begin{figure}[ht]
\begin{center}
\psfrag{s}{$s$}
\psfrag{x}{$x$}
\psfrag{y}{$y$}
\psfrag{m}{$\sqrt{\mu}$}
\psfrag{li}{$\lambda_0$}
\psfrag{lk}{$\lambda_{K-1}$}
\psfrag{R}{\small $\{R_k\}_{k=0}^{M'}$}
\subfigure[Encoder]{\includegraphics[width=12cm]{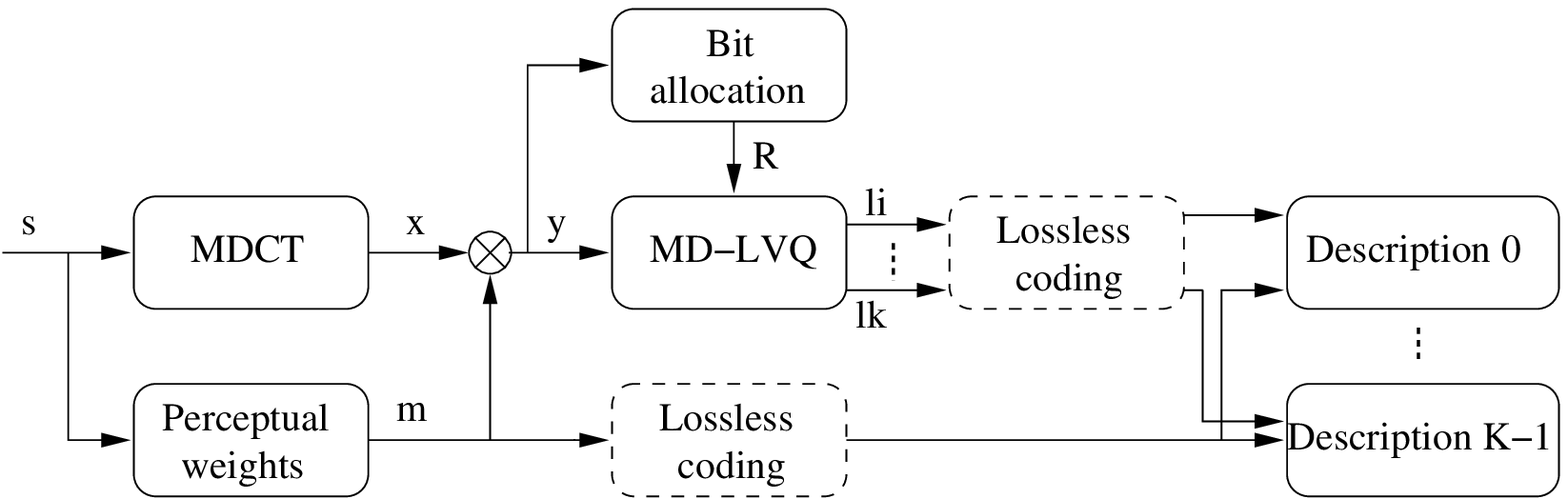}\label{fig:encoder}}
\psfrag{sh}{$\hat{s}$}
\psfrag{xh}{$\hat{x}$}
\psfrag{yh}{$\hat{y}$}
\psfrag{mu}{$\frac{1}{\sqrt{\mu}}$}
\psfrag{li}{$\lambda_0$}
\psfrag{lk}{$\lambda_{K-1}$}
\psfrag{Sum}{$\frac{1}{\kappa}\sum\lambda_i$}
\subfigure[Decoder]{\includegraphics[width=7.5cm]{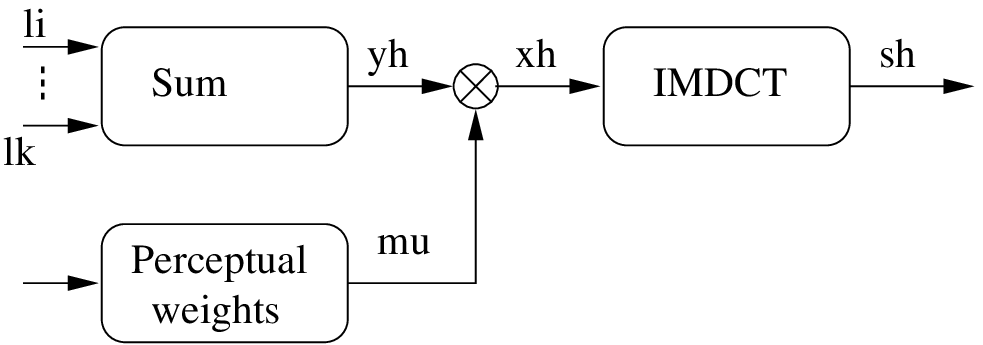}\label{fig:decoder}}
\caption{Encoder and decoder.}
\label{fig:encoderdecoder}
\end{center}
\end{figure}

In this work we apply MD-LVQ on the normalized coefficients of an MDCT to obtain a desired degree of robustness when transmitting encoded audio over a lossy network. The encoder and decoder of the complete scheme are shown in Figs.~\ref{fig:encoder} and~\ref{fig:decoder}, respectively. In the following we describe how the encoding and decoding is performed.

\subsection{Encoder}
By $s$ we denote the current block, which has been obtained by blocking the input signal into overlapping blocks each containing $2M$ samples. The $M$ MDCT coefficients are obtained by applying an $M$-channel MDCT on $s$ and is represented by the vector $x$. It is worth mentioning that we allow for the possibility to use a flexible time segmentation in order to better match the time-varying nature of typical audio signals, cf.\cite{niamut:2006}. 
Each block is encoded into $K$ descriptions independent of previous blocks in order to avoid that the decoder is unable to successfully reconstruct due to previous description losses. 

\begin{figure}[ht]
\begin{center}
\psfrag{hXk}{$\scriptstyle \bar{h}(Y_k)$}
\psfrag{Rk}{$\scriptstyle R_k$}
\mbox{%
\subfigure[Differential entropies]{\includegraphics[width=6cm]{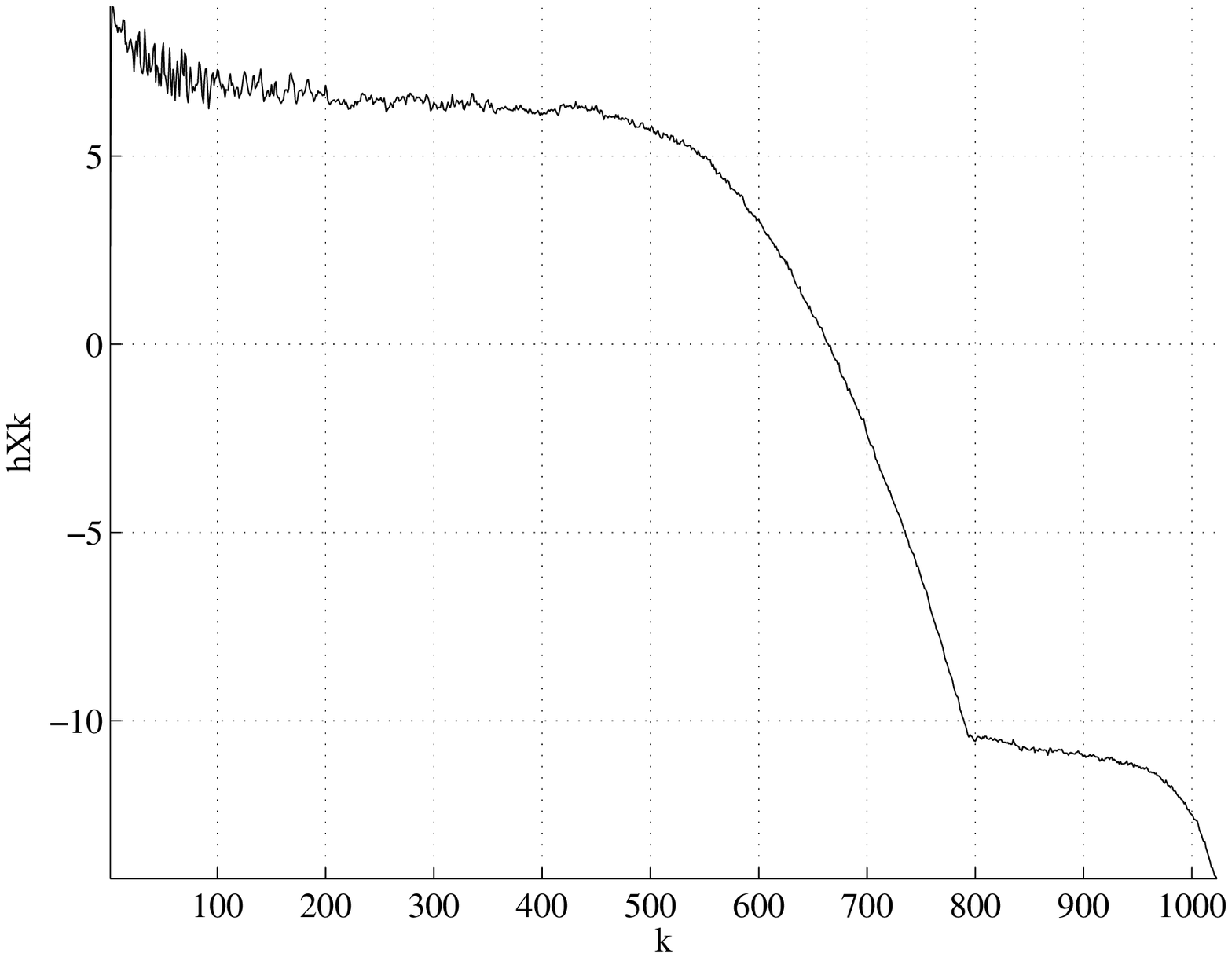}\label{fig:diff_entropies}}
\subfigure[Discrete entropies]{\includegraphics[width=6cm]{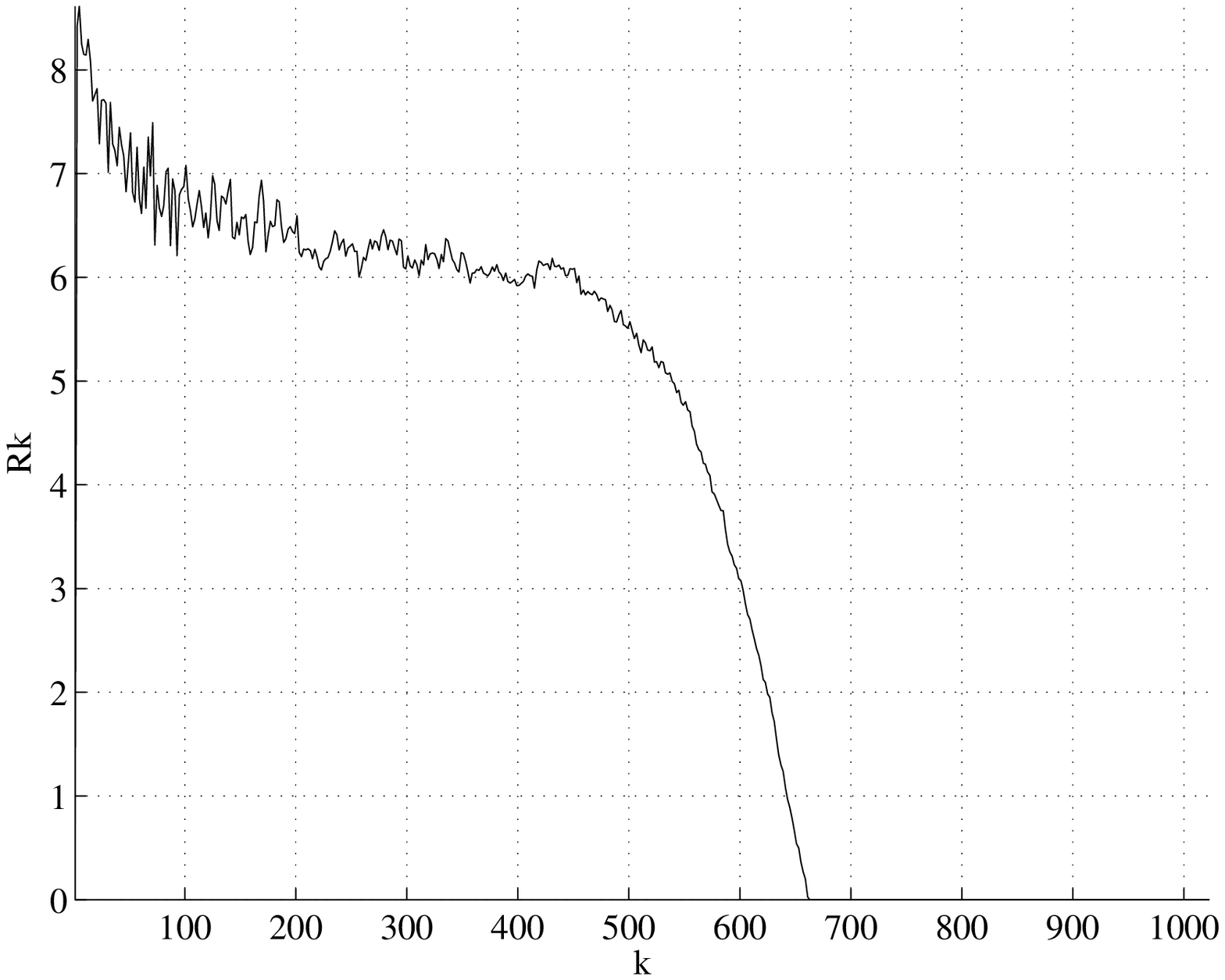}\label{fig:discrete_entropies}}}
\caption{Differential and discrete entropies for the normalized MDCT coefficients (expressed in bit/dim.).}
\label{fig:entropies}
\end{center}
\end{figure}

As discussed in Section~\ref{sec:bitalloc} it is infeasible to jointly encode the entire set of $M$ MDCT coefficients and instead we split $x$ into $M'$ disjoint subsets. The MDCT coefficients are then normalized by the perceptual weights $\mu$ in order to make sure that they are approximately equally sensitive to distortion and moreover to make sure that we operate in an $\ell_2$ domain where it is known that lattice vector quantizers are among the set of good quantizers.
Based on the differential entropies of the normalized MDCT coefficients $y$ and the target entropy $R^*$ we find the individual entropies $R_k, k=0,\dots,M'-1$ by using the algorithm described in Table~\ref{tab:bitallocalg}. Fig.~\ref{fig:diff_entropies} shows an example of the distribution of differential entropies $\bar{h}(Y)$ in a 1024-channel MDCT. In this example a 10 sec.\ audio signal (jazz music) sampled at 48 kHz was input to the MDCT. Fig.~\ref{fig:discrete_entropies} shows the corresponding discrete entropies assigned to each of the 1024 bands when the target entropy is set to $R^*=88$ kbps.
%
%

It may be noticed from Fig.~\ref{fig:discrete_entropies} that the bit budget is mainly spent on the lower part of the normalized MDCT spectrum. This behavior is typical for the audio signals we have encountered. The reason is partly that the audio signals have most of their energy concentrated in the low frequency region but also that the high frequency part is deemphasized by the perceptual weight. 
The perceptual weight is approximately proportional to the inverse of the masking curve and at the high frequency region the steep positive slope of the threshold in quiet\index{threshold in quiet} dominates the masking curve. We remark that the bit allocation effectively limits the band width of the source signal since high frequency bands are simply discarded and it might therefore prove beneficial (perceptually) to use some kind of spectral band replication\index{spectral band replication} at the decoder in order to recover some of the lost high frequency components.

The entropy $R_k$ describes the total entropy assigned to the $k^{th}$ band (or the $k^{th}$ subset of bands if vector quantization is applied). If the number of descriptions is $K$ then each side description operates at an entropy of $R_k/K$ bit/dim. 
Knowledge of $R_k$, the differential entropy $\bar{h}(Y_k)$, the number of descriptions $K$ and the packet-loss probability $p$ makes it possible to find the scaling factors $\nu_k$ and $N_k$ of the central and side quantizers, respectively by use of~(\ref{eq:optnu}) and~(\ref{eq:optN}). 
This in turn completely specify a MD-LVQ scheme having $K$ descriptions. Each normalized MDCT coefficient or vector of coefficients $y_k$ is then first quantized with the central quantizer $Q_k(y_k)=\lambda_{c_k}$ after which index assignments $\alpha(\lambda_{c_k})=\{\lambda_{0_k},\dots,\lambda_{K_k-1}\}$ are performed in order to find the codewords of the side quantizers. The codewords of the side quantizers are losslessly coded and put into $K$ individual packets. Each packet then contains $M'$ encoded codewords. 

It is required that the perceptual weight $\mu$ is somehow transmitted to the decoder in order to be able to reconstruct. Since the $K$ packets have an equal chance of getting lost we need the perceptual weight in all packets, which leads to a certain degree of overhead. In the case where more than one packet is received we therefore waste bits. It might be possible to apply some sort of MD coding on the perceptual weight in order to decrease the amount of side information which needs to be duplicated in all packets. However, it is outside the scope of this chapter to investigate the many aspects of perceptual lossless coding of $\mu$ and we refer the readers to the work of~\cite{niemeyer:2005} for more details. In the following we will simply assume that the perceptual weight can be perceptually lossless coded at 4 kbps, hence if the target entropy is $R^*=96$ kpbs and two packets are to be used, the entropy we can actually use for MD-LVQ is then only $88$ kbps, since $8$ kbps ($4$ kbps in each packet) are used for the weight. If a greater number of packets is desired the overhead for transmitting $\mu$ increases even further.

\subsection{Decoder}
At the receiving side an estimate $\hat{y}$ of the normalized MDCT spectrum is first obtained by simply taking the average of the received descriptions, i.e.\ $\hat{y}_k=\frac{1}{\kappa'}\sum_{i\in l'} \lambda_{i_k}$, where $l'$ denotes the indices of the received descriptions and $\kappa'=|l'|$. This estimate is then denormalized in order to obtain $\hat{x}$, i.e.\ $\hat{x}_k=\hat{y}_k/\sqrt{\mu_k}$. Finally the inverse MDCT (including overlap-add) is applied in order to obtain an approximation $\hat{s}$ of the time domain signal $s$. The decoding procedure is shown in Fig.~\ref{fig:decoder}.

\section{Results}
In this section we compare numerical simulations with theoretical results and in addition we show the results of a subjective listening test\index{listening test}. We first show results related to the expected distortion based on the packet-loss probabilities and then we show results for the case of scalable coding. In both cases we assume a symmetric setup.

\subsection{Expected Distortion Results}
For the objective test we use four short audio clips of different genres (classical jazz music, German male speech, pop music, rock music) each having a duration between 10 and 15 sec.\ and a sampling frequency of $48$ kHz.
We refer to these fragments as ``jazz'', ``speech'' , ``pop'' and ``rock''. We set the target entropy to $96$ kbps (as was done in~\cite{schuller:2005}) which corresponds to $2$ bit/dim.\ since the sampling frequency is $48$ kHz. 
We do not encode the perceptual weight but simply assume that it can be transmitted to the receiver at an entropy of $4$ kbps. 
Since the weight must be included in all of the $K$ descriptions we deduct $4K$ kbps from the total entropy, hence the effective target entropy $R^*_e$ is given by $R^*_e= R^*-4K$ so that a single description system has $R^*_e=92$ kbps whereas a four description system has $R^*_e=80$ kbps (i.e.\ 20 kbps for each side description). 
For simplicity we furthermore assume that the sources are stationary processes so that we can measure the statistics for each vector of MDCT coefficients upfront. However, since audio signals in general have time varying statistics we expect that it will be possible to reduce the bit rate by proper adaptation to the source. 
Since for this particular test we are merely interested in the performance of the proposed audio coder with a varying number of descriptions we will not address the issue of efficient entropy coding but simply assume that the quantized variables can be losslessly coded arbitrarily close to their discrete entropies. Table~\ref{tab:rates} shows the discrete entropies of the quantized normalized MDCT coefficients for the four test fragments.  
\begin{table}[ht]
\begin{center}
\begin{tabular}{lcccccc}\hline
 &  $K=2$ &$K=2$ &$K=3$ & $K=3$ &$K=4$&  $K=4$\\
 & kbps &  bit/dim. &kbps &  bit/dim. &kbps &  bit/dim. \\ \hline
jazz &  96.22& 1.00 & 97.09& 0.67 & 96.87& 0.51\\
speech &  93.48& 0.98 & 96.00& 0.67 & 96.47& 0.50\\
pop & 93.35& 0.98 & 95.25& 0.66 & 95.57& 0.50\\
rock & 93.76& 0.98 & 95.38& 0.66 & 95.60& 0.50\\ \hline
\end{tabular}
\caption{Numerical measured output entropies in kilobits per second (kbps) and bit/dim.\ per description. The target entropy is $R^*=96$ kbps or $2$ bit/dim.}
\label{tab:rates}
\end{center}
\end{table}

\begin{figure}[ht]
\begin{center}
\mbox{%
\subfigure[jazz]{\includegraphics[width=6cm]{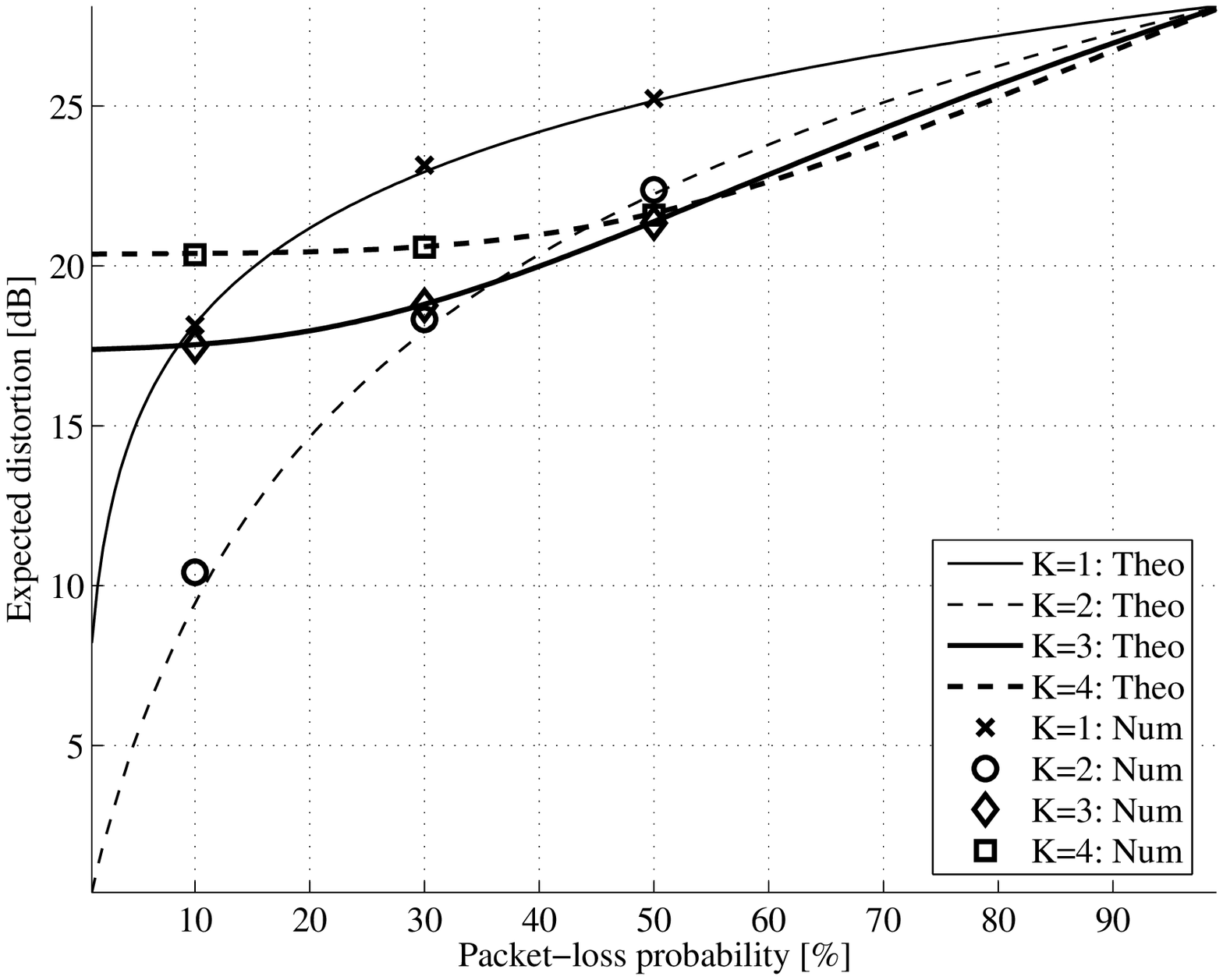}\label{fig:T2}}\quad
\subfigure[speech]{\includegraphics[width=6cm]{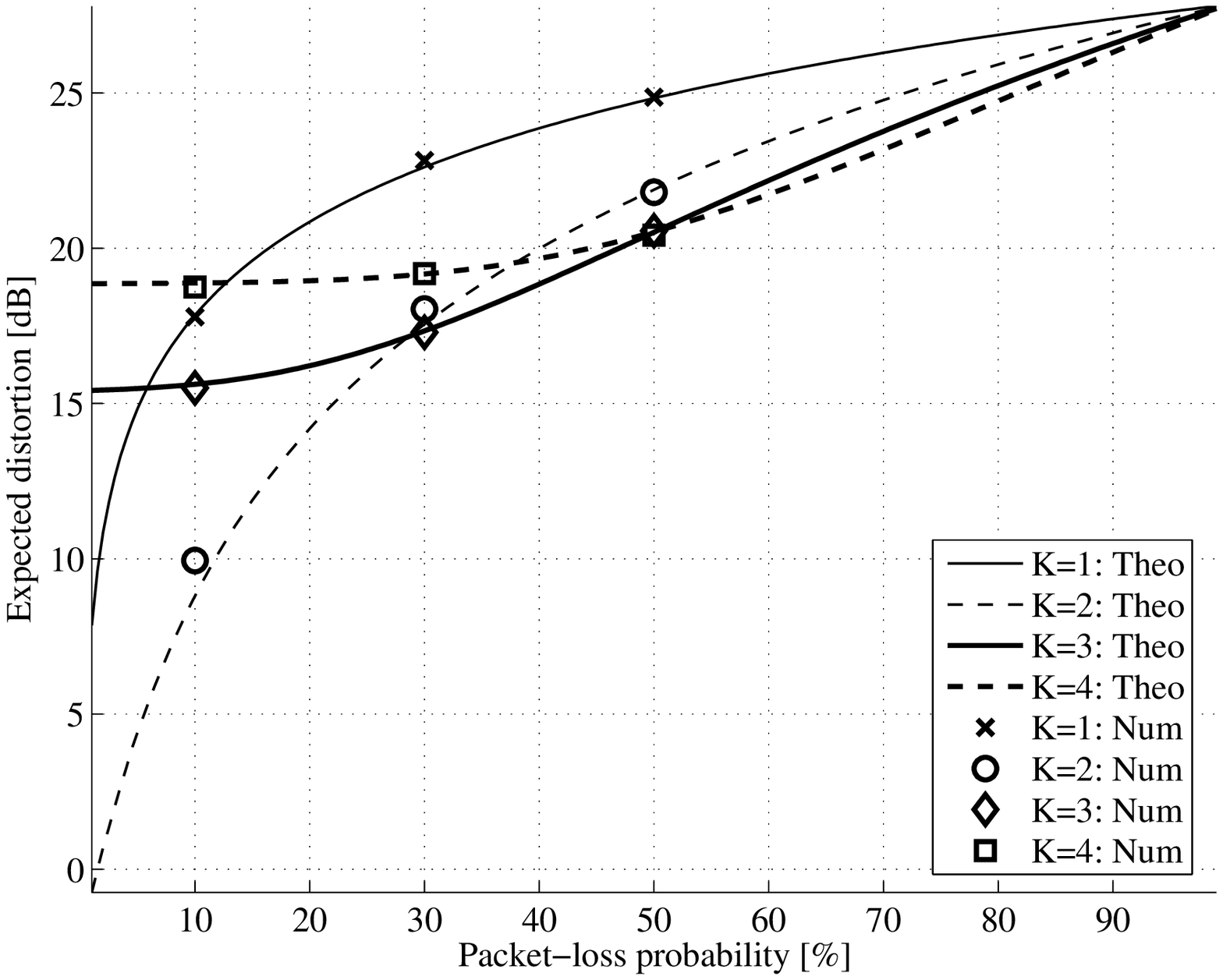}\label{fig:T7}}}
\caption{The expected distortion as a function of packet-loss probabilities for MD-LVQ when operating at a target entropy of 96 kbps.}
\label{fig:T2T7}
\end{center}
\end{figure}

We block the normalized MDCT spectrum into vectors of length two and use the $Z^2$ lattice vector quantizer. 
Because of the short duration of the test fragments 
the resulting expected distortions depend upon the realizations of the packet loss patterns. This phenomena has been noted by other authors, cf.~\cite{arean:2000}. We therefore decided to average the distortion results over three different loss patterns obtained by using different seeds to the random number generator. The numerically obtained expected distortions are shown in Tables~\ref{tab:distK1}--\ref{tab:distK4} and Figs.~\ref{fig:T2} and~\ref{fig:T7}. 
\begin{table}[ht]
\begin{center}
\begin{tabular}{lccc}\hline
$K=1$ & $p=10\%$ & $p=30\%$ & $p=50\%$ \\ \hline
jazz & 18.17 (18.15) & 22.94 (23.12) & 25.16 (25.23) \\
speech & 17.84 (17.79) & 22.61 (22.82) & 24.83 (24.86) \\
pop & 17.89 (17.83) & 22.66 (22.83) & 24.88 (24.91) \\
rock & 18.20 (18.20) & 22.97 (23.12) & 25.18 (25.23) \\ \hline
\end{tabular}
\caption{Theoretical (numerical) expected distortions expressed in dB for $K=1$ and $p=10,30$ and $50\%$. The target entropy is $R^*=96$ kbps or $2$ bit/dim.}
\label{tab:distK1}
\end{center}
\end{table}

\begin{table}[ht]
\begin{center}
\begin{tabular}{lccc}\hline
$K=2$ & $p=10\%$ & $p=30\%$ & $p=50\%$ \\ \hline
jazz & 9.44 (10.42) & 17.96 (18.33) & 22.24 (22.38) \\
speech & 8.80 (9.94)  & 17.55 (18.04) & 21.88 (21.80) \\
pop& 9.04 (10.32) & 17.62 (18.22) & 21.94 (22.11) \\
rock& 9.70 (10.66) & 18.00 (18.36) & 22.27 (22.39) \\ \hline
\end{tabular}
\caption{Theoretical (numerical) expected distortions expressed in dB for $K=2$ and $p=10,30$ and $50\%$. The target entropy is $R^*=96$ kbps or $2$ bit/dim.}
\label{tab:distK2}
\end{center}
\end{table}

\begin{table}[ht]
\begin{center}
\begin{tabular}{lccc}\hline
$K=3$ & $p=10\%$ & $p=30\%$ & $p=50\%$ \\ \hline
jazz & 17.54 (17.49) & 18.80 (18.76) & 21.39 (21.34) \\
speech & 15.62 (15.50) & 17.34 (17.29) & 20.51 (20.56) \\
pop & 16.38 (16.28) & 17.85 (17.75) & 20.76 (20.66) \\
rock & 17.44 (17.33) & 18.75 (18.63) & 21.38 (21.29) \\ \hline
\end{tabular}
\caption{Theoretical (numerical) expected distortions expressed in dB for $K=3$ and $p=10,30$ and $50\%$. The target entropy is $R^*=96$ kbps or $2$ bit/dim.}
\label{tab:distK3}
\end{center}
\end{table}

\begin{table}[ht]
\begin{center}
\begin{tabular}{lccc}\hline
$K=4$ & $p=10\%$ & $p=30\%$ & $p=50\%$ \\ \hline
jazz & 20.39 (20.35) & 20.61 (20.59) & 21.65 (21.59) \\
speech & 18.88 (18.75) & 19.17 (19.18) & 20.52 (20.42) \\
pop & 19.14 (19.08) & 19.41 (19.46) & 20.70 (20.71) \\
rock & 20.27 (20.18) & 20.50 (20.44) & 21.58 (21.50) \\ \hline
\end{tabular}
\caption{Theoretical (numerical) expected distortions expressed in dB for $K=4$ and $p=10,30$ and $50\%$. The target entropy is $R^*=96$ kbps or $2$ bit/dim.}
\label{tab:distK4}
\end{center}
\end{table}

%
%

\begin{figure}[ht]
\begin{center}
\psfrag{P=10\%}{\small $p=10\%$}
\psfrag{P=30\%}{\small $p=30\%$}
\psfrag{P=50\%}{\small $p=50\%$}
\includegraphics[width=11cm]{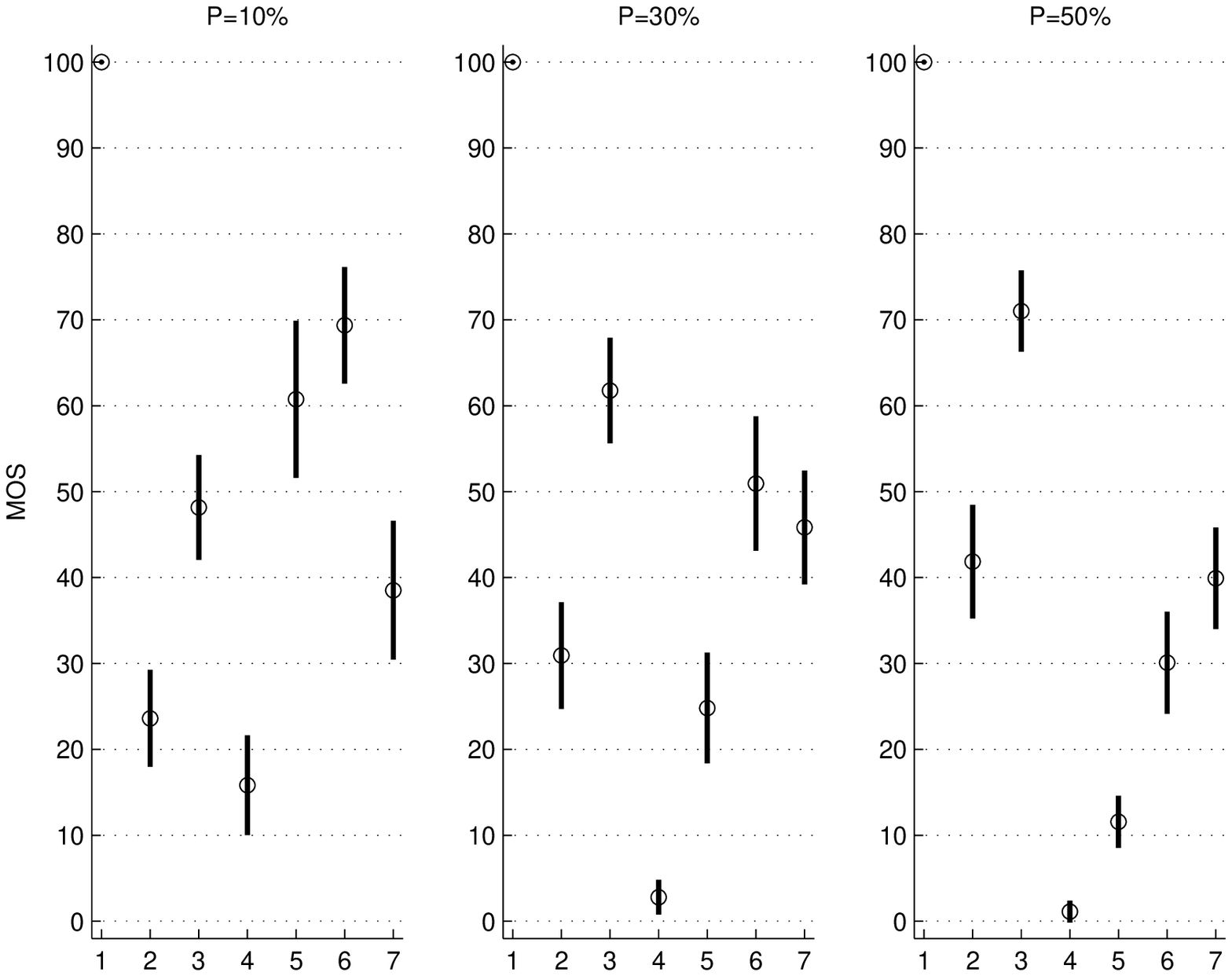} 
\caption{MUSHRA test results averaged over all three audio clips for $p=0.1,0.3$ and $p=0.5$. The seven signals appear in the following order: Hidden ref., 3.5 kHz, 7 kHz, $K=1, K=2, K=3$ and $K=4$.}
\label{fig:mushrares}
\end{center}
\end{figure}

As can be seen in Figs.~\ref{fig:T2} and~\ref{fig:T7} the expected distortions depend not only on the packet-loss rate but also upon the number of descriptions. At high packet-loss rates it is advantageous to use a higher number of packets. To verify these findings we performed an additional subjective comparison test. 
We chose three different fragments (jazz, speech and rock) and three different packet-loss rates ($p=0.1$, $p=0.3$ and $p=0.5$). We then performed a standard MUSHRA test~\cite{MUSHRA}\index{MUSHRA test}. At each packet-loss rate the original signals were encoded using $K=1,2,3$ and $4$ descriptions. Also included in each test were the hidden reference and two anchor signals\index{anchor signals} (3.5 kHz and 7 kHz lowpass filtered signals). We used nine (non-experts) listeners in the listening test and the results are shown in Figs.~\ref{fig:mushra_t2}--\ref{fig:mushra_t19} in Appendix~\ref{app:nac_results} for the individual fragments averaged over the nine participants. The circles in the figures denote mean values and the bars describe 95\% confidence intervals. Fig.~\ref{fig:mushrares} shows the result when averaging over participants and fragments. Notice that for $p=0.3$ and $p=0.5$ there is a significant preference for using more than two descriptions.

%
%

The results of the subjective listening tests\index{listening test} show generally no significant difference between the two and three packet versions for a packet-loss rate of $p=0.1$, cf.\ Figs.~\ref{fig:mushra_t2}(a)--~\ref{fig:mushra_t19}(a). 
However, the results based on the perceptual distortion measure reveals that at $p=0.1$ it is beneficial to use two packets instead of three, cf.\ Figs.~\ref{fig:T2} and~\ref{fig:T7}. In fact, a reduction in distortion of about 7 dB can be achieved. This discrepancy can be partly explained by our implementation of the the bit-allocation strategy outlined in Section~\ref{sec:bitalloc}. 
To avoid assigning a too small rate to a given frequency band (which then would violate the high-resolution assumptions) we have, in the experiments described above, excluded MDCT bands which were assigned a rate lower than 3 bit/dim.\ per description.\footnote{If the numerically measured discrete entropy is, for example, 0.1 bit/dim.\ greater than the specified theoretical entropy, then, since the sampling frequency is 48 kHz, the resulting bit rate is 4.8 kbps above the target entropy. Furthermore, if this 0.1 bit/dim.\ gap is per description, then, in a three-description setup, the resulting rate would exceed the target rate by 14.4 kbps. Practical experiments have shown that at 3 bit/dim.\ per description, the numerically measured discrete entropy is off by less than 0.03 bit/dim.\ per description for a range of index values.} The effect of this is that the high-resolution approximations are good so that theoretical and numerical results agree but the downside is that the input signal is severely lowpass filtered. 
The contribution of the high frequency bands to the total distortion is therefore high, hence, the reception of more than two descriptions does not improve the quality of the reconstructed signal much. 
In addition we would like to emphasize two important factors which might also contribute to the inconsistency between subjective listening tests and the perceptual distortion measure. First of all, the perceptual distortion measure is based upon a single block at a time and therefore the continuity of the signal over time is not addressed.\footnote{The listeners agreed that the ``hick-ups'' resulting from time gaps due to packet losses were the most annoying artifacts present in the coded signals. The overlapping nature of the MDCT is, however, able to reduce the impact of isolated packet losses.} Secondly, the distortion measure is defined in the MDCT domain and since the MDCT is not an orthogonal transform the distortion in the MDCT domain is not equivalent to the distortion in the time domain. 

As previously mentioned we have in these tests excluded MDCT bands where the rate assignment is less than 3 bit/dim.\ per description to make sure that the high-resolution assumptions are valid. 
Such an approach excludes a great amount of MDCT bands (especially those representing the high frequency contents of the signal) and the coded signal sounds muffled (lowpass filtered). The reasoning behind this choice is that a ``lowpass'' filtered version of the signal (without time gaps) is often preferable over a full bandwidth signal (with time gaps). Alternatively, we may take into account that the practical rate becomes too high for the bands that are assigned a too low theoretical rate. Thus, we can heuristically assign a lower target rate for the MDCT coefficients representing the higher frequency bands. Since we encode two-dimensional vectors there are 512 bands in total but only about the first 300 of these are assigned a positive rate. We then modify the scale factor $\nu_k$ for the $k^{th}$ band by the following rule
\begin{equation}
\nu_k =
\begin{cases}
1.0\cdot\nu_k, & 0\leq k   \leq 50, \\ 
1.3\cdot\nu_k, & 51\leq k  \leq 100,\\
1.4\cdot\nu_k, & 101\leq k \leq 200,\\
1.5\cdot\nu_k, & 201\leq k \leq 250,\\
2.0\cdot\nu_k, & 251\leq k \leq 300. 
\end{cases}
\end{equation}

For each different fragment we set the target $R^*$ such that the practical rate is very close to $96$ kbit/sec.\ (incl.\ $4$ kbit/sec.\ per packet for the masking curve). These rates are shown in Table~\ref{tab:practicalrates}.\footnote{In this experiment we have included an additional audio fragment ``harpsi'', which consists of ``music'' from a Harpsichord.}
\begin{table}
\begin{center}
\begin{tabular}{lcc}
Fragment & $R^*$ [kbit/sec.]& $H(Y)$ [kbit/sec.] \\ \hline
jazz & 92.16 & 95.92 \\
harpsi & 86.40 & 96.34 \\
speech & 91.92 & 96.03 \\
pop & 89.76 & 96.29 \\
rock & 94.08 & 96.09 \\ \hline
\end{tabular}
\caption{The target rate $R^*$ is set lower than $96$ kbit/sec.\ which leads to a practical rate $H(Y)$ close to $96$ kbit/sec.}
\label{tab:practicalrates}
\end{center}
\end{table}

\begin{figure}[ht]
\begin{center}
\mbox{%
\subfigure[jazz fragment]{\includegraphics[width=6cm]{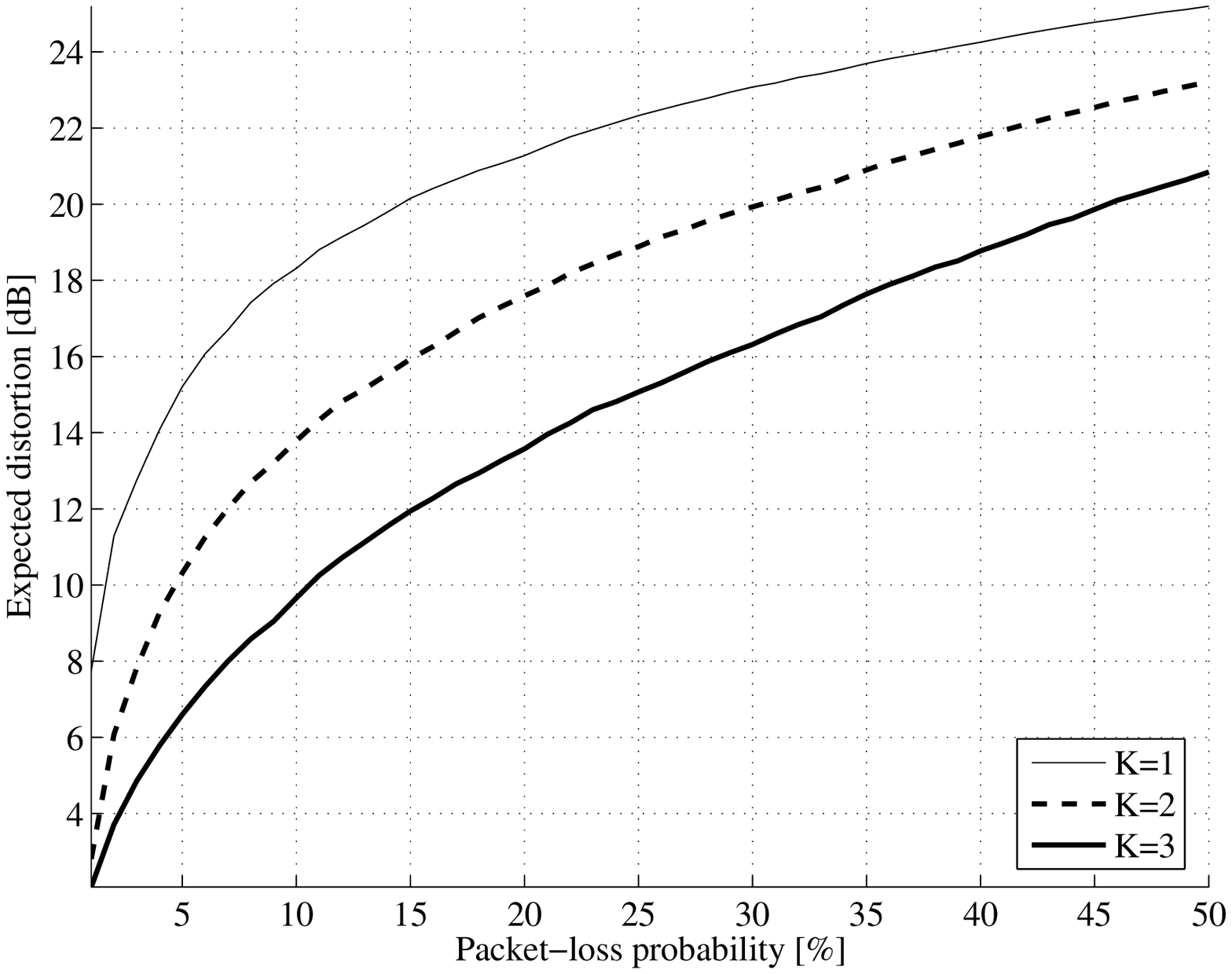}\label{fig:pracexpdistjazz}}
\subfigure[speech fragment]{\includegraphics[width=6cm]{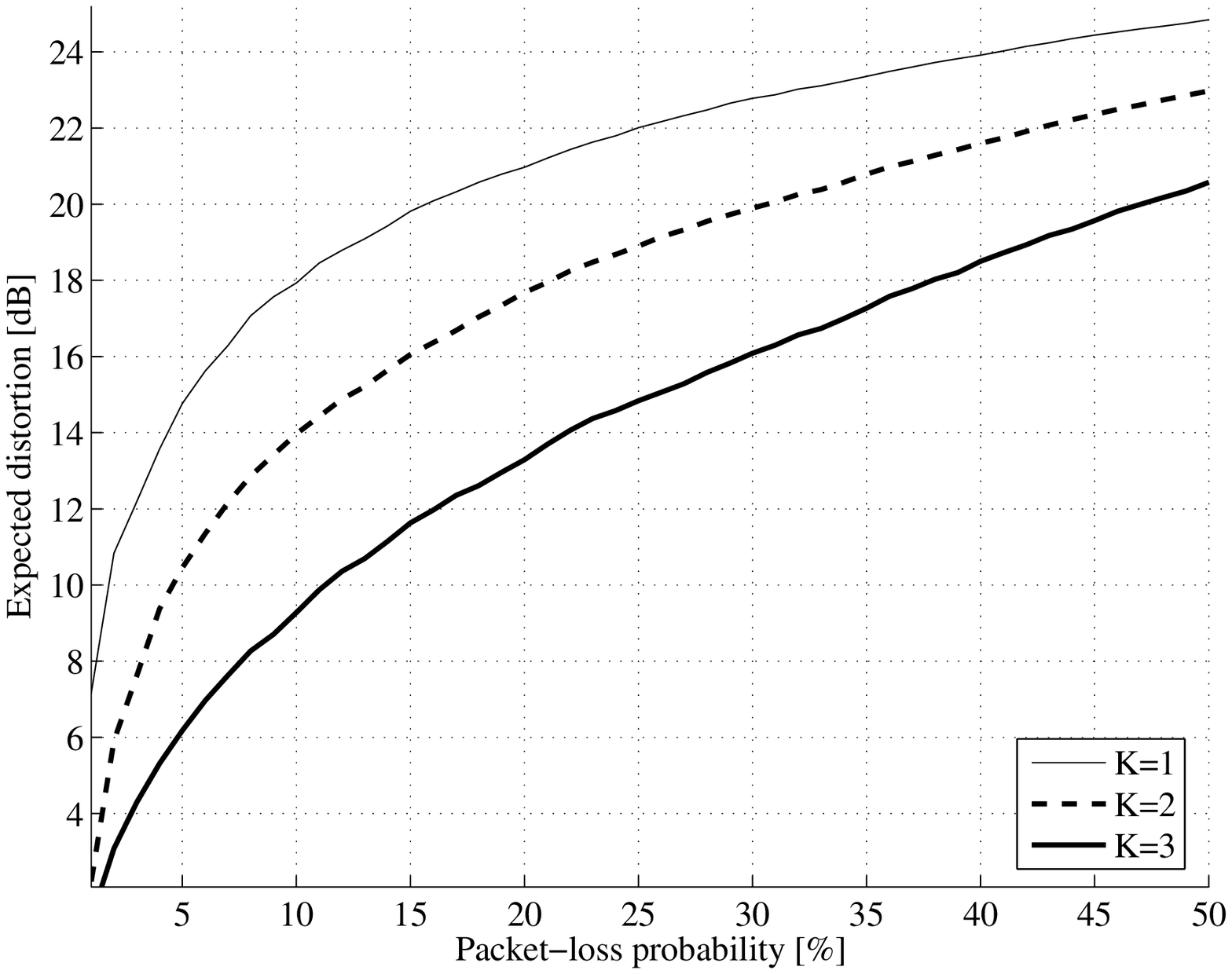}\label{fig:pracexpdistspeech}}
}
\caption{Expected distortion as a function of packet-loss probabilities.}
\label{fig:pracexpdist}
\end{center}
\end{figure}

The numerically measured expected distortions based on the packet-loss probabilities are shown in Fig.~\ref{fig:pracexpdistjazz} for the jazz fragment. We have swept the packet-loss probability between 1\% and 50\% in steps of 1\%. Each test is repeated 10 times to reduce the influence of a particular loss pattern. Notice that already at packet-loss probabilities as low as one percent it becomes advantageous to use three descriptions instead of two descriptions. Fig.~\ref{fig:pracexpdistspeech} shows the results of a similar experiment for the speech fragment.
%
%

\subsection{Scalable Coding Results}
\begin{figure}[ht]
\begin{center}
\includegraphics[width=8cm]{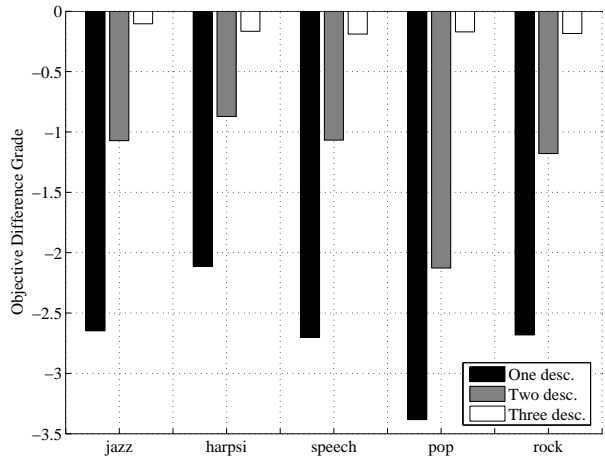}
\caption{ODGs for the reception of one to three packets out of three for different test fragments.}
\label{fig:barplot}
\end{center}
\end{figure}

We now assess the improvement of audio quality as more packets are received. This is a form of scalable coding\index{scalable coding}, where some receivers have access to more information (descriptions) than others. 
In this case no description losses occur. 
Instead of using the expected distortion we will use the Objective Difference Grade\index{objective difference grade} (ODG) based on the Matlab implementation by Kabal et al.~\cite{peaq-kabal} of the PEAQ\index{PEAQ} standard~\cite{peaq:1998}. The ODGs are related to the standard ITU-R 5-grade impairment scale\index{impairment scale} as shown in Table~\ref{tab:SDG}. 
Tables~\ref{tab:odg1}--\ref{tab:odg3} show the ODGs for the five different test fragments. The last column of Tables~\ref{tab:odg1} and~\ref{tab:odg2} show the mean ODGs when averaged over the three different combinations of descriptions. These average ODGs as well as the results of Table~\ref{tab:odg3} are also shown in the bar diagram in Fig.~\ref{fig:barplot}.

From the tables it may be observed that the perceptual distortion is approximately symmetric, i.e.\ the ODG is essentially independent of which packet is received. In addition, it can be seen that as more packets are received a substantial improvement in quality can be expected.
\begin{table}
\begin{center}
\begin{tabular}{lcc}
Impairment & ITU-R Grade & ODG \\ \hline
Imperceptible & 5.0 & 0.0 \\
Perceptiple, but not annoying & 4.0 & -1.0 \\
Slightly annoying & 3.0 & -2.0 \\
Annoying & 2.0 & -3.0 \\
Very annoying & 1.0 & -4.0 \\ \hline
\end{tabular}
\caption{Relationship between the ITU-R 5-grade impairment scale and the ODGs~\cite{bosi:2003}.}
\label{tab:SDG}
\end{center}
\end{table}

\begin{table}
\begin{center}
\begin{tabular}{lcccc}
Fragment & $(\lambda_0)$ & $(\lambda_1)$ & $(\lambda_2)$ & Avg. \\ \hline
jazz   & -2.652 &  -2.571 &  -2.720 & -2.647 \\
harpsi & -1.976 &  -1.757 &  -2.606 & -2.113 \\
speech & -2.649 &  -2.492 &  -2.961 & -2.701 \\
pop    & -3.328 &  -3.375 &  -3.445 & -3.383 \\
rock & -2.699 &  -2.556 &  -2.787 & -2.681 \\ \hline
\end{tabular}
\caption{ODGs when receiving a single description out of three.}
\label{tab:odg1}
\end{center}
\end{table}

\begin{table}
\begin{center}
\begin{tabular}{lcccc}
Fragment & $(\lambda_0,\lambda_1)$ & $(\lambda_0,\lambda_2)$ & $(\lambda_1,\lambda_2)$ & Avg. \\ \hline
jazz   & -1.033 &  -1.162 &  -1.021 & -1.072 \\
harpsi & -0.729 &  -0.993 &  -0.893 & -0.872 \\
speech & -0.994 &  -1.171 &  -1.040 & -1.068 \\
pop    & -1.897 &  -2.401 &  -2.082 & -2.127 \\
rock& -1.125 &  -1.284 &  -1.128 & -1.179 \\ \hline
\end{tabular}
\caption{ODGs when receiving two descriptions out of three.}
\label{tab:odg2}
\end{center}
\end{table}

\begin{table}
\begin{center}
\begin{tabular}{lc}
Fragment & $(\lambda_0,\lambda_1,\lambda_2)$ \\ \hline
jazz   &  -0.104  \\
harpsi &  -0.166  \\
speech &  -0.189  \\
pop    &  -0.171  \\
rock&  -0.184  \\ \hline
\end{tabular}
\caption{ODGs when receiving all three descriptions.}
\label{tab:odg3}
\end{center}
\end{table}

%
%

\section{Conclusion}
We combined MD-LVQ with transform coding in order to obtain a perceptually robust audio coder\index{audio coding}. Previous approaches to this problem were restricted to the case of only two descriptions.
In this work we used $K$-Channel MD-LVQ, which allowed for the possibility of using more than two descriptions.
For a given packet-loss probability we found the number of descriptions
and the bit allocation between transform coefficients, which minimizes a perceptual distortion measure subject to an entropy constraint. 
The optimal MD lattice vector quantizers were presented in closed form, thus avoiding any iterative quantizer design procedures.
The theoretical results were verified with numerical computer simulations using audio signals and it was shown that in environ\-ments with excessive packet losses it is advantageous to use more than two descriptions.
We verified in subjective listening tests that using more than two descriptions
lead to signals of perceptually higher quality.

\chapter{Conclusions and Discussion}

\section{Summary of Results}
We presented an index-assignment based design of $K$-channel MD-LVQ. Where pre\-vious designs have been limited to two descriptions we considered the general case of $K$ descriptions.
Exact rate-distortion results were derived for the case of $K\leq 3$ descriptions and high resolution conditions for smooth stationary sources and the squared error distortion measure. In the asymptotic case of large lattice vector quantizer dimension and high resolution conditions, it was shown that existing rate-distortion MD bounds can be achieved in the quadratic Gaussian case. These results were conjectured to hold also for $K>3$ descriptions. 

In the two-description asymmetric case it was shown that the performance was superior to existing state-of-the-art asymmetric schemes in finite lattice vector quantizer dimensions greater than one. In one and infinite dimensions as well as in the symmetric case (for all dimensions), the performance is identical to existing state-of-the-art sche\-mes.

In the three-description symmetric and asymmetric cases for finite lattice vector quantizer dimen\-sions, the rate loss of the proposed design is superior to that of existing schemes. 

The optimal amount of redundancy in the system was shown to be independent of the source distribution, target rate and type of lattices used for the side quantizers. Basically, the channel conditions (expressed through a set of packet-loss probabilities) describe the required amount of redundancy in the system. 
Thus, for given channel conditions, the optimal index-assignment map can be found and adapting to time-varying source distributions or bit rate requirements amounts to a simple scaling of the central and side lattice vector quantizers.

We proposed an entropy-constrained design where either the side description rates or their sum rate\index{sum rate} are subject to entropy contraints. In the case of a single sum rate entropy contraint, we showed that the optimal bit allocation across descriptions is not unique, but in fact consists of a set of solutions, which all lead to minimal expected distortion.

On the practical side it was shown that the optimal $K$-channel MD lattice vector quantizers can be found in closed-form, hence avoiding any iterative (e.g.\ generalized Lloyd-like) design algorithms. 
Furthermore, we combined MD-LVQ with transform coding in order to obtain a perceptually robust audio coder. Previous approaches to this problem were restricted to the case of only two descriptions.
For a given packet-loss probability we found the number of descriptions
and the bit allocation between transform coefficients, which minimizes a perceptual distortion measure subject to an entropy constraint. 
The theoretical results were verified with numerical computer simulations using audio signals and it was shown that in environments with excessive packet losses it is advantageous to use more than two descriptions.
We verified in subjective listening tests that using more than two descriptions
leads to signals of perceptually higher quality.

\section{Future Research Directions}
In this thesis we considered index-assignment based MD schemes at high resolution conditions, which provide a partial solution to the $K$-channel MD problem. However, more work is needed before the general MD problem is solved. 
Besides the information theoretic open problems discussed in Chapter~\ref{chap:md_theory} there are many unsolved problems related to MD-LVQ. Below we list a few of these.
\begin{itemize}
\item Proving the conjectures of this thesis, i.e.\ proving the rate-distortion results for $K>3$ descriptions.
\item Extending the results to general resolution. To the best of the authors knowledge, the only case where exact rate-distortion expressions (in non high-resolution cases) have been presented for index-assignment based MD schemes, is the two-channel scalar scheme by Frank-Dayan and Zamir~\cite{dayan:2002}.
\item It is an open problem of how to construct practical MD-LVQ schemes that comes arbitrarily close to the known MD bounds. Such schemes require high-dimensional lattice vector quantizers and large index values. 
However, solving the linear assignment problem can become computationally infeasible for large index values. For the symmetric case and certain low dimensional lattices, some progress have been made in reducing this complexity by the design of Huang and Wu~\cite{huang:2006}. A construction for high-dimensional nested lattice codes was recently presented by Zamir et al.~\cite{zamir:2002}. No index-assignment methods have, however, been presented for the nested lattice code design and the problem is therefore not solved.
\item Constructing functional MD schemes for existing applications in real environ\-ments and assessing their performance. For example for real-time application, even if the packet-loss rate of a network is very low, the delay might occasionally be high, which then means that (at least for real-time applications) a delayed packet is considered lost (at least for the current frame) and the use of MD coding might become beneficial. 
\end{itemize}

\appendix

\renewcommand{\chaptermark}[1]{\markboth{\emph{(Appendix \thechapter)}\hspace{1cm} \MakeUppercase{#1}}{}}


\chapter{Quaternions}\label{app:quaternions}
\setcounter{section}{1}
We will here briefly define the Quaternions\index{Quaternions} and describe a few important properties that we will use in this work. For a comprehensive treatment of the Quaternions we refer the reader to~\cite{ward:1997,conway:2003,kantor:1989}. 

The Quaternions, which were discovered in the middle of the $19^{th}$ century by Hamilton~\cite{kantor:1989}, is in some sense a generalization of the complex numbers. Just as $1$ and $i$ denote unit vectors of the complex space $\mathbb{C}$ we define $1,i,j$ and $k$ to be unit vectors in Quaternion space $\mathbb{H}$. The set of numbers defined as $\{a+bi+cj+dk : a,b,c,d\in \mathbb{R}\}$ are then called Quaternion numbers or simply Quaternions. Addition of two Quaternions $q=a+bi+cj+dk$ and $q'=a'+b'i+c'j+d'k$ is defined as 
\begin{equation}\label{eq:quaternionadd}
q+q' = (a+a')+(b+b')i+(c+c')j+(d+d')k,
\end{equation}
and multiplication follows by first defining a multiplication rule for pairs of Quaternion units, that is
\begin{equation}
\begin{split}
i^2&=j^2=k^2=-1, \\
ij&=k,\quad ji=-k, \\
jk&=i,\quad kj=-1, \\
ki&=j,\quad ik=-j,
\end{split}
\end{equation}
which leads to
\begin{equation}\label{eq:quaternionmult}
\begin{split}
qq' &= (aa' - bb' - cc' -dd') + (ab' + ba' + cd' - dc')i \\
&\quad +  (ac'+ca'+db'-bd')j+(ad'+da'+bc'-cb')k.
\end{split}
\end{equation}

\begin{definition}
The skew field\index{skew field} $\mathbb{H}$ of Quaternions is defined as the set $\{z=a+bi+cj+dk : a,b,c,d\in \mathbb{R}\}$ combined with two maps (addition and multiplication) given by~(\ref{eq:quaternionadd}) and~(\ref{eq:quaternionmult}), respectively, that satisfies field properties except that multiplication is non commutative.
\end{definition}

\begin{definition}
Quaternionic conjugation\index{conjugation!Quaternionic} of $q=a+bi+cj+dk \in \mathbb{H}$ is given by $q^*=a-bj-cj-dk$ and we denote by $\bs{q}^{\dagger}$ Quaternionic conjugation and vector transposition where $\bs{q}\in \mathbb{H}^L$.
\end{definition}

\begin{definition}
The real part of a Quaternion $q=a+bi+cj+dk$ is given by $\Re(q)=a$ and the complex part (also called the vector part) is given by $\Im(q)=bi+cj+dk$.
\end{definition}

\begin{lemma}[\cite{baylis:1989}]
Two Quaternions $q_0$ and $q_1$ commute, i.e.\ $q_0q_1=q_1q_0$ if their vector parts are proportional (i.e.\ linear dependent) or, in other words, if the cross product $\Im(q_0)\times\Im(q_1)=0$.
\end{lemma}

\begin{lemma}[\cite{conway:1999}]
The norm\index{norm!Quaternion} $\|q\|$ of a Quaternion $q\in \mathbb{H}$ is given by $\|q\|=\sqrt{q^{*}q}$ and satisfies the usual vector norm, i.e.\ $\|q\|=\sqrt{a^2+b^2+c^2+d^2}$.
\end{lemma}

\begin{definition}[\cite{calderbank:2005}]\label{def:quaternionmatrix}
The Quaternions can be represented in terms of matrices. 
 The isomorphic map $\phi_L\colon (\mathbb{H},+,\cdot) \to (H_{4\times 4},\oplus,\otimes)$ between the space $\mathbb{H}$ of Quaternions and the space $H_{4\times 4}$ of $4\times 4$ matrices over the real numbers $\mathbb{R}$ defined by
\begin{equation}
\phi_L(a+bi+cj+dk)\mapsto
\begin{bmatrix}
a & -b & -c & -d \\
b & \phantom{-}a & -d & \phantom{-}c \\
c & \phantom{-}d & \phantom{-}a & -b \\
d & -c & \phantom{-}b & \phantom{-}a
\end{bmatrix},
\end{equation}
describes left multiplication by the Quaternion $q$. Similar we define right multiplication by the map $\phi_R\colon (\mathbb{H},+,\cdot) \to (H_{4\times 4},\oplus,\otimes)$ given by
\begin{equation}
\phi_R(a+bi+cj+dk)\mapsto
\begin{bmatrix}
a & -b & -c & -d \\
b & \phantom{-}a & \phantom{-}d & -c \\
c & -d & \phantom{-}a & \phantom{-}b \\
d & \phantom{-}c & -b & \phantom{-}a
\end{bmatrix}.
\end{equation}
\end{definition}

It follows from Definition~\ref{def:quaternionmatrix} that addition and multiplication of two Quaternions can be done by use of the usual matrix addition $\oplus$ and matrix multiplication $\otimes$. Furthermore, Quaternionic conjugation\index{conjugation!Quaternionic} can easily be done in $H_{4\times 4}$ space where it is simply the matrix transpose.

\chapter{Modules}\label{app:modules}
In this appendix we give a brief introduction to the theory of algebraic modules\index{module!algebraic}. For more information we refer the reader to the following textbooks~\cite{adkins:1992,keating:1998,grillet:1999,conway:1999}.

\section{General Definitions}
\begin{definition}
Let $\mathcal{J}$ be a ring which is not necessarily commutative with respect to multiplication. Then an Abelian (commutative) group\index{group!commutative} $S$ is called a left $\mathcal{J}$-module or a left module\index{module!left} over $\mathcal{J}$ with respect to a mapping (scalar multiplication on the left which is simply denoted by juxtaposition) $\mathcal{J}\times S\to S$ such that for all $a,b\in \mathcal{J}$ and $g,h\in S$,
\begin{enumerate}
\item[1)] $a(g+h) = ag+ah$,
\item[2)] $(a+b)g = ag+bg$,
\item[3)] $(ab)g = a(bg)$.
\end{enumerate}
\end{definition}

\begin{remark}
For simplicity we have used the same notations for addition/multiplication in the group as well as in the ring.
\end{remark}

\begin{remark}
A right module\index{module!right} is defined in a similar way but with multiplication on the right. In fact if the ring $\mathcal{J}$ is commutative then every left $\mathcal{J}$-module is also a right $\mathcal{J}$-module~\cite{grillet:1999}.
\end{remark}

\begin{definition}
If $\mathcal{J}$ has identity $1$ and if $1a = a$ for all $a\in S$, then $S$ is called a unitary\index{module!unitary} or unital\index{module!unital} $\mathcal{J}$-module. 
\end{definition}

\begin{remark}
In this work all modules have an identity.
\end{remark}

\begin{definition}[\cite{keating:1998}]
A subset $S'=\{m_1,\dots,m_n\}\subset S$ of the $\mathcal{J}$-module $S$ is linearly independent over $\mathcal{J}$ if, for $x_i\in \mathcal{J}$, $x_1m_1+\cdots+x_nm_n=0$ only if $x_1=\cdots=x_n=0$. If in addition $S'$ generates $S$ then $S'$ is a basis for $S$.
\end{definition}

\begin{example}
The set $S=\{2,3\}$ is finite and generates $\mathbb{Z}$, considered as a $\mathbb{Z}$-module over itself~\cite{keating:1998}. However, $S$ is not a linearly independent set. Further, neither element of $S$ can be omitted to give a generating set with one member. Hence, $S$ is not a basis of $\mathbb{Z}$.
\end{example}

\begin{definition}[\cite{keating:1998}]\label{def:free}
A $\mathcal{J}$-module that has a basis is called a free $\mathcal{J}$-module.
\end{definition}

\begin{definition}[\cite{keating:1998}]
The number of elements in a basis of a $\mathcal{J}$-module $S$ is called the rank or dimension of $S$.\end{definition}

\begin{remark}
Not all modules have a basis~\cite{keating:1998}. Let $\mathbb{Z}_m=\mathbb{Z}/m\mathbb{Z}$ be the residue ring of integers\index{integer!residue ring}, i.e.\ $\mathbb{Z}_m=\{[0],\dots,[m-1]\}$, where $[s]=[r]$ in $\mathbb{Z}_m$ implies $s\equiv r\pmod{m}$. Notice that $\mathbb{Z}_m$ contains no linear independent subsets, since $mx=0$ for any $x\in \mathbb{Z}$, hence $\mathbb{Z}_m$ has no basis and is therefore not a free module.
\end{remark}

\begin{definition}[\cite{adkins:1992}]
Let $\mathcal{J}$ be a ring and let $S,S'$ be left $\mathcal{J}$-modules. A function $f: S\rightarrow S'$ is a $\mathcal{J}$-module homomorphism if
\begin{enumerate}
\item $f(m_1+m_2)=f(m_1)+f(m_2)$ for all $m_1,m_2\in S$, and
\item $f(am)=af(m)$ for all $a\in \mathcal{J}$ and $m\in S$.
\end{enumerate}
\end{definition}

\begin{definition}[\cite{adkins:1992}]
The set of all $\mathcal{J}$-module homomorphisms\index{homomorphism} from $S$ to $S'$ is denoted $\text{Hom}(S,S')$. If $S=S'$ then we write $\text{End}(S)$ where elements of $\text{End}(S)$ are called endomorphisms\index{endomorphism}. 
If $f\in \text{End}(S)$ is invertible, then it is called an automorphism\index{automorphism}. 
The group\index{group!of automorphisms} of all automorphisms is denoted $\text{Aut}(S)$. 
\end{definition}

\begin{definition}[\cite{keating:1998}]
Let $\mathcal{J}$ be a ring and $S$ a $\mathcal{J}$-module. Then an annihilator\index{annihilator} of an element $g\in S$ is the set
\begin{equation}
\text{Ann}(g)=\{ h\in \mathcal{J} : hg = 0\}.
\end{equation}
An element $g\in S$ is said to be a torsion\index{torsion} element of $S$ if $\text{Ann}(g)\neq 0$, that is, there is some non-zero element $a\in \mathcal{J}$ with $ag=0$.
\end{definition}

\begin{definition}[\cite{keating:1998}]\label{def:torsionfree}
A $\mathcal{J}$-module $S$ is said to be torsion-free\index{torsion!free} if the only torsion element in $S$ is 0.
\end{definition}

\begin{definition}
Let $S'$ be a finite group\index{group!orbits} and let $S$ be a left $\mathcal{J}$-module. 
The orbit\index{orbits} under the action\index{module!action} of $m\in S$ is obtained by left multiplication, i.e.\ $S'm=\{gm : g\in S'\}$.
\end{definition}

\section{Submodule Related Definitions}
\begin{definition}
Let $S$ be a $\mathcal{J}$-module and $S'$ a nonempty subset of $S$. Then $S'$ is called a submodule of $S$ if $S'$ is a subgroup\index{group!sub} of $S$ and for all $g\in S, h\in S'$, we have $gh\in S'$.
\end{definition}

\begin{definition}[\cite{grillet:1999}]
A cyclic submodule\index{cyclic!submodule} is a submodule which is generated by a single element. For example in a left $\mathcal{J}$-module $S$, a cyclic submodule can be generated by $m\in S$ in the following ways $\mathcal{J}m=\{xm: x\in \mathcal{J}\}$ or $m\mathcal{J}=\{mx: x\in \mathcal{J}\}$.
\end{definition}

\begin{definition}[\cite{grillet:1999}]
In general the submodule $S'$ of the $\mathcal{J}$-module $S$ generated by a finite subset $\{m_1,\dots,m_n\}\subset S$ is the set
\begin{equation}
\mathcal{J}m_1+\cdots+\mathcal{J}m_n = \{x_1m_1+\cdots+x_nm_n:x_1,\dots,x_i\in \mathcal{J}\}
\end{equation}
of all linear combinations of $m_1,\dots,m_n$. Such submodules are called finitely generated. If $S'=S$ then $\{m_1,\dots,m_n\}$ is a set of generators for $S$.
\end{definition}

\begin{lemma}[\cite{keating:1998}]
Let $S$ and $S'$ be submodules of a $\mathcal{J}$-module $S$. Then their sum
\begin{equation}
S+S' = \{l+n : l\in S, n\in S'\},
\end{equation}
is also a submodule. Moreover, $S + S' = S \Longleftrightarrow S'\subseteq S$.
\end{lemma}

\begin{lemma}[\cite{keating:1998}]
Let $S$ and $S'$ be submodules of a $\mathcal{J}$-module $S$. Then their intersection
\begin{equation}
S\cap S' = \{x : x\in S\ \text{and}\ x\in S'\},
\end{equation}
is also a submodule. Moreover, $S\cap S' = S \Longleftrightarrow S\subseteq S'$.
\end{lemma}

\begin{remark}
Submodules of a vector space are its subspaces. 
\end{remark}

\begin{proposition}[\cite{keating:1998}]
Let $S'$ be a submodule of the $\mathcal{J}$-module $S$. 
Let $m,n\in S$ and define a relation on $S$ by the rule that $m\equiv n \Longleftrightarrow m-n \in S'$. The equivalence class of an element $m\in S$ is given by the set
\begin{equation}
[m]=m+S'=\{m+l:l\in S'\}.
\end{equation}
The quotient module\index{module!quotient} (or factor module) $S/S'$ is defined to be the set of all such equivalence classes, with addition given by
\begin{equation}
[m]+[n]=[m+n], [m],[n]\in S/S',
\end{equation}
and multiplication by $r\in \mathcal{J}$ is given by
\begin{equation}
r[m]=[rm], [m]\in S/S'.
\end{equation}
\end{proposition}

\begin{definition}[\cite{keating:1998}]
The module homomorphism  $\pi : S\rightarrow S/S'$ defined by $\pi(m)=[m]$ is called the natural map (or canonical homomorphism) from $S$ to $S/S'$.
\end{definition}

\begin{definition}
Let $S'$ and $S''$ be arbitrary submodules of the $\mathcal{J}$-module $S$. If $S'+S''\neq S$ and $S'\cap S''\neq 0$ their configuration or relationship is often expressed in the diagram shown in Fig.~\ref{fig:diagramextended}. If $S$ is the direct sum\index{direct sum} of $S'$ and $S''$ we have $S'\cap S'' =0$ and $S'+S''=S$ which lead to the simpler diagram shown in Fig.~\ref{fig:diagramsimple}.
\begin{figure}[ht]
\begin{center}
\psfrag{L}{$S'$}
\psfrag{N}{$S''$}
\psfrag{M}{$S$}
\psfrag{LcN}{$S'\cap S''$}
\psfrag{L+N}{$S'+S''$}
\psfrag{0}{$0$}
\mbox{%
\subfigure[]{\includegraphics{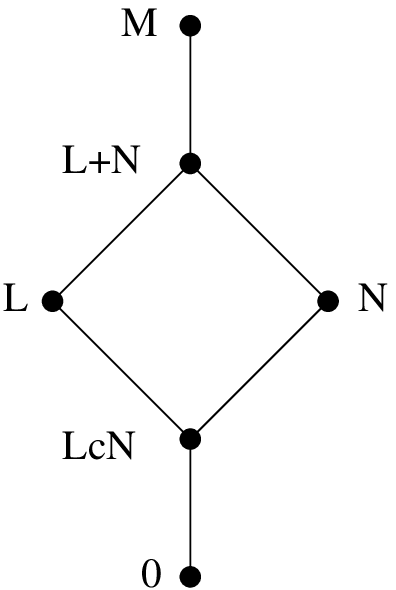}\label{fig:diagramextended}}\qquad%
\subfigure[]{\raisebox{13mm}{\includegraphics{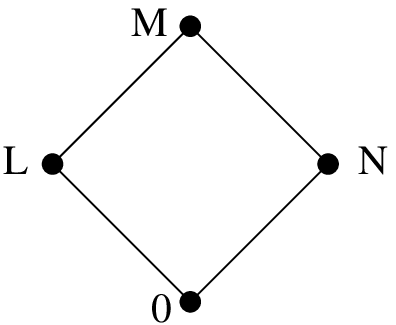}}\label{fig:diagramsimple}}}
\caption{(a) The module $S$ is not a direct sum\index{direct sum} of the submodules $S'$ and $S''$. (b) The module $S$ is a direct sum of $S'$ and $S''$ and therefore $S'\cap S''=0$ and $S'+S''=S$.}
\label{fig:diagramextsim}
\end{center}
\end{figure}
\end{definition}

\section{Quadratic Forms}

\begin{definition}[\cite{adkins:1992}]
A conjugation\index{conjugation!function} on $\mathcal{J}$ is a function $c: \mathcal{J}\to \mathcal{J}$ satisfying
\begin{enumerate}
\item $c(c(\xi)) = \xi,\quad  \forall \xi\in \mathcal{J}$
\item $c(\xi_1+\xi_2) = c(\xi_1) + c(\xi_2),\quad \forall \xi_1,\xi_2\in \mathcal{J}$
\item $c(\xi_1\xi_2) = c(\xi_1)c(\xi_2),\quad  \forall \xi_1,\xi_2\in\mathcal{J}$
\end{enumerate}
\end{definition}

\begin{definition}[\cite{adkins:1992}]\label{def:bilinearform}
Let $S$ be a free $\mathcal{J}$-module. A bilinear\index{bilinear form} form on $S$ is a function $\phi : S\times S \to \mathcal{J}$ satisfying
\begin{enumerate}
\item $\phi(\xi_1 x_1 + \xi_2 x_2, y) = \xi_1\phi(x_1,y) + \xi_2 \phi(x_2,y)$
\item $\phi(x, \xi_1 y_1 + \xi_2 y_2) = \xi_1\phi(x,y_1) + \xi_2 \phi(x,y_2)$
\end{enumerate}
for all $x_1,x_2,y_1,y_2 \in S$ and $\xi_1,\xi_2\in \mathcal{J}$.
\end{definition}

\begin{definition}[\cite{adkins:1992}]\label{def:sesquilinearform}
Let $S$ be a free $\mathcal{J}$-module. A sesquilinear\index{sesquilinear form} form on $S$ is a function $\phi : S\times S \to \mathcal{J}$ satisfying
\begin{enumerate}
\item $\phi(\xi_1 x_1 + \xi_2 x_2, y) = \xi_1\phi(x_1,y) + \xi_2 \phi(x_2,y)$
\item $\phi(x, \xi_1 y_1 + \xi_2 y_2) = c(\xi_1)\phi(x,y_1) + c(\xi_2)\phi(x,y_2)$
\end{enumerate}
for all $x_1,x_2,y_1,y_2 \in S$ and $\xi_1,\xi_2\in \mathcal{J}$ for a non-trivial conjugation $\xi\mapsto c(\xi)$ on $\mathcal{J}$.
\end{definition}

\begin{definition}[\cite{conway:1999}]
Let $\Lambda$ be a $\mathcal{J}$-lattice in $\mathbb{R}^L$ having basis vectors $\zeta_1,\dotsc,\zeta_L \in \mathbb{R}^L$ whose transposes form the rows of the generator matrix $M$. Then any lattice point $\lambda\in \Lambda$ may be written on generic form as $\lambda=\xi^TM$ where $\xi=(\xi_1,\dotsc,\xi_L)^T$ and where $\xi_i\in \mathcal{J}$. Let us define the following function of $\lambda$ (i.e.\ a squared norm)
\begin{equation}\label{eq:squarednorm}
\begin{split}
\tau(\lambda) &= \sum_{i=1}^{L}\sum_{j=1}^{L}(\xi_i\zeta_i)^T\xi_j\zeta_j  \\
 &= \xi^{T}MM^T\xi.
\end{split}
\end{equation}
The function $\tau$ in~(\ref{eq:squarednorm}) is referred to as the quadratic form\index{quadratic!form} associated with the lattice $\Lambda$. If $\Lambda$ has full rank, then $MM^T$ is a positive definite matrix and the associated quadratic form is called a positive definite form. If we extend $\tau$ to $\tau(\lambda_1,\lambda_2)=\xi_1^TMM^T\xi_2$ we obtain the bilinear\index{bilinear form} form of Definition~\ref{def:bilinearform} and if the underlying field is non real we get the sesquilinear\index{sesquilinear form} form of Definition~\ref{def:sesquilinearform} where the conjugation function\index{conjugation!function} $c$ depends on the field.
\end{definition}

\begin{remark}
In this work we will not make explicitely use of quadratic forms. Instead we equip the underlying field with an inner product $\langle \cdot,\cdot \rangle$ which satisfies De\-fi\-nition~\ref{def:bilinearform} and is therefore a bilinear form. \end{remark}
\begin{example}
Let $\Lambda\subset \mathbb{R}^L$, i.e.\ $\Lambda$ is a lattice embedded in the field $\mathbb{R}^L$. Then we can define the usual vector norm $\|\lambda \|^2\triangleq \langle \lambda,\lambda \rangle$, where $\lambda\in \Lambda$. Notice that here the conjugation is simply the identity. See Appendix~\ref{app:lattice_defs} for more examples.
\end{example}

\chapter{Lattice Definitions}\label{app:lattice_defs}
In this appendix we present a number of lattice-related definitions and properties which are used throughout the thesis.

Let $\mathbb{V}$ be a vector space over the field $\mathbb{K}$ and let $\mathbb{V}$ be equipped with an inner product $\langle \cdot , \cdot \rangle$. 
If $\mathbb{K}^L=\mathbb{R}^L$ then $\mathbb{V}$ is the traditional vector space over the Cartesian product of the reals.\footnote{Recall that, in a vector space over $\mathbb{R}^L$, addition and subtraction is with respect to vectors of $\mathbb{R}^L$ whereas multiplication is defined as multiplications of vectors in $\mathbb{R}^L$ with scalar elements of $\mathbb{R}$. Thus, a vector space is a $\mathcal{J}$-module where $\mathcal{J}$ is a field, i.e.\ a ring where all elements (except 0) have inverses.}
As such, $(\mathbb{V}, \langle\cdot,\cdot\rangle)$ is an inner-product space.
An inner-product space induces a norm\index{norm!inner product} $\|\cdot\|$ defined as $\|\cdot\|^2 \triangleq \langle \cdot, \cdot \rangle $. 
If $\mathbb{K}^L=\mathbb{R}^L$ we use the $\ell_2$-norm defined as $\|x\|^2\triangleq x^Tx$ whereas if $\mathbb{K}^L=\mathbb{C}^L$ we have $\|x\|^2\triangleq x^Hx$ where $^H$ denotes Hermitian transposition (i.e.\ the conjugate transpose\index{conjugation!Hermitian}). If $\mathbb{K}^L=\mathbb{H}^L$ then $\|x\|^2\triangleq x^\dagger x$ where $^\dagger$ denotes Quaternionic conjugation\index{conjugation!Quaternionic} and transposition. For more information about inner-product spaces we refer the reader to the widely used textbook by Luenberger~\cite{luenberger:1997}.

\section{General Definitions}
\begin{definition}[\cite{conway:1999}]
A lattice $\Lambda\subset\mathbb{K}^L$ consists of all possible integral linear combinations of a set of basis vectors, or, more formally
\begin{equation}
\Lambda = \left\{ \lambda\in \mathbb{K}^L : \lambda = \sum_{i=1}^{L} \xi_i\zeta_i,\ \forall\xi_i \in \mathcal{J}\right\},
\end{equation}
where $\zeta_i\in \mathbb{K}^L$ are the basis vectors also known as generator vectors of the lattice and $\mathcal{J}\subset \mathbb{K}$ is a well defined ring of integers.
\end{definition}

\begin{remark}
It should be noted that it is often convenient to use $L'>L$ basis vectors to form an $L$-dimensional lattice embedded in $\mathbb{K}^{L'}$. 
\end{remark}

\begin{definition}
Let $M$ be a generator matrix of the lattice $\Lambda$. Then the rows of $M$ are given by the tranposes of the column vectors $\zeta_i^T, i=1,\dotsc,L$, where we actually do not require that $M$ is square.
\end{definition}

\begin{definition}
The square matrix $A=MM^T$ is called the Gram matrix.
\end{definition}

\begin{definition}
A fundamental region\index{fundamental!region} of a lattice is a closed region which contains a single lattice points and tessellate\index{tessellating!region} the underlying space. 
\end{definition}

\begin{lemma}[\cite{conway:1999}]
All fundamental regions have the same volume.
\end{lemma}

\begin{lemma}[\cite{conway:1999}]
The fundamental volume\index{fundamental!volume} $\nu$ of $\Lambda$ is given by $\nu=\sqrt{\det(A)}$, sometimes written as $\nu=\det(\Lambda)$\index{determinant of a lattice}. If $M$ is a square generator matrix then $\nu=|\det(M)|$.
\end{lemma}

\begin{definition}[\cite{coxeter:1973}]
An $L$-dimensional polytope\index{polytope} is a finite convex region in $\mathbb{K}^L$ enclosed by a finite number of hyperplanes.
\end{definition}

\begin{definition}[{\cite{ebeling:1994}}]\label{def:compactquotient}
The quotient\index{quotient!compact} $\mathbb{K}^L/\Lambda$ is the $L$-dimensional torus\index{torus} obtained by combining opposite faces of the fundamental parallelotope $\{a_1\zeta_1 + \dotsc + a_L\zeta_L | 0\leq a_i \leq 1\}$. 
\end{definition}

\begin{definition}\label{def:cartproductlattice}
The Cartesian product $\otimes$ of two lattices $\Lambda_1$ and $\Lambda_2$ is obtained by pairing all points in $\Lambda_1$ with every point in $\Lambda_2$, i.e.\
\begin{equation}
\Lambda_1\otimes\Lambda_2 = \{ (\lambda_1,\lambda_2)| \lambda_1\in\Lambda_1, \lambda_2 \in \Lambda_2\}.
\end{equation}
It follows that the dimension of $\Lambda=\Lambda_1\otimes\Lambda_2$ is equal to the sum of the dimensions of the two lattices $\Lambda_1$ and $\Lambda_2$.
\end{definition}

\begin{definition}[\cite{conway:1999}]
The automorphism\index{automorphism!group} group Aut($\Lambda$) of a lattice $\Lambda$ is the set of distance-preserving transformations\index{automorphism} (or isometries) of the space that fix the origin and takes the lattice to itself.
\end{definition}

\begin{theorem}[\cite{conway:1999}]
For a lattice in ordinary Euclidean space $\mathbb{R}^L$, Aut($\Lambda$) is finite and the transformations in Aut($\Lambda$) may be represented by orthogonal matrices. Let $\Lambda$ have generator matrix $M$. Then an orthogonal matrix $B$ is in Aut($\Lambda$) if and only if there is an integral matrix\index{integral matrix} $U$ with determinant $\pm 1$ such that
\begin{equation}
UM = MB.
\end{equation}
This implies $U=MBM^T A^{-1}$, where $A^{-1}$ is the Gram matrix of $\Lambda$.
\end{theorem}

\begin{remark}
Aut($\Lambda=Z^L$) consists of all sign changes of the $L$ coordinates $(=2^L)$ and all permutations $(=L!)$. Hence, $|\text{Aut}(Z^L)|=2^L L!$~\cite{conway:1999}.
\end{remark}

\begin{definition}[\cite{conway:1999}]
The dual lattice\index{dual lattice} $\tilde{\Lambda}$ of the lattice $\Lambda$ is given by
\begin{equation}
\tilde{\Lambda} = \{ x\in \mathbb{R}^L | x^T\lambda\in \mathbb{Z}\ \text{for all}\ \lambda\in \Lambda\}.
\end{equation}
Alternatively, if $M$ is a square generator matrix of $\Lambda$ then $\tilde{\Lambda}$ can be constructed by use of the generator matrix $\tilde{M}=(M^{-1})^T$.
\end{definition}

\begin{theorem}[\cite{ebeling:1994}]\label{theo:compactlattice}
If $\Lambda\subset \mathbb{R}^L$ is a discrete subgroup with compact quotient\index{quotient!compact} $\mathbb{R}^L/\Lambda$ then $\Lambda$ is a lattice.
\end{theorem}

\begin{definition}[\cite{conway:1999}]
The coefficients $B_i$ of the Theta series\index{theta series} $\Theta_\Lambda(z)\triangleq\sum_iB_iq^i$ of a lattice $\Lambda$ describe the number of points at squared distance $i$ from an arbitrary point in space (which is usually taken to be the origin). The indeterminate $q$ is sometimes set to $q=\exp(i\pi z)$, where $z\in \mathbb{C}$ and $\Im(z)>0$.
\end{definition}

\section{Norm Related Definitions}

\begin{definition}[\cite{conway:1999}]
Let $\Lambda\subset \mathbb{K}^L$ be a lattice. The nearest neighbor region\index{nearest neighbor!region} of $\lambda\in \Lambda$ is defined as 
\begin{equation}
V(\lambda) \triangleq \{ x\in \mathbb{K}^L : \| x - \lambda\|^2 \leq \|x-\lambda'\|^2,\, \forall\, \lambda' \in \Lambda \}.
\end{equation}
\end{definition}

\begin{definition}
The nearest neighbor regions of a lattice are also called Voronoi cells\index{Voronoi cell}, Voronoi regions or Dirichlet regions. In this work we will use the name Voronoi cells.
\end{definition}

\begin{definition}
Voronoi cells of a lattice are congruent polytopes\index{polytope!congruent}, hence they are similar in size and shape and may be seen as translated versions of a fundamental region, e.g.\ $V_0=V(0)$, i.e.\ the Voronoi cell around the origin. 
\end{definition}

\begin{definition}[\cite{conway:1999}]
The dimensionless normalized second moment of inertia\index{second-moment of inertia!of a lattice} $G(\Lambda)$ of a lattice $\Lambda$ is defined by
\begin{equation}\label{eq:Gapp}
G(\Lambda) \triangleq\frac{1}{L\nu^{1+2/L}}\int_{V_0}\|x\|^2dx.
\end{equation}
\end{definition}

\begin{remark}
Applying any scaling or orthogonal transform, e.g.\ rotation\index{rotation!transform} or reflection\index{reflection} on $\Lambda$ will not change $G(\Lambda)$, which makes it a good figure of merit when comparing different lattices (quantizers). In other words, $G(\Lambda)$ depends only upon the shape of the fundamental region, and in general, the more sphere-like shape, the smaller normalized second-moment~\cite{conway:1999}.
\end{remark}

\begin{definition}[\cite{forney:1993}]
The minimum squared distance\index{lattice!minimum squared distance} $d^2_{\text{min}}(\Lambda)$ between lattice points is the minimum non-zero norm of any lattice point $\lambda\in \Lambda$, i.e.\ 
\begin{equation}
d^2_{\text{min}}(\Lambda)\triangleq \min_{\substack{\lambda\in\Lambda\\ \lambda\neq 0}} \|\lambda\|^2.
\end{equation}
\end{definition}

\begin{definition}[\cite{forney:1993}]
The packing radius\index{lattice!packing radius} $\rho_p(\Lambda)$ of the lattice $\Lambda\subset \mathbb{R}^L$ is the radius of the greatest $L$-dimensional sphere that can be inscribed within $V_0$. We then have
\begin{equation}
\rho_p(\Lambda) \triangleq d_{\text{min}}(\Lambda)/2.
\end{equation}
\end{definition}

\begin{definition}[\cite{forney:1993}]
The covering radius\index{lattice!covering radius} $\rho_c(\Lambda)$ of the lattice $\Lambda\subset \mathbb{R}^L$ is the radius of the least $L$-dimensional sphere that contains $V_0$, i.e.
\begin{equation}
\rho_c(\Lambda)\triangleq \max_{x\in V_0}\|x\|.
\end{equation}
\end{definition}

\begin{definition}[\cite{forney:1993}]
The kissing\footnote{The terminology kissing number was introduced by N.\ J.\ A.\ Sloane who drew an analogy to billiards, where two balls are said to kiss if they touch each other, see for example the interview with N.\ J.\ A.\ Sloane by R. Calderbank, which can be found online at http://www.research.att.com/$\tilde{\mbox{ }}$njas/doc/interview.html.} number\index{kissing-number} $\mathfrak{K}(\Lambda)$ is the number of nearest neighbors to any lattice point, which is also equal to the number of lattice points of squared norm $d^2_{\text{min}}(\Lambda)$, i.e.
\begin{equation}
\mathfrak{K}(\Lambda) \triangleq |\{\lambda\in\Lambda : \|\lambda\|^2 = d^2_{\text{min}}(\Lambda)\}|.
\end{equation}
\end{definition}

\begin{definition}
The space-filling loss\index{space-filling loss} of a lattice $\Lambda$ with dimensionless normalized second moment\index{second-moment of inertia} $G(\Lambda)$ is given by
\begin{equation}
D_{\text{Loss}} = 10\log_{10}\left(2\pi eG(\Lambda)\right) \text{dB}.
\end{equation}
\end{definition}

\section{Sublattice Related Definitions}
\begin{definition}
A sublattice $\Lambda'\subseteq \Lambda$ is a subset of the elements of $\Lambda$ that is itself a lattice.
\end{definition}

\begin{definition}[\cite{diggavi:2002}]
A sublattice $\Lambda'\subset \Lambda$ is called clean\index{clean sublattice} if no point of $\Lambda$ lies on the boundary of the Voronoi cells\index{Voronoi cell} of $\Lambda'$.
\end{definition}

\begin{definition}
If $\Lambda'$ is a sublattice of $\Lambda$ then $N=|\Lambda/\Lambda'|$ denotes the index or order of the quotient\index{quotient!lattice} $\Lambda/\Lambda'$. 
\end{definition}

\begin{definition}
If $\Lambda'$ is a clean sublattice\index{clean sublattice} of $\Lambda$ then the index value\index{index!admissible} $N=|\Lambda/\Lambda'|$ is called an admissible index value.
\end{definition}

\begin{definition}
The $L^{th}$ root of the index $N$ is called the nesting ratio\index{index!nesting ratio} $N'$, i.e.\ $N'=N^{1/L}$.
\end{definition}

\begin{definition}[\cite{diggavi:2002}]\label{def:strictlysimilar}
Let $\Lambda$ be an $L$ dimensional lattice with square generator matrix $M$. A sublattice $\Lambda'\subseteq \Lambda$ is geometrically strictly similar\index{geometrically!strictly similar} to $\Lambda$ if and only if the following holds
\begin{enumerate}
\item There is an invertible $L\times L$ matrix $U_1$ with integer entries
\item a non-zero scalar $c_1\in \mathbb{R}$
\item an orthogonal $L\times L$ matrix $K_1$ with determinant 1,
\end{enumerate}
such that a generator matrix $M_1$ for $\Lambda_1$ can be written as
\begin{equation}\label{eq:sublattice}
M_1=U_1M=c_1MK_1.
\end{equation}
If~(\ref{eq:sublattice}) holds then the index $N_1$ of $\Lambda'$ is equal to
\begin{equation}
N_1 = |\Lambda/\Lambda'| = \frac{\det(\Lambda')}{\det(\Lambda)} = \left|\frac{\det(M_1)}{\det(M)}\right| = \det(U_1) = c_1^L.
\end{equation}
Furthermore, $\Lambda'$ has Gram matrix
\begin{equation}
A_1 = M_1M_1^T = U_1MM^TU_1^T = U_1AU_1^T = c_1^2A,
\end{equation}
where $A=MM^T$ is a Gram matrix for $\Lambda$.
\end{definition}

\begin{definition}
If in Definition~\ref{def:strictlysimilar} the determinant of $K_1$ is allowed to be $\pm 1$, i.e.\ $K_1$ can be either a rotation\index{rotation!transform} or a reflection\index{reflection} operator, then the sublattice $\Lambda'$ is said to be geometrically similar\index{geometrically!similar} to $\Lambda$.
\end{definition}

\chapter{Root Lattices}\label{app:rootlattices}
This appendix describes some properties of the root\footnote{The term root lattice refers to a lattice which can be generated by the roots of specific reflection\index{group!reflection} groups~\cite{conway:1999,ericson:2001}.}
lattices considered in this thesis.

\section{$Z^1$}\label{app:z1}
The scalar uniform lattice\index{lattice!$Z^1$} also called $Z^1$ partitions the real line into intervals of equal lengths. Table~\ref{tab:z1constants} outlines important constants related to the $Z^1$ lattice.
\begin{table}[ht]
\begin{center}
\begin{tabular}{|p{4cm}|c|c|}\hline
Description & Notation & Value \\ \hline\hline
Dimension & $L$ & 1 \\ \hline
Fundamental volume & $\nu$ & $1$ \\ \hline
Packing radius & $\rho_p$ & $1/2$ \\ \hline
Covering radius & $\rho_c$ & $1/2$\\ \hline
Space-filling loss & $D_{\text{loss}}$ &  1.5329 dB \\ \hline
Space-filling gain over $Z^1$ & $D_{\text{gain}}$ &  0 dB \\ \hline
Kissing-number\index{kissing-number} & $\mathfrak{K}$ & 2 \\ \hline
Minimal squared distance & $d^2_\text{min}$ & 1 \\ \hline
Dimensionless normalized second moment & \raisebox{-2.5mm}{$G(\Lambda)$} & \raisebox{-2.5mm}{$1/12$} \\ \hline
\end{tabular}
\caption{Relevant constants for the $Z^1$ lattice.}
\label{tab:z1constants}
\end{center}
\end{table}

The set of admissible index values for $Z^1$ is the set of all odd integers~\cite{diggavi:2002} and the coefficients of the Theta series\index{theta series} are given by $B_0=1$ and $B_i = 2, i>0$.

\section{$Z^2$}\label{app:z2}
A generator matrix for $Z^2$ (also known as the square lattice\index{lattice!$Z^2$}) is given by
\begin{equation}
M=\begin{bmatrix}
1 & 0 \\
0 & 1 
\end{bmatrix}.
\end{equation}
The Gram matrix $A$ is identical to the generator matrix, i.e.\ $A=M$. Table~\ref{tab:z2constants} gives an overview of important constants related to the $Z^2$ lattice.
\begin{table}[ht]
\begin{center}
\begin{tabular}{|p{4cm}|c|c|}\hline
Description & Notation & Value \\ \hline\hline
Dimension & $L$ & 2 \\ \hline
Fundamental volume & $\nu$ & $1$ \\ \hline
Packing radius & $\rho_p$ & 1/2 \\ \hline
Covering radius & $\rho_c$ & $\rho_p\sqrt{2}$\\ \hline
Space-filling loss & $D_{\text{loss}}$ &  1.5329 dB \\ \hline
Space-filling gain over $Z^1$ & $D_{\text{gain}}$ &  0 dB \\ \hline
Kissing-number\index{kissing-number} & $\mathfrak{K}$ & 4 \\ \hline
Minimal squared distance & $d^2_\text{min}$ & 1 \\ \hline
Dimensionless normalized second moment & \raisebox{-2.5mm}{$G(\Lambda)$} & \raisebox{-2.5mm}{$1/12$} \\ \hline
\end{tabular}
\caption{Relevant constants for the $Z^2$ lattice.}
\label{tab:z2constants}
\end{center}
\end{table}

\begin{figure}[ht]
\begin{center}
\includegraphics[width=8cm]{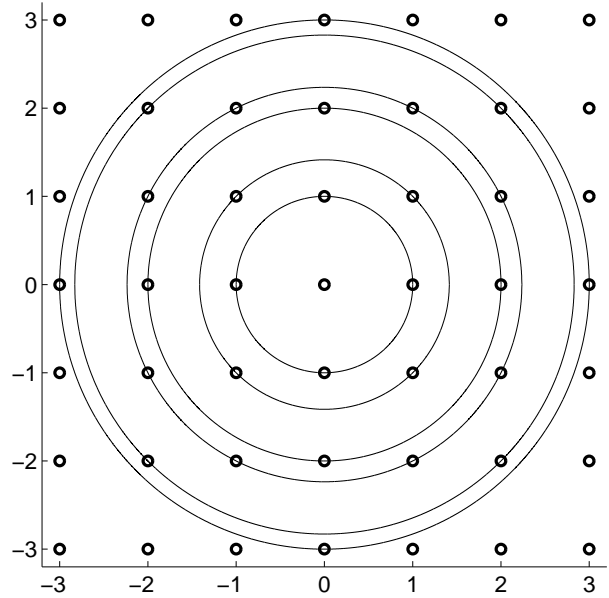}
\caption{The first 7 non-zero shells of $Z^2$ is here shown as large circles (incl.\ the one at the origin). Notice that the number of points lying on each circle agrees with the corresponding coefficient of its Theta series.}
\label{fig:theta_Z2}
\end{center}
\end{figure}

The first 50 coefficients of the Theta series\index{theta series}, i.e.\ the number of points in each of the 50 first shells of $\Lambda$ are shown in Table~\ref{tab:theta_Z2} and the first seven shells are shown in Fig.~\ref{fig:theta_Z2}.
\begin{table}[ht]
\begin{center}
\begin{tabular}{l}\hline
1,4,4,0,4,8,0,0,4,4,8,0,0,8,0,0,4,8,4,0,8,0,0,0,0,12,  \\
8,0,0,8,0,0,4,0,8,0,4,8,0,0,8,8,0,0,0,8,0,0,0,4 \\ \hline
\end{tabular}
\end{center}
\caption{The first 50 coefficients of the Theta series\index{theta series} with starting point at zero for the $Z^2$ lattice.}
\label{tab:theta_Z2}
\end{table}

Let $\Lambda$ be the square lattice represented in the scalar complex domain, i.e.\ $\Lambda=\mathcal{G}$. Then a sublattice $\Lambda'=\xi\Lambda$, where $\xi=a+ib$ and $\xi\in\mathcal{G}$ is clean if and only if $N=a^2+b^2$ is odd~\cite{diggavi:2002}. Equivalently an integer $N$ is an admissible index value if it can be written as a product of primes congruent to $1\!\pmod{4}$ and/or a product of primes congruent to $3\!\pmod{4}$~\cite{conway:1999b,diggavi:2002}.  
This set is given by integer sequence A057653~\cite{sloane:integerseq}, see also Table~\ref{tab:intseqZ2}.
\begin{table}[ht]
\begin{center}
\begin{tabular}{l} \hline
1,5,9,13,17,25,29,37,41,45,49,53,61,65,73,81,85,89,97,101,109,113, \\
117,121,125,137,145,149,153,157,169,173,181,185,197,205,221,225, \\
229,233,241,245,257,261,265,269,277,281,289,293,305,313,317,325 \\
333,337,\dots \\ \hline
\end{tabular}
\caption{Admissible index values for $Z^2$.}
\label{tab:intseqZ2}
\end{center}
\end{table}

A subgroup\index{subgroup} $\Gamma_4\subset \text{Aut}(\Lambda=\mathbb{Z}^2)$ of order 4 is given by~(\ref{eq:gamma4}).

\section{$A_2$}\label{app:a2}
The hexagonal lattice\index{lattice!$A_2$} (also known as $A_2$) can be represented in the complex field where it is identical to $\mathcal{E}$. When represented in $\mathbb{R}^2$ a possible generator matrix is 
\begin{equation}
M=\begin{bmatrix}
1 & 0 \\
-1/2 & \sqrt{3}/2
\end{bmatrix}.
\end{equation}
Its Gram matrix is given by
\begin{equation}
A=\begin{bmatrix}
1 & -1/2 \\
-1/2 & 1
\end{bmatrix}.
\end{equation}
Table~\ref{tab:a2constants} summarizes important constants related to the $A_2$ lattice.
\begin{table}[ht]
\begin{center}
\begin{tabular}{|p{4cm}|c|c|}\hline
Description & Notation & Value \\ \hline\hline
Dimension & $L$ & 2 \\ \hline
Fundamental volume & $\nu$ & $\sqrt{3}/2$ \\ \hline
Packing radius & $\rho_p$ & 1/2\\ \hline
Covering radius & $\rho_c$ & $2\rho_p\sqrt{3}$\\ \hline
Space-filling loss & $D_{\text{loss}}$ &  1.3658 dB \\ \hline
Space-filling gain over $Z^1$ & $D_{\text{gain}}$ &  0.1671 dB \\ \hline
Kissing-number\index{kissing-number} & $\mathfrak{K}$ & 6 \\ \hline
Minimal squared distance & $d^2_\text{min}$ & 1 \\ \hline
Dimensionless normalized second moment & \raisebox{-2.5mm}{$G(\Lambda)$} & \raisebox{-2.5mm}{$5/(36\sqrt{3})$} \\ \hline
\end{tabular}
\caption{Relevant constants for the $A_2$ lattice.}
\label{tab:a2constants}
\end{center}
\end{table}

Let $\Lambda$ be the hexagonal lattice represented in the scalar complex domain, i.e.\ $\Lambda=\mathcal{E}$. Then a sublattice $\Lambda'=\xi\Lambda$, where $\xi=a+\omega b$ and $\xi\in\mathcal{E}$ is clean if and only if 
$a$ and $b$ are relative prime or equivalently if and only if $N$ is a product of primes congruent to $1\pmod{6}$~\cite{diggavi:2002}. This set is given by integer sequence A004611~\cite{sloane:integerseq}, see also Table~\ref{tab:intseqA2}.

\begin{table}[ht]
\begin{center}
\begin{tabular}{l}\hline
1,7,13,19,31,37,43,49,61,67,73,79,91,97,103,109,127,133,139, \\
151,157,163,169,181,193,199,211,217,223,229,241,247,259,271, \\
277,283,301,307,313,331,337,343,349,361,367,373,379,397,403, \\
409,421,427,433,439,457,\dots  \\ \hline
\end{tabular}
\caption{Admissible index values for $A_2$.}
\label{tab:intseqA2}
\end{center}
\end{table}

A subgroup\index{subgroup} $\Gamma_6\subset \text{Aut}(\Lambda=\mathcal{E})$ of order 6 is given by the rotational group\index{group!rotational}
\begin{equation}
\Gamma_6=\{\exp(ik\pi/6), k=0,\dots,5\}.
\end{equation}

\begin{figure}[ht]
\begin{center}
\includegraphics[width=8cm]{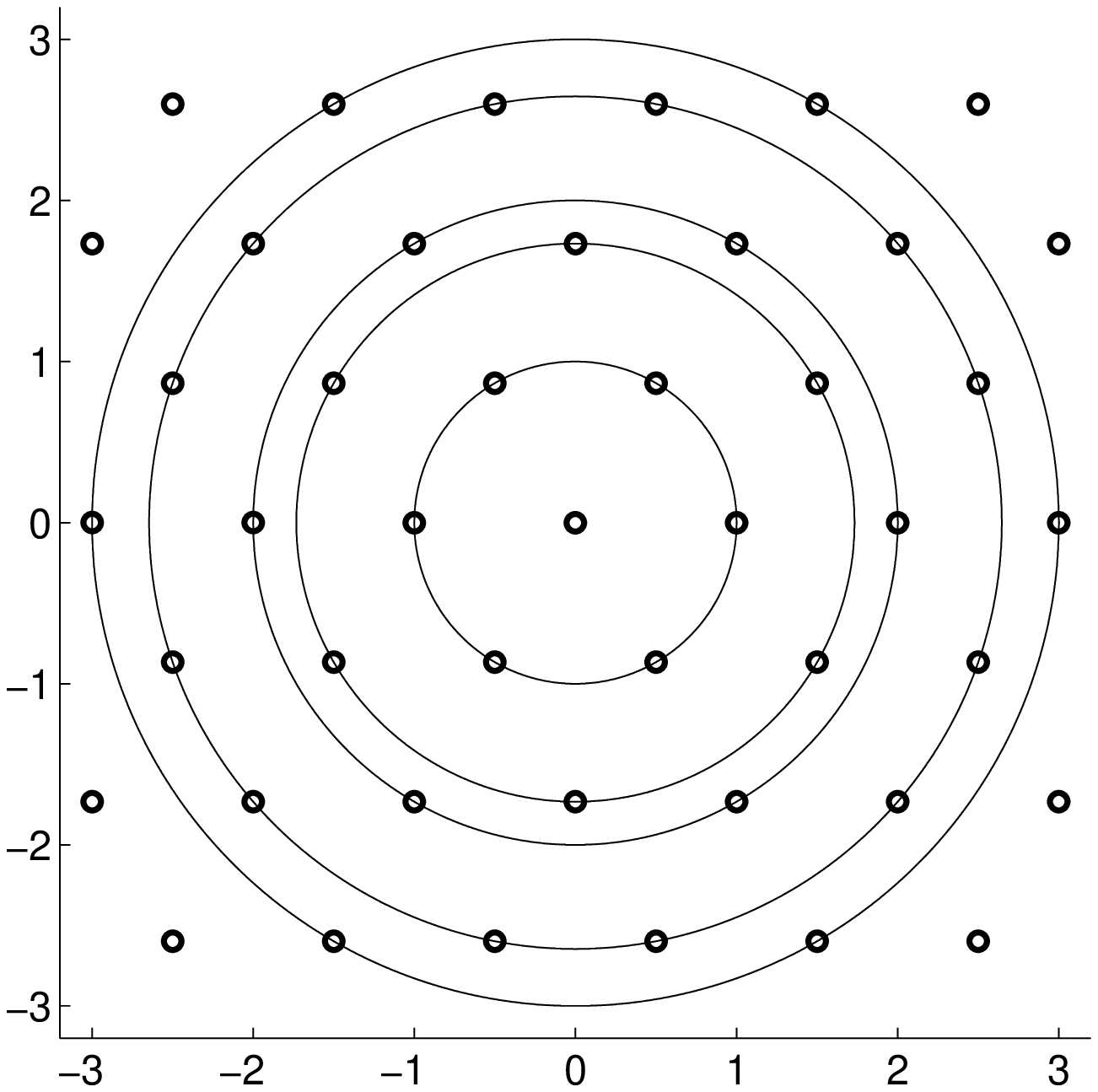}
\caption{The first 6 non-zero shells of $A_2$ is here shown as large circles (incl.\ the one at the origin). Notice that the number of points lying on each circle agrees with the corresponding coefficient of the Theta series.}
\label{fig:theta_A2}
\end{center}
\end{figure}

The first 50 coefficients of the Theta series\index{theta series} for $A_2$ are shown in Table~\ref{tab:theta_A2} and Fig.~\ref{fig:theta_A2}
\begin{table}[ht]
\begin{center}
\begin{tabular}{l} \hline
1,6,0,6,6,0,0,12,0,6,0,0,6,12,0,0,6,0,0,12,0,12,0,0,0,6,0,6, \\
12,0,0,12,0,0,0,0,6,12,0,12,0,0,0,12,0,0,0,0,6,18 \\ \hline
\end{tabular}
\caption{The first 50 coefficients of the Theta series with starting point at zero for the $A_2$ lattice.}
\label{tab:theta_A2}
\end{center}
\end{table}

\section{$Z^4$}\label{app:z4}
The hypercubic lattice\index{lattice!$Z^4$} $Z^4$ is generated by
\begin{equation}
M=\begin{bmatrix}
1 & 0 & 0 & 0\\
0 & 1 & 0 & 0\\
0 & 0 & 1 & 0\\
0 & 0 & 0 & 1
\end{bmatrix}.
\end{equation}
The Gram matrix $A$ is identical to the generator matrix, i.e.\ $A=M$. Table~\ref{tab:z4constants} gives an overview of important constants related to the $Z^4$ lattice.
\begin{table}[ht]
\begin{center}
\begin{tabular}{|p{4cm}|c|c|}\hline
Description & Notation & Value \\ \hline\hline
Dimension & $L$ & 4 \\ \hline
Fundamental volume & $\nu$ & $1$ \\ \hline
Packing radius & $\rho_p$ & 1/2\\ \hline
Covering radius & $\rho_c$ & 1 \\ \hline
Space-filling loss & $D_{\text{loss}}$ &  1.5329 dB \\ \hline
Space-filling gain over $Z^1$ & $D_{\text{gain}}$ &  0 dB \\ \hline
Kissing-number\index{kissing-number} & $\mathfrak{K}$ & 8 \\ \hline
Minimal squared distance & $d^2_\text{min}$ & 1 \\ \hline
Dimensionless normalized second moment & \raisebox{-2.5mm}{$G(\Lambda)$} & \raisebox{-2.5mm}{$1/12$} \\ \hline
\end{tabular}
\caption{Relevant constants for the $Z^4$ lattice.}
\label{tab:z4constants}
\end{center}
\end{table}

$Z^4$ has a geometrically-similar and clean sublattice of index $N$ if and only if $N$ is odd and of the form $a^2$ for some integer $a$~\cite{diggavi:2002}.
The set of admissible index values is given by integer sequence A016754~\cite{sloane:integerseq}, see also Table~\ref{tab:intseq_Z4}.
\begin{table}[ht]
\begin{center}
\begin{tabular}{l} \hline
1,9,25,49,81,121,169,225,289,361,441,529,625,729,841,961, \\
1089,1225,1369,1521,1681,1849,2025,2209,2401,2601,2809, \\
3025,3249,3481,3721,3969,4225,4489,4761,5041,5329,5625, \\
5929,6241,6561,\dots \\ \hline
\end{tabular}
\caption{Admissible index values for $Z^4$.}
\label{tab:intseq_Z4}
\end{center}
\end{table}

The first 50 coefficients of the Theta series\index{theta series} for $Z^4$ are given in Table~\ref{tab:theta_Z4}.
\begin{table}[ht]
\begin{center}
\begin{tabular}{l}\hline
1,8,24,32,24,48,96,64,24,104,144,96,96,112,192,192,24,144, \\
312,160,144, 256,288,192,96,248,336,320,192,240,576,256,24, \\
384,432,384,312,304,480,448,144,336,768,352,288,624,576, \\
384,96,456 \\ \hline
\end{tabular}
\caption{The first 50 coefficients of the Theta series with starting point at zero for the $Z^4$ lattice.}
\label{tab:theta_Z4}
\end{center}
\end{table}

A subgroup\index{subgroup} $\Gamma_8\subset \text{Aut}(\Lambda=\mathbb{Z}^4)$ of order 8 is given by~\cite{vaishampayan:2001}
\begin{equation}\label{eq:gamma8}
 \setlength{\arraycolsep}{.5\arraycolsep}
\Gamma_8\!=\!\left\{\! \pm I_4,\!%
\pm\!\begin{pmatrix}
0 & -1 & 0 & 0\\
1 & 0 & 0 & 0 \\
0 & 0& 0& 1\\
0 & 0 & -1 & 0
\end{pmatrix}\!,
\pm\!\begin{pmatrix}
0 & 0 & -1 & 0 \\
0 & 0 & 0 & -1 \\
1 & 0 & 0 & 0 \\
0 & 1 & 0 & 0
\end{pmatrix}\!,
\pm\!\begin{pmatrix}
0 & 0 & 0 & -1 \\
0 & 0 & 1 & 0 \\
0 & -1 & 0 & 0 \\
1 & 0 & 0 & 0
\end{pmatrix}\!
\right\}.
\end{equation}

\section{$D_4$}
The $D_4$ lattice\index{lattice!$D_4$} (also known as the Schl\"{a}fli lattice\index{Schl\"{a}fli lattice} or checkerboard lattice\index{checkerboard lattice}) consists of all points of $Z^4$ that have even squared norms~\cite{conway:1999}. A possible generator matrix is given by
\begin{equation}
M=\begin{bmatrix}
1 & 1 & 0 & 0\\
-1 & 0 & 1 & 0\\
0 & -1 & 0 & 1\\
0 & -1 & 0 & -1
\end{bmatrix}.
\end{equation}
The Gram matrix is given by
\begin{equation}
A=\begin{bmatrix}
2 & -1 & -1 & -1 \\
-1&  2 &  0 &  0 \\
-1&  0 &  2 &  0 \\
-1&  0 &  0 &  2
\end{bmatrix},
\end{equation}
and a subgroup\index{subgroup} $\Gamma_8\subset \text{Aut}(\Lambda=D_4)$ of order 8 is given by~(\ref{eq:gamma8}).
See Table~\ref{tab:d4constants} for a an overview of important constants related to the $D_4$ lattice.
\begin{table}[ht]
\begin{center}
\begin{tabular}{|p{4cm}|c|c|}\hline
Description & Notation & Value \\ \hline\hline
Dimension & $L$ & 4 \\ \hline
Fundamental volume & $\nu$ & $2$ \\ \hline
Packing radius & $\rho_p$ & $1/\sqrt{2}$\\ \hline
Covering radius & $\rho_c$ & $\rho_p\sqrt{2}$ \\ \hline
Space-filling loss & $D_{\text{loss}}$ &  1.1672 dB \\ \hline
Space-filling gain over $Z^1$ & $D_{\text{gain}}$ &  0.3657 dB \\ \hline
Kissing-number\index{kissing-number} & $\mathfrak{K}$ & 24 \\ \hline
Minimal squared distance & $d^2_\text{min}$ & 2 \\ \hline
Dimensionless normalized second moment & \raisebox{-2.5mm}{$G(\Lambda)$} & \raisebox{-2.5mm}{$0.076603$} \\ \hline
\end{tabular}
\caption{Relevant constants for the $D_4$ lattice.}
\label{tab:d4constants}
\end{center}
\end{table}

If $a$ is 7 or a product of primes congruent to $1\pmod{4}$ then $D_4$ has a geometrically-similar and clean sublattice of index $N=a^2$~\cite{diggavi:2002}. This is the set 7 and integer sequence A004613~\cite{sloane:integerseq}, see also Table~\ref{tab:intseq_D4}.
\begin{table}[ht]
\begin{center}
\begin{tabular}{l}\hline
1,5,7,13,17,25,29,37,41,53,61,65,73,85,89,97,101,109, \\ 
113,125,137,145,149,157,169,173,181,185,193,197,205, \\
221,229,233,241,257,265,269,277,281,289,293,305,313, \\
317,325,337,349,353,365,373,377,389,397,401,409,421,\dots \\ \hline
\end{tabular}
\caption{Admissible index values for $D_4$.}
\label{tab:intseq_D4}
\end{center}
\end{table}

The first 50 coefficients of the Theta series\index{theta series} for $D_4$ are shown in Table~\ref{tab:theta_D4}.
\begin{table}[ht]
\begin{center}
\begin{tabular}{l}\hline
1,0,24,0,24,0,96,0,24,0,144,0,96,0,192,0,24,0,312,0,144,\\
0,288,0,96,0,336,0,192,0,576,0,24,0,432,0,312,0,480,0, \\
144,0,768,0,288,0,576,0,96,0 \\ \hline
\end{tabular}
\caption{The first 50 coefficients of the Theta series with starting point at zero for the $D_4$ lattice.}
\label{tab:theta_D4}
\end{center}
\end{table}

\chapter{Proofs for Chapter~\ref{chap:lattice_theory}}\label{app:proofslatticetheory}

\begin{proof}[Proof of Lemma~\ref{lem:productindex}]
Most of the work towards proving the lemma has already been done in~\cite{conway:1999b} and~\cite{diggavi:2002} and we only need some simple extensions of their results.
For $Z^1$ the proof is trivial, since any odd integer is an admissible index value~\cite{diggavi:2002} and the product of odd integers yields odd integers. Now let $a,b\in \mathbb{Z}^+$ be odd integers that can be written as the product of primes from a certain set $s$.
It is clear that the product $ab$ is also odd an can be written as the product of primes of $s$.
For $Z^2$ an integer is an admissible index value if it can be written as a product of a set of primes which are congruent to $1\!\pmod{4}$ and/or congruent to \mbox{$3\!\pmod{4}$}~\cite{conway:1999b,diggavi:2002}. 
For $A_2$ an integer is an admissible index value if and only if it is a product of primes which are congruent to $1\!\pmod{6}$~\cite{conway:1999b,diggavi:2002} and if $m$ is a product of primes which are congruent to $1\!\pmod{4}$ then $m^2$ is an an admissible integer for $D_4$~\cite{diggavi:2002}.\footnote{We have excluded the index value obtained for $m=7$, since this particular index value cannot be written as a product of primes mod 4 but is a special case found in~\cite{diggavi:2002}.}
It follows that the lemma holds for the lattices mentioned above. 
Finally, for $Z^L$ and $L=4k$, where $k\geq 1$, an integer is an admissible index value if it is odd and can be written on the form $m^{L/2}$ for some integer $m$~\cite{diggavi:2002}. Let $a=m^{L/2}$ and $b=(m')^{L/2}$ we then have that $ab=m^{L/2}(m')^{L/2}=(m'')^{L/2}$, where $m''=mm'$ is odd and therefore an admissible index value for $Z^L$.
\end{proof}

\begin{proof}[Proof of Lemma~\ref{lem:productlattice}]
The cyclic submodule $\Lambda_0=\xi_0\Lambda$ is closed under multiplication by elements of $\Lambda$ so for any $\xi'=\xi_1\xi_2\cdots \xi_{K-1}\in \Lambda$ and any $\lambda_0\in\Lambda_0$ it is true that $\xi'\lambda_0 \in \Lambda_0$ which further implies that $\Lambda_\pi\subseteq \Lambda_0$ since $\xi'\lambda_0\in\Lambda_\pi$. Moreover, multiplication is commutative in $\mathbb{Z}, \mathcal{G}$ and $\mathcal{E}$ so the order of the set of elements $\xi_0,\dotsc,\xi_{K-1}$ when forming $\Lambda_\pi$ is irrelevant. Thus, $\Lambda_\pi\subseteq \Lambda_i$ and it is therefore a product lattice.
\end{proof}

\begin{proof}[Proof of Lemma~\ref{lem:productlattice_int}]
Since the rings considered are unique factorization rings there must be an element $\xi'\in\Lambda$ such that $\xi_0\xi'=\xi_\cap$, where $\xi'$ is unique up to multiplication by units of the respective rings. However, a unit $u\in \Lambda$ belongs to $\text{Aut}(\Lambda)$ and multiplication by $u$ is therefore an isometric operation which takes a lattice to itself. It follows that $\xi_0\xi'\in \Lambda_0$ for any $\xi_0\in\Lambda_0$ which implies that $\Lambda_\pi'\subseteq \Lambda_0=\xi_0\Lambda$. Once again we invoke the fact that $\mathbb{Z}, \mathcal{G}$ and $\mathcal{E}$ are multiplicative commutative rings from which it is clear that $\Lambda_\pi'\subseteq \Lambda_i=\xi_i\Lambda,\ i=0,\dotsc,K-1$.
\end{proof}

\begin{proof}[Proof of Lemma~\ref{lem:lipshitz}]
Follows trivially from the fact that Gaussian integers commute and Lipschitz integers include Gaussian integers as a special case where the $j^{th}$ and $k^{th}$ elements are both zero. 
\end{proof}

\chapter{Estimating $\psi_L$}
\label{app:estimating_psi}
In this appendix we present a method to numerically estimate $\psi_L$ for any $L$ and $K$. 

\section{Algorithm}
In Chapter~\ref{chap:symmetric} we presented closed-form expressions for $\psi_L$ for the case of $K=3$ and $L=2$ or odd as well as for the asymptotic case of $L\rightarrow\infty$. In order to extend these results to $K>3$ it follows from the proof of Theorem~\ref{theo:psiLK3} that we need closed-form expressions for the volumes of all the different convex regions that can be obtained by $K-1$ overlapping spheres. With such expressions it should be straightforward to find $\psi_L$ for any $K$. However, we will take a different approach here. 

Let $\tilde{\nu}$ be the volume of the sphere $\tilde{V}$, which contains the exact number of sublattice points required to construct $N$ distinct $K$-tuples, where the elements of each $K$-tuple satisfy $\|\lambda_i-\lambda_j\|\leq r$, where $r$ is the radius of $\tilde{V}$. Notice that $\tilde{V}$ is the expanded sphere. Thus, the volume $\tilde{\nu}$ of $\tilde{V}$ is $\psi_L^L$ times larger than the lower bound of~(\ref{eq:vtilde}). Now let $\tilde{\nu}' = \tilde{\nu}/\psi_L^L$ denote the volume of a sphere that achieves the lower bound~(\ref{eq:vtilde}) so that $N=(\tilde{\nu}'/\nu_s)^{K-1}$ (at least this is true for large $N$). But this implies that asymptotically as the number of lattice points in $\tilde{V}$ goes to infinity we have
\begin{equation}
N=\left( \frac{\tilde{\nu}/\psi_L^L}{\nu_s}\right)^{K-1},
\end{equation}
which leads to
\begin{equation}\label{eq:estimatepsiL}
\psi_L = \left(\frac{\omega_L r^L}{\nu_s N^{1/K-1}}\right)^{1/L},
\end{equation}
where, without loss of generality, we can assume that $\nu_s=1$ (simply a matter of scaling). For a given $r$ in~(\ref{eq:estimatepsiL}) we can numerically estimate $N$, which then leads to an estimate of $\psi_L$. To numerically estimate $N$ it follows that we need to find the set of lattice points within a sphere $\tilde{V}$ of radius $r$. For each of these lattice points we center another sphere of radius $r$ and find the set of lattice points which are within the intersection of the two spheres. This procedure continues $K-1$ times. In the end we find $N$ by adding the number of lattice points within each intersection, i.e.
\begin{equation}
N =
\sum_{\tilde{\Lambda}_1}
\sum_{\tilde{\Lambda}_2}
\dots
\sum_{\tilde{\Lambda}_{K-2}}
|\Lambda_s \cap \tilde{V}(\lambda_{K-2})\cap \dots \cap \tilde{V}(\lambda_0)|,
\end{equation}
where
\begin{equation}
\begin{split}
\tilde{\Lambda}_1 &= \{ \lambda_1 : \lambda_1\in \Lambda_s\cap \tilde{V}(\lambda_0) \}, \\
\tilde{\Lambda}_2 &= \{ \lambda_2 : \lambda_2\in \Lambda_s\cap \tilde{V}(\lambda_1)\cap \tilde{V}(\lambda_0)\}, \\
&\hspace{2mm}\vdots \\
\tilde{\Lambda}_{K-2} &= \{\lambda_{K-2} : \lambda_{K-2} \in \Lambda_s \cap
\tilde{V}(\lambda_{K-3})\cap \dots \cap \tilde{V}(\lambda_0) \}.
\end{split}
\end{equation}
As $r$ gets large the estimate gets better. For example for $K=4, \Lambda=Z^2$ and $r=10,20,50$ and $70$ then using the algorithm outlined above we find $\psi_2\approx 1.1672, 1.1736,$ $1.1757$ and $1.1762$, respectively.

\chapter{Assignment Example}
\label{app:assignment_example}
\begin{figure}[ht]
\begin{center}
\subfigure[]{\includegraphics[width=8cm]{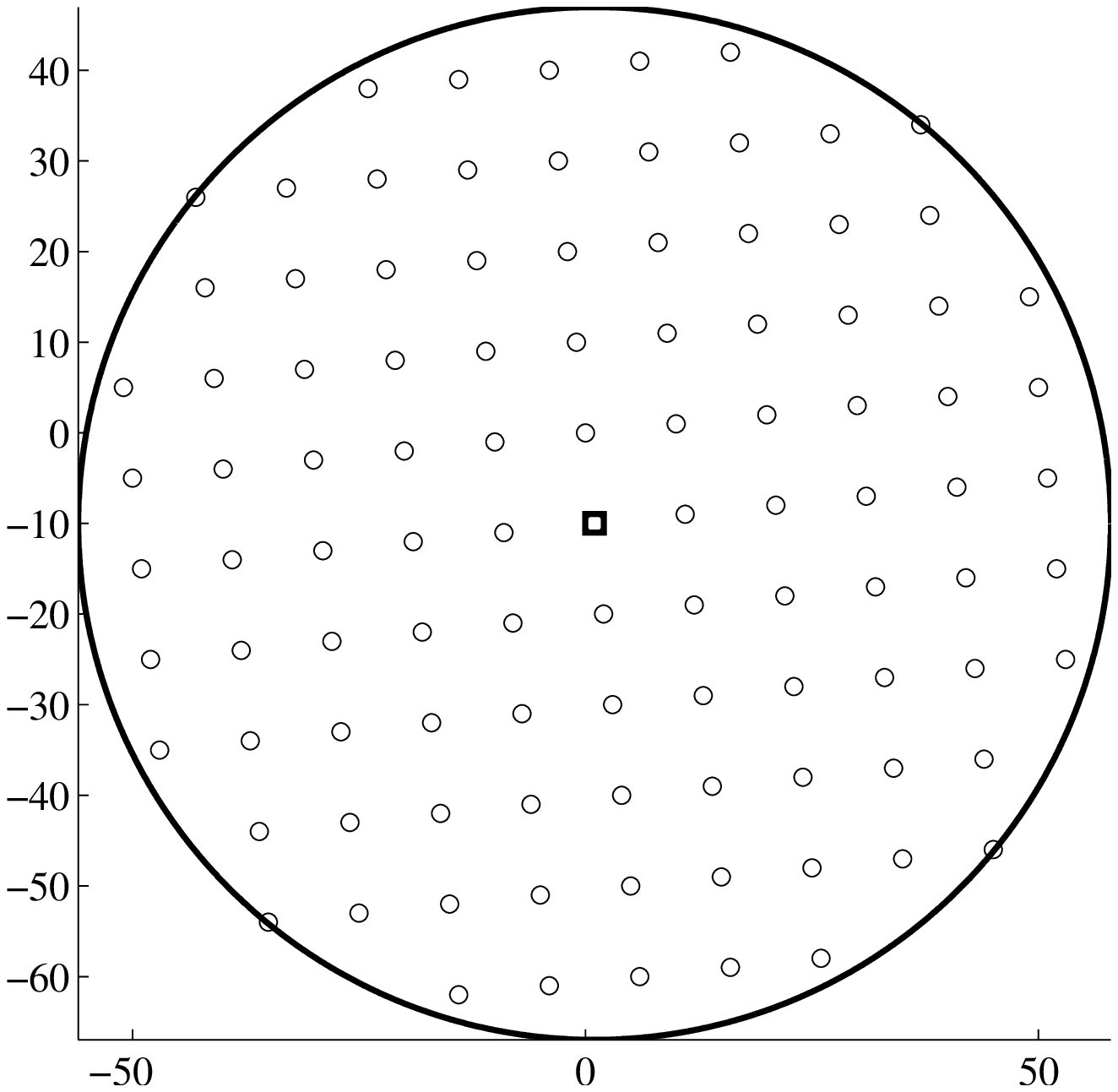}\label{fig:VtildeN101_2}
}\par
\subfigure[]{\includegraphics[width=8cm]{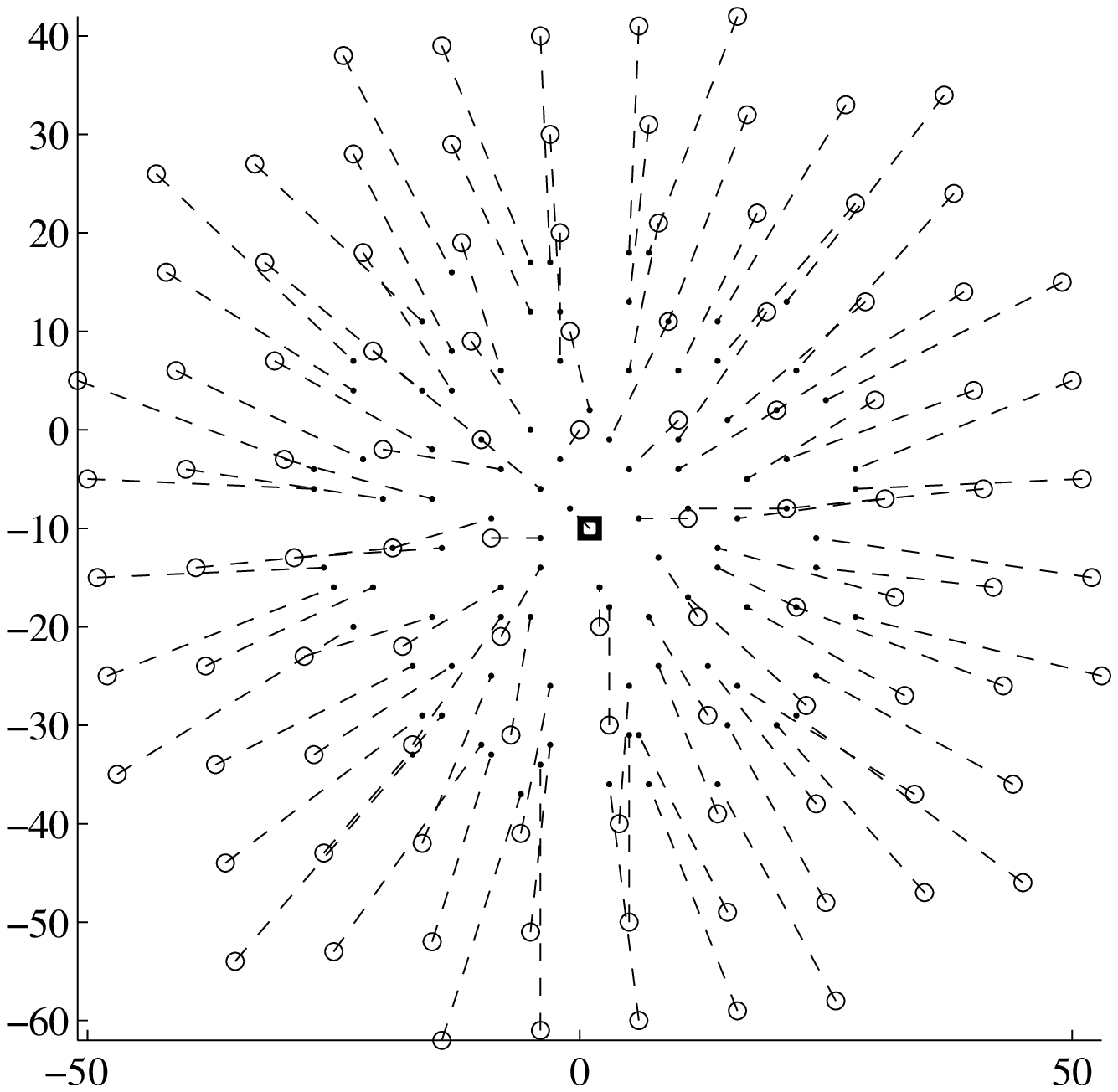}\label{fig:VtildeN101_1}
}

\caption{The square marks the sublattice point $\lambda_0=(1,-10)$ and the small circles illustrate the 101 sublattice points which are associated with $\lambda_0$. (a) The large circle emphasize that the sublattice points are approximately spherically distributed around $\lambda_0$. (b) The assignments are illustrated with dashed lines and the small dots represent central lattice points.}
\end{center}
\end{figure}

In this appendix we give an example of part of a complete assignment. We let $\Lambda=Z^2, K=2$ and $N=101$ and construct 2-tuples as outlined in Sections~\ref{sec:ktuples} and~\ref{sec:summarize_ktuples}. These 2-tuples are then assigned to central lattice points in $V_\pi(0)$. Since $N=101$ then (at least theoretically) each sublattice points will be used 101 times. Furthermore, for a given sublattice point, say $\lambda_0\in\Lambda_s$, the $N$ associated sublattice points, i.e.\ the set of sublattice points representing the second coordinate of the 2-tuples having $\lambda_0$ as first coordinate, will be approximately spherically distributed around $\lambda_0$ (since $\tilde{V}$ forms a sphere). 
Fig.~\ref{fig:VtildeN101_2} shows the set of $N$ sublattice points given by
\begin{equation}
\{\lambda_1\in \Lambda_s : \lambda_1=\alpha_1(\lambda_c)\ \text{and}\ \alpha_0(\lambda_c)=(1,-10), \lambda_c\in \Lambda_c\},
\end{equation}
which represent the set of second coordinates of the $N$ 2-tuples all having $\lambda_0=(1,-10)$ as first coordinate. Each 2-tuple is assigned to a central lattice point. This assignment is illustrated in Fig.~\ref{fig:VtildeN101_1}. Here a dashed line connects a given 2-tuple (represented by its second coordinate $\lambda_1$) with a central lattice point. 
These $N$ assignments are also shown in Table~\ref{tab:101assignments}.
%
%

\begin{table}[ht]
\begin{center}
\begin{tabular}{|ll|ll|ll|} \hline
$\lambda_c\in\Lambda$ & $\alpha_1(\lambda_c)\in \Lambda_s$ & 
$\lambda_c\in\Lambda$ & $\alpha_1(\lambda_c)\in \Lambda_s$ &
$\lambda_c\in\Lambda$ & $\alpha_1(\lambda_c)\in \Lambda_s$ \\ \hline
(-27,-6) & (-50,-5) & (21,-8) & (41,-6) & (-3,-32) & (-5,-51) \\ 
(-27,-4) & (-51,5) & (20,2) & (39,14) & (-10,-32) & (-25,-53) \\ 
(-26,-14) & (-49,-15) & (17,-5) & (30,3) & (6,-31) & (15,-49) \\ 
(-25,-16) & (-48,-25) & (16,-9) & (31,-7) & (5,-31) & (5,-50) \\ 
(-23,-20) & (-47,-35) & (15,1) & (29,13) & (20,-30) & (35,-47) \\ 
(-23,4) & (-42,16) & (14,11) & (27,33) & (15,-30) & (25,-48) \\ 
(-23,7) & (-43,26) & (14,7) & (28,23) & (22,-29) & (45,-46) \\ 
(-22,-3) & (-41,6) & (14,-12) & (32,-17) & (-14,-29) & (-26,-43) \\ 
(-21,-16) & (-38,-24) & (11,-8) & (21,-8) & (-16,-29) & (-36,-44) \\ 
(-20,-7) & (-40,-4) & (10,6) & (18,22) & (16,-26) & (34,-37) \\ 
(-19,-12) & (-39,-14) & (10,-1) & (19,12) & (5,-26) & (4,-40) \\ 
(-16,4) & (-32,17) & (10,-4) & (20,2) & (-3,-26) & (-6,-41) \\ 
(-16,11) & (-33,27) & (5,-4) & (10,1) & (24,-25) & (44,-36) \\ 
(-15,-7) & (-30,-3) & (3,-1) & (9,11) & (-9,-25) & (-16,-42) \\ 
(-15,-2) & (-31,7) & (5,18) & (6,41) & (13,-24) & (24,-38) \\ 
(-14,-12) & (-29,-13) & (7,18) & (16,42) & (8,-24) & (14,-39) \\ 
(-13,4) & (-22,18) & (-5,17) & (-14,39) & (-13,-24) & (-27,-33) \\ 
(-13,8) & (-23,28) & (-3,17) & (-4,40) & (-17,-24) & (-37,-34) \\ 
(-10,-1) & (-21,8) & (-13,16) & (-24,38) & (7,-19) & (13,-29) \\ 
(-9,-9) & (-19,-12) & (5,13) & (7,31) & (-5,-19) & (-7,-31) \\ 
(-8,-4) & (-20,-2) & (-5,12) & (-13,29) & (-8,-19) & (-17,-32) \\ 
(-8,6) & (-12,19) & (-2,12) & (-3,30) & (-15,-19) & (-28,-23) \\ 
(-5,0) & (-11,9) & (9,11) & (17,32) & (17,-18) & (33,-27) \\ 
(28,-4) & (50,5) & (-2,7) & (-2,20) & (3,-18) & (3,-30) \\ 
(28,-6) & (51,-5) & (5,6) & (8,21) & (11,-17) & (23,-28) \\ 
(28,-19) & (53,-25) & (1,2) & (-1,10) & (2,-16) & (2,-20) \\ 
(25,3) & (49,15) & (-6,-37) & (-14,-62) & (-8,-16) & (-18,-22) \\ 
(24,-11) & (52,-15) & (14,-36) & (26,-58) & (14,-14) & (22,-18) \\ 
(24,-14) & (42,-16) & (7,-36) & (16,-59) & (-4,-14) & (-8,-21) \\ 
(22,6) & (38,24) & (3,-36) & (6,-60) & (8,-13) & (12,-19) \\ 
(22,-18) & (43,-26) & (-4,-34) & (-4,-61) & (-4,-11) & (-9,-11) \\ 
(21,13) & (37,34) & (-9,-33) & (-15,-52) & (6,-9) & (11,-9) \\ 
(21,-3) & (40,4) & (-17,-33) & (-35,-54) & (-1,-8) & (1,-10) \\
(-4,-6) & (-10,-1) & (-2,-3) & (0,0) & &  \\ \hline
\end{tabular}
\end{center}
\caption{The assignments of the $N=101$ 2-tuples which all have $\lambda_0=(1,-10)$ as first coordinate, i.e.\ $\alpha_0(\lambda_c)=(1,-10)$.}
\label{tab:101assignments}
\end{table}

\chapter{Proofs for Chapter~\ref{chap:symmetric}}
\label{app:proof_symmetric}
For notational convenience we will in this appendix use the shorter notation $\mathcal{L}$ instead of $\mathcal{L}^{(K,\kappa)}$.

\section{Proof of Theorem~\ref{theo:sums}}
\label{app:sums}
In order to prove Theorem~\ref{theo:sums}, we need the following results.

\begin{lemma}\label{lem:r2}
For $1\leq \kappa \leq K$ we have
\begin{equation*}
\begin{split}
\sum_{l\in\mathcal{L}}
\left\langle \lambda_c,\sum_{j=0}^{\kappa-1} \lambda_{l_j}\right\rangle 
&= 
\frac{\kappa}{K}\binom{K}{\kappa}
\left\langle \lambda_c, \sum_{i=0}^{K-1} \lambda_i\right\rangle.
\end{split}
\end{equation*}
\end{lemma}

\begin{proof}
Expanding the sum on the left-hand-side leads to $\binom{K}{\kappa}\kappa$ different terms of the form $\langle \lambda_c, \lambda_i\rangle$, where $i\in \{0,\dots, K-1\}$. There are $K$ distinct $\lambda_i$'s so the number of times each $\lambda_i$ occur is $\binom{K}{\kappa}\kappa/K$.
\end{proof}

\begin{lemma}\label{lem:r1}
For $1\leq \kappa \leq K$ we have
\begin{equation*}
\begin{split}
\sum_{l\in\mathcal{L}}\left\| \sum_{j=0}^{\kappa-1} \lambda_{l_j}\right\|^2
=
\frac{\kappa}{K}\binom{K}{\kappa}\sum_{i=0}^{K-1}\|\lambda_i\|^2
+\frac{2\kappa(\kappa-1)}{K(K-1)}\binom{K}{\kappa}
\sum_{i=0}^{K-2}\sum_{j=i+1}^{K-1}\langle \lambda_i,\lambda_j \rangle.
\end{split}
\end{equation*}
\end{lemma}

\begin{proof}
There are $\binom{K}{\kappa}$ distinct ways of adding $\kappa$ out of $K$ elements. Squaring a sum of $\kappa$ elements leads to $\kappa$ squared elements and $2\binom{\kappa}{2}$ cross products (product of two different elements). This gives a total of $\binom{K}{\kappa}\kappa$ squared elements, and $2\binom{K}{\kappa}\binom{\kappa}{2}$ cross products. Now since there are $K$ distinct elements, the number of times each squared element occurs is given by
\begin{equation}\label{eq:lemnum1}
\#_{\|\lambda_i\|^2} =\binom{K}{k}\frac{\kappa}{K}.
\end{equation}
 There are $\binom{K}{2}$ distinct cross products, so the number of times each cross product occurs is given by
\begin{equation}\label{eq:lemnum2}
\#_{\langle\lambda_i,\lambda_j\rangle} = 
\binom{K}{\kappa}\frac{2\binom{\kappa}{2}}{\binom{K}{2}}=
\displaystyle\frac{2\kappa(\kappa-1)}{K(K-1)}\displaystyle\binom{K}{\kappa}.
\end{equation}
\rspace
\end{proof}

\begin{lemma}\label{lem:r3}
For $K\geq 1$ we have
\begin{equation}\label{eq:r3}
\begin{split}
(K-1)\sum_{i=0}^{K-1}\|\lambda_i\|^2 - 2\sum_{i=0}^{K-2}\sum_{j=i+1}^{K-1}\langle \lambda_i , \lambda_j \rangle
= \sum_{i=0}^{K-2}\sum_{j=i+1}^{K-1} \|\lambda_i - \lambda_j\|^2.
\end{split}
\end{equation}
\end{lemma}
\begin{proof}
Expanding the right-hand-side of~(\ref{eq:r3}) yields
\begin{equation}
\begin{split}
\sum_{i=0}^{K-2}\sum_{j=i+1}^{K-1}\|\lambda_i - \lambda_j\|^2
=\sum_{i=0}^{K-2}\sum_{j=i+1}^{K-1}\left( \|\lambda_i\|^2 + \|\lambda_j\|^2 -
2\langle \lambda_i, \lambda_j \rangle \right).
\end{split}
\end{equation}
We also have
\begin{equation}
\begin{split}
\sum_{i=0}^{K-2}\sum_{j=i+1}^{K-1}\left( \|\lambda_i\|^2 + \|\lambda_j\|^2\right)
&= 
\sum_{i=0}^{K-2}(K-1-i)\|\lambda_i\|^2 + \sum_{i=0}^{K-2}\sum_{j=i+1}^{K-1}\|\lambda_j\|^2 \\
&= \sum_{i=0}^{K-2}(K-1-i)\|\lambda_i\|^2 + \sum_{j=1}^{K-1}j\|\lambda_j\|^2 \\
&= \sum_{i=0}^{K-1}(K-1-i)\|\lambda_i\|^2 + \sum_{j=0}^{K-1}j\|\lambda_j\|^2 \\
&= \sum_{i=0}^{K-1}(K-1)\|\lambda_i\|^2 - \sum_{i=0}^{K-1}i\|\lambda_i\|^2 + \sum_{j=0}^{K-1}j\|\lambda_j\|^2 \\
&= (K-1)\sum_{i=0}^{K-1}\|\lambda_i\|^2,
\end{split}
\end{equation}
which completes the proof.
\end{proof}

We are now in a position to prove the following result.
\begin{proposition}{\label{prop:sums}}
For $1\leq \kappa \leq K$ we have
\begin{equation*}
\begin{split}
\sum_{l\in\mathcal{L}}\left\| \lambda_c - \frac{1}{\kappa}
\sum_{j=0}^{\kappa-1} \lambda_{l_j}\right\|^2 &= 
\binom{K}{\kappa}
\Bigg(
\left\| \lambda_c - 
\frac{1}{K}\sum_{i=0}^{K-1}\lambda_i\right\|^2
\\
&\quad+\left(
\frac{K-\kappa}{K^2\kappa(K-1)}\right)
\sum_{i=0}^{K-2}\sum_{j=i+1}^{K-1}\| \lambda_i - \lambda_j\|^2
\Bigg).
\end{split}
\end{equation*}
\end{proposition}

\begin{proof}
We have
\begin{equation*}\label{eq:costfunctional4}
\begin{split}
\left\| \lambda_c - \frac{1}{\kappa}\sum_{j=0}^{\kappa-1} \lambda_{l_j}\right\|^2
=\|\lambda_c\|^2- 2\left\langle \lambda_c,\frac{1}{\kappa}\sum_{j=0}^{\kappa-1} \lambda_{l_j}\right\rangle +
\frac{1}{\kappa^2}\left\| \sum_{j=0}^{\kappa-1} \lambda_{l_j}\right\|^2.
\end{split}
\end{equation*}
Hence, by use of Lemmas~\ref{lem:r2} and~\ref{lem:r1}, we have that
\begin{equation*}
\begin{split}
\sum_{l\in\mathcal{L}}
\left\| \lambda_c - \frac{1}{\kappa}\sum_{j=0}^{\kappa-1} \lambda_{l_j}\right\|^2 &=
\binom{K}{\kappa}\Bigg(
\|\lambda_c\|^2 
- \frac{2}{K}\left\langle \lambda_c,\sum_{i=0}^{K-1}\lambda_i\right\rangle
+
\frac{1}{K\kappa}\sum_{i=0}^{K-1}\|\lambda_i\|^2 \\
&\quad+\frac{2(\kappa-1)}{K(K-1)\kappa}
\sum_{i=0}^{K-2}\sum_{j=i+1}^{K-1}\langle \lambda_i,\lambda_j \rangle
\Bigg) \\
&=
\binom{K}{\kappa}\Bigg(
\left\| \lambda_c - 
\frac{1}{K}\sum_{i=0}^{K-1}\lambda_i\right\|^2
-
\frac{1}{K^2}
\left\|\sum_{i=0}^{K-1}\lambda_i\right\|^2\\
&\quad+
\frac{1}{K\kappa}\sum_{i=0}^{K-1}\|\lambda_i\|^2
+\frac{2(\kappa-1)}{K(K-1)\kappa}
\sum_{i=0}^{K-2}\sum_{j=i+1}^{K-1}\langle \lambda_i,\lambda_j \rangle
\Bigg) \\
&=
\binom{K}{\kappa}\Bigg(
\left\| \lambda_c - 
\frac{1}{K}\sum_{i=0}^{K-1}\lambda_i\right\|^2
+
\left(\frac{1}{K\kappa} - \frac{1}{K^2}\right)
\sum_{i=0}^{K-1}\|\lambda_i\|^2\\
&\quad+
\left(\frac{2(\kappa-1)}{K(K-1)\kappa}-\frac{2}{K^2}\right)
\sum_{i=0}^{K-2}\sum_{j=i+1}^{K-1}\langle \lambda_i,\lambda_j \rangle
\Bigg) \\
&=
\binom{K}{\kappa}\Bigg(
\left\| \lambda_c - 
\frac{1}{K}\sum_{i=0}^{K-1}\lambda_i\right\|^2
+
\left(\frac{K-\kappa}{K^2\kappa}\right)
\sum_{i=0}^{K-1}\|\lambda_i\|^2 \\
&\quad-
\left(\frac{K-\kappa}{K^2\kappa(K-1)}\right)
2\sum_{i=0}^{K-2}\sum_{j=i+1}^{K-1}\langle \lambda_i,\lambda_j \rangle
\Bigg)
\intertext{so that, by Lemma~\ref{lem:r3}, we finally have that}
&=
\binom{K}{\kappa}\Bigg(
\left\| \lambda_c - 
\frac{1}{K}\sum_{i=0}^{K-1}\lambda_i\right\|^2 \\
&\quad+
\left(\frac{K-\kappa}{K^2\kappa(K-1)}\right)
\sum_{i=0}^{K-2}\sum_{j=i+1}^{K-1}\| \lambda_i-\lambda_j \|^2
\Bigg),
\end{split}
\end{equation*}
which completes the proof.
\end{proof}

\begin{theo_empty}{\ref{theo:sums}.}
For $1\leq \kappa \leq K$ we have
\begin{equation*}
\begin{split}
\sum_{\lambda_c}\sum_{l\in\mathcal{L}}
\left\| \lambda_c - \frac{1}{\kappa}\sum_{j=0}^{\kappa-1} \lambda_{l_j}\right\|^2
&= \sum_{\lambda_c}\binom{K}{\kappa}\Bigg(\left\| \lambda_c - 
\frac{1}{K}\sum_{i=0}^{K-1}\lambda_i\right\|^2\\
&\quad+
\left(
\frac{K-\kappa}{K^2\kappa(K-1)}\right)
\sum_{i=0}^{K-2}\sum_{j=i+1}^{K-1}\| \lambda_i - \lambda_j\|^2\Bigg).
\end{split}
\end{equation*}
\end{theo_empty}

\begin{proof}
Follows trivially from Proposition~\ref{prop:sums}.
\end{proof}

\section{Proof of Theorem~\ref{theo:psiLK3}}\label{app:theo:psiLK3}
\begin{theo_empty}{\ref{theo:psiLK3}.}
For the case of $K=3$ and any odd $L$, the dimensionless expansion factor\index{$\psi_L$} is given by 
\begin{equation}
\psi_L=\left(\frac{\omega_L}{\omega_{L-1}}\right)^{1/2L}\left(\frac{L+1}{2L}\right)^{1/2L}\beta_L^{-1/2L},
\end{equation}
where $\beta_L$ is given by
\begin{equation}\label{eq:betaL1}
\begin{split}
\beta_L&=
\sum_{n=0}^{\frac{L+1}2}\binom{\frac{L+1}2}{n}2^{\frac{L+1}2-n}(-1)^n 
\sum_{k=0}^{\frac{L-1}2} \frac{\left(\frac{L+1}2\right)_k \left(\frac{1-L}2\right)_k}{\left(\frac{L+3}2\right)_k\, k!}\\
&\quad\times\sum_{j=0}^k\binom{k}{j}\left(\frac{1}2\right)^{k-j}(-1)^j\left(\frac{1}{4}\right)^j \frac{1}{L+n+j}.
\end{split}
\end{equation}
\end{theo_empty}

\begin{proof}
In the following we consider the case of $K=3$. For a specific $\lambda_0\in \Lambda_s$ we need to construct $N$ $3$-tuples all having $\lambda_0$ as the first coordinate. To do this we first center a sphere $\tilde{V}$ of radius $r$ at $\lambda_0$, see Fig.~\ref{fig:tuples}. For large $N$ and small $\nu_s$ this sphere contains approximately $\tilde{\nu}/\nu_s$ lattice points from $\Lambda_s$. Hence, it is possible to construct $(\tilde{\nu}/\nu_s)^2$ distinct 3-tuples using lattice points inside $\tilde{V}$. However, the maximum distance between $\lambda_1$ and $\lambda_2$ points is greater than the maximum distance between $\lambda_0$ and $\lambda_1$ points and also between $\lambda_0$ and $\lambda_2$ points. To avoid this bias towards $\lambda_0$ points we only use 3-tuples that satisfy $\|\lambda_i-\lambda_j\|\leq r$ for $i,j=0,1,2$. However, with this restriction we can no longer form $N$ 3-tuples. 
In order to make sure that exactly $N$ 3-tuples can be made we expand $\tilde{V}$ by the factor $\psi_L$.
It is well known that the number of lattice points at exactly squared distance $l$ from $c$, for any $c \in \mathbb{R}^L$ is given by the coefficients of the Theta series of the lattice $\Lambda$~\cite{conway:1999}. Theta series depend on the lattices and also on $c$~\cite{conway:1999}. Instead of working directly with Theta series we will, in order to be lattice and displacement independent, consider the $L$-dimensional \emph{hollow} sphere $\bar{\mathcal{C}}$ obtained as $\bar{\mathcal{C}}=S(c,m)-S(c,m-1)$ and shown in Fig.~\ref{fig:circles}(a). The number of lattice points $a_m$ in $\bar{\mathcal{C}}$ is given by $|\bar{\mathcal{C}}\cap \Lambda|$ and asymptotically as $\nu_s\rightarrow 0$ (and independent of $c$)
\begin{equation}\label{eq:am}
a_m = \text{Vol}(\bar{\mathcal{C}})/\nu_s  = \frac{\omega_L}{\nu_s}\big( m^{L}-(m-1)^L).
\end{equation}

The following construction makes sure that we have $\|\lambda_1-\lambda_2\|\leq r$. For a specific $\lambda_1\in \tilde{V}(\lambda_0)\cap\Lambda_s$ we center a sphere $\tilde{V}$ at $\lambda_1$ and use only $\lambda_2$ points from $\tilde{V}(\lambda_0)\cap\tilde{V}(\lambda_1)\cap\Lambda_s$. In Fig.~\ref{fig:circles}(b) we have shown two overlapping spheres where the first one is centered at some $\lambda_0$ and the second one is centered at some $\lambda_1\in\tilde{V}(\lambda_0)$ which is at distance $m$ from $\lambda_0$, i.e.\ $\|\lambda_0-\lambda_1\|=m$. Let us by $\mathcal{C}$ denote the convex region obtained as the intersection of the two spheres, i.e.\ $\mathcal{C}=\tilde{V}(\lambda_0)\cap \tilde{V}(\lambda_1)$. Now let $b_m$ denote the number of lattice points in $\mathcal{C}\cap \Lambda_s$. With this we have, asymptotically as $\nu_s\rightarrow 0$, that $b_m$ is given by
\begin{equation}\label{eq:bm}
b_m = \text{Vol}(\mathcal{C})/\nu_s.
\end{equation}
\begin{figure}
\begin{center}
\psfrag{Vt}{$\tilde{V}$}
\psfrag{m}{$m$}
\psfrag{am}{$\bar{\mathcal{C}}$}
\psfrag{bm}{$\mathcal{C}$}
\psfrag{r}{$r$}
\psfrag{l0}{$\scriptstyle\lambda_0$}
\mbox{%
\subfigure[]{\includegraphics{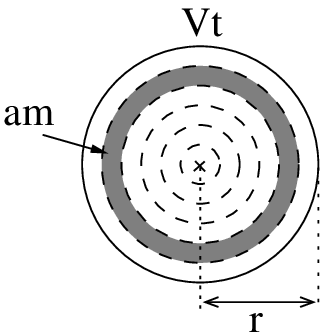}}\quad\qquad
\subfigure[]{\includegraphics{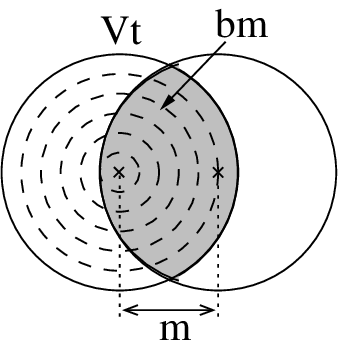}}}
\caption{The number of lattice points in the shaded region in (a) given by $a_m=\text{Vol}(\bar{\mathcal{C}})/\nu_s$ and in (b) it is given by $b_m=\text{Vol}(\mathcal{C})/\nu_s$.}
\label{fig:circles}
\end{center}
\end{figure}
It follows that the number $T$ of distinct 3-tuples which satisfy $\|\lambda_i-\lambda_j\|\leq r$ is given by
\begin{equation}\label{eq:T}
\lim_{\nu_s\rightarrow 0} T = \sum_{m=1}^r a_m b_m.
\end{equation}

We now proceed to find a closed-form expression for the volume of $\mathcal{C}$, which eventually will lead to a simple expression for $b_m$. Let ${}_2\mathcal{F}_1(\cdot)$ denote the Hypergeometric function\index{hypergeometric function} defined by~\cite{rainville:1960}
\begin{equation}\label{eq:hypergeom}
{}_2\mathcal{F}_1\left(a,b;c;z\right)= \sum_{k=0}^{\infty} \frac{(a)_k (b)_k}{(c)_k\, k!}z^k,
\end{equation}
where $(\cdot)_k$ is the Pochhammer symbol\index{Pochhammer symbol} defined as
\begin{equation}\label{eq:pochhammer}
(a)_k=\begin{cases}
1 & k=0 \\
a(a+1)\cdots(a+k-1) & k\geq 1.
\end{cases}
\end{equation}

\begin{lemma}
The volume of an $L$-dimensional ($L$ odd) spherical cap\index{volume!spherical cap}\index{spherical cap} $V_\text{cap}$ is given by 
\begin{equation}
\begin{split}
\text{Vol}(V_\text{cap}) &= \frac{2\omega_{L-1}}{L+1} r^{(L-1)/2} (2r-m)^{(L+1)/2}\\
&\quad\times {}_2\mathcal{F}_1\left(\frac{L+1}{2},\frac{1-L}{2};\frac{L+3}{2};\frac{2r-m}{4r} \right),
\end{split}
\end{equation}
\end{lemma}

\begin{proof}
This is a special case of what was proven in~\cite{jeanlouis:2005} and we can therefore use the same technique with only minor modifications. Let $h=m/2$ and let $u$ be a unit vector of $\mathbb{R}^L$. Furthermore, let $H_{h,u}$ be the affine hyperplane\index{affine hyperplane} $\{z+hu | z\in \mathbb{R}^L, z\cdot u =0\}$ of $\mathbb{R}^L$ which contains the intersection of two spheres of equal radii $r$ and with centers at distance $m\leq r$ apart, see Fig.~\ref{fig:balls}. 
\begin{figure}[ht]
\psfrag{H}{$H_{h,u}$}
\psfrag{C}{$\mathcal{C}_{r,h,u}$}
\psfrag{S}{$\mathcal{S}_{r,h,u}$}
\psfrag{r}{$r$}
\psfrag{h}{$h$}
\psfrag{m}{$m$}
\psfrag{x}{$x$}
\begin{center}
\includegraphics[width=6cm]{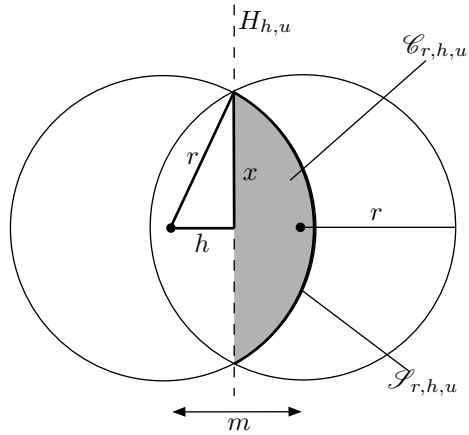}
\caption{Two balls in $\mathbb{R}^2$ of equal radii $r$ and distance $m$ apart.}
\label{fig:balls}
\end{center}
\end{figure}

We define the spherical cap as  
\begin{equation}
\mathcal{C}_{r,h,u} = \{z\in B(0,r) | z\cdot u \geq h\},
\end{equation}
and its surface is described by
\begin{equation}
\mathcal{S}_{r,h,u} = \{z\in S(0,r) | z\cdot u \geq h\},
\end{equation}
where $B(0,r) \in \mathbb{R}^L$ and $S(0,r) \in \mathbb{R}^L$ denote the ball\footnote{In this proof we redefine the concept of a sphere to be in line with~\cite{jeanlouis:2005}. As such, the term sphere denotes the surface of a ball\index{ball}, hence, a sphere has no interior. This terminology is only needed in this proof. Elsewhere we define the sphere to be a solid sphere (i.e.\ a ball and its surface) as is customary in the lattice literature.} respectively the sphere of radius $r$ and centered at the origin.

The sphere $H_{h,u}\cap \mathcal{S}_{r,h,u}$ has radius $x=\sqrt{r^2-h^2}$ and it is clear that $h=\sqrt{r^2-x^2}$. Moreover, any point of $\mathcal{S}_{r,h,u}$ which is at distance\footnote{By distance we mean the length of the shortest straight line that can be drawn between $H_{h,u}$ and $\mathcal{S}_{r,h,u}$. It is clear that this line is perpendicular to $H_{h,u}$.} $t$ from $H_{h,u}$ is at distance $(x^2-t^2-2th)^{1/2}$ from the real line $\mathbb{R}u$ (i.e.\ the span of $u$).  Hence, the volume $\text{Vol}(\mathcal{C}_{r,h,u})$ of $\mathcal{C}_{r,h,u}$ is given by
\begin{equation}\label{eq:Vctmp}
\begin{split}
\text{Vol}(\mathcal{C}_{r,h,u}) &=\int_{0}^{r-h}\omega_{L-1}(x^2-t^2-2th)^{(L-1)/2}\, dt \\
&=\omega_{L-1}\int_{0}^{r-h}\left((r-h-t)(r+h+t)\right)^{(L-1)/2}\, dt \\
&= \omega_{L-1}\int_{0}^{\alpha} (\alpha-t)^{\gamma}(t-\beta)^{\gamma}\, dt,
\end{split}
\end{equation}
where $\alpha=r-h, \beta=-r-h$ and $\gamma=(L-1)/2$. The last integral in~(\ref{eq:Vctmp}) can be shown to be equal to~\cite[Eq.\ 3.196.1]{gradshteyn:2000}
\begin{equation}
\int_{0}^{\alpha} (\alpha-t)^{\gamma}(t-\beta)^{\gamma}\, dt = \frac{\alpha^{\gamma+1}(-\beta)^{\gamma}}{\gamma+1} {}_2\mathcal{F}_1\left(1,-\gamma;\gamma+2;\frac{\alpha}{\beta} \right),
\end{equation}
which by use of~(\ref{eq:hypergeomtrans}) can be written as
\begin{equation}\label{eq:partialhyper}
\begin{split}
\frac{\alpha^{\gamma+1}(-\beta)^{\gamma}}{\gamma+1} &{}_2\mathcal{F}_1\left(1,-\gamma;\gamma+2;\frac{\alpha}{\beta} \right)= \\
&\left(1-\frac{\alpha}{\beta}\right)^{\gamma}\frac{\alpha^{\gamma+1}(-\beta)^{\gamma}}{\gamma+1} {}_2\mathcal{F}_1\left(\gamma+1,-\gamma;\gamma+2;\frac{\alpha}{\alpha-\beta} \right).
\end{split}
\end{equation}
The volume $\text{Vol}(\mathcal{C}_{r,h,u})$ follows by inserting~(\ref{eq:partialhyper}) in~(\ref{eq:Vctmp}), that is 
\begin{equation}\label{eq:Vc}
\begin{split}
\text{Vol}(\mathcal{C}_{r,h,u}) &= \omega_{L-1} \left(1-\frac{\alpha}{\beta}\right)^{\gamma}\frac{\alpha^{\gamma+1}(-\beta)^{\gamma}}{\gamma+1} {}_2\mathcal{F}_1\left(\gamma+1,-\gamma;\gamma+2;\frac{\alpha}{\alpha-\beta} \right) \\
&= \frac{\omega_{L-1}}{L+1} r^{(L-1)/2} (2r-m)^{(L+1)/2} \\
&\quad\times {}_2\mathcal{F}_1\left(L/2+1/2,1/2-L/2;L/2+3/2;\frac{2r-m}{4r} \right),
\end{split}
\end{equation}
which completes the proof.
\end{proof}

The region $\mathcal{C}$ consists of two equally sized spherical caps. 
Inserting~(\ref{eq:am}) and~(\ref{eq:bm}) into~(\ref{eq:T}) leads to\footnote{In this asymptotic analysis we assume that all $\lambda_1$ points within a given $\bar{\mathcal{C}}$ is at exact same distance from the center of $\tilde{V}$ (i.e.\ from $\lambda_0$). The error due to this assumption is neglectable, since any constant offset from $m$ will appear inside $\mathcal{O}(\cdot)$.} (asymptotically as $\nu_s\rightarrow 0$)
\begin{equation}\label{eq:Tapprox1}
\begin{split}
T&= \sum_{m=1}^ra_m b_m \\
&=\frac{2\omega_L\omega_{L-1}}{\nu_s^2(L+1)}\sum_{m=1}^r (m^L-(m-1)^L)
r^{(L-1)/2}\\
&\quad\times (2r-m)^{(L+1)/2} {}_2\mathcal{F}_1\left(\frac{L+1}2,\frac{1-L}2;\frac{L+3}2;\frac{2r-m}{4r} \right) \\
&\overset{(a)}{=}\frac{2\omega_L\omega_{L-1}}{\nu_s^2(L+1)}r^{\frac{L-1}2}
\sum_{n=0}^{\frac{L+1}2}\binom{\frac{L+1}2}{n}(2r)^{\frac{L+1}2-n}(-1)^n \\
&\quad\times\sum_{k=0}^{\frac{L-1}2} \frac{\left(\frac{L+1}2\right)_k \left(\frac{1-L}2\right)_k}{\left(\frac{L+3}2\right)_k\, k!}\sum_{j=0}^k\binom{k}{j}\left(\frac{1}2\right)^{k-j}(-1)^j\left(\frac{1}{4r}\right)^j\\
&\quad\times\sum_{m=1}^r (m^L-(m-1)^L)m^nm^j\\
&\overset{(b)}{=} \frac{2\omega_L\omega_{L-1}}{\nu_s^2(L+1)}r^{\frac{L-1}2}
\sum_{n=0}^{\frac{L+1}2}\binom{\frac{L+1}2}{n}(2r)^{\frac{L+1}2-n}(-1)^n \\
&\quad\times\sum_{k=0}^{\frac{L-1}2} \frac{\left(\frac{L+1}2\right)_k \left(\frac{1-L}2\right)_k}{\left(\frac{L+3}2\right)_k\, k!}\sum_{j=0}^k\binom{k}{j}\left(\frac{1}2\right)^{k-j}(-1)^j\left(\frac{1}{4r}\right)^j\\
&\quad\times \left(L\sum_{m=1}^r m^{L-1+n+j} + \mathcal{O}(m^{L-2+n+j})\right).
\\
&\overset{(c)}{=} \frac{2\omega_L\omega_{L-1}}{\nu_s^2(L+1)}r^{\frac{L-1}2}
\sum_{n=0}^{\frac{L+1}2}\binom{\frac{L+1}2}{n}(2r)^{\frac{L+1}2-n}(-1)^n \\
&\quad\times\sum_{k=0}^{\frac{L-1}2} \frac{\left(\frac{L+1}2\right)_k \left(\frac{1-L}2\right)_k}{\left(\frac{L+3}2\right)_k\, k!}\sum_{j=0}^k\binom{k}{j}\left(\frac{1}2\right)^{k-j}(-1)^j\left(\frac{1}{4r}\right)^j\\
&\quad\times\left(\frac{L}{L+n+j}r^{L+n+j} + \mathcal{O}\left(r^{L-1+n+j}\right)\right),
\end{split}
\end{equation}
where $(a)$ follows by use of the binomial series\index{binomial series expansion} expansion~\cite[p.162]{graham:1994}, i.e.\ $(x+y)^k = \sum_{n=0}^k\binom{k}{n}x^{k-n}y^{n}$, which in our case leads to 
\begin{equation}
(2r-m)^{\frac{L+1}2} = \sum_{n=0}^{\frac{L+1}2}\binom{\frac{L+1}2}{n}(2r)^{\frac{L+1}2-n}(-1)^n m^n
\end{equation}
and
\begin{equation}
\left(\frac{2r-m}{4r}\right)^k = \sum_{j=0}^k\binom{k}{j}\left(\frac{1}2\right)^{k-j}(-1)^j\left(\frac{m}{4r}\right)^j.
\end{equation}
$(b)$ is obtained  by once again applying the binomial series expansion, that is
\begin{equation}
(m-1)^L = m^L-Lm^{L-1} + \mathcal{O}(m^{L-2}),
\end{equation}
and $(c)$ follows from the fact that $\sum_{m=1}^r m^L = \frac{1}{L+1}r^{L+1}+\mathcal{O}(r^{L})$.

Next we let $r\rightarrow \infty$ so that the number of \emph{hollow} spheres inside $\tilde{V}$ goes to infinity.\footnote{We would like to emphasize that this is equivalent to keeping $r$ fixed, say $r=1$, and then let the number of \emph{hollow} spheres inside $\tilde{V}$ go to infinity. To see this let $M\rightarrow\infty$ and then rewrite~(\ref{eq:am}) as
\begin{equation}\label{eq:amprime}
a_{m/M} = \text{Vol}(\bar{\mathcal{C}})/\nu_s  = \frac{\omega_L}{\nu_s}\left( \left(\frac{m}{M}\right)^{L}-\left(\frac{m-1}{M}\right)^L\right), \quad 1\leq m\leq M.
\end{equation}
A similar change applies to~(\ref{eq:bm}). Hence, the asymptotic expression for $T$ is also valid within a localized region of $\mathbb{R}^L$ which is a useful property we exploit when proving Proposition~\ref{prop:riemann2}.}
From~(\ref{eq:Tapprox1}) we see that, asymptotically as $\nu_s\rightarrow 0$ and $r\rightarrow \infty$, we have
\begin{equation}\label{eq:Tapprox}
T = 2\frac{\omega_L\omega_{L-1}}{\nu_s^2}\frac{L}{L+1}\beta_L r^{2L},
\end{equation}
where $\beta_L$ is constant for fixed $L$ and given by~(\ref{eq:betaL1}).

We are now in a position to find an expression for $\psi_L$. Let $\bar{\nu}$ be equal to the lower bound~(\ref{eq:vtilde}), i.e.\ $\bar{\nu}=\nu_s\sqrt{N}$ and let $\bar{r}$ be the radius of the sphere having volume $\bar{\nu}$. Then $\psi_L$ is given by the ratio of $r$ and $\bar{r}$, i.e.\ $\psi_L=r/\bar{r}$, where $r$ is the radius of $\tilde{V}$. Using this in~(\ref{eq:Tapprox}) leads to
\begin{equation}\label{eq:r}
r=\left( \frac{T\nu_s(L+1)}{2\omega_L\omega_{L-1}L\beta_L} \right)^{1/2L}.
\end{equation}
Since the radius $\bar{r}$ of an $L$-dimensional sphere\index{radius of $L$-sphere} of volume $\bar{\nu}$ is given by
\begin{equation}\label{eq:rbar}
\bar{r}=\left(\frac{\bar{\nu}}{\omega_L}\right)^{1/L},
\end{equation}
we can find $\psi_L$ by dividing~(\ref{eq:r}) by~(\ref{eq:rbar}), that is
\begin{equation}\label{eq:psiL1}
\psi_L=\frac{r}{\bar{r}}=\left(\frac{T\nu_s^2(L+1)}{2\omega_L\omega_{L-1}L\beta_L}\right)^{1/2L}\left(\frac{\bar{\nu}}{\omega_L}\right)^{-1/L}.
\end{equation}
Since we need to obtain $N$ 3-tuples we let $T=N$ so that with $\bar{\nu}=\sqrt{N}\nu_s$ we can rewrite~(\ref{eq:psiL1}) as
\begin{equation}\label{eq:psiL}
\psi_L=\left(\frac{\omega_L}{\omega_{L-1}}\right)^{1/2L}\left(\frac{L+1}{2L}\right)^{1/2L}\beta_L^{-1/2L}.
\end{equation}
This completes the proof. 
\end{proof}

\section{Proof of Theorem~\ref{theo:psiLinf}}\label{app:theo:psiLinf}
\begin{lemma}\label{lem:omegaratio}
For $L\rightarrow \infty$ we have
\begin{equation}
\left(\frac{\omega_{L}}{\omega_{L-1}}\right)^{1/2L}=1.
\end{equation}
\end{lemma}
\begin{proof}
The volume\index{sphere!volume}\index{volume!unit-sphere} $\omega_L$ of an $L$-dimensional unit hypersphere is given by $\omega_L=\pi^{L/2}/(L/2)!$ so we have that
\begin{equation}
\begin{split}
\lim_{L\rightarrow \infty}&\left(\frac{\pi^{L/2}}{(L/2)!}\frac{(L/2-1/2)!}{\pi^{L/2-1/2}}\right)^{1/2L}\\
&= \lim_{L\rightarrow \infty} \pi^{1/4L} \left(\mathcal{O}(L^{-1})\right)^{1/2L}\\
&= 1.
\end{split}
\end{equation}
\rspace
\end{proof}

\begin{lemma}\label{lem:betalimit}
For $L\rightarrow \infty$ we have 
\begin{equation}
\frac{1}{\beta_L^{1/2L}} = \left(\frac{4}{3}\right)^{1/4}.
\end{equation}
\end{lemma}
\begin{proof}
The inner sum in~(\ref{eq:betaL}) may be well approximated by using that $\frac{1}{L+c}\approx \frac{1}{L}$ for $L\gg c$, which leads to
\begin{equation}\label{eq:approxfracL}
\begin{split}
&\sum_{j=0}^{k}\binom{k}{j}\left(\frac{1}{2}\right)^{k-j}(-1)^j\left(\frac{1}4\right)^j\frac{1}{L+n+j}\\
&\qquad\approx
\sum_{j=0}^{k}\binom{k}{j}\left(\frac{1}{2}\right)^{k-j}(-1)^j\left(\frac{1}4\right)^j\frac{1}{L} \\
&\qquad= \frac{1}L  \left(\frac{1}{4}\right)^k.
\end{split}
\end{equation}
We also have that 
{\allowdisplaybreaks
\begin{align}\notag
\sum_{k=0}^{\frac{L-1}2}&\frac{\left(\frac{L+1}2\right)_k \left(\frac{1-L}2\right)_k}{\left(\frac{L+3}2\right)_k\, k!} \left(\frac{1}4\right)^k\!\!=\!
{}_2\mathcal{F}_1\left(\frac{L+1}2, \frac{1-L}2;\frac{L+3}2; \frac{1}4\right) \\ \notag
&\overset{(a)}{=} (1-1/4)^{(-1+L)/2}{}_2\mathcal{F}_1\left(1, \frac{1-L}{2};\frac{L+3}2; -\frac{1}3\right) \\ \notag
&= (3/4)^{(-1+L)/2} \sum_{k=0}^{L/2-1/2} \frac{k!}{k!}\frac{(1/2-L/2)_k }{(3/2+L/2)_k} (-1/3)^k \\ \notag
&=(3/4)^{(-1+L)/2}\\ \notag 
&\quad\times\sum_{k=0}^{L/2-1/2}\left(\frac{(-L/2)^k}{(L/2)^k + \mathcal{O}(L^{k-1})} + \mathcal{O}(L^{-1})\right) (-1/3)^k \\ \label{eq:approxhyper}
&\approx (3/4)^{(-1+L)/2} \sum_{k=0}^{L/2-1/2} (1/3)^k,
\end{align}}
where $(a)$ follows from the following hypergeometric transformation~\cite{rainville:1960}
\begin{equation}\label{eq:hypergeomtrans}
{}_2\mathcal{F}_1\left(a,b;c;z\right) = (1-z)^{-b}{}_2\mathcal{F}_1\left(c-a,b;c;\xi\right),
\end{equation}
where $\xi=\frac{z}{z-1}$. Finally, it is true that
\begin{equation}\label{eq:sumbinomn}
\sum_{n=0}^{L/2+1/2}\binom{L/2+1/2}{n} 2^{L/2+1/2-n}(-1)^n = 1.
\end{equation}
Inserting~(\ref{eq:approxfracL}),~(\ref{eq:approxhyper}) and~(\ref{eq:sumbinomn}) into~(\ref{eq:betaL1}) leads to
\begin{equation}
\beta_L \approx (3/4)^{(-1+L)/2} \frac{1}{L}\sum_{k=0}^{L/2-1/2} (1/3)^k,
\end{equation}
where since $\sum_{k=0}^{\infty} (1/3)^k = 3/2$,
we get
\begin{equation}
\begin{split}
\lim_{L\rightarrow \infty}\frac{1}{\beta_L^{1/2L}} &= \lim_{L\rightarrow \infty} (4/3)^{1/4} (4/3)^{-1/4L} L^{1/2L} (2/3)^{1/2L} \\
&= (4/3)^{1/4},
\end{split}
\end{equation}
which proves the Lemma.
\end{proof}

We are now in a position to prove the following theorem.
\begin{theo_empty}{\ref{theo:psiLinf}.}
For $K=3$ and $L\rightarrow \infty$ the dimensionless expansion factor\index{$\psi_L$} $\psi_L$ is given by
\begin{equation}
\psi_{\infty} = \left(\frac{4}{3}\right)^{1/4}.
\end{equation}
\end{theo_empty}
\begin{proof}
The proof follows trivially by use of Lemma~\ref{lem:omegaratio} and Lemma~\ref{lem:betalimit} in~(\ref{eq:psiL}).
\end{proof}

\section{Proof of Proposition~\ref{prop:riemann2}}\label{app:riemann2}
Let $T_i = \{ \lambda_i : \lambda_i = \alpha_i(\lambda_c),\ \lambda_c \in V_\pi(0) \}$, i.e.\ the set of $N^2$ sublattice points $\lambda_i\in\Lambda_s$ associated with the $N^2$ central lattice points within $V_\pi(0)$.
Furthermore, let $T'_i \subseteq T_i$ be the set of unique elements of $T_i$, where $|T_i'|\approx N$. 
Finally, let 
\begin{equation}
T_j(\lambda_i) = \{ \lambda_j : \lambda_j = \alpha_j(\lambda_c)\ \text{and}\ \lambda_i = \alpha_i(\lambda_c),\  \lambda_c \in V_\pi(0) \},
\end{equation}
and let $T'_j(\lambda_i)\subseteq T_j(\lambda_i)$ be the set of unique elements. That is, $T_j(\lambda_i)$ contains all the elements $\lambda_j\in \Lambda_s$ which are in the $K$-tuples that also contains a specific $\lambda_i$. We will also make use of the notation $\#_{\lambda_j}$ to indicate the number of occurrences of a specific $\lambda_j$ in $T_j(\lambda_i)$. 

For the pair $(i,j)$ we have
\begin{equation*}
\sum_{\lambda_c\in V_\pi(0)}\| \alpha_i(\lambda_c) - \alpha_j(\lambda_c)\|^2 =
\sum_{\lambda_i\in T'_i}\sum_{\lambda_j\in T_j(\lambda_i)} \|\lambda_i - \lambda_j \|^2.
\end{equation*}
Given $\lambda_i\in T'_i$, we have
\begin{equation}\label{eq:riemannassump}
\begin{split}
\sum_{\lambda_j\in T_j(\lambda_i)} \|\lambda_i - \lambda_j \|^2 \nu_s 
&=\sum_{\lambda_j\in T'_j(\lambda_i)} \#_{\lambda_j} \|\lambda_i-\lambda_j\|^2\nu_s \\
&\overset{(a)}{\approx}
\frac{N}{\tilde{N}}\sum_{\lambda_j\in T'_j(\lambda_i)}\|\lambda_i-\lambda_j\|^2\nu_s \\
&\approx \frac{N}{\tilde{N}} \int_{\tilde{V}(\lambda_i)}\|\lambda_i-x\|^2\, dx \\
&\approx \frac{N}{\tilde{N}} \tilde{\nu}^{1+2/L}L G(S_L)\\
&\overset{(b)}{=} N\nu_s\tilde{\nu}^{2/L} LG(S_L),
\end{split}
\end{equation}
where $(a)$ follows by assuming (see the discussion below for the case of $K=3$) that $\#_{\lambda_j}=N/\tilde{N}$ for all $\lambda_j\in T_j(\lambda_i)$ and $(b)$ follows since $\tilde{\nu}=\tilde{N}\nu_s$. Hence, with $\tilde{\nu}=\tilde{N}\nu_s=\psi N^{1/(K-1)} \nu_s$ and $\nu_s=N\nu$, we have
\begin{equation*}
\begin{split}
\frac{1}{L}\sum_{\lambda_j\in T_j(\lambda_i)} \|\lambda_i - \lambda_j \|^2 \nu_s 
&\approx
N\nu_s \psi_L^{2} \nu^{2/L}N^{2/L} N^{2/L(K-1)} G(S_L)\\ 
&=
\nu_s\psi_L^{2} N^{1+2K/L(K-1)}\nu^{2/L} G(S_L),
\end{split}
\end{equation*}
which is independent of $\lambda_i$, so that
\begin{equation*}
\begin{split}
\frac{1}{L}\sum_{\lambda_i\in T'_i}\sum_{\lambda_j\in T_j(\lambda_i)}\| \lambda_i - \lambda_j\|^2 
&\approx \frac{N}{L} \sum_{\lambda_j\in T_j(\lambda_i)}\| \lambda_i - \lambda_j\|^2 \\
&\approx \psi_L^{2} N^{2+2K/L(K-1)}\nu^{2/L} G(S_L).
\end{split}
\end{equation*}

In~(\ref{eq:riemannassump}) we used the approximation $\#_{\lambda_j}\approx N/\tilde{N}$ without any explanation. For the case of $K=2$ and as $N\rightarrow \infty$ we have that $T'_i = T_i$ and $N=\tilde{N}$, hence the approximation becomes exact, i.e.\ $\#_{\lambda_j}= 1$. This proves the Proposition for $K=2$. We will now consider the case of $K=3$ and show that asymptotically, as $L\rightarrow \infty$, the following approximation becomes exact.
\begin{equation}\label{eq:g}
\frac{1}{L}\sum_{\lambda_j\in T_j(\lambda_i)} \|\lambda_i-\lambda_j\|^2 \approx N\tilde{\nu}^{2/L}G(S_L).
\end{equation}
To prove this we use the same procedure as when deriving closed-form expressions for $\psi_L$ leads to the following asymptotic expression 
\begin{equation}\label{eq:exactK3}
\sum_{\lambda_j\in T_j(\lambda_i)} \|\lambda_i-\lambda_j\|^2 = \sum_{m=1}^r a_m b_m m^2,
\end{equation}
where we without loss of generality assumed that $\lambda_i=0$ and used the fact that we can replace $\|\lambda_j\|^2$ by $m^2$ for the $\lambda_j$ points which are at distance $m$ from $\lambda_i=0$. It follows that we have
\begin{equation}\label{eq:f1}
\frac{1}{L}\sum_{\lambda_j\in T_j(\lambda_i)} \|\lambda_i-\lambda_j\|^2 = 2\frac{\omega_L\omega_{L-1}}{\nu_s^2}\frac{1}{L+1}\beta_L'r^{2L+2},
\end{equation}
where
\begin{equation}
\begin{split}
\beta_L'&=
\sum_{n=0}^{\frac{L+1}2}\binom{\frac{L+1}2}{n}2^{\frac{L+1}2-n}(-1)^n 
\sum_{k=0}^{\frac{L-1}2} \frac{\left(\frac{L+1}2\right)_k \left(\frac{1-L}2\right)_k}{\left(\frac{L+3}2\right)_k\, k!}\\
&\quad\times\sum_{j=0}^k\binom{k}{j}\left(\frac{1}2\right)^{k-j}(-1)^j\left(\frac{1}{4}\right)^j \frac{1}{L+n+j+2}.
\end{split}
\end{equation}

Since $\tilde{\nu}=\omega_L r^L = \psi_L^L\sqrt{N}\nu_s$ we can rewrite~(\ref{eq:f1}) as
\begin{equation}\label{eq:f}
\begin{split}
&\sum_{\lambda_j\in T_j(\lambda_i)} \|\lambda_i-\lambda_j\|^2 = 2\frac{\omega_L\omega_{L-1}}{\nu_s^2}\frac{1}{L+1}\beta_L' \tilde{\nu}^{2+2/L}\frac{1}{\omega_L^{2+2/L}} \\ 
&\quad= 2\frac{\omega_{L-1}}{\omega_L^{1+2/L}}\frac{1}{L+1}\beta_L' \tilde{\nu}^{2/L}\psi_L^{2L} N \\
&\quad\overset{(a)}{=} 2\frac{\omega_{L-1}}{\omega_L^{1+2/L}}\frac{1}{L+1}\beta_L' \tilde{\nu}^{2/L}N\left(\frac{\omega_L}{\omega_{L-1}}\right)\left(\frac{L+1}{2L}\right)\frac{1}{\beta_L} \\
&\quad=\frac{1}{\omega_L^{2/L}}\frac{1}{L}\tilde{\nu}^{2/L}N\frac{\beta_L'}{\beta_L},
\end{split}
\end{equation}
where $(a)$ follows by inserting~(\ref{eq:psiL}). Dividing~(\ref{eq:f}) by~(\ref{eq:g}) leads to
\begin{equation}
\frac{1}{\omega_L^{2/L}}\frac{1}{L}\frac{1}{G(S_L)}\frac{\beta_L'}{\beta_L}
= \frac{L+2}{L}\frac{\beta_L'}{\beta_L}.
\end{equation}
Hence, asymptotically as $L\rightarrow \infty$ we have that
\begin{equation}
\lim_{L\rightarrow \infty}\frac{L+2}{L}\frac{\beta_L'}{\beta_L} = 1,
\end{equation}
which proves the Proposition.

\begin{remark}\label{rem:optfiniteL}
Proposition~\ref{prop:riemann2} considered the asymptotic case of $L\rightarrow\infty$. Exact distortion expressions for the case of $K=3$ and finite $L$ follow by replacing~(\ref{eq:riemannassump}) with~(\ref{eq:exactK3}).
\end{remark}

\begin{remark}
For $K>3$ it is very likely that similar equations can be found for $\psi_L$ which can then be used to verify the goodness of the approximations for any $K$. Moreover, in Appendix~\ref{app:growthriemann2} we show that the rate of growth of~(\ref{eq:riemannassump}) is unaffected if we replace $\#_{\lambda_j}$ by either $\min_{\lambda_j} \{\#_{\lambda_j}$\} or $\max_{\lambda_j} \{\#_{\lambda_j}\}$ which means that the error by using the approximation $N/\tilde{N}$ instead of the true $\#_{\lambda_j}$ is constant (i.e.\ it does not depend on $N$) for fixed $K$ and $L$. It remains to be shown whether this error term tends to zero as $L\rightarrow \infty$ for $K>3$. However, based on the discussion above we conjecture that Proposition~\ref{prop:riemann2} is true for any $K$ asymptotically as $N,L\rightarrow\infty$ and $\nu_s\rightarrow 0$. In other words, the side distortion of a $K$-channel MD-LVQ system can be expressed through the normalized second moment of a sphere as the dimension goes to infinity.
\end{remark}

\section{Proof of Proposition~\ref{prop:growthriemann2}}\label{app:growthriemann2}
Before proving Proposition~\ref{prop:growthriemann2} we need to lower and upper bound $\#_{\lambda_j}$ (see Ap\-pendix~\ref{app:riemann2} for an introduction to this notation).
As previously mentioned the $\lambda_j$ points which are close (in Euclidean sense) to $\lambda_i$ occur more frequently than $\lambda_j$ points farther away. To see this observe that the construction of $K$-tuples can be seen as an iterative procedure that first picks a $\lambda_0\in \Lambda_s \cap V_\pi(0)$ and then any $\lambda_1\in \Lambda_s$ is picked such that $\|\lambda_0 -\lambda_1\|\leq r$, hence $\lambda_1\in \Lambda_s\cap \tilde{V}(\lambda_0)$.  
The set of $\lambda_{K-1}$ points that can be picked for a particular $(K-1)$-tuple e.g.\ $(\lambda_0,\dots,\lambda_{K-2})$ is then given by $\{ \lambda_{K-1} : \lambda_{K-1}\in\Lambda_s\cap \tilde{V}(\lambda_{K-2})\cap\cdots \cap \tilde{V}(\lambda_0)\}$. It is clear that $\|\lambda_i - \lambda_j\|\leq r$ where $(\lambda_i,\lambda_j) = (\alpha_i(\lambda_c),\alpha_j(\lambda_c)), \forall \lambda_c \in \Lambda_c$ and any $i,j \in \{0,\dots,K-1\}$. 

\begin{figure}[ht]
\psfrag{l0}{$\lambda_0$}
\psfrag{l1}{$\lambda_1$}
\psfrag{l2}{$\lambda_2$}
\psfrag{C(s)}{$\mathcal{C}(s)$}
\begin{center}
\includegraphics[width=5cm]{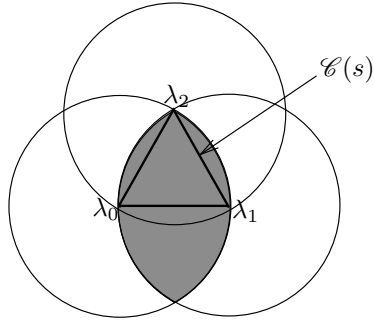}
\caption{Three spheres of equal radius are here centered at the set of points $s=\{\lambda_0,\lambda_1,\lambda_2\}$. The shaded area describes the intersection of two spheres. The equilateral triangle describes the convex hull\index{convex hull} $\mathcal{C}(s)$ of $s$.}
\label{fig:3balls}
\end{center}
\end{figure}

Let $T_\text{min}(\lambda_i,\lambda_j)$ denote the minimum number of times the pair $(\lambda_i,\lambda_j)$ is used. The minimum $T_\text{min}$ of $T_\text{min}(\lambda_i,\lambda_j)$ over all pairs $(\lambda_i,\lambda_j)$ lower bounds $N/\tilde{N}$. We will now show that $T_\text{min}$ is always bounded away from zero. To see this notice that 
the minimum overlap between two spheres of radius $r$ centered at $\lambda_0$ and $\lambda_1$, respectively, is obtained when $\lambda_0$ and $\lambda_1$ are are maximally separated, i.e.\ when $\|\lambda_0-\lambda_1\|=r$. This is shown by the shaded area in Fig.~\ref{fig:3balls} for $L=2$. For three spheres the minimum overlap is again obtained when all pairwise distances are maximized, i.e.\ when $\|\lambda_i-\lambda_j\|=r$ for $i,j\in \{0,1,2\}$ and $i\neq j$. 
It is clear that the volume of the intersection of three spheres is less than that of two spheres, hence the minimum number of $\lambda_2$ points is greater than the minimum number of $\lambda_3$ points. However, by construction it follows that when centering $K$ spheres at the set of points $s=\{\lambda_0,\dots,\lambda_{K-1}\} = \{\alpha_0(\lambda_c),\dots,\alpha_{K-1}(\lambda_c)\}$ each of the points in $s$ will be in the intersection $\cap_s$ of the $K$ spheres. 
Since the intersection of an arbitrary collection of convex sets leads to a convex set~\cite{rockafellar:1970},  the convex hull $\mathcal{C}(s)$ of $s$ will also be in $\cap_s$. Furthermore, for the example in Fig.~\ref{fig:3balls}, it can be seen that $\mathcal{C}(s)$ (indicated by the equilateral triangle) will not get smaller for $K\geq 3$ and this is true in general since points are never removed from $s$ as $K$ grows.
For $L=3$ the regular tetrahedron~\cite{coxeter:1973}\index{tetrahedron} consisting of four points with a pairwise distance of $r$ describes a regular convex polytope which lies in $\cap_s$. In general the regular $L$-simplex~\cite{coxeter:1973}\index{simplex!regular} lies in $\cap_s$ and the volume $\text{Vol}(L)$ of a regular $L$-simplex\index{volume!regular simplex} with side length $r$ is given by~\cite{buchholz:1992}
\begin{equation}
\text{Vol}(L)=\frac{r^L}{L!}\sqrt{\frac{L+1}{2^L}} = c_L r^L,
\end{equation}
where $c_L$ depends only on $L$. It follows that the minimum number of $K$-tuples that contains a specific $(\lambda_i,\lambda_j)$ pair is lower bounded by $\text{Vol}(L)^{K-2}/\nu_s^{K-2}$. Since the volume $\tilde{\nu}$ of $\tilde{V}$ is given by $\tilde{\nu}=\omega_L r^L$ we get
\begin{equation}\label{eq:lowerboundballs}
\left(\frac{\text{Vol}(L)}{\nu_s}\right)^{K-2} = \left(\frac{c_L}{\omega_L}\right)^{K-2}\left(\frac{\tilde{\nu}}{\nu_s}\right)^{K-2}.
\end{equation}
Also, by construction we have that $N\leq (\tilde{\nu}/\nu_s)^{K-1}$ and that $\tilde{N}=\tilde{\nu}/\nu_s$ so an upper bound on $N/\tilde{N}$ is given by
\begin{equation}\label{eq:upperboundballs}
\frac{N}{\tilde{N}}\leq \left( \frac{\tilde{\nu}}{\nu_s}\right)^{K-2},
\end{equation}
which differs from the lower bound in~(\ref{eq:lowerboundballs}) by a multiplicative constant.

We are now in a position to prove Proposition~\ref{prop:growthriemann2}.

\begin{prop_empty}{\ref{prop:growthriemann2}}
For $N\rightarrow \infty$ and $2\leq K<\infty$ we have
\begin{equation}\label{eq:o}
\frac{\sum_{\lambda_c\in V_\pi(0)}\left\| \lambda_c - \frac{1}{K}\sum_{i=0}^{K-1}\lambda_i\right\|^2}
{\sum_{\lambda_c\in V_\pi(0)}\sum_{i=0}^{K-2}\sum_{j=i+1}^{K-1}\| \lambda_i - \lambda_j\|^2} \rightarrow 0.
\end{equation}
\end{prop_empty}

\begin{proof}
The numerator describes the distance from a central lattice point to the mean vector of its associated $K$-tuple. This distance is upper bounded by the covering radius\index{lattice!covering radius} of the sublattice $\Lambda_s$. The rate of growth of the covering radius is proportional to $\nu_s^{1/L}=(N\nu)^{1/L}$, hence
\begin{equation}\label{eq:o1}
\sum_{\lambda_c\in V_\pi(0)}\left\|\lambda_c - \frac{1}K\sum_{i=0}^{K-1}\lambda_i\right\|^2= \mathcal{O}\left(N^2N^{2/L}\nu^{2/L}\right).
\end{equation}
Since the approximation $N/\tilde{N}$ used in Proposition~\ref{prop:riemann2} is sandwiched between the lower and upper bounds (i.e.\ Eqs.~(\ref{eq:lowerboundballs}) and~(\ref{eq:upperboundballs})) we can write
\begin{equation}
\begin{split}
&\sum_{\lambda_c\in V_\pi(0)}
\sum_{i=0}^{K-2}\sum_{j=i+1}^{K-1}
\|\alpha_i(\lambda_c) - \alpha_j(\lambda_c)\|^2\\
&\quad=
\sum_{i=0}^{K-2}\sum_{j=i+1}^{K-1}
\sum_{\lambda_c\in V_\pi(0)}
\|\alpha_i(\lambda_c) - \alpha_j(\lambda_c)\|^2 \\
&\quad\approx \frac{L}2K(K-1)
G(S_L)\psi_L^{2}N^2N^{2K/L(K-1)}\nu^{2/L},
\end{split}
\end{equation}
so that, since $\lambda_i=\alpha_i(\lambda_c)$,
\begin{equation}\label{eq:o2}
\sum_{\lambda_c\in V_\pi(0)}\sum_{i=0}^{K-2}\sum_{j=i+1}^{K-1}\| \lambda_i - \lambda_j\|^2
= \mathcal{O}\left(N^2N^{2K/L(K-1)}\nu^{2/L}\right).
\end{equation}
Comparing~(\ref{eq:o1}) to~(\ref{eq:o2}) we see that (\ref{eq:o}) grows as $\mathcal{O}\left(N^{-K/(K-1)}\right) \rightarrow 0$ for $N\rightarrow \infty$ and $K<\infty$.
\end{proof}

\chapter{Proofs for Chapter~\ref{chap:asym}}
For notational convenience we will in this appendix use the shorter notations $\mathcal{L}, \mathcal{L}_i$ and $\mathcal{L}_{i,j}$ instead of $\mathcal{L}^{(K,\kappa)}, \mathcal{L}_i^{(K,\kappa)}$ and $\mathcal{L}_{i,j}^{(K,\kappa)}$. 
 
\section{Proof of Theorem~\ref{theo:sums_asym}}\label{app:proofsums_asym}
To prove Theorem~\ref{theo:sums_asym} we need the following results.

\begin{lemma}\label{lem:sumLj}
For $1 \leq \kappa \leq K$ and any $i\in \{0,\dots, K-1\}$ we have
\begin{equation*}
\sum_{\substack{j=0\\j\neq i}}^{K-1}p(\mathcal{L}_j)=\kappa p(\mathcal{L})-p(\mathcal{L}_i).
\end{equation*}
\end{lemma}
\begin{proof}
Since $|\mathcal{L}_j| = \binom{K-1}{\kappa-1}$ the sum $\sum_{j=0}^{K-1}p(\mathcal{L}_j)$ contains $K\binom{K-1}{\kappa-1}$ terms. However, the number of distinct terms is $|\mathcal{L}|=\binom{K}{\kappa}$ and each individual term occurs $\kappa$ times in the sum, since
\begin{equation*}
\frac{K\binom{K-1}{\kappa-1}}{\binom{K}{\kappa}} = \kappa.
\end{equation*}
Subtracting the terms for $j=i$ proves the lemma.
\end{proof}

\begin{lemma}\label{lem:sumLij}
For $1 \leq \kappa \leq K$ and any $i,j\in \{0,\dots, K-1\}$ we have
\begin{equation*}
\sum_{j=0}^{K-1}p(\mathcal{L}_{i,j})=\kappa p(\mathcal{L}_i).
\end{equation*}
\end{lemma}
\begin{proof}
It is true that $\mathcal{L}_{i,i}=\mathcal{L}_i$ and since $|\mathcal{L}_{i}|=\binom{K-1}{\kappa-1}$ and $|\mathcal{L}_{i,j}|=\binom{K-2}{\kappa-2}$ the sum $\sum_{j=0}^{K-1}p(\mathcal{L}_{i,j})$ contains $(K-1)\binom{K-2}{\kappa-2} + \binom{K-1}{\kappa-1}$ terms. However, the number of distinct $l\in \mathcal{L}_i$ terms is $|\mathcal{L}_i|=\binom{K-1}{\kappa-1}$ and each term occurs $\kappa$ times in the sum, since
\begin{equation*}
\frac{(K-1)\binom{K-2}{\kappa-2} + \binom{K-1}{\kappa-1}}{\binom{K-1}{\kappa-1}} = \kappa.
\end{equation*}
\rspace
\end{proof}

\begin{lemma}\label{lem:suminnerprod}
For $1 \leq \kappa \leq K$ we have
\begin{equation*}
\sum_{l\in\mathcal{L}} p(\mathcal{L})\left\langle\lambda_c,\frac{1}{\kappa}\sum_{i\in l} \lambda_i \right\rangle = \left\langle \lambda_c, \frac{1}{\kappa}\sum_{i=0}^{K-1}\lambda_i p(\mathcal{L}_i)\right\rangle.
\end{equation*}
\end{lemma}
\begin{proof}
We have that
\begin{equation*}
\begin{split}
\sum_{l\in\mathcal{L}} p(\mathcal{L})\left\langle\lambda_c,\frac{1}{\kappa}\sum_{i\in l} \lambda_i \right\rangle &=
\left\langle\lambda_c, \frac{1}{\kappa}\sum_{l\in\mathcal{L}}p(l)\sum_{i\in l}\lambda_i \right\rangle \\
&= \left\langle \lambda_c, \frac{1}{\kappa}\sum_{i=0}^{K-1}\lambda_i p(\mathcal{L}_i)\right\rangle,
\end{split}
\end{equation*}
where the last equality follows since $\mathcal{L}_i$ denotes the set of all $l$-terms that contains the index $i$.
\end{proof}

\begin{lemma}\label{lem:normsum0}
For $1 \leq \kappa \leq K$ we have
\begin{equation*}
\begin{split}
\sum_{i=0}^{K-2}\sum_{j=i+1}^{K-1}p(\mathcal{L}_i)p(\mathcal{L}_j)\|\lambda_i-\lambda_j\|^2 &=
\sum_{i=0}^{K-1}p(\mathcal{L}_i)\left(\kappa p(\mathcal{L})-p(\mathcal{L}_i)\right)\|\lambda_i\|^2\\
&\quad- 2\sum_{i=0}^{K-2}\sum_{j=i+1}^{K-1}p(\mathcal{L}_i)p(\mathcal{L}_j)\langle \lambda_i,\lambda_j\rangle.
\end{split}
\end{equation*}
\end{lemma}
\begin{proof}
We have that
\begin{equation*}
\begin{split}
\sum_{i=0}^{K-2}\sum_{j=i+1}^{K-1}p(\mathcal{L}_i)p(\mathcal{L}_j)\|\lambda_i-\lambda_j\|^2 &=\sum_{i=0}^{K-2}\sum_{j=i+1}^{K-1}p(\mathcal{L}_i)p(\mathcal{L}_j)(\|\lambda_i\|^2 + \|\lambda_j\|^2)\\
&\quad - 2\sum_{i=0}^{K-2}\sum_{j=i+1}^{K-1}p(\mathcal{L}_i)p(\mathcal{L}_j)\langle \lambda_i,\lambda_j\rangle.
\end{split}
\end{equation*}
Furthermore, it follows that
{\allowdisplaybreaks
\begin{align*}
\sum_{i=0}^{K-2}\sum_{j=i+1}^{K-1}&p(\mathcal{L}_i)p(\mathcal{L}_j)(\|\lambda_i\|^2 + \|\lambda_j\|^2) \\
&= 
\sum_{i=0}^{K-2}p(\mathcal{L}_i)\|\lambda_i\|^2\sum_{j=i+1}^{K-1}p(\mathcal{L}_j) + \sum_{j=1}^{K-1}p(\mathcal{L}_j)\|\lambda_j\|^2 \sum_{i=0}^{j-1}p(\mathcal{L}_i) \\
&= \sum_{i=0}^{K-1}p(\mathcal{L}_i)\|\lambda_i\|^2\underbrace{\sum_{j=i+1}^{K-1}p(\mathcal{L}_j)}_{0\ \text{for}\ i=K-1} + \sum_{j=0}^{K-1}p(\mathcal{L}_j)\|\lambda_j\|^2\underbrace{\sum_{i=0}^{j-1}p(\mathcal{L}_i)}_{0\ \text{for}\ j=0} \\
&= \sum_{i=0}^{K-1}p(\mathcal{L}_i)\|\lambda_i\|^2\left( \sum_{j=0}^{i-1}p(\mathcal{L}_j)+\sum_{j=i+1}^{K-1}p(\mathcal{L}_j)\right) \\
&= \sum_{i=0}^{K-1}p(\mathcal{L}_i)\|\lambda_i\|^2\sum_{\substack{j=0\\j\neq i}}^{K-1}p(\mathcal{L}_j) \\
&= \sum_{i=0}^{K-1}p(\mathcal{L}_i)\|\lambda_i\|^2\left( \kappa p(\mathcal{L}) - p(\mathcal{L}_i)\right),
\end{align*}}
where the last equality follows by use of Lemma~\ref{lem:sumLj}.
\end{proof}

\begin{lemma}\label{lem:normsum}
For $1 \leq \kappa \leq K$ we have
\begin{equation*}
\begin{split}
\sum_{i=0}^{K-2}\sum_{j=i+1}^{K-1}p(\mathcal{L}_{i,j})\|\lambda_i-\lambda_j\|^2 &=
(\kappa -1)\sum_{i=0}^{K-1}p(\mathcal{L}_i)\|\lambda_i\|^2\\
&\quad -
2\sum_{i=0}^{K-2}\sum_{j=i+1}^{K-1}p(\mathcal{L}_{i,j})\langle \lambda_i,\lambda_j\rangle.
\end{split}
\end{equation*}
\end{lemma}
\begin{proof}
We have that
\begin{equation*}
\begin{split}
\sum_{i=0}^{K-2}\sum_{j=i+1}^{K-1}p(\mathcal{L}_{i,j})\|\lambda_i-\lambda_j\|^2 &= \sum_{i=0}^{K-2}\sum_{j=i+1}^{K-1}p(\mathcal{L}_{i,j})(\|\lambda_i\|^2\\
&\quad + \|\lambda_j\|^2) - 2\sum_{i=0}^{K-2}\sum_{j=i+1}^{K-1}p(\mathcal{L}_{i,j})\langle \lambda_i,\lambda_j\rangle.
\end{split}
\end{equation*}
Furthermore, it follows that
{\allowdisplaybreaks
\begin{align*}
\sum_{i=0}^{K-2}\sum_{j=i+1}^{K-1}p(\mathcal{L}_{i,j})(\|\lambda_i\|^2 &+ \|\lambda_j\|^2)\\
& =
\sum_{i=0}^{K-2}\sum_{j=i+1}^{K-1}p(\mathcal{L}_{i,j})\|\lambda_i\|^2 +
\sum_{i=0}^{K-2}\sum_{j=i+1}^{K-1}p(\mathcal{L}_{i,j})\|\lambda_j\|^2 \\
&= 
\sum_{i=0}^{K-2}\|\lambda_i\|^2\sum_{j=i+1}^{K-1}p(\mathcal{L}_{i,j})
+\sum_{j=1}^{K-1}\sum_{i=0}^{j-1}p(\mathcal{L}_{i,j})\|\lambda_j\|^2 \\
&= \sum_{i=0}^{K-1}\|\lambda_i\|^2\underbrace{\sum_{j=i+1}^{K-1}p(\mathcal{L}_{i,j})}_{0\ \text{for}\ i=K-1}
+\sum_{j=0}^{K-1}\|\lambda_j\|^2\underbrace{\sum_{i=0}^{j-1}p(\mathcal{L}_{i,j})}_{0\ \text{for}\ j=0} \\
&=\sum_{i=0}^{K-1}\|\lambda_i\|^2\left(\sum_{j=0}^{i-1}p(\mathcal{L}_{i,j})+\sum_{j=i+1}^{K-1}p(\mathcal{L}_{i,j})\right) \\
&=\sum_{i=0}^{K-1}\|\lambda_i\|^2\left(\sum_{j=0}^{K-1}p(\mathcal{L}_{i,j})-p(\mathcal{L}_{i})\right) \\
&\overset{(a)}{=}\sum_{i=0}^{K-1}\|\lambda_i\|^2\left(\kappa p(\mathcal{L}_{i})-p(\mathcal{L}_{i})\right) \\
&=(\kappa-1)\sum_{i=0}^{K-1}\|\lambda_i\|^2p(\mathcal{L}_{i}),
\end{align*}}
where $(a)$ follows by use of Lemma~\ref{lem:sumLij}.
\end{proof}

\begin{lemma}\label{lem:normsum2}
For $1 \leq \kappa \leq K$ we have
\begin{equation*}
\sum_{l\in\mathcal{L}}p(l)\left\|\sum_{i\in l}\lambda_i\right\|^2 =
\kappa \sum_{i=0}^{K-1}p(\mathcal{L}_i)\|\lambda_i\|^2 - \sum_{i=0}^{K-2}\sum_{j=i+1}^{K-1}p(\mathcal{L}_{i,j})\|\lambda_i-\lambda_j\|^2.
\end{equation*}
\end{lemma}
\begin{proof}
The set of all elements $l$ of $\mathcal{L}$ that contains the index $i$ is denoted by $\mathcal{L}_i$. Similarly the set of all elements that contains the indices $i$ and $j$ is denoted by $\mathcal{L}_{i,j}$. From this we see that
\begin{equation*}
\begin{split}
\sum_{l\in\mathcal{L}}p(l)\left\|\sum_{i\in l}\lambda_i\right\|^2 &=
\sum_{l\in\mathcal{L}}p(l)\left(\sum_{i\in l}\|\lambda_i\|^2 + 2\sum_{i=0}^{\kappa-2}\sum_{j=i+1}^{\kappa-1}\langle \lambda_{l_i},\lambda_{l_j}\rangle\right) \\
&= \sum_{i=0}^{K-1}p(\mathcal{L}_i)\|\lambda_i\|^2 + 2\sum_{i=0}^{K-2}\sum_{j=i+1}^{K-1}p(\mathcal{L}_{i,j})\langle \lambda_{i},\lambda_{j}\rangle.
\end{split}
\end{equation*}
By use of Lemma~\ref{lem:normsum} it follows that
\begin{equation*}
\begin{split}
\sum_{l\in\mathcal{L}}p(l)\left\|\sum_{i\in l}\lambda_i\right\|^2 &=
\sum_{i=0}^{K-1}p(\mathcal{L}_i)\|\lambda_i\|^2 + (\kappa-1)\sum_{i=0}^{K-1}p(\mathcal{L}_i)\|\lambda_i\|^2\\
&\quad - \sum_{i=0}^{K-2}\sum_{j=i+1}^{K-1}p(\mathcal{L}_{i,j})\| \lambda_{i}-\lambda_{j}\|^2 \\
&=\kappa\sum_{i=0}^{K-1}p(\mathcal{L}_i)\|\lambda_i\|^2 - \sum_{i=0}^{K-2}\sum_{j=i+1}^{K-1}p(\mathcal{L}_{i,j})\| \lambda_{i}-\lambda_{j}\|^2
\end{split}
\end{equation*}
\rspace
\end{proof}

We are now in a position to prove the following result.
\begin{proposition}\label{prop:lcmean}
For $1 \leq \kappa \leq K$ we have
\begin{equation}\label{eq:lcmean0}
\begin{split}
\sum_{l\in\mathcal{L}}p(l)\left\|\lambda_c-\frac{1}{\kappa}\sum_{i\in l}\lambda_i\right\|^2 &=
p(\mathcal{L})\left\|\lambda_c-\frac{1}{\kappa p(\mathcal{L})}\sum_{i=0}^{K-1}p(\mathcal{L}_i) \lambda_i\right\|^2 \\ 
& + \frac{1}{\kappa^2}\sum_{i=0}^{K-2}\sum_{j=i+1}^{K-1}\left(\frac{p(\mathcal{L}_i)p(\mathcal{L}_j)}{p(\mathcal{L})}-p(\mathcal{L}_{i,j})\right)\|\lambda_i-\lambda_j\|^2.
\end{split}
\end{equation}
\end{proposition}
\begin{proof}
Expansion of the norm on the left-hand-side in~(\ref{eq:lcmean0}) leads to 
{\allowdisplaybreaks
\begin{align*}
\sum_{l\in\mathcal{L}}p(l)\bigg\|\lambda_c&-\frac{1}{\kappa}\sum_{i\in l}\lambda_i\bigg\|^2\!\! =\! \sum_{l\in\mathcal{L}}p(l)\!\!\left(
\|\lambda_c\|^2 - 2\left\langle\lambda_c,\frac{1}{\kappa}\sum_{i\in l}\lambda_i\right\rangle + \frac{1}{\kappa^2}\left\|\sum_{i\in l}\lambda_i\right\|^2\right) \\
&\overset{(a)}{=} p(\mathcal{L})\|\lambda_c\|^2 - 2\left\langle \lambda_c , \frac{1}{\kappa}\sum_{i=0}^{K-1}p(\mathcal{L}_i)\lambda_i\right\rangle
+ \frac{1}{\kappa^2}\sum_{l\in \mathcal{L}}p(l)\left\|\sum_{i\in l}\lambda_i\right\|^2\\
&= p(\mathcal{L})\left\| \lambda_c - \frac{1}{\kappa p(\mathcal{L})}\sum_{i=0}^{K-1}p(\mathcal{L}_i)\lambda_i\right\|^2 - \frac{1}{\kappa^2p(\mathcal{L})}\left\|\sum_{i=0}^{K-1}p(\mathcal{L}_i)\lambda_i\right\|^2\\
&\quad+ \frac{1}{\kappa^2}\sum_{l\in \mathcal{L}}p(l)\left\|\sum_{i\in l}\lambda_i\right\|^2\\
&=p(\mathcal{L})\left\| \lambda_c - \frac{1}{\kappa p(\mathcal{L})}\sum_{i=0}^{K-1}p(\mathcal{L}_i)\lambda_i\right\|^2 + \frac{1}{\kappa^2}\sum_{l\in \mathcal{L}}p(l)\left\|\sum_{i\in l}\lambda_i\right\|^2\\
&\quad - \frac{1}{\kappa^2p(\mathcal{L})}\left( \sum_{i=0}^{K-1}p(\mathcal{L}_i)^2\|\lambda_i\|^2 + 2\sum_{i=0}^{K-2}\sum_{j=i+1}^{K-1}p(\mathcal{L}_i)p(\mathcal{L}_j)\langle \lambda_i,\lambda_j\rangle\right)\\
&\overset{(b)}{=}p(\mathcal{L})\left\| \lambda_c - \frac{1}{\kappa p(\mathcal{L})}\sum_{i=0}^{K-1}p(\mathcal{L}_i)\lambda_i\right\|^2 + 
\frac{1}{\kappa}\sum_{i=0}^{K-1}p(\mathcal{L}_i)\|\lambda_i\|^2\\
&\quad - \frac{1}{\kappa^2}\sum_{i=0}^{K-2}\sum_{j=i+1}^{K-1}p(\mathcal{L}_{i,j})\|\lambda_i-\lambda_j\|^2 -\frac{1}{\kappa^2p(\mathcal{L})}\sum_{i=0}^{K-1}p(\mathcal{L}_i)^2\|\lambda_i\|^2 \\
&\quad + \frac{1}{\kappa^2 p(\mathcal{L})}\bigg(
\sum_{i=0}^{K-2}\sum_{j=i+1}^{K-1}p(\mathcal{L}_i)p(\mathcal{L}_j)\| \lambda_i-\lambda_j\|^2\\
&\quad -\sum_{i=0}^{K-1}p(\mathcal{L}_i)(\kappa p(\mathcal{L})-p(\mathcal{L}_i))\|\lambda_i\|^2 \bigg)\\
&=p(\mathcal{L})\left\| \lambda_c - \frac{1}{\kappa p(\mathcal{L})}\sum_{i=0}^{K-1}p(\mathcal{L}_i)\lambda_i\right\|^2 \\
&\quad+\frac{1}{\kappa^2}\sum_{i=0}^{K-2}\sum_{j=i+1}^{K-1}\left(\frac{p(\mathcal{L}_i)p(\mathcal{L}_j)}{p(\mathcal{L})}- p(\mathcal{L}_{i,j})\right)\|\lambda_i-\lambda_j\|^2,
\end{align*}}
where $(a)$ follows by use of Lemma~\ref{lem:suminnerprod} and $(b)$ by use of Lemmas~\ref{lem:normsum0} and~\ref{lem:normsum2}.
\end{proof}

\begin{theorem}
For $1 \leq \kappa \leq K$ we have
\begin{equation}\label{eq:lcmean1}
\begin{split}
\sum_{\lambda_c}\sum_{l\in\mathcal{L}}p(l)&\left\|\lambda_c-\frac{1}{\kappa}\sum_{i\in l}\lambda_i\right\|^2 =\sum_{\lambda_c}\bigg(
p(\mathcal{L})\left\|\lambda_c-\frac{1}{\kappa p(\mathcal{L})}\sum_{i=0}^{K-1}p(\mathcal{L}_i) \lambda_i\right\|^2 \\ 
&\qquad + \frac{1}{\kappa^2}\sum_{i=0}^{K-2}\sum_{j=i+1}^{K-1}\left(\frac{p(\mathcal{L}_i)p(\mathcal{L}_j)}{p(\mathcal{L})}-p(\mathcal{L}_{i,j})\right)\|\lambda_i-\lambda_j\|^2\bigg).
\end{split}
\end{equation}
\end{theorem}
\begin{proof}
Follows trivially from Proposition~\ref{prop:lcmean}.
\end{proof}

\section{Proof of Proposition~\ref{prop:riemann2_asym}}
\label{app:riemann2_asym}
\begin{prop_empty}{\ref{prop:riemann2_asym}}
For $K=2$ and asymptotically as $N_i\rightarrow \infty, \nu_i\rightarrow 0$ as well as for $K=3$ and asymptotically as $N_i,L\rightarrow \infty$ and $\nu_i\rightarrow 0$, we have
for any pair of sublattices, $(\Lambda_i,\Lambda_j),\ i,j=0,\dots,K-1,\ i\neq j$,
\begin{equation*}
\frac{1}{L}\sum_{\lambda_c\in V_\pi(0)}
\| \alpha_i(\lambda_c)-\alpha_j(\lambda_c)\|^2 = \psi_L^{2}\nu^{2/L} G(S_L) N_\pi\prod_{m=0}^{K-1}N_m^{2/L(K-1)}.
\end{equation*}
\end{prop_empty}

\begin{proof}
Let $T_i = \{ \lambda_i : \lambda_i = \alpha_i(\lambda_c),\ \lambda_c \in V_\pi(0) \}$, i.e.\
the set of $N_\pi$ sublattice points $\lambda_i \in \Lambda_i$ associated with the $N_\pi$ central lattice points within $V_\pi(0)$.
Furthermore, let $T'_i \subseteq T_i$ be the set of unique elements of $T_i$. Since (for large $N_i$) all the lattice points of $\Lambda_i$ which are contained within $V_\pi(0)$ are used in some $K$-tuples, it follows that $|T'_i|\approx \tilde{\nu}/\nu_i = N_\pi/N_i$. 
Finally, let $T_j(\lambda_i) = \{ \lambda_j : \lambda_j = \alpha_j(\lambda_c)\ \text{and}\ \lambda_i = \alpha_i(\lambda_c),\  \lambda_c \in V_\pi(0) \}$ and let $T'_j(\lambda_i)\subseteq T_j(\lambda_i)$ be the set of unique elements. That is, $T_j(\lambda_i)$ contains all the elements $\lambda_j\in \Lambda_j$ which are in the $K$-tuples that also contains a specific $\lambda_i \in \Lambda_i$. We will also make use of the notation $\#_{\lambda_j}$ to indicate the number of occurrences of a specific $\lambda_j$ in $T_j(\lambda_i)$. For example for $K=2$ we have $\#_{\lambda_j}=1, \forall \lambda_j$ whereas for $K>2$ we have $\#_{\lambda_j}\geq 1$. We will show later that using the approximation $\#_{\lambda_j}\approx N_i/\tilde{N}_j$ is asymptotically good for $K=3, L\rightarrow \infty$ and $N_n\rightarrow\infty, \forall n$. Furthermore, we conjecture this to be the case for $K>3$ as well.

For sublattice $\Lambda_i$ and $\Lambda_j$ we have
\begin{equation*}
\sum_{\lambda_c\in V_\pi(0)}\| \alpha_i(\lambda_c) - \alpha_j(\lambda_c)\|^2 =
\sum_{\lambda_i\in T'_i}\sum_{\lambda_j\in T_j(\lambda_i)} \|\lambda_i - \lambda_j \|^2.
\end{equation*}
Given $\lambda_i\in T'_i$, we have
\begin{equation}
\begin{split}
\sum_{\lambda_j\in T_j(\lambda_i)} \|\lambda_i - \lambda_j \|^2 \nu_j
&= \sum_{\lambda_j\in T'_j(\lambda_i)}\#_{\lambda_j}\|\lambda_i-\lambda_j\|^2\nu_j \\
&\approx
\frac{N_i}{\tilde{N}_j}\sum_{\lambda_j\in T'_j(\lambda_i)}\|\lambda_i-\lambda_j\|^2\nu_j \\
&\approx \frac{N_i}{\tilde{N}_j} \int_{\tilde{V}(\lambda_i)}\|\lambda_i-x\|^2\, dx \\
&= \frac{N_i}{\tilde{N}_j} \tilde{\nu}^{1+2/L} LG(S_L) \\ \label{eq:f_asym}
&= N_i \nu_j \tilde{\nu}^{2/L} LG(S_L) 
\end{split}
\end{equation}
since $\tilde{N}_j=\tilde{\nu}/\nu_j$. Hence, with $\tilde{\nu}=\psi_L^L\nu\prod_{m=0}^{K-1}N_m^{1/(K-1)}$, we have
\begin{equation*}
\begin{split}
\frac{1}{L}\sum_{\lambda_j\in T_j(\lambda_i)} \|\lambda_i - \lambda_j \|^2 \nu_j 
&\approx
N_i\nu_j \psi_L^{2} \nu^{2/L}G(S_L)\prod_{m=0}^{K-1}N_m^{2/L(K-1)},
\end{split}
\end{equation*}
which is independent of $\lambda_i$, so that
\begin{equation*}
\begin{split}
\frac{1}{L}\sum_{\lambda_i\in T'_i}\sum_{\lambda_j\in T_j(\lambda_i)}\| \lambda_i - \lambda_j\|^2 
&\approx \frac{1}{L}\frac{N_\pi}{N_i}\sum_{\lambda_j\in T_j(\lambda_i)}\| \lambda_i - \lambda_j\|^2 \\
&\approx \psi_L^{2} \nu^{2/L} G(S_L) N_\pi\prod_{m=0}^{K-1}N_m^{2/L(K-1)},
\end{split}
\end{equation*}
which completes the first part of the proof. We still need to show that for $K=3$ and $L\rightarrow\infty$ as well as $N_m\rightarrow\infty, \forall m$ the approximation $\#_{\lambda_j}\approx N_i/\tilde{N}_j$ is good. That this is so can be deduced from the proof of Proposition~\ref{prop:riemann2} (the last part where $K=3$) by using the fact that $\tilde{\nu}= \psi_L^L\nu\prod{N_m^{1/(K-1)}}$ in order to prove that
\begin{equation}
\frac{1}{L}\sum_{\lambda_j\in T_j(\lambda_i)}\|\lambda_i-\lambda_j\|^2 = N_i\tilde{\nu}^{2/L}G(S_L),
\end{equation}
which shows that~(\ref{eq:f_asym}) is asymptotically true for $K=3, L\rightarrow\infty$ and $N_n\rightarrow \infty, \forall n$.
\end{proof}

\section{Proof of Proposition~\ref{prop:growthriemann2_asym}}
\label{app:growthriemann2_asym}

\begin{prop_empty}{\ref{prop:growthriemann2_asym}}
For $N_i\rightarrow \infty$ we have
\begin{equation}\label{eq:o_asym}
\frac{\sum_{\lambda_c\in V_\pi(0)}\left\| \lambda_c - \frac{1}{\kappa p(\mathcal{L})}\sum_{i=0}^{K-1}p(\mathcal{L}_i)\lambda_i\right\|^2}
{\sum_{\lambda_c\in V_\pi(0)}
\sum_{i=0}^{K-2}\sum_{j=i+1}^{K-1}\left(\frac{p(\mathcal{L}_i)p(\mathcal{L}_j)}{p(\mathcal{L})}-p(\mathcal{L}_{i,j})\right)\|\lambda_i-\lambda_j\|^2}
\rightarrow 0.
\end{equation}
\end{prop_empty}

\begin{proof}
The numerator describes the distance from a central lattice point to the weighted centroid of its associated $K$-tuple. Let us choose $\Lambda_0$ such that $N_0\leq N_i, \forall i$. Then, since by construction there is no bias towards any of the sublattices, the weighted centroids will be evenly distributed around $\lambda_0$ points. Hence, the distance from central lattice points to the centroids can be upper bounded by the covering radius of $\Lambda_0$. This is a conservative\footnote{The number of distinct centroids per unit volume is larger than the number of points of $\Lambda_0$ per unit volume.}
 upper bound but will suffice for the proof. The rate of growth of the covering radius is proportional to $\nu_0^{1/L}=(N_0\nu)^{1/L}$, hence
\begin{equation}\label{eq:o1_asym}
\sum_{\lambda_c\in V_\pi(0)}\left\|\lambda_c - \frac{1}{\kappa p(\mathcal{L})}\sum_{i=0}^{K-1}p(\mathcal{L}_i)\lambda_i\right\|^2= \mathcal{O}\left(N_\pi N_0^{2/L}\nu^{2/L}\right).
\end{equation}
By use of Proposition~\ref{prop:riemann2_asym} we have\footnote{The approximation of $\#_{\lambda_j}$ in Proposition~\ref{prop:riemann2_asym} does not influence this analysis. To see this we refer the reader to Appendix~\ref{app:growthriemann2}.}
\begin{equation*}
\begin{split}
\frac{1}{L}&\sum_{\lambda_c\in V_\pi(0)}
\sum_{i=0}^{K-2}\sum_{j=i+1}^{K-1}
\left(\frac{p(\mathcal{L}_i)p(\mathcal{L}_j)}{p(\mathcal{L})}-p(\mathcal{L}_{i,j})\right)
\|\alpha_i(\lambda_c) - \alpha_j(\lambda_c)\|^2 \\
&\quad=
\frac{1}{L}\sum_{i=0}^{K-2}\sum_{j=i+1}^{K-1}
\left(\frac{p(\mathcal{L}_i)p(\mathcal{L}_j)}{p(\mathcal{L})}-p(\mathcal{L}_{i,j})\right)
\sum_{\lambda_c\in V_\pi(0)}\|\alpha_i(\lambda_c) - \alpha_j(\lambda_c)\|^2 
\\
&\quad\approx
\psi_L^{2}\nu^{2/L}G(S_L)N_\pi\prod_{m=0}^{K-1}N_m^{2/L(K-1)}
\sum_{i=0}^{K-2}\sum_{j=i+1}^{K-1}
\left(\frac{p(\mathcal{L}_i)p(\mathcal{L}_j)}{p(\mathcal{L})}-p(\mathcal{L}_{i,j})\right),
\end{split}
\end{equation*}
so that, since $\lambda_i=\alpha_i(\lambda_c)$, we get by use of Proposition~\ref{prop:riemann2_asym}\footnote{In this case we actually lower bound the expression and as such the order operator $\mathcal{O}$ is in fact $\Omega$. Recall that we say that $f(n)=\mathcal{O}(g(n))$ if $0< f(n) \leq c_1g(n)$ and $f(n)=\Omega(g(n))$ if $f(n)\geq c_0g(n),$ for $c_0,c_1>0$ and some large $n$. Furthermore, $f(n)=\Theta(g(n))$ if $c_0g(n)\leq f(n) \leq c_1g(n)$, cf.~\cite{graham:1994}.}
\begin{equation}\label{eq:o2_asym}
\begin{split}
\sum_{\lambda_c\in V_\pi(0)}\sum_{i=0}^{K-2}\sum_{j=i+1}^{K-1}
\left(\frac{p(\mathcal{L}_i)p(\mathcal{L}_j)}{p(\mathcal{L})}-p(\mathcal{L}_{i,j})\right)&
\| \lambda_i - \lambda_j\|^2 \\
&\quad= \Omega\left(N_\pi\nu^{2/L}\prod_{m=0}^{K-1}N_m^{2/L(K-1)}\right).
\end{split}
\end{equation}
Comparing~(\ref{eq:o1_asym}) to~(\ref{eq:o2_asym}) we see that (\ref{eq:o_asym}) grows as $\Theta\left(N_0^{2/L}/N_\pi^{2/L(K-1)}\right) \rightarrow 0$ for $N_i\rightarrow \infty$.
\end{proof}

\section{Proof of Lemmas}\label{app:lemmas}
\begin{proof}[Proof of Lemma~\ref{lem:nobias}]
For simplicity (and without any loss of generality) we assume that $V_\pi(0)$ forms the shape of a sphere, see Fig.~\ref{fig:Vpi}. 
The $K$-tuples are constructed by centering a sphere $\tilde{V}$ of volume $\tilde{\nu}$ around each $\lambda_0 \in V_\pi(0)$ and taking all combinations of lattice points within this region (keeping $\lambda_0$ as first coordinate). From Fig.~\ref{fig:Vpi} it may be seen that any $\lambda_0$ which is contained in the region denoted $\mathcal{A}$ will always be combined with sublattice points that are also contained in $V_\pi(0)$. 
On the other hand, any $\lambda_0$ which is contained in region $\mathcal{B}$ will occasionally be combined with points outside $V_\pi(0)$. Therefore, we need to show that the volume $V_\mathcal{A}$ of $\mathcal{A}$ approaches the volume of $V_\pi(0)$ as $N_i\rightarrow \infty$ or equivalently that the ratio of $V_\mathcal{B}/V_\mathcal{A}\rightarrow 0$ as $N_i\rightarrow \infty$, where $V_\mathcal{B}$ denotes the volume of the region $\mathcal{B}$.
\begin{figure}
\psfrag{Vpi}{$V_\pi(0)$}
\psfrag{Vtil}{$\tilde{V}$}
\psfrag{r0}{$r_0$}
\psfrag{r1}{$r_1$}
\psfrag{A}{$\mathcal{A}$}
\psfrag{B}{$\mathcal{B}$}
\begin{center}
\includegraphics[]{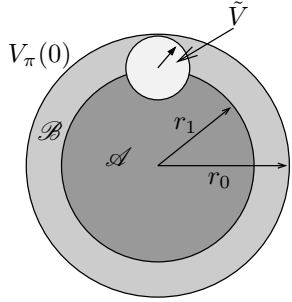}
\caption{The complete sphere consisting of the regions $\mathcal{A}$ and $\mathcal{B}$ describe $V_\pi(0)$. The radius of $V_\pi(0)$ is $r_0$. The small bright sphere describe $\tilde{V}$. When $\tilde{V}$ is centered at $\lambda_0$ points within the sphere $\mathcal{A}$ of radius $r_1$ it will be completely contained within $V_\pi(0)$.}
\label{fig:Vpi}
\end{center}
\end{figure}

Let $\omega_L$ denote the volume of an $L$-dimensional unit sphere. Then $V_\mathcal{A}=\omega_Lr_1^L$ and $V_\mathcal{B}=\nu_\pi-V_\mathcal{A}$, where $\nu_\pi$ is the volume of $V_\pi(0)$. The radius $r_1$ of $\mathcal{A}$ can be expressed as the difference between the radius $r_0$ of $V_\pi(0)$ and the radius of $\tilde{V}$, that is
\begin{equation}
r_1 = (\nu_\pi/\omega_L)^{1/L} - (\tilde{\nu}/\omega_L)^{1/L}.
\end{equation}
Since, $\nu_\pi=\nu\prod N_i=\nu N_{\pi}$ and $\tilde{\nu}=\psi_L^L\nu \prod N_i^{1/(K-1)}=\psi_L^L\nu N_{\pi}^{1/(K-1)}$ we can write $V_\mathcal{A}$ as
\begin{equation}
\begin{split}
V_\mathcal{A}&=\omega_Lr_1^L \\
&= \omega_L\left( \left(\frac{\nu N_{\pi}}{\omega_L}\right)^{1/L} - \left(\frac{\psi_L^L\nu N_{\pi}^{1/(K-1)}}{\omega_L}\right)^{1/L}\right)^L \\
&= \nu\left( N_{\pi}^{1/L} - \psi_L N_{\pi}^{1/L(K-1)} \right)^{L}.
\end{split}
\end{equation}
The volume of $\mathcal{B}$ can be expressed through the volume of $\mathcal{A}$ as
\begin{equation}
V_\mathcal{B}=\nu_\pi-V_\mathcal{A},
\end{equation}
so that their ratio is given by
\begin{equation}
\frac{V_\mathcal{B}}{V_\mathcal{A}} = \frac{N_{\pi}}{\left(N_{\pi}^{1/L} - N_{\pi}^{1/L(K-1)}\right)^{L}} - 1.
\end{equation}
Clearly, for $K>2$ we have 
\begin{equation}
\lim_{N_\pi\rightarrow \infty} \frac{N_{\pi}}{\left(N_{\pi}^{1/L} - N_{\pi}^{1/L(K-1)}\right)^{L}} = 1,
\end{equation}
which proves the claim.
\end{proof}

\begin{proof}[Proof of Lemma~\ref{lem:equivktuples}]
We only need to prove Lemma~\ref{lem:equivktuples} for $\Lambda_0$ and $\Lambda_1$. Then by symmetry it must hold for any pair.
Let $\mathcal{S}_{\lambda_0}$ denote the set of $K$-tuples constructed by centering $\tilde{V}$ at some $\lambda_0\in V_\pi(0)\cap\Lambda_0$.
Hence, $s\in \mathcal{S}_{\lambda_0}$ has $\lambda_0$ as first coordinate and the distance between any two elements of $s$ is less than $r$, the radius of $\tilde{V}$. We will assume\footnote{This is always the case if $r \geq \max_i r(\Lambda_i)$ where $r(\Lambda_i)$ is the covering radius of the $i^{th}$ sublattice. The covering radius depends on the lattice and is maximized if $\Lambda_i$ is geometrically similar to $Z^L$, in which case we have\cite{conway:1999}
\begin{equation*}
r(\Lambda_i)=\frac{1}2\sqrt{2}\nu^{1/L}N_i^{1/L}.
\end{equation*}
Since $r=\psi_L \nu^{1/L} N_\pi^{1/L(K-1)}/\omega_L^{1/L}$ it follows that in order to make sure that $\mathcal{S}_{\lambda_0}\neq\emptyset$ the index values must satisfy
\begin{equation*}\label{eq:Niupperbound}\tag{*}
N_i\leq (\sqrt{2}\psi_L)^L\omega_L N_\pi^{1/(K-1)},\quad i=0,\dots, K-1.
\end{equation*}
Throughout this work we therefore require (and implicitly assume) that~(\ref{eq:Niupperbound}) is satisfied.%
}
that $S_{\lambda_0}\neq \emptyset, \forall \lambda_0$.

Similarly, define the set $\mathcal{S}_{\lambda_1}\neq \emptyset$ by centering $\tilde{V}$ at some $\lambda_1\in V_\pi(0)\cap\Lambda_1$. 
Assume\footnote{This is asymptotically true according to Lemma~\ref{lem:nobias} since we at this point do not consider the cosets of the $K$-tuples. Furthermore, the cosets are invariant to which lattice is used for the construction of $K$-tuples as long as all elements of the $K$-tuples are within $V_\pi(0)$.} all elements of the $K$-tuples are in $V_\pi(0)$. Then it must hold that for any $s\in \mathcal{S}_{\lambda_1}$ we have $s\in \bigcup_{\lambda_0\in V_\pi\cap\Lambda_0}\mathcal{S}_{\lambda_0}$. But it is also true that for any $s'\in \mathcal{S}_{\lambda_0}$ we have $s'\in \bigcup_{\lambda_1\in V_\pi\cap\Lambda_1}\mathcal{S}_{\lambda_1}$. 
Hence, we deduce that $\bigcup_{\lambda_0\in V_\pi\cap\Lambda_0}\mathcal{S}_{\lambda_0} \equiv \bigcup_{\lambda_1\in V_\pi\cap\Lambda_1}\mathcal{S}_{\lambda_1}$. 
Furthermore, $|V_\pi(0)\cap \Lambda_0|=N_\pi/N_0, |\mathcal{S}_{\lambda_0}|=N_0, \forall \lambda_0\in V_\pi(0)\cap\Lambda_0$ and 
$\mathcal{S}_{\lambda'_0}\cap\mathcal{S}_{\lambda''_0}=\emptyset, \lambda'_0\neq\lambda''_0$, 
which implies that $|\bigcup_{\lambda_0\in V_\pi\cap\Lambda_0}\mathcal{S}_{\lambda_0}|=N_\pi$. 
\end{proof}

\section{Proof of  Theorem~\ref{theo:sidedist}}
\label{app:sidedist}
Before proving Theorem~\ref{theo:sidedist} we need the following results.


\begin{lemma}\label{lem:sumjside}
For $1\leq \kappa \leq K$ and any $l \in \mathcal{L}$ we have
\begin{equation*}
\left\|\sum_{j\in l} \lambda_j\right \|^2=
\kappa\sum_{j\in l}\|\lambda_j\|^2 - \sum_{i=0}^{\kappa-2}\sum_{j=i+1}^{\kappa-1}\|\lambda_{l_j}-\lambda_{l_i}\|^2.
\end{equation*}
\end{lemma}
\begin{proof}
We can write
\begin{equation*}
\left\|\sum_{j\in l} \lambda_j\right \|^2=
\sum_{j\in l}\|\lambda_j\|^2 + 2\sum_{i=0}^{\kappa-2}\sum_{j=i+1}^{\kappa-1}\langle \lambda_{l_j}, \lambda_{l_i} \rangle,
\end{equation*}
which by use of Lemma~\ref{lem:r3} leads to
\begin{equation*}
\begin{split}
\left\|\sum_{j\in l} \lambda_j\right \|^2 &=
\sum_{j\in l}\|\lambda_j\|^2 + (\kappa-1)\sum_{j\in l}\|\lambda_j\|^2 -
\sum_{i=0}^{\kappa-2}\sum_{j=i+1}^{\kappa-1}\| \lambda_{l_j}- \lambda_{l_i}\|^2 \\
&= \kappa\sum_{j\in l}\|\lambda_j\|^2 -
\sum_{i=0}^{\kappa-2}\sum_{j=i+1}^{\kappa-1}\| \lambda_{l_j}- \lambda_{l_i}\|^2.
\end{split}
\end{equation*}
\rspace
\end{proof}

\begin{lemma}\label{lem:innerprod1}
For $1 \leq \kappa \leq K$ and any $l\in \mathcal{L}$ we have
\begin{equation*}
\begin{split}
2\left\langle \sum_{j\in l}\lambda_j , \sum_{i=0}^{K-1}p(\mathcal{L}_i)\lambda_i \right\rangle &=
p(\mathcal{L})\kappa\sum_{j\in l}\|\lambda_j\|^2\\
&\quad + \kappa\sum_{i=0}^{K-1}p(\mathcal{L}_i)\|\lambda_i\|^2  - \sum_{j\in l}\sum_{i=0}^{K-1}p(\mathcal{L}_i)\|\lambda_j - \lambda_i\|^2.
\end{split}
\end{equation*}
\end{lemma}
\begin{proof}
\begin{equation*}
\begin{split}
2\left\langle \sum_{j\in l}\lambda_j , \sum_{i=0}^{K-1}p(\mathcal{L}_i)\lambda_i \right\rangle &=
2\sum_{j\in l}\sum_{i=0}^{K-1}p(\mathcal{L}_i) \langle \lambda_j ,\lambda_i \rangle   \\
&= -\sum_{j\in l}\sum_{i=0}^{K-1}p(\mathcal{L}_i)\|\lambda_j-\lambda_i\|^2 \\
&\quad+ \sum_{j\in l}\sum_{i=0}^{K-1}p(\mathcal{L}_i)\left( \|\lambda_j\|^2 + \|\lambda_i\|^2\right)
\end{split}
\end{equation*}
where by use of Lemma~\ref{lem:sumLj} we obtain
\begin{equation*}
\begin{split}
2\left\langle \sum_{j\in l}\lambda_j , \sum_{i=0}^{K-1}p(\mathcal{L}_i)\lambda_i \right\rangle &=
-\sum_{j\in l}\sum_{i=0}^{K-1}p(\mathcal{L}_i)\|\lambda_j-\lambda_i\|^2 \\
&\quad+ \kappa p(\mathcal{L})\sum_{j\in l}\|\lambda_j\|^2 + \kappa\sum_{i=0}^{K-1}p(\mathcal{L}_i)\|\lambda_i\|^2.
\end{split}
\end{equation*}
\rspace
\end{proof}

\begin{proposition}\label{prop:normdiff2}
For $0< \kappa \leq K\leq 3$, $N_i\rightarrow \infty, \nu_i \rightarrow 0$ and any $l\in \mathcal{L}$ we have
\begin{equation*}
\begin{split}
\frac{1}{L}\sum_{\lambda_c\in V_{\pi}(0)}\left\|\frac{1}{\kappa}\sum_{j\in l}\lambda_j - \frac{1}{\kappa p(\mathcal{L})}\sum_{i=0}^{K-1}p(\mathcal{L}_i)\lambda_i\right\|^2&\\
&\hspace{-2cm}= \omega^{(K,l)}
\psi_L^{2}\nu^{2/L} G(S_L)N_\pi\prod_{m=0}^{K-1}N_m^{2/L(K-1)},
\end{split}
\end{equation*}
where
\begin{equation*}
\begin{split}
\omega^{(K,l)}&=\frac{1}{p(\mathcal{L})^2\kappa^2}\bigg(p(\mathcal{L})^2\kappa^2-p(\mathcal{L})^2\binom{\kappa}{2} - p(\mathcal{L})\sum_{j\in l} p(\mathcal{L}_j)\\
&\quad -
\sum_{i=0}^{K-2}\sum_{j=i+1}^{K-1}p(\mathcal{L}_i)p(\mathcal{L}_j)\bigg),
\end{split}
\end{equation*}
where $\binom{\kappa}{2}=0$ for $\kappa=1$.
\end{proposition}

\begin{proof}
We have that
\begin{equation*}
\begin{split}
\bigg\|\frac{1}{\kappa}\sum_{j\in l}\lambda_j - \frac{1}{\kappa p(\mathcal{L})}\sum_{i=0}^{K-1}& p(\mathcal{L}_i)\lambda_i\bigg\|^2\\
&= \frac{1}{p(\mathcal{L})^2\kappa^2}\bigg(p(\mathcal{L})^2\left\|\sum_{j\in l}\lambda_j\right\|^2
+ \left\|\sum_{i=0}^{K-1}p(\mathcal{L}_i)\lambda_i \right\|^2 
 \\
&\quad -
2p(\mathcal{L})\left\langle \sum_{j\in l}\lambda_j, \sum_{i=0}^{K-1} p(\mathcal{L}_i) \lambda_i \right\rangle \bigg),
\end{split}
\end{equation*}
which by use of Lemmas~\ref{lem:sumjside} and~\ref{lem:innerprod1} leads to
{\allowdisplaybreaks
\begin{align}\notag
\bigg\|\frac{1}{\kappa}\sum_{j\in l}\lambda_j &- \frac{1}{p(\mathcal{L})\kappa}\sum_{i=0}^{K-1} p(\mathcal{L}_i)\lambda_i\bigg\|^2 \\ \notag
&=\frac{1}{p(\mathcal{L})^2\kappa^2}\bigg(
p(\mathcal{L})^2\kappa\sum_{j\in l}\|\lambda_j\|^2 - p(\mathcal{L})^2\sum_{i=0}^{\kappa-2}\sum_{j=i+1}^{\kappa-1}\|\lambda_{l_i}-\lambda_{l_j}\|^2\\ \notag
&\quad + 
p(\mathcal{L})\kappa\sum_{i=0}^{K-1} p(\mathcal{L}_i)\|\lambda_i\|^2 
-\sum_{i=0}^{K-2}\sum_{j=i+1}^{K-1}p(\mathcal{L}_i)p(\mathcal{L}_j)\|\lambda_i - \lambda_j\|^2 \\ \notag
&\quad -p(\mathcal{L})^2\kappa\sum_{j\in l}\|\lambda_j\|^2 - p(\mathcal{L})\kappa\sum_{i=0}^{K-1}p(\mathcal{L}_i)\|\lambda_i\|^2 \\ \notag
&\quad  + p(\mathcal{L})\sum_{j\in l}\sum_{i=0}^{K-1}p(\mathcal{L}_i)\|\lambda_j - \lambda_i\|^2 \bigg) \\ \notag
&= 
\frac{1}{p(\mathcal{L})^2\kappa^2}\bigg(
p(\mathcal{L})\sum_{j\in l}\sum_{i=0}^{K-1} p(\mathcal{L}_i)\|\lambda_j - \lambda_i\|^2 \\ \notag
&\quad-p(\mathcal{L})^2\sum_{i=0}^{\kappa-2}\sum_{j=i+1}^{\kappa-1}\|\lambda_{l_i}-\lambda_{l_j}\|^2 \\ \label{eq:sideperm}
&\quad-\sum_{i=0}^{K-2}\sum_{j=i+1}^{K-1}p(\mathcal{L}_i)p(\mathcal{L}_j)\|\lambda_i - \lambda_j\|^2  \bigg).
\end{align}}
It follows from Proposition~\ref{prop:riemann2_asym}, (\ref{eq:sideperm}) and Lemma~\ref{lem:sumLj} that we can write
\begin{equation*}
\begin{split}
&\frac{1}{L}\sum_{\lambda_c\in V_{\pi}(0)}\bigg\|\frac{1}{\kappa}\sum_{j\in l}\lambda_j - \frac{1}{p(\mathcal{L})\kappa}\sum_{i=0}^{K-1}p(\mathcal{L}_i)\lambda_i \bigg\|^2 \\
&\approx
\frac{1}{ p(\mathcal{L})^2\kappa^2}\bigg(
p(\mathcal{L})\sum_{j\in l}\sum_{\substack{i=0\\i\neq j}}^{K-1} p(\mathcal{L}_i)- p(\mathcal{L})^2\sum_{i=0}^{\kappa-2}\sum_{j=i+1}^{\kappa-1}
-\sum_{i=0}^{K-2}\sum_{j=i+1}^{K-1} p(\mathcal{L}_i)p(\mathcal{L}_j) \bigg)\\
&\quad
\times\psi_L^{2}\nu^{2/L} G(S_L)N_\pi\prod_{m=0}^{K-1}N_m^{2/L(K-1)}\\
&=
\frac{1}{p(\mathcal{L})^2\kappa^2}\bigg(
p(\mathcal{L})^2\kappa^2 - p(\mathcal{L})\sum_{j\in l} p(\mathcal{L}_j)
-p(\mathcal{L})^2\binom{\kappa}{2} \\
&\quad-\sum_{i=0}^{K-2}\sum_{j=i+1}^{K-1}p(\mathcal{L}_i)p(\mathcal{L}_j) \bigg)
\psi_L^{2}\nu^{2/L} G(S_L)N_\pi\prod_{m=0}^{K-1}N_m^{2/L(K-1)}.
\end{split}
\end{equation*}
This completes the proof.
\end{proof}

\begin{proposition}\label{prop:normdiff}
For any $1\leq \kappa \leq K$ and $l\in \mathcal{L}$ we have
\begin{equation*}
\sum_{\lambda_c\in V_\pi(0)}\left\|\frac{1}{\kappa}\sum_{j\in l}\lambda_j - \frac{1}{p(\mathcal{L})\kappa}\sum_{i=0}^{K-1}p(\mathcal{L}_i)\lambda_i \right\| =
\mathcal{O}\left(\nu^{1/L}N_\pi\prod_{m=0}^{K-1}N_m^{1/L(K-1)} \right)
\end{equation*}
\end{proposition}

\begin{proof}
Recall that the sublattice points $\lambda_i$ and $\lambda_j$ satisfy $\|\lambda_i-\lambda_j\|\leq r$, where $r=(\tilde{\nu}/\omega_L)^{1/L}$ is the radius of $\tilde{V}$. Hence, without loss of generality, we let $\lambda_j=r$ and $\lambda_i=0$, which leads to
\begin{equation*}
\begin{split}
\sum_{\lambda_c\in V_\pi(0)}\left\|\frac{1}{\kappa}\sum_{j\in l}\lambda_j - \frac{1}{p(\mathcal{L})\kappa}\sum_{i=0}^{K-1} p(\mathcal{L}_i)\lambda_i\right\| &\leq
rN_\pi \\
&= \mathcal{O}\left(\nu^{1/L}N_\pi\prod_{m=0}^{K-1}N_m^{1/L(K-1)}\right),
\end{split}
\end{equation*}
since $\tilde{\nu}=\psi_L^L\nu\prod_{m=0}^{K-1}N_m^{1/(K-1)}$.
\end{proof}

\begin{proposition}\label{prop:lclmean}
For $1\leq \kappa \leq K\leq 3$, $l\in \mathcal{L}$, $N_i\rightarrow \infty$ and $\nu_i\rightarrow 0$ we have
\begin{equation*}
\sum_{\lambda_c\in V_\pi(0)}\left\| \frac{1}{\kappa}\sum_{j\in l}\lambda_j - \lambda_c\right\|^2 = \sum_{\lambda_c\in V_\pi(0)}\left\| \frac{1}{\kappa}\sum_{j\in l}\lambda_j - \frac{1}{p(\mathcal{L})\kappa}\sum_{i=0}^{K-1} p(\mathcal{L}_i)\lambda_i\right\|^2.
\end{equation*}
\end{proposition}
\begin{proof}
Let $\bar{\lambda} = \frac{1}{p(\mathcal{L})\kappa}\sum_{i=0}^{K-1}p(\mathcal{L}_i)\lambda_i$ and $\lambda' = \frac{1}{\kappa}\sum_{j\in l}\lambda_j$. We now follow~\cite[Eqs. (67) -- (72)]{diggavi:2002} and obtain the following inequalities:
\begin{equation*}\label{eq:lambdacmean1}
\begin{split}
\|\lambda' - \lambda_c\|^2 &= \|\lambda' - \bar{\lambda} + \bar{\lambda} - \lambda_c\|^2 \\
&= \|\lambda' - \bar{\lambda}\|^2 + \|\bar{\lambda} - \lambda_c\|^2 + 2\langle(\lambda'-\bar{\lambda}),(\bar{\lambda}-\lambda_c)\rangle,
\end{split}
\end{equation*}
from which we can establish the inequality
\begin{equation*}
\begin{split}
\|\lambda' - \bar{\lambda}\|^2 + \|\bar{\lambda} - \lambda_c\|^2 &- 2|\langle(\lambda'-\bar{\lambda}),(\bar{\lambda}-\lambda_c)\rangle|
\leq \|\lambda' - \lambda_c \|^2 \\
&\leq \|\lambda' - \bar{\lambda}\|^2 + \|\bar{\lambda} - \lambda_c\|^2 + 2|\langle(\lambda'-\bar{\lambda}),(\bar{\lambda}-\lambda_c)\rangle|.
\end{split}
\end{equation*}
Using the Cauchy-Schwartz inequality\index{inequality!Cauchy-Schwartz} we get
\begin{equation*}
\begin{split}
\|\lambda' - \bar{\lambda}\|^2 + \|\bar{\lambda} - \lambda_c\|^2 &- 2\|(\lambda'-\bar{\lambda})\|\|(\bar{\lambda}-\lambda_c)\|
\leq \|\lambda' - \lambda_c \|^2 \\
&\leq \|\lambda' - \bar{\lambda}\|^2 + \|\bar{\lambda} - \lambda_c\|^2 + 2\|(\lambda'-\bar{\lambda})\|\|(\bar{\lambda}-\lambda_c)\|.
\end{split}
\end{equation*}
which can be rewritten as
\begin{align*}
\|\lambda' - \bar{\lambda}\|^2 \left(1- \frac{\|\bar{\lambda} - \lambda_c\|}{\|\lambda' - \bar{\lambda}\|}\right)^2 
&\leq \|\lambda' - \lambda_c \|^2 \\
&\leq\|\lambda' - \bar{\lambda}\|^2 \left(1+ \frac{\|\bar{\lambda} - \lambda_c\|}{\|\lambda' - \bar{\lambda}\|}\right)^2 
\end{align*}
Summing over $\lambda_c \in V_\pi(0)$ and observing that $\|\bar{\lambda} - \lambda_c\|^2 \geq 0$, we get
\begin{equation*}
\begin{split}
\sum_{\lambda_c\in V_\pi(0)}&\left( \|\lambda' - \bar{\lambda}\|^2 -
2\|\lambda'-\bar{\lambda}\|\|\bar{\lambda}-\lambda_c\| \right)\leq 
\sum_{\lambda_c\in V_\pi(0)}\|\lambda_c - \lambda'\|^2 \\
&\leq \sum_{\lambda_c\in V_\pi(0)}\left( \|\lambda' - \bar{\lambda}\|^2 +
\|\bar{\lambda} - \lambda_c\|^2 +
2\|\lambda'-\bar{\lambda}\|\|\bar{\lambda}-\lambda_c\| \right),
\end{split}
\end{equation*}
which can be rewritten as
\begin{align}\label{eq:cauchy1}
&\hspace{-5mm}\left( \sum_{\lambda_c\in V_\pi(0)} \|\lambda' - \bar{\lambda}\|^2\right)
\left( 1 - 2\frac{\sum_{\lambda_c\in V_\pi(0)}\|\lambda' - \bar{\lambda}\|\|\bar{\lambda}-\lambda_c\|}
{\sum_{\lambda_c\in V_\pi(0)}\|\lambda' - \bar{\lambda}\|^2}\right)\\ 
&\quad\leq \sum_{\lambda_c\in V_\pi(0)}\|\lambda_c - \lambda'\|^2 \\ \notag
&\quad\leq
\left( \sum_{\lambda_c\in V_\pi(0)} \|\lambda' - \bar{\lambda}\|^2\right)\\
&\quad\times
\left( 1 + \frac{\sum_{\lambda_c\in V_\pi(0)}\|\bar{\lambda}-\lambda_c\|^2}
{\sum_{\lambda_c\in V_\pi(0)}\|\lambda' - \bar{\lambda}\|^2}
+2\frac{\sum_{\lambda_c\in V_\pi(0)}\|\lambda' - \bar{\lambda}\|\|\bar{\lambda}-\lambda_c\|}
{\sum_{\lambda_c\in V_\pi(0)}\|\lambda' - \bar{\lambda}\|^2}\right) \label{eq:cauchy2}.
\end{align}
By use of~(\ref{eq:o1_asym}) and Proposition~\ref{prop:normdiff} it is possible to upper bound the numerator of the fraction in~(\ref{eq:cauchy1}) by
\begin{equation*}
\begin{split}
\sum_{\lambda_c\in V_\pi(0)}\|\lambda' - \bar{\lambda}\|\|\bar{\lambda}-\lambda_c\| 
&= \mathcal{O}\left((N_k\nu)^{1/L} N_\pi\nu^{1/L}\prod_{m=0}^{K-1}N_m^{1/L(K-1)}\right),
\end{split}
\end{equation*}
since the covering radius of the $k^{th}$ sublattice is proportional to $(N_k\nu)^{1/L}$, where $N_k$ is the minimum of $N_i, i=0,\dots, K-1$.

By use of Proposition~\ref{prop:normdiff2} it is easily seen that the denominator in~(\ref{eq:cauchy1}) grows as
\begin{equation*}
\sum_{\lambda_c \in V_\pi(0)}\|\lambda' - \bar{\lambda}\|^2 = \mathcal{O}\left(\nu^{2/L}N_\pi\prod_{m=0}^{K-1}N_m^{2/L(K-1)}\right),
\end{equation*}
hence the fraction in~(\ref{eq:cauchy1}) go to zero for $N_i\rightarrow \infty$. By a similar analysis it is easily seen that the fractions in~(\ref{eq:cauchy2}) also go to zero as $N_i\rightarrow \infty$.

Based on the asymptotic behavior of the fractions in (\ref{eq:cauchy1}) and (\ref{eq:cauchy2}) we see that (asymptotically as $N_i\rightarrow\infty$)
\begin{equation*}
\sum_{\lambda_c\in V_\pi(0)}\|\lambda' - \bar{\lambda}\|^2 \leq
\sum_{\lambda_c\in V_\pi(0)}\|\lambda_c - \lambda'\|^2 \leq
\sum_{\lambda_c\in V_\pi(0)}\|\lambda' - \bar{\lambda}\|^2,
\end{equation*}
hence
\begin{equation*}
\sum_{\lambda_c\in V_\pi(0)}\|\lambda_c - \lambda'\|^2 \approx \sum_{\lambda_c\in V_\pi(0)}\|\lambda' - \bar{\lambda}\|^2,
\end{equation*}
which completes the proof.
\end{proof}

We are now in a position to prove Theorem~\ref{theo:sidedist}.
\begin{theo_empty}{\ref{theo:sidedist}}
The side distortion $D^{(K,l)}$ due to reception of descriptions $\{l\}$, where $l\in \mathcal{L}$ for any $1\leq \kappa\leq K\leq 3$ is, asymptotically as $L,N_i\rightarrow \infty$ and $\nu_i \rightarrow 0$, given by
\begin{equation*}
D^{(K,l)} = \omega^{(K,l)}
\psi_L^{2}\nu^{2/L} G(S_L)\prod_{i=0}^{K-1}N_i^{2/L(K-1)},
\end{equation*}
where
\begin{equation*}
\begin{split}
\omega^{(K,l)}&=\frac{1}{p(\mathcal{L})^2\kappa^2}\bigg(p(\mathcal{L})^2\kappa^2- p(\mathcal{L})^2\binom{\kappa}{2} - p(\mathcal{L})\sum_{j\in l} p(\mathcal{L}_j) \\
&\quad -
\sum_{i=0}^{K-2}\sum_{j=i+1}^{K-1}p(\mathcal{L}_i)p(\mathcal{L}_j)\bigg),
\end{split}
\end{equation*}
where $\binom{\kappa}{2}=0$ for $\kappa=1$.
\end{theo_empty}

\begin{proof}
By use of~(\ref{eq:di1}) we can write the distortion as 
\begin{equation}\label{eq:dldist}
\begin{split}
D^{(K,l)} &= \frac{1}{L}E\left\|\frac{1}{\kappa}\sum_{j\in l}\lambda_j - X\right\|^2 \\
&\approx D_c + \frac{1}{L}\frac{1}{N_\pi}\sum_{\lambda_c\in V_{\pi}(0)}\left\|\frac{1}{\kappa}\sum_{j\in l}\lambda_j - \lambda_c\right\|^2 \\
&\approx \frac{1}{L}\frac{1}{N_\pi}\sum_{\lambda_c\in V_{\pi}(0)}\left\|\frac{1}{\kappa}\sum_{j\in l}\lambda_j - \lambda_c\right\|^2,
\end{split}
\end{equation}
where the second approximation follows from that fact that as $N_i\rightarrow \infty$, the distortion due to the index assignment is dominating. Furthermore, by use of Propositions~\ref{prop:lclmean} and~\ref{prop:normdiff2} in~(\ref{eq:dldist}) we are able to write
\begin{equation*}
\begin{split}
D^{(K,l)} &\approx \frac{1}{L}\frac{1}{N_\pi}\sum_{\lambda_c\in V_{\pi}(0)}\left\|\frac{1}{\kappa}\sum_{j\in l}\lambda_j - \frac{1}{p(\mathcal{L})\kappa}\sum_{i=0}^{K-1} p(\mathcal{L}_i) \lambda_i\right\|^2 \\
&\approx \omega^{(K,l)}
\psi_L^{2}\nu^{2/L}G(S_L)\prod_{m=0}^{K-1}N_m^{2/L(K-1)},
\end{split}
\end{equation*}
which completes the proof.
\end{proof}

\chapter{Proofs for Chapter~\ref{chap:comparison}}
\label{app:proofs_comparison}
This appendix contains proofs of the Lemmas and Theorems presented in Chapter~\ref{chap:comparison}. 

\section{Proofs of Lemmas}\label{app:comparison_proofs_lemmas}
\begin{proof}[Proof of Lemma~\ref{lem:31scecdist}]
For $K=3,k=1$ and $R_s\rightarrow\infty$ we see from~(\ref{eq:sigmanorm}) that 
\begin{equation}\label{eq:approxsigmaq2}
\begin{split}
\sigma_q^{2} &= \left((1-\rho_q)2^{2R_s}\left(\frac{1+2\rho_q}{1-\rho_q}\right)^{1/3} - 1\right)^{-1}  \\
&\approx (1-\rho_q)^{-2/3}(1+2\rho_q)^{-1/3} 2^{-2R_s},
\end{split}
\end{equation}
where the approximation follows from the high resolution assumption which implies that $2^{2R_s}\gg 1$. 
With this, we can write the optimal single-channel distortion of a $(3,1)$ SCEC, which is given by~(\ref{eq:MMSE}), as
\begin{equation}\label{eq:D1approx}
\begin{split}
D^{(3,1)} &= \frac{\sigma_q^2}{\sigma_q^2 + 1} \\
&\approx  \sigma_q^2,
\end{split}
\end{equation}
where the approximation follows since $\sigma_q^2\ll 1$.
We now equalize the single-channel distortion of three-channel MD-LVQ (or (3,1) MD-LVQ) and (3,1) SCECs (i.e.\ we set (\ref{eq:sidedist1}) equal to~(\ref{eq:D1approx})) so that we can express $\rho_q$ as a function of $N'$. This leads to
\begin{equation}\label{eq:rhoN}
1+2\rho_q = \left(\frac{3}{\psi_\infty^2 N'}\right)^3(1-\rho_q)^{-2}.
\end{equation}
Using~(\ref{eq:MMSE}) we rewrite the two-channel distortion of (3,1) SCECs as
{\allowdisplaybreaks
\begin{align} \notag
D^{(3,2)} &= \frac{\sigma_q^2(1+\rho_q)}{\sigma_q^2(1+\rho_q) + 2} \\ \notag
&\overset{(a)}{\approx} \frac{1}{2}\sigma_q^2(1+\rho_q) \\ \notag
&\overset{(b)}= \frac{1+\rho_q}6\psi_\infty^2N'2^{-2R_s} \\ \label{eq:opt2channel}
&\overset{(c)}{\approx} \frac{1}{12}\psi_\infty^2N'2^{-2R_s}
\end{align}}%
where $(a)$ is true at high resolution since $\sigma_q^2\ll 1$, $(b)$ follows by replacing $\sigma_q^2$ with~(\ref{eq:approxsigmaq2}) and inserting~(\ref{eq:rhoN}) and $(c)$ is valid for large $N'$ since $N'\gg 1$ implies that $\rho_q\approx -1/2$.
Similarly, by using~(\ref{eq:rhoN}) in~(\ref{eq:MMSE}) the optimal three-channel distortion can be written as
\begin{equation}\label{eq:opt3channel}
\begin{split}
D^{(3,3)} &= \frac{\sigma_q^2(1+2\rho_q)}{\sigma_q^2(1+2\rho_q) + 3} \\
&\approx \frac{1}{3}\sigma_q^2(1+2\rho_q) \\
&= 3\psi_\infty^{-4}(1-\rho_q)^{-2}\left(\frac{1}{N'}\right)^{2}2^{-2R_s} \\
&\approx \left(\frac{1}{N'}\right)^{2}2^{-2R_s},
\end{split}
\end{equation}
where the first approximation is valid when $\sigma_q^2\ll 1$ and the second follows since $\rho_q\approx -1/2$. Comparing~(\ref{eq:opt2channel}) and~(\ref{eq:opt3channel}) to~(\ref{eq:sidedist2}) and~(\ref{eq:centraldist1}) shows that three-channel MD-LVQ reach the achievable rate-distortion region of a $(3,1)$ SCEC in the quadratic Gaussian case at high resolution.
\end{proof}

\begin{proof}[Proof of Lemma~\ref{lem:boundCi}]
Let $\Ae$ denote the set of epsilon-typical sequences~\cite{cover:1991} and note that $\Ae$ must have bounded support\index{bounded support} since, for any $L$, $f_X(x_0,\dots,x_{L-1})>2^{-L(h(X)+\epsilon)}$ for $x\in\Ae$ and 
$$\int_{\Ae}f_X(x_0,\dots,x_{L-1})dx \leq 1.$$
Let the side quantizers of an MD-LVQ system be SD entropy-constrained lattice vector quantizers. An SD lattice vector quantizer designed for an output entropy of, say $R_i$, for the $L$-dimensional uniform source with bounded support (in fact matched to the support of \Ae) has a finite number of codewords given by $2^{LR_i}$. The distortion performance of a lattice vector quantizer is, under high-resolution assumptions, inde\-pendent of the source pdf~\cite{linder:1994,conway:1999}. Therefore, using this quantizer for $\Ae$ instead of a truly uniformly distributed source will not affect the distortion performance but it might affect the rate. However, since the bounded uniform distribution is entropy maximizing it follows that $R_i$ upper bounds the rate of the quantizer.
\end{proof}

\begin{proof}[Proof of Lemma~\ref{lem:32scecdist}]
The two-channel distortion of a $(3,2)$ SCEC is given by
\begin{equation}\label{eq:D2bin2}
D^{(3,2)} \approx \frac{1}{2}\sigma_q^2(1+\rho_q),
\end{equation}
where from~(\ref{eq:sigmanorm}) we see that
\begin{equation}\label{eq:sigmaqbin2}
\sigma_q^2 = 2 (1-\rho_q)^{-1/3}(1+2\rho_q)^{-2/3}2^{-4R_b}.
\end{equation}
Inserting~(\ref{eq:sigmaqbin2}) into~(\ref{eq:D2bin2}) and setting the result equal to~(\ref{eq:D2barbin}), i.e.\ we normalize such that the two-channel distortion of (3,2) SCECs is equal to that of (3,2) MD-LVQ. This leads to 
\begin{equation}
(1+\rho_q)(1-\rho_q)^{-1/3}(1+2\rho_q)^{-2/3} = \frac{1}{12}\psi_\infty^4(N')^2,
\end{equation}
from which we find that
\begin{equation}
(1+2\rho_q)^{1/3} = \left(\frac{1}{12}\psi_\infty^4(N')^2\right)^{-1/2}(1+\rho_q)^{1/2}(1-\rho_q)^{-1/6}.
\end{equation}
It follows that we can write $D^{(3,3)}$ as
\begin{align}\notag
D^{(3,3)} &= \frac{1}{3}\sigma_q^2(1+2\rho_q) \\ \notag
    &= \frac{2}{3}(1-\rho_q)^{-1/3}(1+2\rho_q)^{1/3}2^{-4R_b} \\ \notag
    &= \frac{2}{3}(1-\rho_q)^{-1/3}\sqrt{12}(1+\rho_q)^{1/2}(1-\rho_q)^{-1/6}\psi_\infty^{-2}(N')^{-1}2^{-4R_b} \\ \label{eq:D3bin2}
    &\approx \frac{\psi_\infty^{2}}{N'} 2^{-4R_b},
\end{align}
where the approximation follows by inserting $\rho\approx -1/2$. The proof is now complete since~(\ref{eq:D3bin2}) is identical to~(\ref{eq:Dcbin2a}).
\end{proof}

\section{Proof of Theorem~\ref{theo:scec-lc}}\label{app:theo:scec-lc}
Since $(3,2)$ MD-LVQ is closely related to $(3,2)$ SCECs we can to some extent use the proof techniques of~\cite{pradhan:2004}. However, there are some important differences. We cannot rely on random coding arguments since we are not using random codebooks. For example where~\cite{pradhan:2004} exploit properties of the entropy of subsets, we need to show that certain properties hold for all subsets and not just on average. Furthermore, we consider the asymmetric case where the individual codebook rates $R_i$ and binning rates $R_{b,i}$ are allowed to be unequal whereas in~\cite{pradhan:2004} the symmetric case was conside\-red, i.e.\ only a single codebook rate $R_s$ and a single binning rate $R_b$ was taken into account.

\begin{theo_empty}{\ref{theo:scec-lc}}
Let $X\in \mathbb{R}^L$ be a source vector constructed by blocking an arbitrary i.i.d.\ source with finite differential entropy into sequences of length $L$. Let $J\subseteq\{0,\dots,K-1\}$ and let $\lambda_{J}$ denote the set of codewords indexed by $J$. The set of decoding functions is denoted $g_J\colon \bigotimes_{j\in J}\Lambda_j \to \mathbb{R}^L$. Then, under high-resolution assumptions, if 
\begin{equation*}
E[\rho(X,g_{J}(\lambda_J))] \leq D^{(K,J)},\quad \forall J\in \mathcal{K},
\end{equation*}
where $\rho(\cdot,\cdot)$ is the squared-error distortion measure and for all $S\subseteq J$
\begin{equation}\label{eq:sumbinrateR_proof}
\sum_{i\in S} R_{b,i} > \sum_{i\in S}\gamma_i + \frac{1}{L}\log_2(|\{\lambda_S|\lambda_{J-S}\}|),
\end{equation}
the rate-distortion tuple $(R_{b,0},\dots,R_{b,(K-1)},\{D^{(K,J)}\}_{J\in\mathcal{K}})$ is achievable.
\end{theo_empty}

\begin{proof}[Proof of Theorem~\ref{theo:scec-lc}.]
Define the following error events.
\begin{enumerate}
\item $\mathcal{E}_0:$ $X$ does not belong to $A_\epsilon^{(L)}(X)$.
\item $\mathcal{E}_1:$ There exists no indices $(j_0,\dots,j_{K-1})$ such that $(\lambda_0(j_0),\dots,\lambda_{K-1}(j_{K-1})) = \alpha(\lambda_c)$ for $\lambda_c = Q(X)$.
\item $\mathcal{E}_2:$ Not all channel indices are valid.
\item $\mathcal{E}_3:$ For some $k$ received bin indices there exists another admissible $k$-tuple in the same bins.
\end{enumerate}
As usual we have $\mathcal{E}=\bigcup_{i=0}^{K-1}\mathcal{E}_i$ and the probability of error is bounded from above by the union bound, i.e.\ $P(\mathcal{E})\leq \sum_{i=0}^{K-1} P(\mathcal{E}_i)$. 

\emph{Bounding $P(\mathcal{E}_0)$:}
Applying standard arguments for typical sequences it can be shown that $P(\mathcal{E}_0)\rightarrow 0$ for $L$ sufficiently large~\cite{cover:1991}. We may now assume the event $\mathcal{E}_0^c$, i.e.\ all source vectors belong to the set of typical sequences and hence they are approximately uniformly distributed. 

\emph{Bounding $P(\mathcal{E}_1)$:}
The source vector $X$ is encoded by the central quantizer using a nearest neighbor rule. Since any source vector will have a closest element (which might not be unique) in $\mathcal{C}_c$ and by construction all $\lambda_c\in\mathcal{C}_c$ have an associated $K$-tuple of sublattice points, it follows that $P(\mathcal{E}_1)=0$ for all $L$.

\emph{Bounding $P(\mathcal{E}_2)$:}
We only have to prove this for one of the channels. Then by symmetry it holds for all of them. Furthermore, since the intersection of a finite number of sets of probability 1 is 1 it follows that with probability 1 a codeword $\lambda_i$ given $\lambda_c$ can be found in some bin. In the following we assume $K<\infty$. Let $\lambda_c$ be the codeword associated with $X$ (i.e.\ $X$ is quantized to $\lambda_c$), where $X\in A_\epsilon^{(L)}(X)$. Let $A$ denote the event that $\lambda_0(j_0)$ exists in the codebook $\mathcal{C}_0$, i.e.\ the event $[\lambda_0(j_0)\in \mathcal{C}_0, \lambda_0(j_0)=\alpha_0(\lambda_c)]$. We then have that
\begin{equation*}
\begin{split}
P(f_0(\lambda_0(j_0))\neq \vartheta)&=P(f_0(\lambda_0(j_0))\neq \vartheta | A^c)P(A^c) \\
&\quad+ P(f_0(\lambda_0(j_0))\neq \vartheta | A)P(A),
\end{split}
\end{equation*}
where the first term on the right hand side is zero if we make sure that all $\lambda_c$'s are assigned a (unique) $K$-tuple. Therefore, we only have to look at the second term as was the case in~\cite{pradhan:2004}. We must show that $G= P[f_0(\lambda_0(j_0))\neq \vartheta| A]\rightarrow 1$, i.e.\ 
\begin{equation}\label{eq:notG}
1-G = P[\lambda_{0_j}\neq \lambda_0(j_0), 0\leq j\leq M_0-1 | A],
\end{equation}
where $M_0=\xi_0 2^{LR_{b,0}}=2^{L(R_0+\gamma_0)}$ is the total number of codewords selected for all bins from $\mathcal{C}_0$ and $\lambda_{0_j}$ indicates the $j^{th}$ such selected codeword. Since the codewords $\lambda_{0_j}$ are chosen independently (uniformly) and with replacement they all have the same probability of being equal to $\lambda_0(j_0)$, so we let $j=0$ and rewrite~(\ref{eq:notG}) as
\begin{equation*}
1-G=[P(\lambda_{0_0}\neq \lambda_0(j_0)|A)]^{M_0}.
\end{equation*}
The size of $\mathcal{C}_0$ is $|\mathcal{C}_0|$ and all codewords of $\mathcal{C}_0$ are equally probable so 
\begin{equation}\label{eq:notG1}
1-G=\left(1-\frac{1}{|\mathcal{C}_0|}\right)^{M_0}.
\end{equation}
Taking logs and invoking the log-inequality\footnote{The log-inequality is given by $\log(z)\leq z-1, z>0$, where $\log$ denotes the natural logarithm.}\index{inequality!log}, Eq.~(\ref{eq:notG1}) can be rewritten as
\begin{equation*}
\log(1-G)\leq -\frac{M_0}{|\mathcal{C}_0|} 
= -\frac{2^{L(R_0+\gamma_0)}}{|\mathcal{C}_0|},
\end{equation*}
which goes to $-\infty$ for $L\rightarrow \infty$ if $R_0+\gamma_0 > \frac{1}L\log_2(|\mathcal{C}_0|)$. By use of Lemma~\ref{lem:boundCi} we have $|C_0|=2^{LR_0}$ so that $1-G\rightarrow 0$ for $L\rightarrow \infty$ if $\gamma_0>0$.

\emph{Bounding $P(\mathcal{E}_3)$:}
Assume we receive $k$ bin indices from the encoder. We then need to show that there is a unique set of codewords (one from each bin) which form an admissible $k$-tuple. Let $J=\{i_0,\dots,i_{k-1}\}$. Along the lines of~\cite{pradhan:2004} we define the following error event for any $S\subseteq J:$
\begin{equation*}
\begin{split}
\mathcal{E}'_S &: \exists j_i' \neq j_i, \forall i \in S, f_S(\lambda_S(j_S')) =
f_S(\lambda_S(j_S)),\\
& (\lambda_S(j_S'),\lambda_{J-S}(j_{J-S}))=\alpha_J(\lambda_c),\quad \lambda_c\in C_c,
\end{split}
\end{equation*}
i.e.\ that there exist more than one admissible $k$-tuple in the given $k$ bins. The event $\mathcal{E}_3$ can be expressed as $\mathcal{E}_3 = \bigcup_{S\subseteq J}\mathcal{E}'_S$.
The probability of the error event $\mathcal{E}'_S$ can be upper bounded by
\begin{equation*}
P(\mathcal{E}'_S)\leq \prod_{i\in S}(\xi_i-1)P[(\lambda_S^*,\lambda_{J-S}(j_{J-S}))=\alpha_J(\lambda_c)],
\end{equation*}
for some $\lambda_c$, where $\lambda_i^*$ is a randomly chosen vector from $\mathcal{C}_i$ for $i\in S$. Let $\{\lambda_S|\lambda_{J-S}\}$ denote the set of admissible $k$-tuples that contains $\lambda_{J-S}$ so that
\begin{equation*}
P[(\lambda_S^*,\lambda_{J-S}(j_{J-S}))=\alpha_J(\lambda_c)] < 
\frac{|\{\lambda_S|\lambda_{J-S}\}|}{\prod_{i\in S}|\mathcal{C}_i|}.
\end{equation*}
We are then able to bound $P(\mathcal{E}_s')$ by
\begin{equation*}
\begin{split}
P(\mathcal{E}_S')&< \prod_{i\in S}\xi_i\frac{|\{\lambda_S|\lambda_{J-S}\}|}{\prod_{i\in S}|\mathcal{C}_i|} \\
&= \prod_{i\in S}2^{L(\gamma_i-R_{b,i})}|\{\lambda_S|\lambda_{J-S}\}|,
\end{split}
\end{equation*}
which goes to zero if
\begin{equation}\label{eq:binningrateR}
\sum_{i\in S}R_{b,i} > \sum_{i\in S}\gamma_i + \frac{1}{L}\log_2(|\{\lambda_S|\lambda_{J-S}\}|).
\end{equation}

Finally, the expected distortion is bounded by $P(\mathcal{E}^c)D_J + P(\mathcal{E})d_\text{max}, \forall J\in \mathcal{K}$ where $P(\mathcal{E}) \rightarrow 0$ for $L\rightarrow \infty$ and assuming that the distortion measure is bounded, i.e.\ $d_\text{max}< \infty$, proves the theorem.\footnote{We here make the assumption, as appears to be customary, that the distortion measure is bounded also for sources with unbounded support.}
\end{proof}

\chapter{Results of Listening Test}\label{app:nac_results}

In this appendix we present the results of the MUSHRA listening test described in Chapter~\ref{chap:nac}.

%
%
\begin{figure}[ht]
\begin{center}
\psfrag{P=10\%}{\small $p=10\%$}
\psfrag{P=30\%}{\small $p=30\%$}
\psfrag{P=50\%}{\small $p=50\%$}
\includegraphics[width=11cm]{nac/figs/mushrares.eps} 
\caption{MUSHRA test results averaged over all three audio clips for $p=0.1,0.3$ and $p=0.5$. The seven signals appear in the following order: Hidden ref., 3.5 kHz, 7 kHz, $K=1, K=2, K=3$ and $K=4$.}
\label{fig:mushrares_app}
\end{center}
\end{figure}

\begin{figure}[ht]
\begin{center}
\subfigure[$p=0.1$]{\includegraphics[width=7cm]{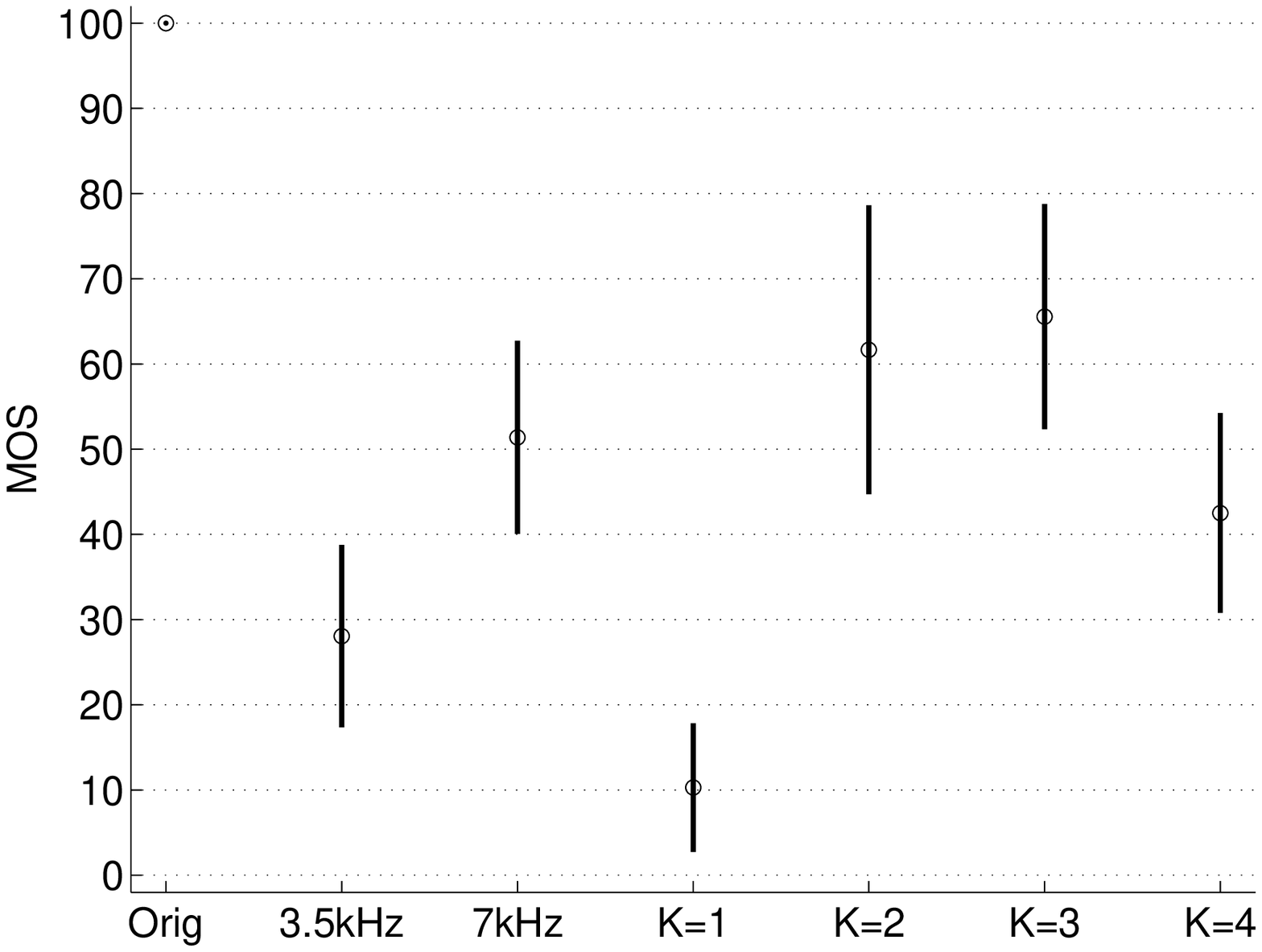}}\par
\subfigure[$p=0.3$]{\includegraphics[width=7cm]{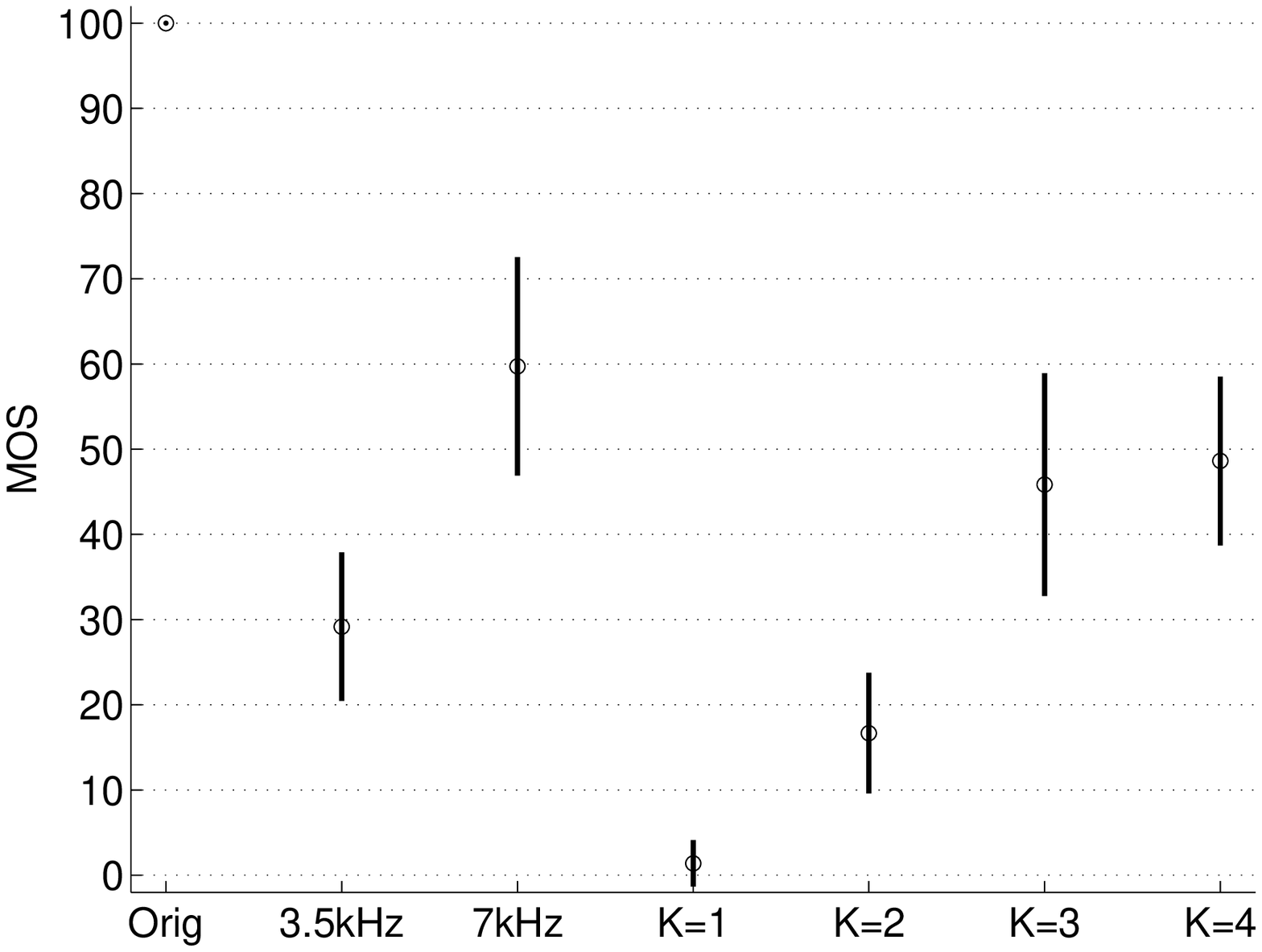}}\par
\subfigure[$p=0.5$]{\includegraphics[width=7cm]{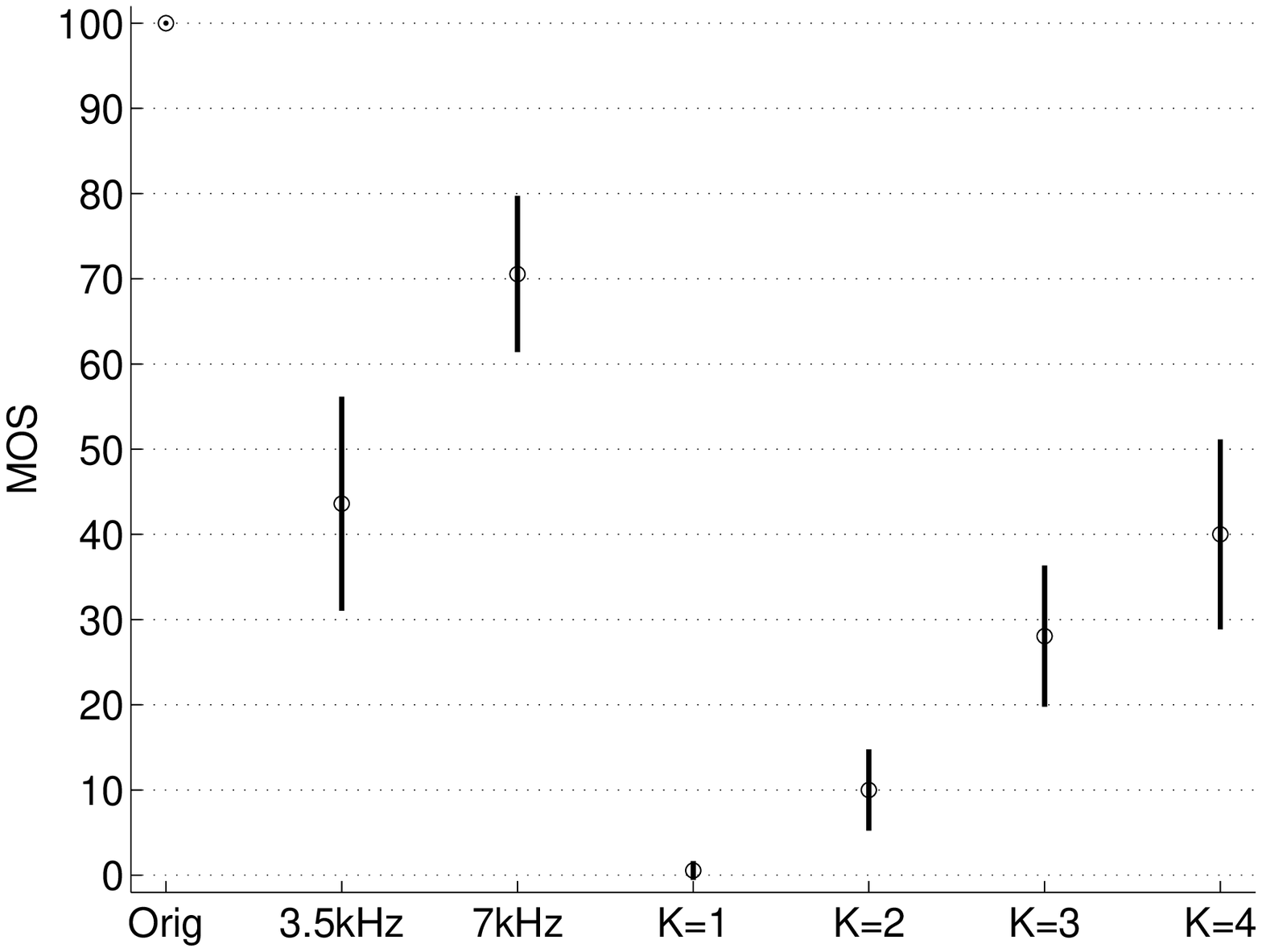}}
\caption{MUSHRA test results for the jazz fragment and $p=0.1,0.3$ and $p=0.5$.}
\label{fig:mushra_t2}
\end{center}
\end{figure}

\begin{figure}[ht]
\begin{center}
\subfigure[$p=0.1$]{\includegraphics[width=7cm]{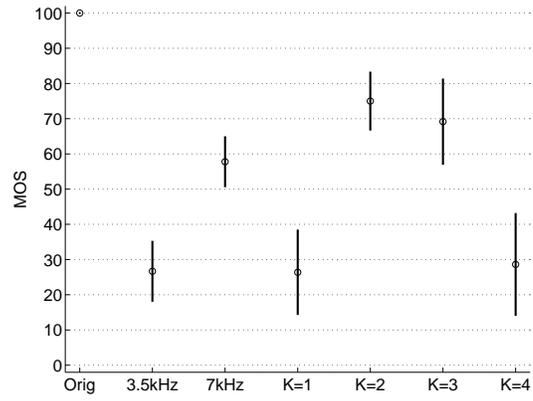}}\par
\subfigure[$p=0.3$]{\includegraphics[width=7cm]{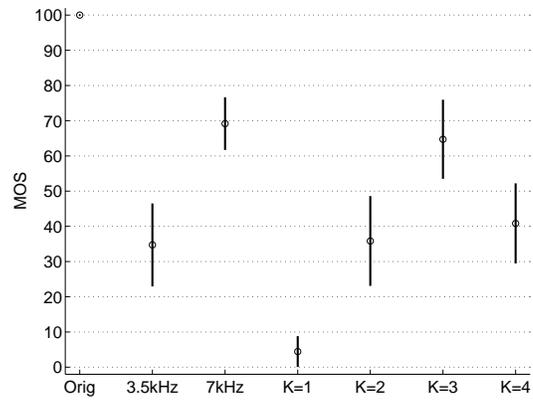}}\par
\subfigure[$p=0.5$]{\includegraphics[width=7cm]{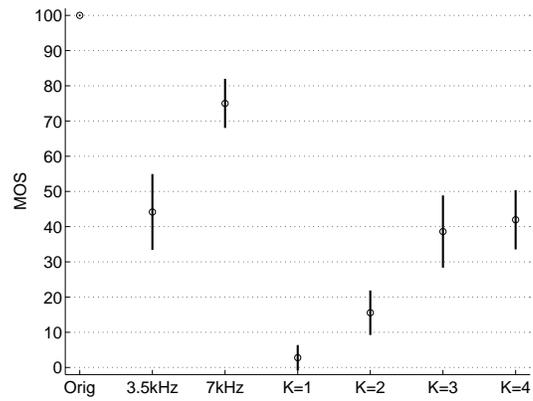}}
\caption{MUSHRA test results for the speech fragment and $p=0.1,0.3$ and $p=0.5$.}
\label{fig:mushra_t7}
\end{center}
\end{figure}

\begin{figure}[ht]
\begin{center}
\subfigure[$p=0.1$]{\includegraphics[width=7cm]{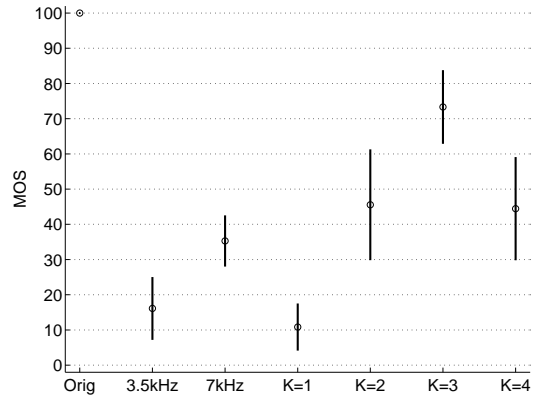}}\par
\subfigure[$p=0.3$]{\includegraphics[width=7cm]{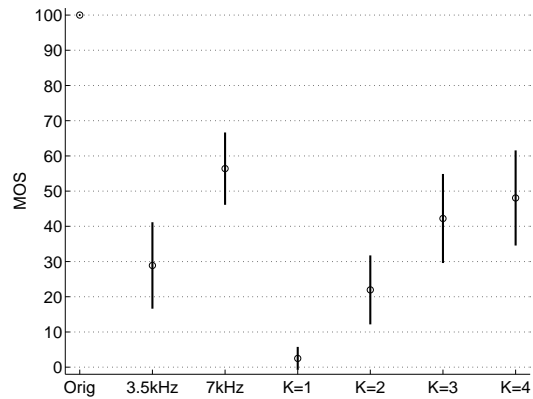}}\par
\subfigure[$p=0.5$]{\includegraphics[width=7cm]{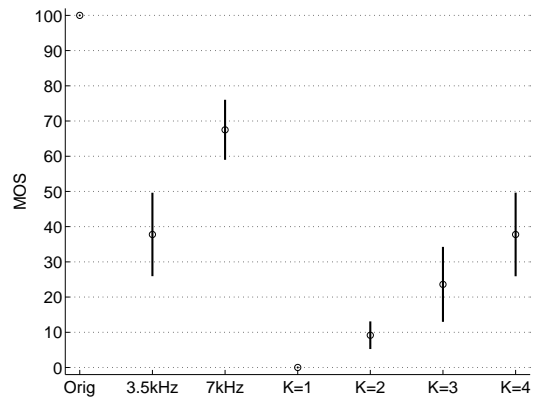}}
\caption{MUSHRA test results for the rock fragment and $p=0.1,0.3$ and $p=0.5$.}
\label{fig:mushra_t19}
\end{center}
\end{figure}

\chapter*{Samenvatting}
\addcontentsline{toc}{chapter}{Samenvatting}
\markboth{Samenvatting}{}
Internetdiensten zoals het voice over Internet-protocol (VoIP) 
en audio/video streaming (b.v. video op verzoek en video vergaderen)  
worden steeds populairder door de recente groei van breedbandnetwerken. 
Dit soort "real-time" diensten vereisen vaak een lage verzendtijd, een hoge bandbreedte en een lage pakket-verlies kans om acceptabele kwaliteit voor de 
eindgebruikers te leveren. De heterogene communicatie infrastructuur van de huidige pakketgeschakelde netwerken verschaffen echter geen 
gegarandeerde prestaties met betrekking tot bandbreedte of verzendtijd en daarom 
wordt de gewenste kwaliteit over het algemeen niet bereikt.

Om een bepaalde mate van robuustheid te bereiken op kanalen waarop fouten kunnen voorkomen, kan multiple-description (MD) coding toegepast worden.
Dit is een methode waar de laatste tijd erg veel aandacht aan is besteed. Het MD probleem is in wezen een gecombineerd bron-kanaal coderingsprobleem dat gaat over (het met verlies) coderen van informatie voor transmissie 
over een onbetrouwbaar $K$-kanalen communicatie systeem. De kanalen kunnen falen, met als resultaat het verlies van een pakket en daardoor een verlies van informatie aan de ontvangende kant. Welke van de $2^K-1$ niet-triviale deelverzamelingen van de $K$ kanalen falen, wordt bekend verondersteld aan de ontvangende kant, maar niet bij de encoder. Het probleem is dan een MD schema te ontwerpen dat, voor gegeven kanaal rate (of een gegeven som rate), de distorsies minimaliseert die een gevolg zijn van reconstruering van de bron, gebruik makend van informatie van willekeurige deelverzamelingen van de kanalen.

Hoewel wij ons in dit proefschrift hoofdzakelijk richten op de informatie theo\-retische aspecten van MD 
codering, zullen we voor de volledigheid ook laten zien hoe het voorge\-stelde MD coderingsschema 
kan worden gebruikt om een perceptueel robuuste audio coder te construeren, die  
geschikt is voor b.v. audio-streaming op pakketgeschakelde netwerken. 

We richten ons op het MD probleem vanuit een bron-codering standpunt en bekijken het 
algemene geval van $K$ pakketten. We maken uitgebreid 
gebruik van lattice vector kwantisatie (LVQ) theorie, hetgeen een goed instrument blijkt, in de zin dat het voorge\-stelde MD-LVQ schema als 
brug tussen theorie en praktijk dient. Voor asymptotische gevallen van hoge 
resolutie en grote lattice vector kwantisator dimensie, tonen wij aan 
dat de beste bekende informatie theoretische rate-distorsie MD grenzen kunnen worden bereikt, terwijl we, in niet asymptotische gevallen van 
eindig-dimensionale lattice vector kwantisators (maar nog onder hoge 
resolutie veronderstelling), praktische MD-LVQ schemas construeren, 
die vergelijkbaar met en vaak superieur zijn aan be\-staande state-of-the-art schemas. 

In het twee-kanaal symmetrische geval is eerder aangetoond dat de zij-representa\-ties van een MD-LVQ schema zij-distorsies toelaten, 
die (bij hoge resolutie voorwaar\-den) identiek zijn aan die van $L$-dimensionale kwantisators met bolvormige Voronoi cellen. In dit geval zeggen wij dat 
de zij-kwantisators de $L$-bol grens bereikt. Een dergelijk resultaat is niet eerder aangetoond voor het twee-kanaal asymmetrische geval. Het voorgestelde MD-LVQ schema is echter in staat de $L$-bol grens te bereiken, bij hoge resolutie voorwaarden, voor zowel het symmetrische geval als het 
asymmetrische geval. 

Het voorgestelde MD-LVQ schema schijnt een van de  
eerste schemas in de litera\-tuur te zijn die het grootst bekende 
hoge resolutie drie-kanaal MD gebied in het kwadratische Gaussische geval 
bereikt. Hoewel de optimaliteit alleen voor $K\leq 3$ wordt bewezen, nemen we aan dat het optimaal is voor willekeurige $K$ representaties. 

We laten gesloten-vorm uitdrukkingen zien voor de rate en distorsie prestaties voor algemene gladde stationaire bronnen en een kwadratische-fout distorsie criterium en voor hoge resolutie voorwaarden (ook voor eindig-dimensionale lattice vector kwanti\-sators). Er wordt aangetoond dat de zij-distorsies in het drie-kanaal geval kan worden uitgedrukt in het dimensieloze, genormaliseerde, tweede moment van een $L$-bol, onaf\-hankelijk van het type lattice dat wordt gebruikt voor de zij-kwantisators. Dit komt overeen met eerdere resultaten voor het geval van twee representaties. 

Het rate verlies wanneer eindig-dimensionale lattice vector kwantisators gebruikt worden is onafhankelijk van het lattice en wordt gegeven door het rate verlies van een $L$-bol en een bijkomende term die de ratio van twee dimensieloze expansie factoren beschrijft. Er wordt aangetoond dat het totale rate verlies superieur is aan bestaande drie-kanaal schemas. Dit resultaat lijkt te gelden voor elk aantal representaties.

\chapter*{Curriculum Vitae}
\addcontentsline{toc}{chapter}{Curriculum Vitae}
\markboth{Curriculum Vitae}{}
Jan \O stergaard was born in Frederikshavn, Denmark, in 1974. He obtained his high school diploma at Frederikshavns Tekniske Skole (HTX) in Frederikshavn in 1994. 
In 1999 he received the M.Sc.\ degree in Electrical Engineering at Aalborg University, Denmark.
From 1999 to 2002 he worked as a researcher in the area of signal analysis and classification at ETI A/S in Aalborg, Denmark. From 2002 to 2003 he worked as a researcher at ETI US in Virginia, United States.
In February 2003, he started as a Ph.D.\ student in the Information and Communication Theory group at Delft University of Technology, Delft, The Netherlands. During the period June 2006 -- September 2006 he was a visiting researcher in
the department of Electrical Engineering-Systems at Tel Aviv University, Tel Aviv, Israel.

\chapter*{Glossary of Symbols and Terms}
\addcontentsline{toc}{chapter}{Glossary of Symbols and Terms}
\markboth{Glossary of Symbols and Terms}{Glossary of Symbols and Terms}

\begin{table}[ht]
\begin{center}
\begin{tabular}{cl}\hline
Symbol & Description \\ \hline
$\mathbb{R}^L$ & $L$-dimensional Euclidean space (real field)\\
$\mathbb{C}^L$ & $L$-dimensional complex field \\
$\mathbb{Z}^L$ & $L$-dimensional set of all rational integers \\
$\mathcal{G}$ & Gaussian integers \\
$\mathcal{Q}$ & Algebraic integers \\
$\mathcal{E}$ & Eisenstein integers \\
$\mathcal{H}_0$ & Lipschitz integers \\
$\mathcal{H}_1$ & Hurwitzian integers \\

$x^H$ & Hermitian transposition (conjugate transposition) \\
$x^\dagger$ & Quaternionic transposition (Quaternionic conjugate transposition) \\
$\mathcal{J}$ & $\mathcal{J}$-module ($\mathcal{J}$-lattice) \\
$\|X\|$ & Vector norm with respect to underlying field \\
$\langle X,X\rangle$ & Inner product \\
\hline
\end{tabular}
\caption{Algebra-related symbols.}
\end{center}
\end{table}

\begin{table}[ht]
\begin{center}
\begin{tabular}{cl}\hline
Symbol & Description \\ \hline
$X$ & Scalar random process or $L$-dimensional random vector $(X\in \mathbb{R}^L)$ \\
$x$ & $L$-dimensional vector (realization of $X$) \\
$f_X$ & Distribution of $X$ \\
$\hat{X}$ & Reconstruction of $X$ \\
$\mathcal{X}$ & Alphabet of $X$ (usually $\mathcal{X}=\mathbb{R}^L$) \\
$\hat{\mathcal{X}}$ & Alphabet of $\hat{X}$ (usually $\hat{\mathcal{X}}\subset \mathbb{R}^L$) \\
$\rho$ & Fidelity criterion (usually squared-error)\\
$R(D)$ & Rate-distortion function \\
$D(R)$ & Distortion-rate function \\
$I(\cdot;\cdot)$ & Mutual information \\
$h(\cdot)$ & Differential entropy  \\
$\bar{h}(\cdot)$ & Differential entropy rate  \\
$H(\cdot)$ & Discrete entropy  \\
$E$ & Statistical expectation operator \\
$R_{\text{SLB}}$ & Shannon lower bound \\
$R_{\text{Loss}}$ & Rate loss \\
$R^*_{\text{red}}$ & Rate redundancy \\
$D_{\text{Loss}}$ & Space-filling loss \\
$\sigma_X^2$ & Variance of $X$ \\
$P_X$ & Entropy power \\
$Q(X)$ & Quantization of $X$ \\ 
\hline
\end{tabular}
\caption{Source-coding related symbols.}
\end{center}
\end{table}

\begin{table}[ht]
\begin{center}
\begin{tabular}{cl}\hline
Symbol & Description \\ \hline
$\Lambda_c$ & Central lattice (central quantizer) \\
$\Lambda_s$ & Sublattice $\Lambda_s\subseteq \Lambda_c$ (side quantizer in symmetric case)\\
$\Lambda_i$ & Sublattice $\Lambda_i\subseteq \Lambda_c$ (side quantizer in asymmetric case) \\
$\Lambda_\pi$& Product lattice $\Lambda_\pi\subset \Lambda_i$ or $\Lambda_\pi\subset \Lambda_s$  \\
$\Lambda_c/\Lambda_\pi$ & Quotient lattice \\
$V_c$ & Voronoi cell of $\Lambda_c$ \\
$V$ & Voronoi cell of $\Lambda_s$ or $\Lambda_i$ \\
$\nu$ & Volume of Voronoi cell of $\Lambda_c$ \\
$\nu_s$ & Volume of Voronoi cell of $\Lambda_s$ \\
$\nu_i$ & Volume of Voronoi cell of $\Lambda_i$ \\
$\nu_\pi$ & Volume of Voronoi cell of $\Lambda_\pi$ \\
$N$ & Index value of sublattice $\Lambda_s$ ($N=|\Lambda_c/\Lambda_s|$) \\
$N_i$ & Index value of sublattice $\Lambda_s$ ($N_i=|\Lambda_c/\Lambda_i|$) \\
$N_\pi$ & Index value of product lattice $\Lambda_\pi$ ($N_\pi=|\Lambda_c/\Lambda_\pi|$) \\
$N'$ & Nesting ratio of $\Lambda_s$ (index per dimension) \\
$G(\Lambda)$ & Dimensionless normalized second moment of $\Lambda$ \\
$G(S_L)$ & Dimensionless normalized second moment of $L$-sphere \\
$\zeta_i$ & Basis vector (lattice generator vector) \\
$M$ & Lattice generator matrix \\
$A$ & Gram matrix \\
$\Gamma_m$ & Multiplicative group of automorphisms of order $m$ \\
$\Lambda_c/\Lambda_\pi/\Gamma_m$ & Set of orbit representatives  \\
$Z^1$ & Scalar lattice (uniform lattice) \\
$Z^2$ & Square lattice \\
$Z^L$ & Hypercubic lattice \\
$A_2$ & Hexagonal two-dimensional lattice \\
$D_4$ & Four dimensional (checker board) lattice \\
$\xi\Lambda_c$ & Sublattice of $\Lambda_c$ (cyclic right submodule) \\
$\Lambda_c\xi$ & Sublattice of $\Lambda_c$ (cyclic left submodule) \\
$\mathfrak{K}(\Lambda)$ & Kissing number of $\Lambda$ \\
\hline
\end{tabular}
\caption{Lattice-related symbols.}
\end{center}
\end{table}

\begin{table}[ht]
\begin{center}
\begin{tabular}{cp{8cm}}\hline
Symbol & Description \\ \hline
$\alpha$ & Index assignment map ($\alpha(\lambda_c)=(\lambda_0,\dotsc,\lambda_{K-1})$)\\
$\alpha^{-1}$ & Inverse index assignment map \\
$\alpha_i$ & Component function ($\lambda_i=\alpha_i(\lambda_c)$)\\
$K$ & Number of descriptions \\
$\kappa$ & Number of received descriptions \\
$\tilde{V}$ & $L$-dimensional sphere \\ 
$\tilde{\nu}$ & Volume of $\tilde{V}$ \\
$\tilde{N}_i$ & Number of lattice points of $\Lambda_i$ within $\tilde{V}$\\
$\psi_L$ & Dimensionless expansion factor \\ 
$\omega_L$ & Volume of unit $L$-sphere \\
$R_s$ & Description rate [bit/dim.] in symmetric setup \\
$R_i$ & Description rate [bit/dim.] of $i^{th}$ description \\
$R_c$ & Rate of central quantizer  \\
$R_T$ & Sum rate ($R_T=\sum R_i$) \\
$D_i$ & Side distortion of $i^{th}$ description \\
$D_c$ & Central distortion \\
$D^{(K,\kappa)}$ & Distortion due to reconstructing using $\kappa$ descriptions out of $K$ \\
$J^{(K)}$ & Cost functional \\
$p$ & Packet-loss probability \\
$\mathcal{L}^{(K,\kappa)}$ & Index set describing all distinct $\kappa$-tuples out of the set $\{0,\dotsc, K-1\}$ \\
$\mathcal{L}^{(K,\kappa)}_i$ & Index set describing all distinct $\kappa$-tuples out of the set $\{0,\dotsc, K-1\}$, which contains the index $i$ \\
$\mathcal{L}^{(K,\kappa)}_{i,j}$ & Index set describing all distinct $\kappa$-tuples out of the set $\{0,\dotsc, K-1\}$, which contains the pair of indices $(i,j)$ \\
$p(\mathcal{L}^{(K,\kappa)}_{i,j})$ & Probability of the set $\mathcal{L}^{(K,\kappa)}_{i,j}$ \\[2mm]
$D_a^{(K,\kappa)}$ & Expected distortion when receiving $\kappa$ out of $K$ descriptions based on the packet-loss probability \\
$D^{(K,l)}$ & Distortion due to reconstructing using the subset of descriptions $l\subseteq \{0,\dotsc,K-1\}$ \\
MD-LVQ & Multiple-description lattice vector quantization \\
SCEC & Source-channel erasure code \\
SPSD & Sum of pairwise squared distances \\
WSPSD & Weighted sum of pairwise squared distances \\
\hline
\end{tabular}
\caption{MD-LVQ related symbols and terms.}
\end{center}
\end{table}

\bibliographystyle{plain}
\bibliography{bibliography/strings,bibliography/IEEEabrv,bibliography/mdc,bibliography/math,bibliography/quantization,bibliography/perceptual,bibliography/ostergaard}

\clearemptydoublepage
\index{information divergence|see{Kullback-Leibler distance}}
\index{labeling function|see{index-assignment (map)}}
\index{continuous entropy|see{entropy (differential)}}
\index{hexagonal lattice|see{lattice ($A_2$)}}
\index{scalar lattice|see{lattice ($Z^1$)}}
\index{square lattice|see{lattice ($Z^2$)}}
\index{hypercubic lattice|see{lattice ($Z^4$)}}
\index{high rate|see{high resolution conditions}}
\index{Dirichlet regions|see{Voronoi cell}}
\index{Lipschitz integer|see{integer (Lipschitz)}}
\index{nesting ratio|see{index (nesting ratio)}}
\index{space-filling gain|see{space-filling loss}}
\index{product lattice|see{lattice (product lattice)}}
\index{expansion factor|see{$\psi_L$}}
\index{isometry|see{isomorphism}}
\index{MSE|see{distortion measure (squared error)}}
\index{input weighted|see{distortion measure (non difference)}}
\index{factor module|see{module (quotient)}}
\index{dimensionless normalized second moment|see{second moment of inertia}}
\index{relative entropy|see{Kullback-Leibler distance}}
\index{entropy!relative|see{Kullback-Leibler distance}}
\index{Lagrangian multipliers|see{Lagrangian weights}}
\index{sublattice|see{lattice (sub)}}
\index{subgroup|see{group (sub)}}
\index{group!Abelian|see{group (commutative)}}
\index{subgroup|see{group (sub)}}
\index{admissible index value|see{index (admissible)}}
\index{AAC|see{MPEG-2}}
\index{MDCT|see{modified discrete cosine transform}}
\index{side rate|see{description rate}}
\index{target entropy|see{target rate}}
\index{inner product|see{norm (inner product)}}
\index{equivalence class|see{coset}}
\index{memoryless process|see{memoryless source}}
\index{pdf|see{probability density function}}
\index{convex closure|see{convex hull}}
\index{SPSD|see{sum of pairwise squared distances}}
\index{WSPSD|see{weighted sum of pairwise squared distances}}
\index{module!cyclic|see{cyclic (submodule)}}
\index{covering radius|see{lattice (covering radius)}}
\index{packing radius|see{lattice (packing radius)}}
\index{SCEC|see{source-channel erasure code}}

\printindex


\end{document}